\documentclass{aa}  

\usepackage{graphicx}
\usepackage{txfonts}
\usepackage{multirow}
\usepackage{xcolor}
\usepackage{ulem}
\usepackage[section]{placeins}
\usepackage{lscape}
\usepackage[colorlinks=true,linkcolor=black,citecolor=blue,urlcolor=blue,draft=false]{hyperref}

\begin{document} 

   \title{Analysing the SEDs of protoplanetary disks with machine learning}
   \author{T. Kaeufer\inst{1}\fnmsep\inst{2}\fnmsep\inst{3}\fnmsep\inst{4}, 
           P. Woitke\inst{1}, 
           M. Min\inst{3}, 
           I. Kamp\inst{2}, \and 
           C. Pinte\inst{5}\fnmsep\inst{6}}
   \institute{
   Space Research Institute, Austrian Academy of Sciences, Schmiedlstrasse 6, A-8042 Graz, Austria \\ \email{till.kaeufer@oeaw.ac.at}
        \and
    Kapteyn Astronomical Institute, University of Groningen, PO Box 800, 9700 AV Groningen, The Netherlands
        \and
   SRON Netherlands Institute for Space Research, Niels Bohrweg 4, 2333CA Leiden, The Netherlands
        \and
        TU Graz, Fakultät für Mathematik, Physik und Geodäsie, Petersgasse 16 8010 Graz, Austria
        \and
   School of Physics and Astronomy, Monash University, Clayton Vic 3800, Australia
        \and
   Universit{\'e} Grenoble-Alpes, CNRS
   Institut de Plan{\'e}tologie et d'Astrophyisque (IPAG),
   F-38000 Grenoble, France}

   \date{Received 14/11/2022; accepted 23/01/2022}

 
  \abstract
   {The analysis of spectral energy distributions (SEDs) of protoplanetary disks to determine their physical properties is known to be highly degenerate. Hence, a full Bayesian analysis is required to obtain parameter uncertainties and degeneracies. The main challenge here is computational speed, as one proper full radiative transfer model requires at least a couple of CPU minutes to compute.}
   {We performed a full Bayesian analysis for 30 well-known protoplanetary disks to determine their physical disk properties, including uncertainties and degeneracies. To circumvent the computational cost problem, we created neural networks (NNs) to emulate the SED generation process.}
   {We created two sets of MCFOST Monte-Carlo radiative transfer disk models to train and test two NNs that predict SEDs for continuous and discontinuous disks, with 18 and 26 free model parameters, respectively. A Bayesian analysis was then performed on $30$ protoplanetary disks with SED data collected by the FP7-Space DIANA project to determine the posterior distributions of all parameters. We ran this analysis twice, (i) with old distances and additional parameter constraints as used in a previous study, to compare results, and (ii) with updated distances and free choice of parameters to obtain homogeneous and unbiased model parameters. We evaluated the uncertainties in the determination of physical disk parameters from SED analysis, and detected and quantified the strongest degeneracies.}
   {The NNs are able to predict SEDs within $\sim\!1\,$ms with uncertainties of about $5\%$ compared to the true SEDs obtained by the radiative transfer code. We find parameter values and uncertainties that are significantly different from previous values obtained by $\chi^2$~fitting. Comparing the global evidence for continuous and discontinuous disks, we find that $26$ out of $30$ objects are better described by disks that have two distinct radial zones. The analysed sample shows a significant trend for massive disks to have small scale heights, which is consistent with lower midplane temperatures in massive disks. We find that the frequently used analytic relationship between disk dust mass and millimetre-flux systematically underestimates the dust mass for high-mass disks (dust mass $\geq\!10^{-4}\,\rm M_\odot$). We determine how well the dust mass can be determined with our method for different numbers of flux measurements. As a byproduct, we created an interactive graphical tool\thanks{https://tillkaeufer.github.io/sedpredictor} that instantly returns the SED predicted by our NNs for any parameter combination.}
   {}


   \maketitle
%

\section{Introduction}
Spectral energy distributions (SEDs) are commonly used to determine the physical structure and the dust properties of protoplanetary disks (e.g. \citealp{Andrews2007,Ricci2011}). To determine, for example, the mass and shape of the disk, or the properties of dust grains and polycyclic aromatic hydrocarbons (PAHs), radiative transfer (RT) models are usually applied which, after some fitting procedure, provide the values of these physical parameters.

However, SED analysis can be highly degenerate (e.g. \citealp{Thamm1994,Heese2017}). For example, simultaneous changes in disk mass and dust composition can lead to almost indistinguishable SEDs. Furthermore, dust particle settling and disk flaring may have very similar effects on the SED at (sub-)millimetre wavelengths.  Such degeneracies can result in large uncertainties when the values of single physical parameters are determined. 

One approach to address this problem is to combine SED fitting with other observations, for example, images at millimetre wavelengths to constrain the outer radius, which effectively reduces the number of free parameters. 
However, such a procedure will be very specific to a certain combination of multi-kind data, and it is often done only for one or a selected number of individual objects where that particular data exist (e.g. \citealp{Pinte2008,Tannirkulam2008,Dong2012}).
Analysing larger samples entails substantial computational efforts (e.g. \citealp{Sheehan2022}).

The statistically correct way to take all the degeneracies in an analysis into account is to calculate the posterior distributions for all parameters via a full Bayesian analysis. However, this requires running at least a few million SED models. Due to the high computational cost of state-of-the-art 2D RT disk modelling tools, this process is computationally very demanding. Even though this has been done for single objects using very large computational resources, \citep[e.g][]{Pinte2008}, it becomes computationally unfeasible when considering larger samples of objects.
Another drawback of this brute-force method is that modellers are forced to use SED modelling approaches that are as fast as possible. Hence, they tend to avoid full RT models and fall back on simpler underlying disk models or allow only for a few physical parameters, in which case fewer degeneracies are found \citep[e.g.][]{Liu2015}.

To circumvent the high computational cost, the modelling process can be emulated. \citet{Ribas2020} emulated the D'Alessio irradiated accelerator disk (DIAD) models \citep{Dalessio1998} using neural networks (NNs). DIAD assumes that the disk is geometrically thin, and that therefore the radial energy transport is negligible. Furthermore, the disks are assumed to be in a steady state, with a constant mass accretion rate, and in vertical hydrostatic equilibrium. In an iterative process, the vertical disk structure is solved under these assumptions. The NNs created by \citet{Ribas2020} predicted spectral fluxes with an estimated $1\sigma$ uncertainty of $10\%$ compared to the true SEDs of the DIAD models. This enabled them to fit the SEDs of $23$ objects, using the Markov chain Monte-Carlo method to derive posterior parameter distributions. Their work focusses on confronting the $\alpha$ description \citep{Shakura1973} with observations\, finding high accretion rates and viscosities for many sources.

The uncertainties of disk parameters are expected to increase with fewer observational constraints. Nevertheless, analytical relationships are often used to estimate individual disk parameters from limited sets of observations. In particular, the dust mass is often derived analytically from millimetre flux, assuming certain values for the dust absorption opacity and mean temperature, assuming the disk to be optically thin, and assuming the dust temperatures to be high enough to emit in the Rayleigh limit \citep{Woitke2019}. 
The disk dust masses derived this way have been shown to disagree with the masses derived from the full RT models, even when using the correct average dust temperature and opacity from the models \citep{Ballering2019,Woitke2019}.  
The Bayesian analysis carried out in this paper will allow us, for example, to quantify the inherent uncertainties in this disk mass determination method, and to confront these results with the frequently applied analytical relationship. 

To do so, we generated a large set of SEDs using the Monte-Carlo continuum RT code MCFOST \citep{Pinte2006,Pinte2009}, and used NNs to emulate this process of creating SEDs from model parameters. The benefit of this process is the drastic decrease in computational cost to generate SEDs. This makes a full Bayesian analysis that is carried out for 30 well-known protoplanetary disks feasible.

This paper is structured as follows. Section~\ref{sec:method} introduces the models, emulating process, and the fitting procedure including data selection, data preparation and the Bayesian framework. We present the results of this paper in Section~\ref{sec:results} including the quality of the SED predictions, the overlap to the DIANA fitting results, the new fits, and the determination of the dust mass. These results are discussed in Section~\ref{sec:discussion} with a focus on parameter trends in the sample, parameter uncertainties, and degeneracies. We conclude this paper in Section~\ref{sec:summary} with a summary of our main findings.

\section{Method\label{sec:method}} 

We created large sets of more than $10^5$ Monte-Carlo continuum RT 2D disk SED models with MCFOST \citep{Pinte2006,Pinte2009}, which cover the expected parameter space in an efficient way. We train a NN to reproduce these results from the values of the model parameters. The predicted SEDs are then used to perform a full Bayesian analysis on multiwavelength photometric and spectroscopic data sets of 30 well-known protoplanetary disks as published by \citet{Dionatos2019}. We compare our new results for the physical disk properties with those previously obtained by \citet{Woitke2019}, who used an evolutionary strategy to optimise $\chi^2$. In the next step, we refit these objects using updated GAIA distances \citep{Gaia2021}, fixing only the stellar parameters, and inclination. As a result, we obtain the most probable, up-to-date disk parameters for 30 well-known disks from the SED data, taking into account all uncertainties and degeneracies.

\subsection{The SED forward model\label{sec:mcfost}}
All SEDs are calculated by the Monte-Carlo radiative transfer code MCFOST 2.20.8 \citep{Pinte2006,Pinte2009}. 
MCFOST solves the radiative transfer problem of a dusty disk being illuminated by a star. The disk is assumed to be passive, that is, we assume that all energy is coming from the central star and that at every point the radiative heating of the dust particles equals their radiative cooling. 
All relevant stellar, disk, and dust parameters for this problem, including disk shape, dust material, size distribution, and settling, are listed in Table \ref{tab:parameters} and are explained in Sect. \ref{sec:free_para}. The meaning of these parameters is described in detail in \citet{Woitke2016}.

MCFOST sets up a 2D density structure with a grid of $150\times 100$ points concerning the radial and vertical directions, respectively. This structure is defined by various disk shape parameters. The material composition of the dust grains is assumed to be constant throughout the disk. A size distribution function is assumed and mapped onto 70 size bins. The opacities of these dust grains are pre-computed using Mie and effective mixing theory. We are using the so-called DIANA standard dust opacities, see \citet{Woitke2016}. The grains in each size bin are then vertically redistributed, using the settling prescription of \cite{Dubrulle1995}, and their opacities are co-added.

Photon packets are propagated in 3D from the stellar surface through the disk, while undergoing interactions with the dust on the 2D grid. The packets are traced until they exit the computational domain. Scattering events are modelled as a change of direction, dependent on the scattering angle, but without changing the wavelength. In case of absorption events, the packets are re-emitted isotropically with a wavelength distribution that is based on the local dust temperature \citep{Bjorkman2001}. During the Monte-Carlo modelling procedure, the grains heat up until radiative equilibrium is achieved. Packets that are absorbed by PAHs are stopped. At the end of the Monte-Carlo run, the PAH temperature distribution is updated, and the absorbed packets are re-emitted. The procedure is iterated until convergence. We use $5\cdot10^6$ photon packages for the temperature determination phase. For more information, please see \citet{Pinte2006} and \citet{Pinte2009}.

Once the temperature structure is computed using the above procedure, the SED is calculated in a subsequent step for a predefined set of 140 wavelength points. For this, a raytracing method is used, integrating the formal solution of the radiative transfer equation. The full wavelength dependent source function has to be known at each location in the disk. The thermal part of the source function is obtained from the temperature and emissivity of the grains. The scattering part of the source function is computed using a monochromatic Monte-Carlo run. For doing so, photon packets are systematically generated by the star and by each cell. At each interaction, the weight of the package is reduced according to the probability of absorption, and only the scattered fraction is traced further after the direction has changed. From this procedure the angle dependent specific intensity is computed at each location in the disk which allows us to compute the contribution of scattering to the local source function. In this way, the raytraced SED is calculated for all inclination angles. The advantage of this separate raytracing step is that the generated SEDs contain only very little remaining Monte-Carlo noise. In our setup, we have used $10^4$ photon packages per wavelength point and 10 inclinations. These inclinations are equidistant in cosine space: $18.19^\circ$, $31.79^\circ$, $41.41^\circ$, $49.46^\circ$, $56.63^\circ$, $63.26^\circ$, $69.51^\circ$, $75.52^\circ$, $81.37^\circ$, and $87.13^\circ$.

\begin{figure}
    \centering
    \includegraphics[width=\linewidth,trim=16 15 4 0,clip]{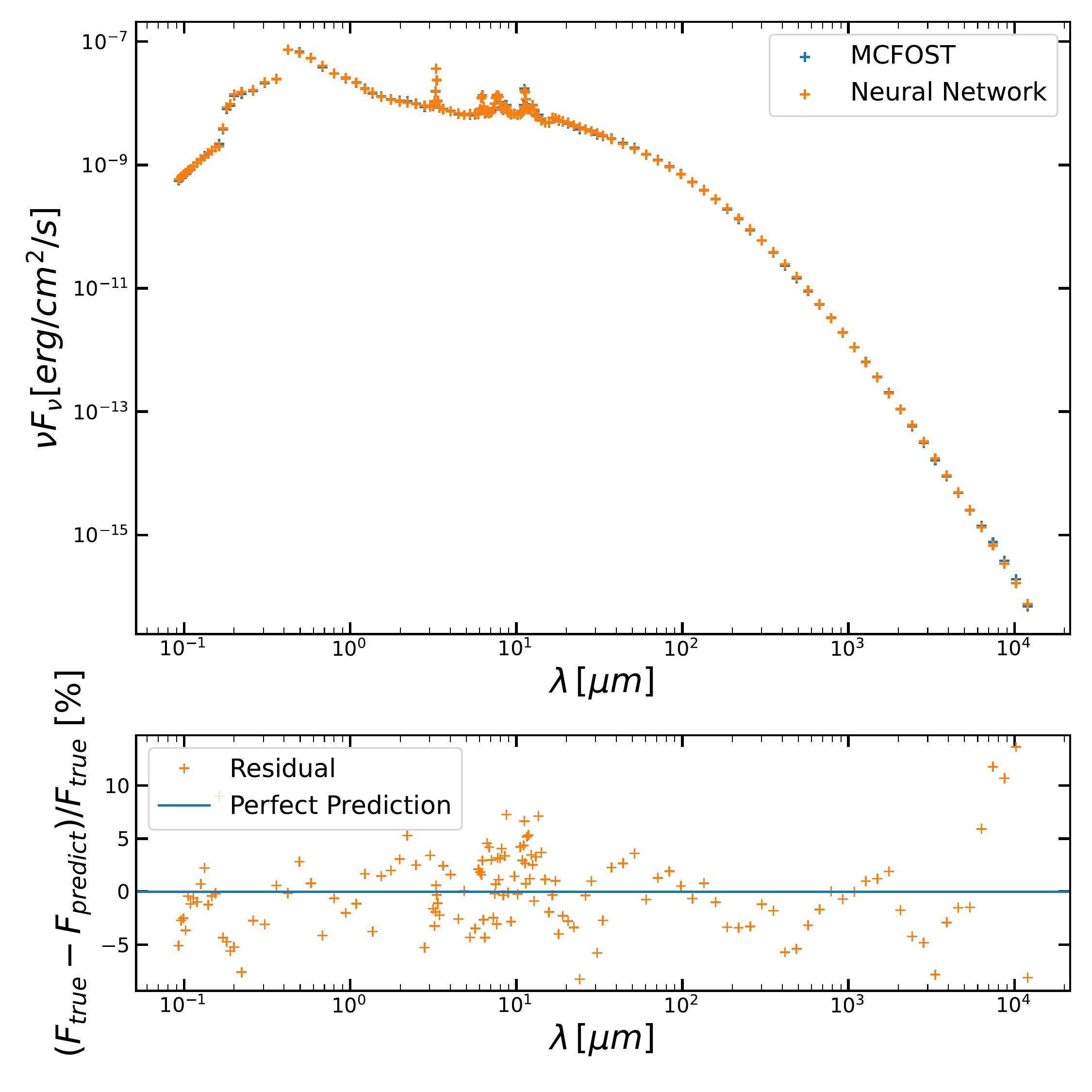}
    \caption{Example of a predicted SED with strong PAH features for a single zone model of a A6 star that is part of the test sample compared to the true SED derived by the full RT code. The orange and blue points in the upper panel show the predicted and the true SEDs, respectively. The residual (orange) is shown in percent in the lower panel.}
    \label{fig:pred_sed}
\end{figure}

The selected wavelengths are evenly distributed in logarithmic space, but with a denser coverage around the 10\,$\mu$m and 20\,$\mu$m silicate emission features, the mid-IR PAH emission features at $3.3\,\mathrm{\mu m}$, $6.25\,\mathrm{\mu m}$, $7.85\,\mathrm{\mu m}$, $8.7\,\mathrm{\mu m}$, $11.3\,\mathrm{\mu m}$ and $12.7\,\mathrm{\mu m}$, and in the UV, see Fig.~\ref{fig:pred_sed}.

In total, the computation of one disk model, which generates ten SEDs, takes about 3-15 minutes on 8 processors. We have used MCFOST in this setup to calculate some $10^5$ disk models, that is, a few $10^6$ SEDs due to the $10$ inclination angles, which are used for training and testing of the NNs, see Sect. \ref{sect:NN}.

\subsection{The free parameters\label{sec:free_para}}

The star is described by four independent parameters $T_{\rm eff}$, $L_\star$, $f_{\rm UV}$ and $p_{\rm UV}$. 
We use pre-main-sequence evolutionary tracks from \citet{Siess2000} to connect $T_{\rm eff}$ and $L_\star$ to the mass $M_\star$ and age of the star.
The photospheric spectrum is interpolated from the PHOENIX stellar atmosphere spectra \citep{Brott2005} as function of effective temperature $T_{\rm eff}$ and surface gravity $\log g$. 
The surface gravity is calculated from the given photospheric luminosity $L_\star$ and the stellar mass $M_\star$. 
At UV wavelengths, a power law spectrum replaces the photospheric spectrum to account for the excess UV emission due to accretion and stellar activity. 
This power law is described by $f_{\rm UV}$ and $p_{\rm UV}$, which are the total UV luminosity shortwards of 250\,nm relative to $L_\star$, and the spectral slope in the UV, respectively.

The dust is described by five free parameters, using the DIANA standard dust opacities (see \citealp{Woitke2016}). 
The dust size distribution function before settling is assumed to be a power law, $f(a)\propto a^{-a_{\rm pow}}$, from a minimum size ($a_{\rm min}$) to a maximum size ($a_{\rm max}$), with power law index $a_{\rm pow}$. 
We assume the dust material to be a mixture of silicate $\rm Mg_{0.7}Fe_{0.3}SiO_3$ \citep{Dorschner1995} and amorphous carbon \citep{Zubko1996} with a fixed porosity of $25\,\%$. 
The opacity computations apply Mie-theory to a distribution of hollow spheres \citep[DHS;][]{Min2005} with a maximum volume fraction of the hollow core of 80\%. The DHS provides a reasonable representation of the spectral properties of irregularly shaped aggregate dust grains, see \citet{Min2016}.
In addition to the three parameters of dust size, we have one additional free parameter for the dust material (the mixing ratio of amorphous carbon) and one additional parameter for dust settling $\alpha_{\rm settle}$.

Polycyclic aromatic hydrocarbons can add multiple emission features to the SED at mid-IR wavelengths, in particular for central stars that are bright in the blue and soft-UV. 
We include PAHs in all models by means of two additional free parameters in the models, namely the PAH-abundance in comparison to the standard abundance in the interstellar medium ($f_{\rm PAH}$) and the ratio of charged to neutral PAHs ($\rm PAH_{\rm charged}$ ). We chose representative PAHs consisting of 54 carbon atoms and 18 hydrogen atoms.
For details, see \citet{Woitke2016}.

\begin{figure}
    \centering
    \includegraphics[width=\linewidth]{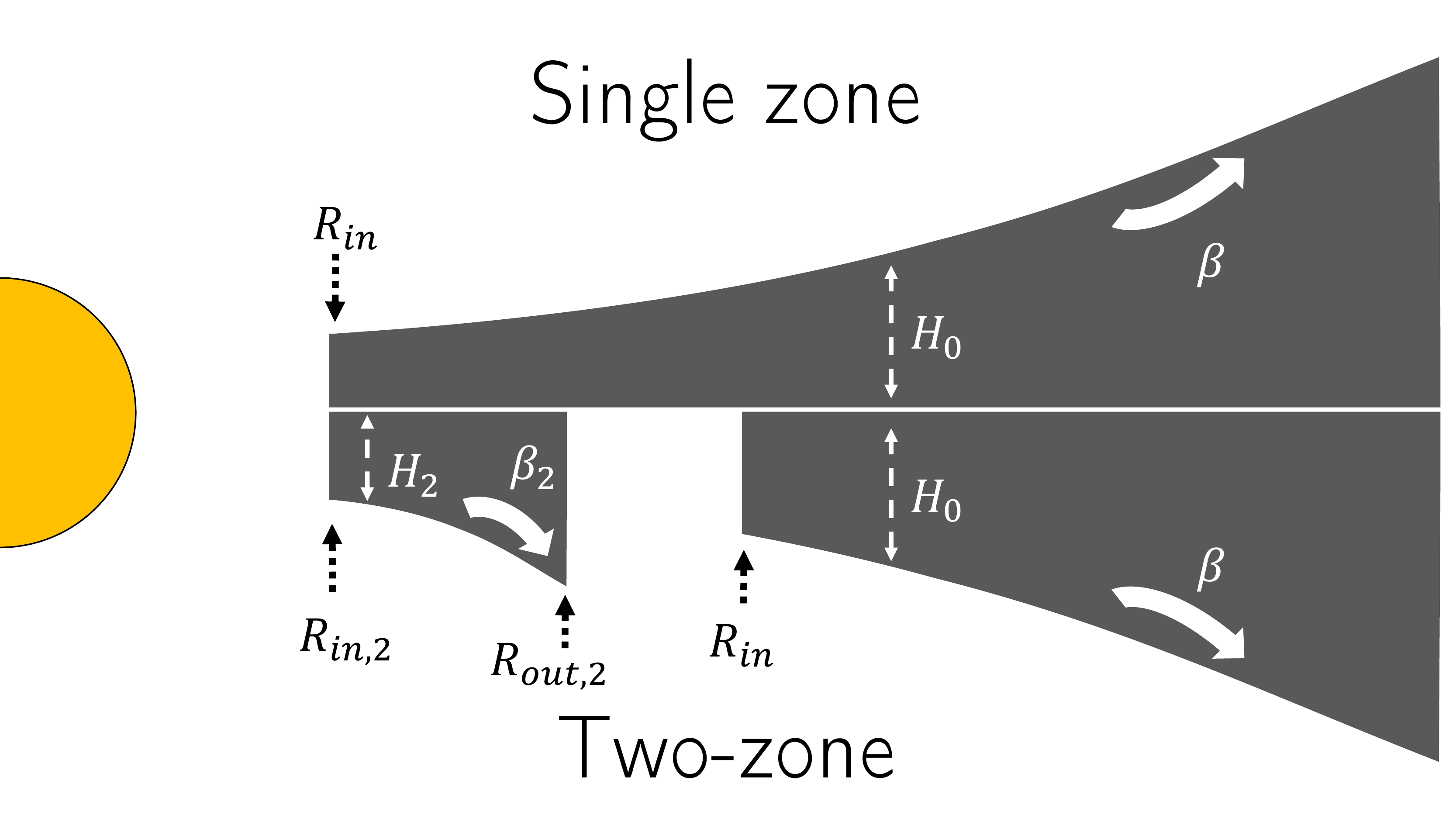}
    \caption{Sketch of a single and two-zone disk structure. The shown parameters are explained in Table~\ref{tab:parameters}.}
    \label{fig:sketch_disk}
\end{figure}

We are using two different types of disk structures (Fig.~\ref{fig:sketch_disk}), single and two-zone models. While single zone models have a continuous radial surface density structure, two-zone models consist of two disk zones that can be connected or have a gap in between. This additional complexity is often used to improve the fit quality of SEDs, see, for example \citet{Woitke2019}. 

There are six free parameters that define the shape of every disk zone.
The column density is assumed to decrease as a radial power law from an inner radius ($R_{\rm in}$) to a tapering-off radius ($R_{\rm taper}$) beyond which the column density deceases with an exponential function, until the hydrogen nuclei column density reaches a threshold value of $10^{20}\rm\,cm^{-2}$ which defines the outer radius, see Eq.\,(1) in \citet{Woitke2016}. 
The exponent in the power law is assumed to be the same as in the exponential ($\gamma=\epsilon$) in all models in this paper.
The vertical extension of the gas density above the midplane is defined by a scale height ($H_0$) at a reference radius and the flaring index ($\beta$), see Eq.\,(2) in \citet{Woitke2016}. 
The gas mass in the disk zone ($M_{\rm disk}$) results from radial integration and is used to set the proportionality constant for the surface-density law. 

The respective parameters for the inner disk zone in two-zone models have the same name, marked with a $2$ (see Fig.~\ref{fig:sketch_disk} and Table~\ref{tab:parameters}). Additionally, the maximum dust size ($a_{\rm max,2}$) and the PAH abundance ($f_{\rm PAH,2}$) in the inner zone are set independently. All other dust and PAH parameters are set globally.

In all disk models, we assume a gas to dust mass ratio of 100 in all disk zones. However, the gas mass influences only the settling efficiency in our models. Therefore, different gas to dust ratios can be simulated by changing the settling parameter accordingly. Additionally, we use an interstellar background radiation field composed of (i) the cosmic microwave background (CMB) modelled with a Planckian of temperature 2.7\,K, plus (ii) a standard UV background \citep{Draine1978,Draine1996} $\chi_{\rm ISM}\!=\!1$. The number of free parameters (stellar, dust, PAHs, disk, and inclination) adds up to $18$ and $26$ for single and two-zone models, respectively.

\subsection{Grid creation\label{sec:grid_creation}}
We created two separate disk model grids to train the two NNs predicting the single zone and two-zone SEDs, respectively. The goal of both grids is to vary all model parameters that we want to use to fit SEDs, so that the NNs can learn the influence of these parameters. To find reasonable distributions for our model parameters, we analysed the parameter values found in the DIANA SED models \citep{Woitke2019} and adjusted our parameter distributions for the grid production accordingly. Both grids are created similarly, but the two-zone grid has additional parameters for an inner disk zone. In this section, we describe the general idea behind the grid creation. A detailed explanation highlighting special parameters is attached in Appendix \ref{App:grid}.

\begin{figure}[t]
    \centering
    \includegraphics[width=0.9\linewidth]{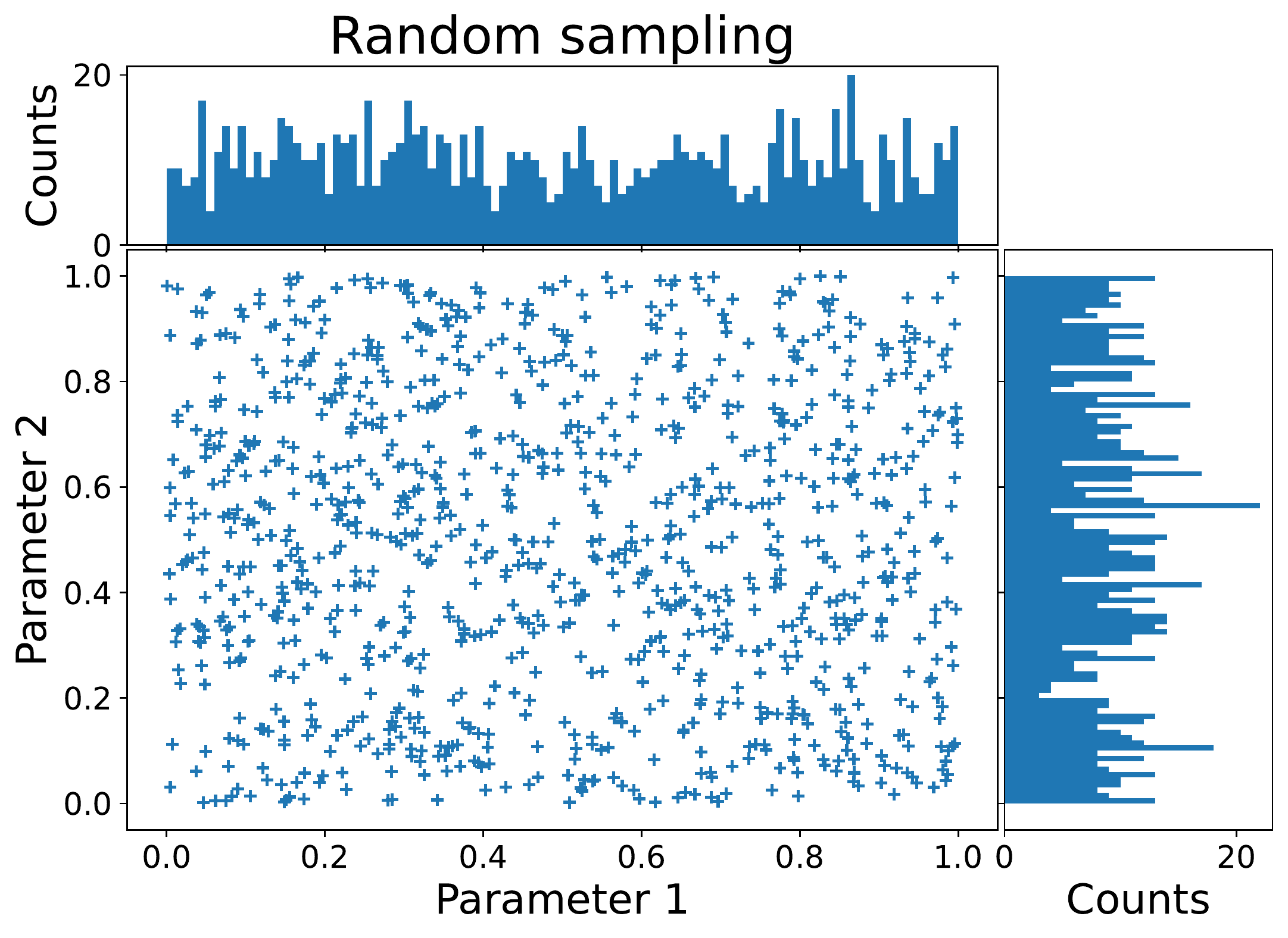}
    \includegraphics[width=0.9\linewidth]{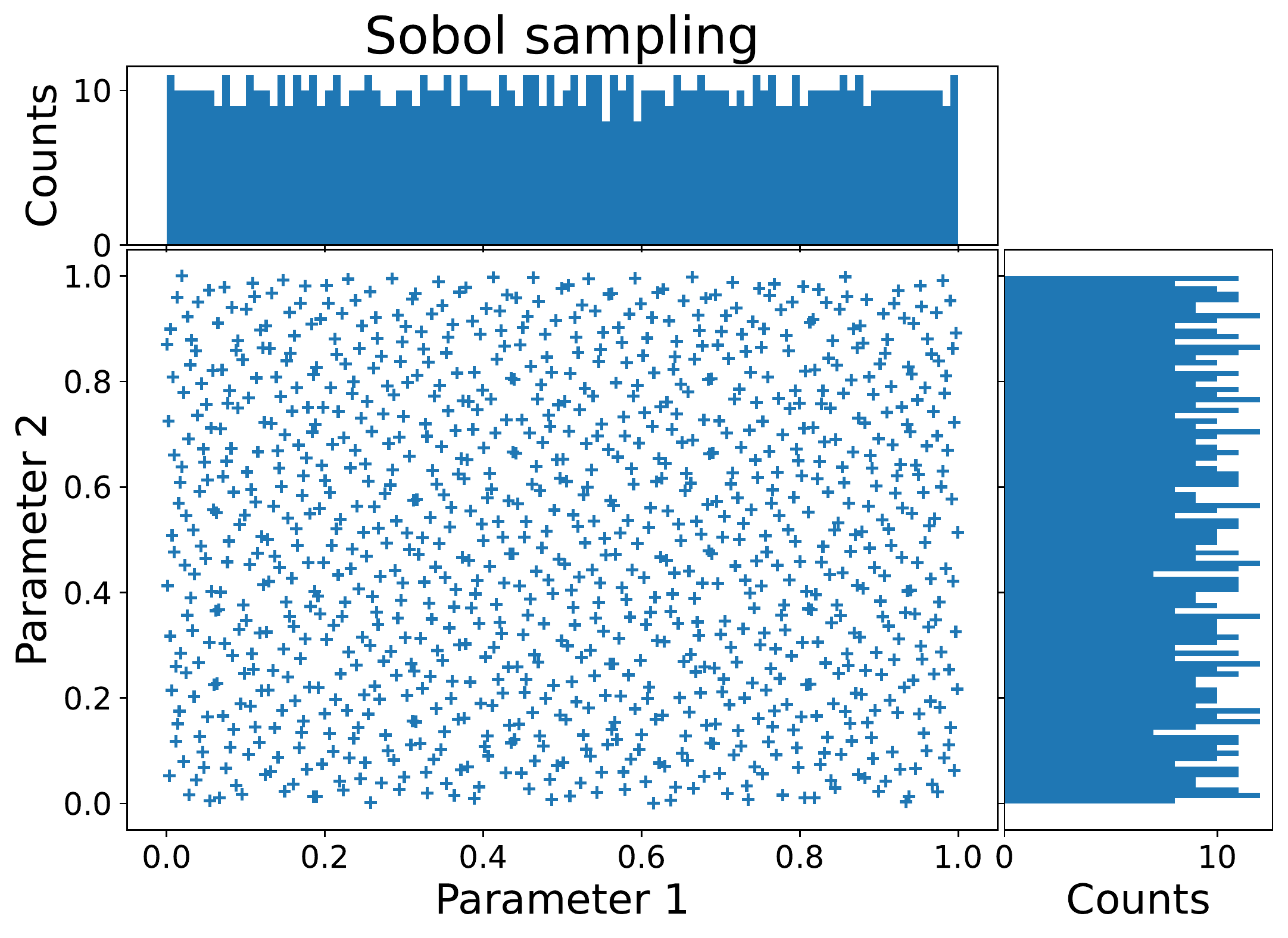}
    \caption{Two parameters sampled with $1,000$ points using random numbers (upper panel) and the Sobol sequence (lower panel). Both parameters are evenly distributed between $0$ and $1$.}
    \label{fig:sobol}
\end{figure}

We use the low-discrepancy Sobol sequence \citep{Sobol1967} to sample the parameter space. This results in a smooth quasi-random coverage, as seen in Fig.~\ref{fig:sobol}. A random distribution shows larger areas without any models and a more uneven distribution for every parameter compared to the Sobol sampling. Therefore, we think that the Sobol sampling improves the NN's ability to learn how the SEDs look like over the whole parameter space.

All parameters that vary for single and two-zone models are listed in Table \ref{tab:parameters}. They are either uniformly or Gaussian distributed between a lower ($x_{\mathrm{min}}$) and upper limit ($x_{\mathrm{max}}$).
The uniform distribution uses these minimal and maximal values to map the Sobol numbers ($x_{\rm sobol}$) that range from $0$ to $1$ to parameter values ($x_{\mathrm{para}}$):
\begin{align}
    x_{\mathrm{para}}=x_{\mathrm{min}}+(x_{\mathrm{max}}-x_{\mathrm{min}})\,x_{\mathrm{sobol}} \label{eq:minmax}.
\end{align}
If we decided to sample a parameter with a Gaussian distribution, we used the inverse error function ($\rm erf^{-1}$) to map the Sobol numbers to the parameter value. The Gaussian distribution is defined by the mean value ($x_{\rm mean}$) and a standard deviation ($x_{\rm std}$):
\begin{align}
    x_{\mathrm{para}}=x_{\mathrm{mean}}+x_{\mathrm{std}}\sqrt{2}\;{\rm erf}^{-1}(1-2\,x_{\mathrm{sobol}}) \label{eq:gauss}.
\end{align}
The Sobol numbers that result in values for $x_{\mathrm{para}}$ outside the limits of $x_{\mathrm{min}}$ and $x_{\mathrm{max}}$ are discarded and uniformly distributed in the allowed range using Eq.\,(\ref{eq:minmax}). The values for $x_{\rm min}$, $x_{\rm max}$, $x_{\rm mean}$, and $x_{\rm std}$ are derived from the DIANA sample. We calculated the respective values of the sample, discarded eventual outliers, and influenced our decisions by knowledge of the physical meaning of all parameters. This knowledge resulted in stricter or looser limits for some parameters or in sampling the logarithm of a parameter instead of its linear value.

All values describing the parameter distributions are listed in Table~\ref{tab:parameters}, including the decisions whether or not to use logarithmic values and whether the sampling is even or Gaussian. In total, the single zone grid consists of $673,378$ SEDs. The two-zone grid consists of $1.195,285$ SEDs.

\begin{sidewaystable*}[p]
\caption{All model parameters for single and two-zone models.}
\label{tab:parameters}
\centering
\begin{tabular}{lllllllll}
\hline
\hline
 & & & & & & & & \\[-1.9ex]
\# & name & description &  sampling\tablefootmark{(1)} & function\tablefootmark{(2)}  & $x_{\rm min}$ & $x_{\rm max}$ & $x_{\rm mean}$ & $x_{\rm std}$\\ 
 & & & & & & & &  \\[-1.9ex] \hline
 & & & & & & & & \\[-1.9ex]
 &  & stellar parameters &  & &  &  &  &    \\
1 & $M_{\star}$ & mass of the star [$M_\odot$]  &$\log_{10}(M_{\star})$ & even\tablefootmark{(3)} & $\log_{10}(0.2)$ & $\log_{10}(2.5)$ &  &   \\
2 & Age & age of the star [Myr] &  $\log_{10}(\rm Age)$ & even\tablefootmark{(3)} & $\log_{10}(0.5)$ & $\log_{10}(20)$ &  &  \\
3 & $T_{\rm eff}$ & effective stellar temperature [K]  &$\log_{10}(T_{\rm eff})$ & even\tablefootmark{(3)} & indirect\tablefootmark{(4)} & indirect\tablefootmark{(4)} &  &    \\
4 & $L_{\star}$ & luminosity of the star [$L_\odot$] & $\log_{10}(L_{\star})$ & even\tablefootmark{(3)} & indirect\tablefootmark{(4)} & indirect\tablefootmark{(4)} &  &    \\
5 & $f_{\rm UV}$ & excess UV $L_{\rm UV}/L_{\star}$  & $\log_{10}(f_{\rm UV})$ & gauss & $-3$ & $-1$ & $\log_{10}(0.08)$ & $\log_{10}(0.12)$ \\
6 & $p_{\rm UV}$ & exponent of excess UV  & $\log_{10}(p_{\rm UV})$ & gauss & $\log_{10}(0.5)$ & $\log_{10}(2)$ & $\log_{10}(1.2)$ & 0.4\\  
 & & & & & & & &  \\[-1.9ex] \hline
 & & & & & & & & \\[-1.9ex]
 &  & dust parameters &  & & &  &  &   \\
7 & $a_{\rm min}$& minimal dust grain size [$\mu m$]& $\log_{10}(a_{\rm min})$ & gauss & $-3$ & $\log_{10}(0.2)$ & $\log_{10}(0.03)$ & $\log_{10}(0.3)$  \\
8 & $a_{\rm max}$ & maximal dust grain size [$\mu m$] & $\log_{10}(a_{\rm max})$ & gauss & $\log_{10}(300)$ & $4$ & $\log_{10}(4000)$ & $\log_{10}(2)$\\
9 & $a_{\rm pow}$ & exponent of dust size distribution & $a_{\rm pow}$ & gauss & 3 & 5 & 3.6 & 0.35 \\
10 & amC-Zubko & amount of amorphous carbon & amC-Zubko & gauss & 0.05 & 0.3 & 0.18 & 0.05 \\
11 & $\alpha_{\rm settle}$ & viscosity parameter &$\log_{10}(\alpha_{\rm settle})$ & gauss & $-5$ & $-1$ & $-3$ & $0.8$ \\  
 & & & & & & & & \\[-1.9ex] \hline
 & & & & & & & & \\[-1.9ex]
 &  & PAH parameters &  &  &  & &  &    \\
12 & $f_{\rm PAH}$ & amount of PAHs relative to the ISM & $\log_{10}(f_{\rm PAH})$ & gauss & $-3.5$ & $0$ & $-1.5$ & $0.9$ \\
13& $\rm PAH_{\rm charged}$  & amount of charged PAHs & $\rm PAH_{\rm charged}$  & even & 0 & 1 &  &  \\  
 & & & & & & & & \\[-1.9ex] \hline
 & & & & & & & &  \\[-1.9ex]
 &  & disk parameters &  &  &  & &  &   \\
14 & $M_{\rm disk}$ & disk mass [$M_{\rm sun}$] & $\log_{10}(M_{\rm disk}/M_{\star})$ & gauss & $-5$ & $0$ & $-2$ & 0.8  \\
15 & $R_{\rm in}$\tablefootmark{(5)} & inner disk radius [AU] & $T_{\rm sub}$\,[K]\tablefootmark{(6)} & gauss & 100 & 1677 & 1333 & 457 \\
 & $R_{\rm in}$\tablefootmark{(7,8)} & inner disk radius [AU]& $R_{\rm in}$& gauss &  1 & 70 & 18 & 19  \\
16 & $R_{\rm taper}$ & taper radius [AU] & $\log_{10}(R_{\rm taper})$ & gauss & $\log_{10}(5)$ & $\log_{10}(350)$ & $\log_{10}(90)$ &$\log_{10}(5)$ \\
17 & $\epsilon$ & column density exponent $N_H(r) \propto r^{-\epsilon}$ & $\epsilon$ & gauss & 0 & 2.5 & 1 & 0.4  \\
18 & $H_0$ & scale height at 100 AU\tablefootmark{(9)} [AU] & $H_0$ & gauss & 3 & 35 & 12 & 7 \\
19 & $\beta$ & flaring index\tablefootmark{(9)}& $\beta$ & gauss & 0.9 & 1.4 & 1.15 & 0.08  \\  
 & & & & & & & & \\[-1.9ex] \hline
 & & & & & & & &  \\[-1.9ex]
 &  & inner disk parameters &  &  &  &  & &    \\
20 & $M_{\rm disk,2}$\tablefootmark{(8)} & disk mass [$M_{\rm sun}$]& $\log_{10}(M_{\rm disk,2}/M_{\rm disk})$ & gauss  & $-7$ & $-0.7$ & $-4$ & $1.3$\\
21 & $\epsilon_2$\tablefootmark{(8)} & column density exponent $N_{H,2}(r) \propto r^{-\epsilon_2}$ & $\epsilon_2$ & gauss   & -1 & 2 & 0.7 & 0.7 \\
22 & $H_2$\tablefootmark{(8)} & scale height at 1 AU\tablefootmark{(10)} [AU] & $H_2$ & gauss   & 0.02 & 0.5 & 0.1 & 0.1 \\
23 & $\beta_2$\tablefootmark{(8)} &  flaring index\tablefootmark{(10)} & $\beta_2$ & gauss  & 0.05 & 1.7 & 1. & 0.26 \\
24 & $R_{\rm in,2}$\tablefootmark{(8)} & inner radius [AU] & $T_{\rm sub}$\,[K]\tablefootmark{(6)} & gauss  & 100 & 1677 & 1368 & 497 \\
25 & $R_{\rm out,2}$\tablefootmark{(8)} & outer radius & $\log_{10}(R_{\rm out,2}/R_{in})$ & gauss   & $-3$ & $0$ & $-0.48$ & $0.8$\\
26 & $a_{\rm max,2}$\tablefootmark{(8)} &  maximal dust size&  $\log_{10}(a_{\rm max,2}/a_{\rm max})$  & gauss  & $-4$ & $0$ & $-0.95$ & 1.6 \\
27 &  $f_{\rm PAH,2}$\tablefootmark{(8)} & amount of PAHs& log($f_{\rm PAH,2}$/$f_{\rm PAH}$) & gauss   & $-3.5$ & $0$ & $-0.99$ & 1.72 \\
\end{tabular}
\tablefoot{\\
\tablefoottext{1}{Sampled scale (linear or logarithmic) for the parameter or the ratio of the parameter to anotherone.}\\
\tablefoottext{2}{Even (Eq. \ref{eq:minmax}) or Gaussian distribution (Eq. \ref{eq:gauss}).}\\
\tablefoottext{3}{All single zone models are evenly distributed in mass and age, while all two-zone models are evenly distributed in luminosity and temperature. The undetermined parameters are derived from evolutionary tracks \citep{Siess2000} by a Neural Network.}\\
\tablefoottext{4}{There are no boundries for the stellar temperature and luminosity, but limits in stellar mass and age. They lead to indirect limits in temperature and luminosity through their relation in the HRD using pre-mainsequence tracks from \cite{Siess2000}.}\\
\tablefoottext{5}{Inner radius for single zone models.}\\
\tablefoottext{6}{Using the stellar parameters and the sublimation temperature defines the radius (For details see Appendix \ref{App:grid}).}\\
\tablefoottext{7}{Inner radius of the outer disk zone.}\\
\tablefoottext{8}{Only applicable to two-zone models.}\\
\tablefoottext{9}{$H(r)= H_0\cdot \left(\frac{r}{100 AU}\right)^\beta$}
}

\end{sidewaystable*}

\subsection{Neural network\label{sect:NN}}
\begin{figure}
    \centering
    \includegraphics[width=\linewidth]{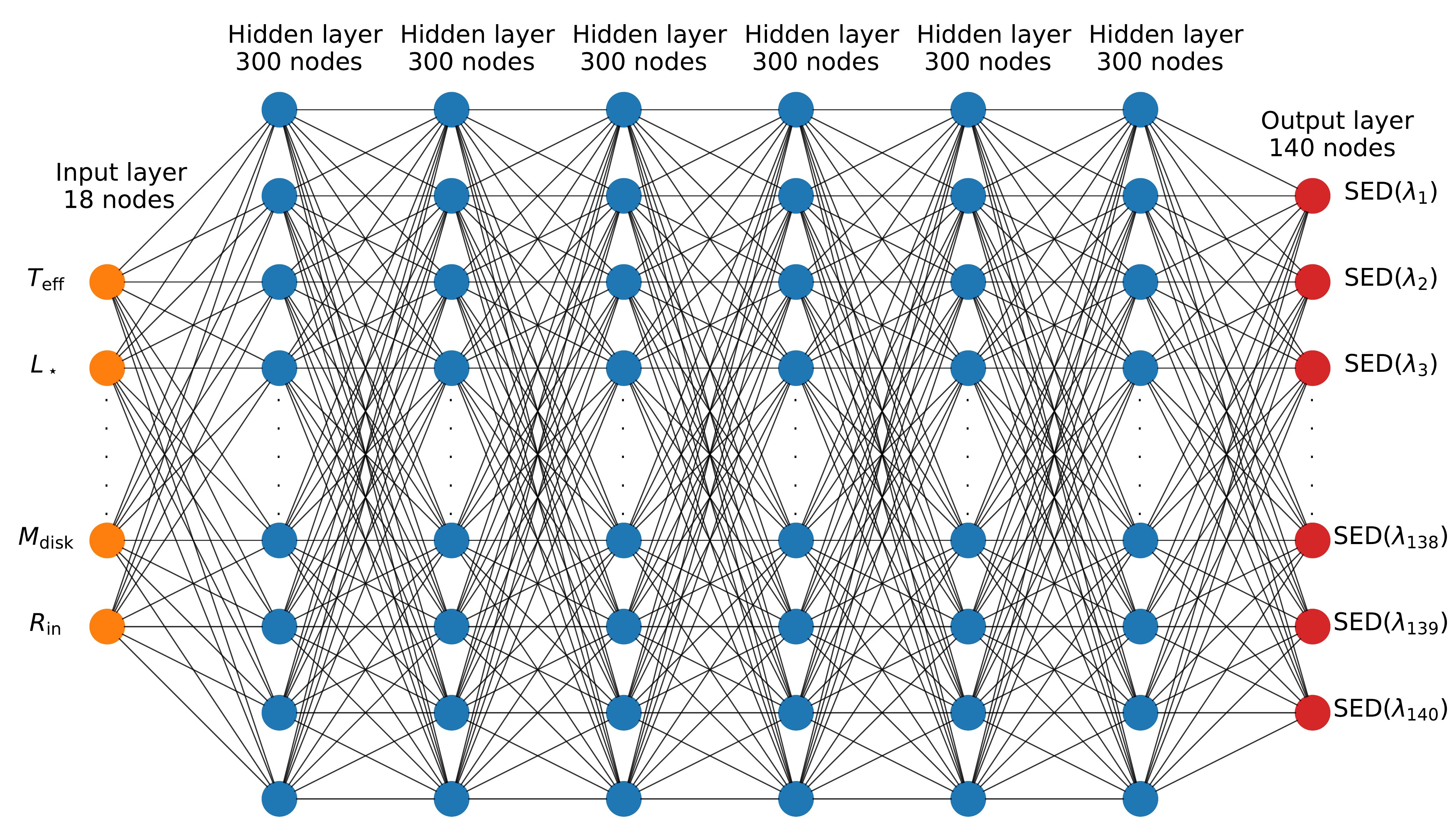}
    \caption{Chosen architecture of the neural network (NN) to predict single zone model's SEDs. The NN consists of an input layer (orange) with $18$ neurons (18 free parameters), $6$ hidden layers (blue circles) with $300$ neurons each, and an output layer (red) with $140$ neurons (the SED values at wavelength $i$ of the 140 points ($\mathrm{SED}(\lambda_i)$). All neurons are connected to every neuron in the adjacent layer. This sums up to $499,340$ trainable parameters.}
    \label{fig:NN_sketch} 
\end{figure}

The NNs used in this paper consist of a number of layers, each with a number of nodes (see, e.g. \citealp{Gangal2021} for an overview of different NNs). Each node or neuron is connected to every node in the adjacent layers (Fig.~\ref{fig:NN_sketch}). The first layer, also called input layer, takes all input parameter values, one for each node. In the final layer, also called output layer, each node produces the spectral flux at a different wavelength. Both layers are connected through a number of hidden layers. Every connection between a neuron $i$ in layer $\ell$ and a neuron $j$ in the next layer $\ell+1$ has a weight $w_{i,j,\ell}$. These weights are optimised during the training process. The sum of the neurons' values $u_{i,\ell}$ in layer $\ell$, multiplied by the respective weights, plus a trainable constant $b_{j,\ell}$, called bias, is taken as input for the activation function $f$ to get the output value of the neuron $u_{j,\ell+1}$ in the next layer:
\begin{align}
    u_{j,\,\ell+1} = f\left(\sum_{i} w_{i,j,\ell} \cdot u_{i,\ell} + b_{j,\ell}\right). 
\end{align}
We use the rectified linear unit (relu) function $f(u) = \max(0,u)$ as the activation function for all neurons in all hidden layers. This function introduces non-linearities to the network, which is essential for the learning process. It enables a NN with just one hidden layer and a finite number of neurons to approximate any continuous function \citep{hornik1989}. Multiple layers are used to reach convergence quicker.

We trained two different networks for the single-zone and the two-zone disk models. As input for both NNs, we take all model parameters. Therefore, the input layer of the NN to predict single zone models consists of $18$ nodes, while the respective NN for two-zone models has $26$ nodes. For all parameters that are logarithmically varied in the grid, we take the logarithm of their values as input for the network. The output predicted by the network is a vector with the logarithmic fluxes for the SED at the $140$ wavelength points. Therefore, there is no knowledge about the relationship between the different wavelengths. Both input and output data are standardised by a scaler derived from the training data. This proved to be more successful than normalising the data (Appendix \ref{app:train}). $70\%$ of the data are taken for training, while the rest is used as a test set to determine the point at which the training does not improve any more. We made sure that all models that differ only in their inclination are part of the same set, to insure the independence of the test set. This is necessary, because the inclination is the only parameter not sampled randomly.

In the training process, the weights are changed using stochastic gradient descent to minimise the loss function. We decided to use the mean squared error between predicted and true logarithmic flux values as our loss function.

We use Keras \citep{chollet2015keras}, which is part of the python package TensorFlow \citep{Abadi2016}, to build the multilayer feedforward NNs. Most hyperparameters are determined using hypergrids, which means running multiple NNs to find good settings. A summary of this process is shown in Appendix \ref{app:train}. We are aware that we have not found the best combination of hyperparameters, but the change in quality turned out to be small for reasonable settings. This process results in an architecture of $6$ hidden layers with $300$ neurons for the single zone NN, a prediction of which is seen in Fig.~\ref{fig:pred_sed}. The two-zone NN has $7$ layers with $400$ neurons. Therefore, both NNs have a number of weights that is comparable to the size of the training sets (single zone: $499,340$ weights and $470,379$ SEDs in the training set; two-zone: $1,029,340$ weights and $1,192,877$ SEDs in the training set). For a detailed description of the training and determination of the hyperparameters, see Appendix \ref{app:train}.

We trained the networks until the test loss function stopped improving for $100$ epochs. Then we selected the state of the NN with the best test predictions. The loss value improves rapidly during the first epochs and converges quickly. Therefore, the test score does not improve any more after epoch $1615$, and the training process is stopped.

 
\subsection{Fitting process}

In this section, we explain how we connect observations and the method to derive model parameters with their uncertainties for a sample of $30$ protoplanetary disks.

\subsubsection{The DIANA Sample\label{sec:sample}}

\begin{table*}[ht]
\caption{Assumed stellar parameters for the analysed sample.}
    \label{tab:objects}     
\centering 
\begin{tabular}{llllllll}
\hline
& & & & & & & \\[-1.9ex]
object & SpTyp\tablefootmark{(1)} & $d\,\rm[pc]$\tablefootmark{2} & Av\tablefootmark{(3)} & $T_{\rm eff}$\,[K]\tablefootmark{(3)}& $L_\star\,[L_{\rm sun}]$\tablefootmark{(4)} & $M_\star\, [M_{\rm sun}]$\tablefootmark{(1)} & age\,[Myr]\tablefootmark{(1)} \\

& & & & & & & \\[-1.9ex]\hline
& & & & & & & \\[-1.9ex]
HD\,97048 & B9 & $184.4\pm0.8$ & 1.28 & 10000 & $45.82$ & 2.6 & 4.5 \\
& & & & & & & \\[-1.9ex]
HD\,100546 & B9\tablefootmark{(5)} & $108.1\pm0.4$ & 0.22 & 10470 & $33.55$ &  2.5\tablefootmark{(5)} &  $>4.8$\tablefootmark{(5)} \\
& & & & & & & \\[-1.9ex]
AB\,Aur & B9 & $155.9\pm0.9$ & 0.42 & 9550 & $49.35$ & 2.6 & 3.9 \\
& & & & & & & \\[-1.9ex]
HD\,95881 & A1\tablefootmark{(6)} & $111.0\pm2.4$ & 0.89 & 9900 & $14.43$ & 2.5\tablefootmark{(6)} & $>5.8$\tablefootmark{(6)} \\
& & & & & & & \\[-1.9ex]
HD\,163296 & A1\tablefootmark{(6)} & $101.0\pm0.4$ & 0.48 & 9000 & $25.17$ & 2.5\tablefootmark{(6)} &  $>5.8$\tablefootmark{(6)}\\
& & & & & & & \\[-1.9ex]
49\,Cet & A2 & $57.23\pm0.18$ & 0.00 & 8770 & $15.58$ & 1.99 & 10 \\
& & & & & & & \\[-1.9ex]
MWC\,480 & A5 & $156.2\pm1.3$ & 0.16 & 8250 & $17.81$ & 1.99 & $4.9\cdot10^2$ \\
& & & & & & & \\[-1.9ex]
HD\,169142 & A7\tablefootmark{(7)} & $114.87\pm0.35$ & 0.06 & 7800 & $6.15$ & 1.74\tablefootmark{(7)} & $>15$\tablefootmark{(7)} \\
& & & & & & & \\[-1.9ex]
CQ\,Tau & A9 & $149.4\pm1.3$ & 2.53 & 7231 & $11.58$ & 1.76 & 9.4 \\
& & & & & & & \\[-1.9ex]
HD\,142666 & F1 & $146.3\pm0.5$ & 0.81 & 7050 & $10.02$ & 1.7 & $1.7\cdot10^3$ \\
& & & & & & & \\[-1.9ex]
HD\,135344B & F3 & $135.0\pm0.4$ & 0.40 & 6620 & $7.07$ & 1.58 & 12 \\
& & & & & & & \\[-1.9ex]
V\,1149\,Sco & F9 & $167.3\pm0.5$ & 0.71 & 6080 & $3.75$ & 1.33 & 16 \\
& & & & & & & \\[-1.9ex]
PDS\,66 &  K0 & $97.89\pm0.12$ & 1.01 & 5205 & $1.42$ &  1.25 &  15 \\
& & & & & & & \\[-1.9ex]
Lk\,Ca\,15 & K3\tablefootmark{(8)} & $157.2\pm0.7$ & 1.7 & 4730 & $1.51$ & 1.5\tablefootmark{(8)} & $>5.5$\tablefootmark{(8)} \\
& & & & & & & \\[-1.9ex]
RY\,Lup & K4 & $153.5\pm1.4$ & 0.29 & 4420 & $1.95$ & 1.23 & 1.9 \\
& & & & & & & \\[-1.9ex]
Usco\,J1604-2130 & K4 & $145.3\pm0.6$ & 1.0 & 4550 & $0.76$ & 1.0 & 10 \\
& & & & & & & \\[-1.9ex]
CI\,Tau & K6 & $160.3\pm0.5$ & 1.77 & 4200 & $1.21$ & 0.90 & 1.9 \\
& & & & & & & \\[-1.9ex]
TW\,Cha & K6 & $183.1\pm0.4$ & 1.61 & 4110 & $0.78$ & 0.82 & 2.8 \\
& & & & & & & \\[-1.9ex]
TW\,Hya & K7 & $60.14\pm0.05$ & 0.20 & 4000 & $0.34$ & 0.75 & 7.8 \\
& & & & & & & \\[-1.9ex]
RU\,Lup & K7 & $157.5\pm1.0$ & 0.00 & 4060 & $1.49$ & 0.73 & 1.1 \\
& & & & & & & \\[-1.9ex]
AA\,Tau & K7 & $134.7\pm1.6$ & 0.99 & 4010 & $0.78$ & 0.71 & 2.3 \\
& & & & & & & \\[-1.9ex]
GM\,Aur & K7 & $158.1\pm1.2$ & 0.30 & 4000 & $0.77$ & 0.70 & 2.3 \\
& & & & & & & \\[-1.9ex]
DN\,Tau & K7 & $128.6\pm0.4$ & 0.71 & 3986 & $0.71$ & 0.69 & 2.4 \\ 
& & & & & & & \\[-1.9ex]
BP\,Tau & K7 & $127.4\pm0.6$ & 0.57 & 3950 & $0.74$ & 0.65 & 2.0 \\
& & & & & & & \\[-1.9ex]
DF\,Tau & K7 & $176\pm16$ & 1.27 & 3900 & $3.89$ & 0.61 & 0.36 \\
& & & & & & & \\[-1.9ex]
DO\,Tau & M0 & $138.5\pm0.7$ & 2.6 & 3800 & $0.90$ & 0.52 & 1.1 \\
& & & & & & & \\[-1.9ex]
DM\,Tau & M0 & $144.0\pm0.5$ & 0.55 & 3780 & $0.25$ & 0.52 & 5.4 \\
& & & & & & & \\[-1.9ex]
CY\,Tau & M1 & $126.33\pm0.33$ & 0.10 & 3640 & $0.29$ & 0.42 & 2.8 \\
& & & & & & & \\[-1.9ex]
FT\,Tau & M3 & $130.2\pm0.4$ & 1.09 & 3400 & $0.26$ & 0.30 & 2.2 \\
& & & & & & & \\[-1.9ex]
RECX\,15 & M3 & $103.4\pm2.4$ & 0.65 & 3400 & $0.11$ & 0.29 & 5.2 \\ 
\end{tabular}
\tablefoot{\\
\tablefoottext{1}{Spectral type, stellar mass and age for every object is derived using the updated luminosity and pre-main-sequence tracks by \cite{Siess2000} for solar-metallicity.}\\
\tablefoottext{2}{Distances are taken from \cite{Gaia2021}.}\\
\tablefoottext{3}{Taken from \cite{Woitke2019}}\\
\tablefoottext{4}{Updated Luminosity: Luminosity used by \cite{Woitke2019} scaled according to new distances.}\\
\tablefoottext{5}{No tracks from \cite{Siess2000} match, values are taken from the closest point at $T_{\rm eff} =9650\,\rm K$ and $L_\star= 42\, L_{\rm sun}$}\\
\tablefoottext{6}{No tracks from \cite{Siess2000} match, values are taken from the closest point at $T_{\rm eff} =9000\,\rm K$ and $L_\star= 30\, L_{\rm sun}$}\\
\tablefoottext{7}{No tracks from \cite{Siess2000} match, values are taken from the closest point at $T_{\rm eff} =7800\,\rm K$ and $L_\star= 9\, L_{\rm sun}$}\\
\tablefoottext{8}{No tracks from \cite{Siess2000} match, values are taken from the closest point at $T_{\rm eff} =4730\,\rm K$ and $L_\star= 1.6\, L_{\rm sun}$}
}
\end{table*}

We apply our machine learning SED-fitting method to a sample of 30 well-studied disks (the so-called DIANA sample, see \citealp{Woitke2019}), for which many multiwavelength observations (photometric fluxes and low-resolution spectra) are available and which have already been fitted using the same RT code (MCFOST). This will allow us to compare our results with a previous study.
$27$ of these objects are introduced in \cite{Woitke2019}. However, here we expand this set by adding $3$ more unpublished SED fits for DN\,Tau, CQ\,Tau and PDS\,66, which have meanwhile been fitted as well using the same approach as described in \cite{Woitke2019}\footnote{https://prodimo.iwf.oeaw.ac.at/models/diana-sedfit}.
In this section, we introduce this sample and explain the modelling done by \cite{Woitke2019} briefly. For details on the data collection, see \cite{Dionatos2019} and \cite{Woitke2019} for details of the modelling.

The observational data for each object comprise photometric fluxes and low-resolution spectra (UV, ISO, Spitzer/IRS, Herschel/PACS, Herschel/SPIRE) that are individually selected. Measurement uncertainties smaller than $5\,\%$ are set to $5\,\%$ to account for calibration problems in the mix of different measurements.

The SED models for the DIANA sample in \cite{Woitke2019} are derived in a two-step approach. First, the stellar parameters are fitted and fixed afterwards. The stellar parameters ($T_{\rm eff}$, $L_{\star}$, $A_V$, $M_{\star}$, age, and spectral type) are consistent with pre-main-sequence evolutionary tracks by \cite{Siess2000}.

In a second modelling step, a disk model was fitted to the SED measurements for every object. This was done using an evolutionary strategy to minimise $\chi^2$. For the modelling of each object, either a single or two-zone setup was selected by hand, which had a number of parameters fitted while other parameters were fixed using literature knowledge. This process resulted in the best fit for every object. We note that for a number of objects, the fits have recently been updated. The most up-to-date versions can be found online$^1$.

For the three objects added to the DIANA sample in this paper, we exactly followed the same procedure, with the most up-to-date distances available at the time.
We introduce these three objects briefly here, while referring to \cite{Woitke2019} for detailed information about the other objects.

\smallskip\noindent
\textit{CQ\,Tau:\ } CQ\,Tau is a F5 star \citep{Manoj2006} at a distance of about $163\,\mathrm{pc}$ \citep{Collaboration2018}, and with an age of about $10\,\mathrm{Myr}$, it is one of the oldest Herbig Ae/Be stars \citep{Chapillon2018}.
We used a stellar luminosity of $13.81\,\rm L_\odot$, an effective temperature of $7231\,\rm K$, a stellar mass of $1.25\,\rm M_\odot$, a visual extinction Av of $2.53$, an age of $8.4\,\rm Myr$, and a spectral type of A9 to fit the selected observations, roughly reproducing the literature values for age and spectral type.
\citet{Gabellini2019} found a cavity between $15\,\mathrm{AU}$ and $25\,\mathrm{AU}$. The disk is very similar to the TW Hydra system, due to its gas-rich disk with dust grains up to a few centimetres \citep{testi2003}.

\smallskip\noindent
\textit{PDS\,66:\ } PDS\,66 is a T Tauri-type K1 star \citep{Pecaut2016} at a distance of about $99\,\mathrm{pc}$ \citep{Collaboration2018}. Depending on its membership to the Lower Centaurus Crux (LCC) subgroup or the eta Cha association, it is thought to have an age between $7$ and $17$\,Myr \citep{Wolff2016}. The stellar fit by \citet{Grafe2013} resulted in an age of $13\,$Myr. Despite the old age of the system, the inner rim of the disk is very close to the star ($0.12"$) in polarised light \citep{Wolff2016}. This is in line with \citet{Cortes2009}, who find an inner radius of the disk consistent with the sublimation radius at $0.1\,\mathrm{AU}$. This is peculiar, because normally the inner part of the disk is cleared at that time.

\smallskip\noindent
\textit{DN\,Tau:\ } DN\,Tau in a classical T Tauri star of spectral type M0 \citep{Long2018} at a distance of about 129\,pc \citep{Collaboration2018}. A fit to our photometric, UV and X-ray data collection (see \citealt{Dionatos2019}) finds an effective temperature of about 3990\,K and a stellar luminosity of about $\rm 0.71\,L_\odot$ with a reddening of $A_V\!\approx\!0.71$, which according to the Siess tracks \citep{Siess2000} corresponds to spectral type K7, a stellar mass of $0.69\,M_\odot$ and an age of 2.4\,Myrs. These stellar properties are close to the ones discussed for DN\,Tau in \citep{Robrade2014}, even though our estimates use a slightly stronger reddening and hence larger luminosity, more in line with \citet{Ingleby2013}. According to the ALMA observations by \citep{Long2018} the dusty disk of DN\,Tau shows a regular Gaussian intensity profile with only little evidence for rings. The apparent radius of the dusty disk at 1.3\,mm is about 56\,au, which makes DN\,Tau one to the smallest targets in the DSHARP sample.

The DIANA project used the observed UV spectra directly, whenever available. However, for the general fitting performed in this paper, we need to assume a power law in the UV with two input parameters. Therefore, we estimated $p_{\rm UV}$ and $f_{\rm UV}$ by an iterative process (resulting values in Table \ref{tab:compare_diana_single} and Table \ref{tab:compare_diana_two}), until the UV power law roughly fits the observed UV data. In case of RU\,Lup, the UV observations do not follow a power law. Hence, we opted for a power law that matches the UV luminosity from the DIANA fit instead. Additionally, there is no power law within the limitations of the grid setup that can reproduce the UV spectrum of DO\,Tau, due to the strong reddening of the object. We therefore opted for a power law that fits the data as good as possible. We do not expect these manipulations to have a large impact on the derived disk parameters, because the UV wavelength points are excluded from our likelihood function.

All model fits use the best distances available at that time. However, since all DIANA objects have meanwhile updated distances \citep{Gaia2021}, we scaled the luminosities to the new distances and used the evolutionary tracks from \cite{Siess2000} to update the stellar mass, age, and spectral type accordingly. A list of all stellar parameters are shown in Table \ref{tab:objects}. The resulting stellar spectra fit well to the observational data. Therefore, a completely new stellar fit is not necessary for the aims of this project.

This means that the reader will find two SED-fits for every object in this paper. The first fit uses old distance measurements and the same parameter constraints used in DIANA to compare the results to the previously published fits. The second fit uses up-to-date distances without any parameter constraints to derive new values with uncertainties for all disk parameters.

\subsubsection{Data preparation}

To compare observations to the SEDs derived with our technique, a few in-between steps are needed.
The NNs return SEDs for models that are at a distance of $100\,\mathrm{pc}$, this has to be changed in post-processing by scaling the flux to the new distance.

Furthermore, the output of the NNs needs to be reddened. We use the reddening law by \cite{Fitzpatrick1999} to derive the wavelength-dependent amount of interstellar extinction using two reddening parameters, $E_{\rm B-V}$ and $R_{\rm V}$. 
In the last step, the reddened and distance-corrected SEDs are interpolated using a spline fit in log-log space to the wavelengths of the observations.

\subsubsection{Bayesian analysis\label{sec:bayesian_analy}}

We use MultiNest \citep{Feroz2008,Feroz2009,Feroz2019} through the PyMultiNest Python package \citep{Buchner2014} to compare the models with observations and determine the posterior distribution for the free parameters. MultiNest is a Bayesian inference tool. It uses a multimodal nested sampling algorithm to compute the Bayesian evidence and posterior distribution from a likelihood function. All details of the algorithm and the numerical implementation are described in the references given above.

We use the observational data selected by \cite{Woitke2019}, which were introduced in Sect. \ref{sec:sample}. 
Since the goal of this study is to fit the disk parameters with reasonable assumptions about the star, we exclude all observations shortwards of $0.5\,\rm \mu m$. 
The uncertainties for all (photometric and spectral) flux data points ($\sigma_{i,obs}$) is set to be at least $5\%$ times the observed values, to account for calibration problems and stellar variability when using multiple instruments and multi-epoch data.
As mentioned before, the model emulation by a NN introduces an error on the model prediction. To account for this, we add a model error of $5\%$ of the predicted flux value (see Sect.~\ref{Sec:acc}).

The likelihood function $\mathcal{L}$ accounts for the deviations of the interpolated model fluxes ($y_{i,model}\!=\!\nu F_{\nu,model}$) from the observational data ($y_{i,obs}\!=\!\nu F_{\nu,obs}$) considering all wavelength points $i\!=\!1\,...\,N_{obs}$ longwards of 0.5\,$\mu$m, with respect to the uncertainties of model ($\sigma_{i,model}$) and observations ($\sigma_{i,obs}$):

\begin{align}
    \mathcal{L} = \prod_{i=1}^{N_{obs}} \frac{1}{\sqrt{2\pi\,(\sigma_{i,model}^2 + \sigma_{i,obs}^2})}\exp{\left(-\frac{\left(y_{i,model} -y_{i,obs}\right)^2}{2\, (\sigma_{i,model}^2 + \sigma_{i,obs}^2)}\right)}.
\end{align}

The low-resolution spectra consist of many more data points than the photometric measurements, which would make them dominate the likelihood value. Therefore, we reduce the influence of the spectra by weighting the logarithmic likelihood of spectral points. The weights are set to $10$ divided by the spectral resolution. Therefore, the weight of every spectral point equals the weight a spectrum with a spectra resolution of $10$ would have.

In addition to the likelihood function, we need to specify the prior probability distribution function for each object and parameter. We use the same parameter distributions that were used for the creation of the model grids as our prior distribution for all objects. We use additional constraints to ensure that the NNs cannot predict parameter values outside of the range used to create the training grid. We solved this problem by assigning a very small likelihood in all such cases. These constraints concern $\log_{10}(M_{\rm disk}/M_\star)$, $\log_{10}(M_{\rm disk,2}/M_{\rm disk})$, $\log_{10}(R_{\rm out,2}/R_{\rm in})$, $\log_{10}(a_{\rm max,2}/a_{\rm max})$,  $\log_{10}(f_{\rm PAH,2}/f_{\rm PAH})$, and $T_{\rm sub}$, which are all checked for values outside of the limits $x_{\rm min}$ and $x_{\rm max}$. 
We also use additional physical constraints to avoid the outer radii of all zones smaller than their respective inner radii, overlapping radial zones, and maximum dust sizes smaller than minimum dust sizes. 

As described in Appendix \ref{App:grid}, we excluded SEDs from the training sample that have disks that occult the star (shielded SEDs). In those cases, the NNs are unable to predict this behaviour even if certain parameter combinations would lead to it. Hence, we also need to exclude these cases from the Bayesian analysis. To do so, we trained two NNs to distinguish between shielded and non-shielded SEDs with $2$ and $4$ hidden layers and $64$ neurons per layer for single and two-zone models, respectively. These NNs were trained on all SEDs created during the grid creation. While all parameters (identical to the main NNs) are used as input, the output is a number between $0$ and $1$ that encodes the probability of this parameter set resulting in a shielded SED. Since we want to make sure that most shielded SEDs are excluded, we set the threshold to distinguish shielded and non-shielded SEDs to $0.0721$ and $0.0577$ for single and two-zone models, respectively. These values are chosen because they result in the detection of more than $99.9\,\%$ of shielded SEDs while incorrectly classifying as non-shielded SEDs in about $1\,\%$ of all test cases. We incorporated this classification into the likelihood in the same manner as previously shown, which hinders the Bayesian analysis to explore these areas of parameter space.

\section{Results\label{sec:results}}

\subsection{SED predictions\label{Sec:acc}}
In this section, we evaluate the quality of the NNs' predictions.
We use the relative deviations of predicted flux ($F_{\rm predict,\lambda}$) at a wavelength $\lambda$ compared to the true flux calculated by the full RT code ($F_{\rm true,\lambda}$) as our measure of quality for a predicted flux:
\begin{align}
    Q_{\rm pred}= \frac{|F_{\rm true,\lambda}-F_{\rm predict,\lambda}|}{F_{\rm true,\lambda}} \label{eq:quality_indi}.
\end{align}
\begin{figure}[t]
    \centering
    \includegraphics[width=\linewidth]{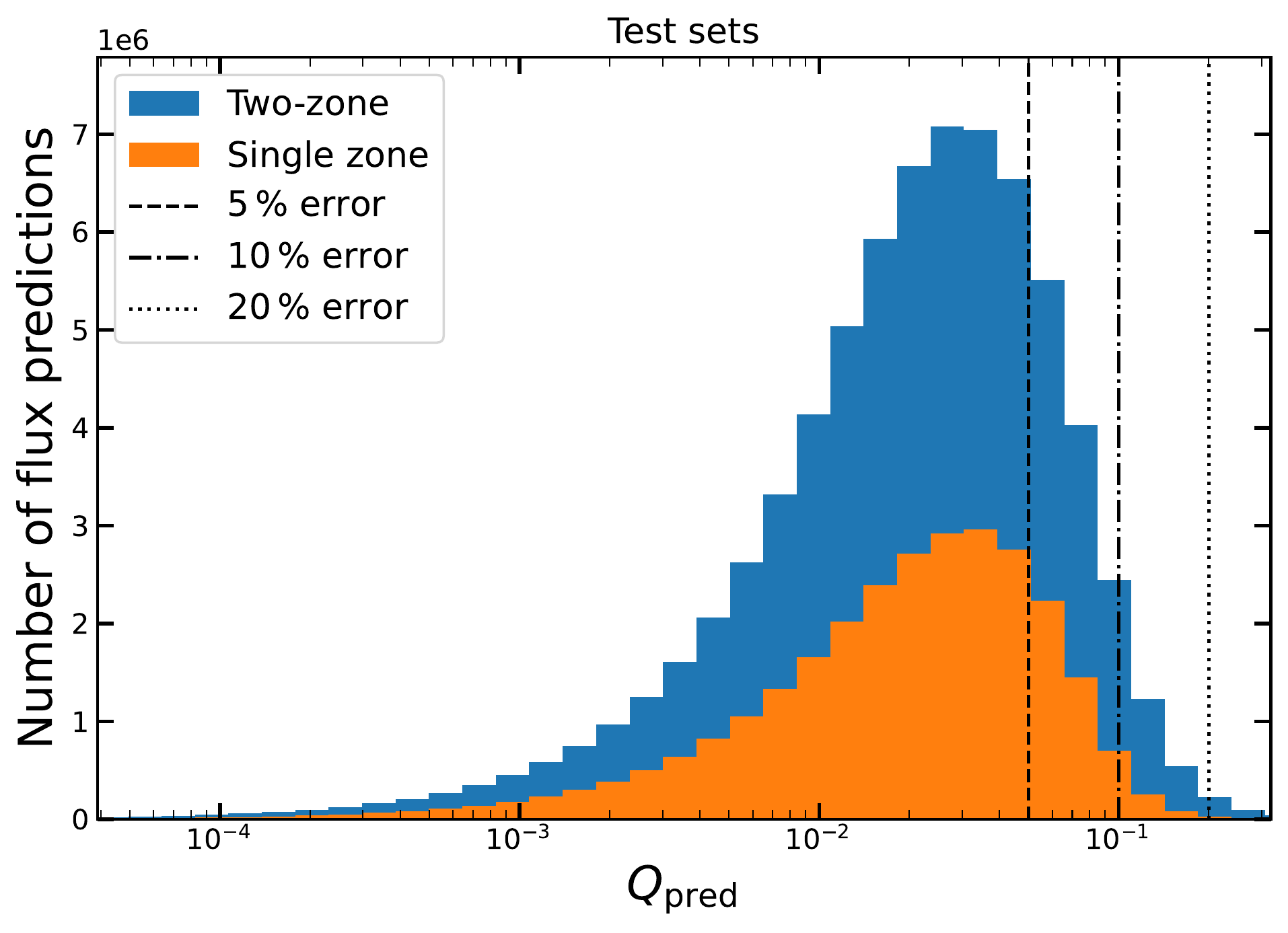}
    \caption{Quality distribution according to Eq.~\ref{eq:quality_indi} of the predictions in the test set for the single zone (orange) and two-zone (blue) NN. The histogram shows how many predictions fall in each quality bin. The dashed, dashed-dotted, and dotted black lines show the $5\,\%$, $10\,\%$, and $20\,\%$ difference, respectively.} 
    \label{fig:errorshisto_NN}
\end{figure}
Figure \ref{fig:errorshisto_NN} shows the quality distribution for every flux prediction of the single and two-zone SEDs in the respective test sets.
The prediction quality for both networks peaks at similar values, with median\,/\,mean values of 2.2\%\,/\,2.9\% (single zone) and 2.3\%\,/\,3.2\% (two-zone).

Most predictions have similar qualities, with 68\% of the predictions ($1\sigma$) being better than 3.4\% and 3.6\% for the single and two-zone NN, respectively. Therefore, we use a conservative value of $5\%$ as the $1\sigma$ uncertainty for the NNs' predictions in the Bayesian Analysis.

We note that both NNs predict a few models poorly, while the overwhelming majority are predicted well, with similar quality. We individually inspected the worst predictions and found common patterns. Individual wavelength points in the UV ($0.1-0.15\,\mu m$ and $\sim0.42\,\rm\mu m$) were sometimes poorly predicted. This can be explained by the many emission lines in the UV. A few wavelength points coincide with these lines. Therefore, the resulting flux at these points are hard to predict. Nevertheless, this does not affect the further analysis since the UV is not used in the Bayesian analysis. The second most common source for errors are the PAH features. Since the variation of the SED at these wavelengths is large, stars that show strong PAH features (high stellar luminosity and temperature) have the potential for slightly worse predictions.

Fig.~\ref{fig:pred_sed} shows an SED prediction for an A-type star from the single zone NN. This prediction has mean prediction quality for the $140$ fluxes of this model of $2.9\,\%$, which is the mean value of the test sample. Therefore, the quality of this prediction is a good representation of the average prediction quality.

\subsection{Reproducing DIANA fits\label{sec:reproduce_DIANA}}
In a first step, we aim to reproduce the DIANA fits with parameter uncertainties. This means setting the same constraints to the model (fixed parameters and complexity) as done for the DIANA fits. 

Regarding the complexity of the models, we opted for single or two-zone models according to the DIANA fits. Two-zone models exist in three settings (gap, no gap, and smooth transition). A gap exists if $R_{\rm out,2} < R_{\rm in}$ and there is no gap if $R_{\rm out,2} = R_{\rm in}$. A smooth transition means that the flaring index and mass of the inner zone are adjusted so that the surface density and scale height are the same for the inner and outer disk zone at $R_{\rm out,2} = R_{\rm in}$. 
The list of fixed parameters always included all stellar parameters ($L_{\star}$, $T_{\rm eff}$, $f_{\rm UV}$, and $p_{\rm UV}$) and the inclinations.

For one object, RECX\,15, the dust-to-gas ratio is not $0.01$ in the DIANA data. Since we cannot account for that with our setup, we adjusted the disk gas mass to a value that corresponds to the needed dust-to-gas ratio. Since the gas mass influences only the settling in our setup, we adjusted the settling parameter of that model to have the same settling efficiency. 

For some objects, DIANA fixed the outer radius to values that do not fit into the relation we enforce between the outer radius and taper radius. Additionally, and exponent of the exponential function which decreases the column density outside $R_{\rm taper}$ was sometimes fixed to values different from $\epsilon$. Both of these effects, cannot be simulated with our neural networks. However, we expect the resulting differences of the SED to be minimal.

In Table \ref{tab:compare_diana_single} (Appendix \ref{sec:tables}) all parameter values for single zone objects are listed with the values from \cite{Woitke2019} for comparison. If no value is given for a parameter for the fit from this study, it was fixed to the value used by DIANA. If no values are given for the PAH parameters in DIANA, this means that they were excluded from the model. In that case, we set $f_{\rm PAH}=0.001$ and $\rm PAH_{\rm charged}=0.0$. Even though this does not completely remove PAHs from our model, it makes them negligible.

In Table~\ref{tab:compare_diana_two} (Appendix \ref{sec:tables}) the parameter values for this work and DIANA are shown for all two-zone models. Again, values that are only listed for DIANA are fixed to that value in our study. For some object the PAH parameters are not listed meaning that PAHs were excluded for the respective zone in DIANA and set to $f_{\rm PAH}=0.001$, $f_{\rm PAH,2}=0.000001$, and $\rm PAH_{\rm charged}=0.0$ in this study to mimic this effect. 

If the transition between zones is smooth, $M_{\rm disk,2}$ and $\beta_2$ are adjusted accordingly. This is noted in Table \ref{tab:compare_diana_two} as 'smooth' at the relevant entries.

\begin{figure}[t]
    \centering
    \includegraphics[width=\hsize]{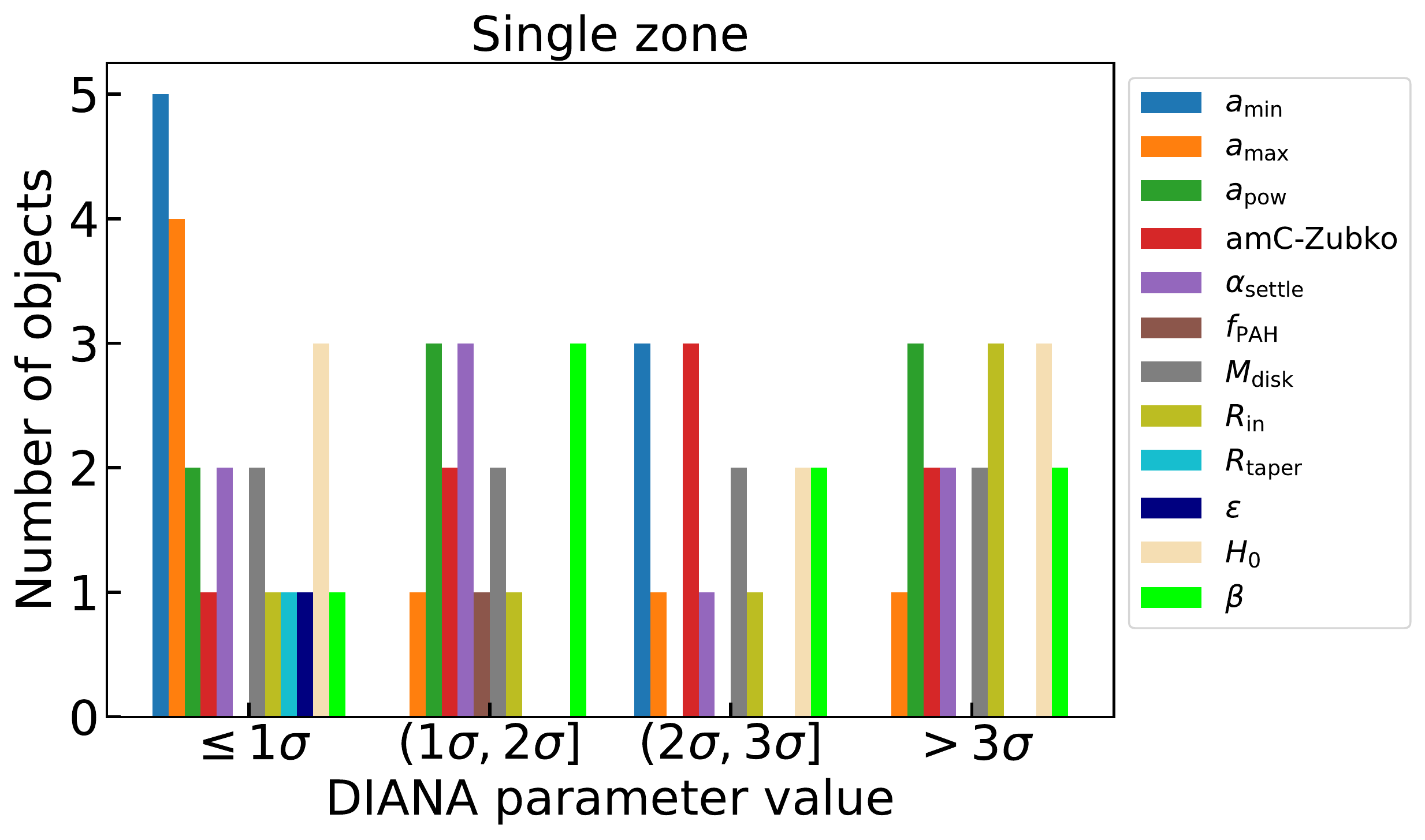}
    \caption{Comparison between resulting parameter posterior distribution and the DIANA fits for single zone objects. The histograms show the fraction of DIANA models that have parameter values within the $1\sigma$, $2\sigma$, and $3\sigma$ contour or outside for that parameter.} 
    \label{fig:hist_single_diana_quality} 
\end{figure}

\begin{figure}[ht]
    \centering
    \includegraphics[width=\hsize]{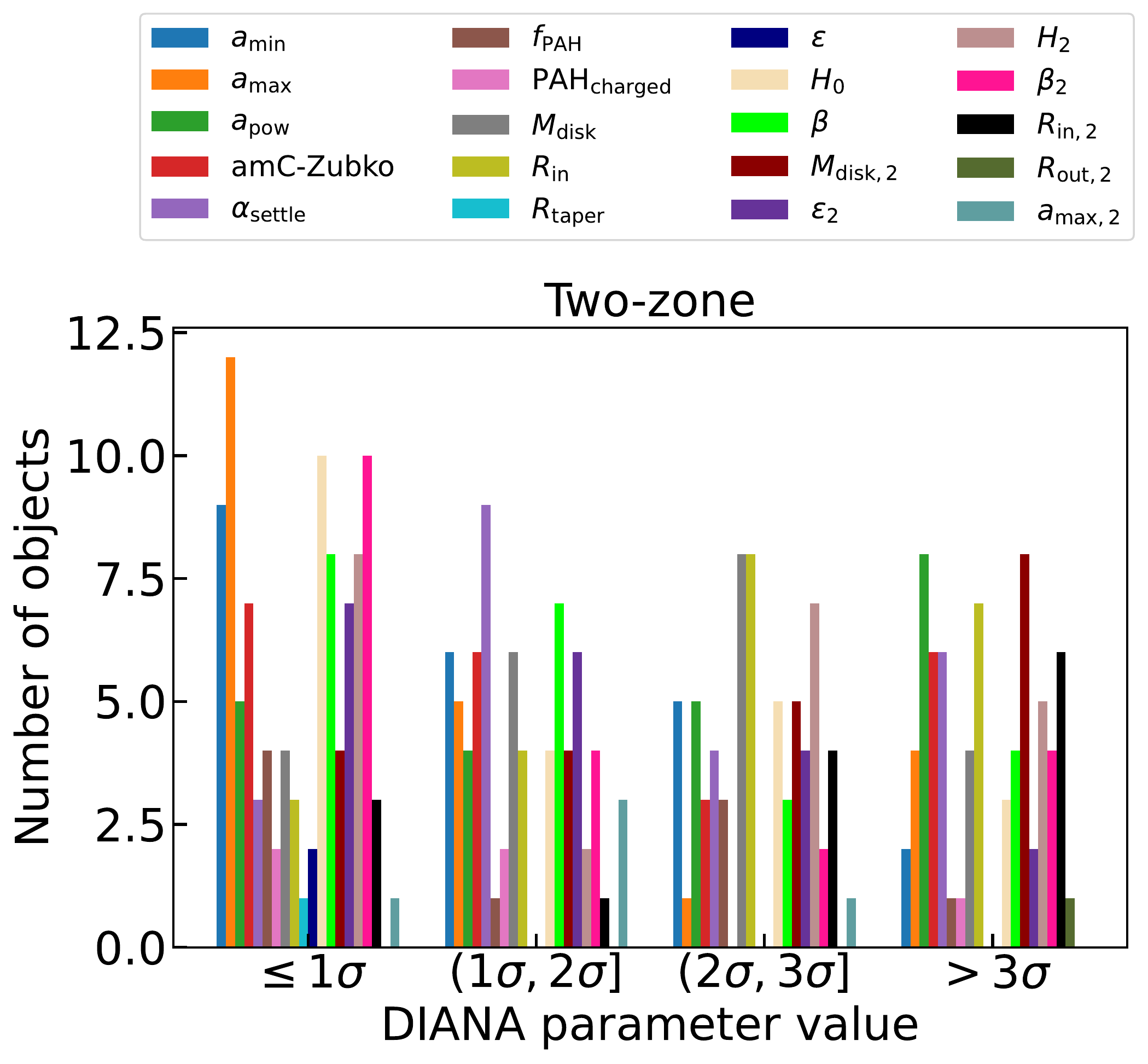}
    \caption{Comparison between resulting parameter posterior distribution and the DIANA fits for two-zone objects. The histograms show the fraction of DIANA models that have parameter values within the $1\sigma$, $2\sigma$, and $3\sigma$ contour or outside for that parameter.} 
    \label{fig:hist_two_diana_quality} 
\end{figure}

Now, we discuss the quantitative comparison of our fits and the DIANA fits.
Fig.~\ref{fig:hist_single_diana_quality} and Fig.~\ref{fig:hist_two_diana_quality} show how often the parameter values found in DIANA fall within different $\sigma$ levels from the posterior for single and two-zone models, respectively.

For single zone models $23$, $16$, and $15$ of the $72$ parameter predictions fall within the $1\sigma$, $2\sigma$, and $3\sigma$ contour, with $18$ DIANA parameter values outside the $3\sigma$ level. For the $317$ two-zone model parameter predictions $103$, $74$, and $68$ predictions fall within the respective contours, while $72$ are outside the $3\sigma$ level.

A good agreement between DIANA and this study can have two causes. The parameter can be easy to constrain which makes it possible for DIANA to find these values, or the parameter uncertainties are very large and therefore many parameter prediction falls within a certain $\sigma$ contour. We test the size of the uncertainties in Sect. \ref{sect:uncertain}.
In general, there is little agreement between the DIANA parameter values and the derived posteriors for both single and two-zone models.

The best overlap for single zone parameters (Fig.~\ref{fig:hist_single_diana_quality}) is found for $a_{\rm min}$, $a_{\rm max}$, $\alpha_{\rm settle}$, with $63\,\%$, $71\,\%$, and $63\,\%$ of the fitted parameters within the $2\sigma$ level, respectively. Additionally, $R_{\rm taper}$ and $\epsilon$ are fitted only one time each with good agreement between the studies (within $1\sigma$). Among the worst fitted parameters are amC-Zubko, $R_{\rm in}$, $H_0$, and $\beta$.

The two-zone fits (Fig.~\ref{fig:hist_two_diana_quality}) in this study find similar overlap compared to the single zone fits. Very often $R_{\rm in}$ values of DIANA are outside the uncertainties of this study. Similarly, there is little overlap for $a_{\rm pow}$ and $M_{\rm disk,2}$.

We constrain the Bayesian Analysis to only explore areas of the parameter space in which the NN's predictions are valid. Therefore, it is not possible to predict values outside these restrictions. Even though, we informed our parameter space by the DIANA predictions, some parameter values that were found in DIANA are so extreme that we excluded them from our grid creation. As an example, it seems that the innermost radius causes some disagreement between DIANA and this study with $3$ ($6$) predictions for single zone (two-zone) models outside the $3\sigma$ contour. $3$ DIANA fit radii have temperatures that we exclude from our study. Therefore, our method does not explore this area of parameters space and these parameters are predetermined to be outside the $3\sigma$ contour.

The different levels of overlap are not equally distributed over all fitted objects. For example, the single zone fits of DN\,Tau and DF\,Tau and the two-zone fit for CQ\,Tau, HD\,135344B and CY\,Tau find no $>3\sigma$ disagreements. On the other hand, the single zone DIANA fit for 49\,Cet does end up in a completely different area of parameter space, resulting in $>3\sigma$ differences between both studies. The largest number of parameter disagreements for two-zone fits is found for AB\,Aur with $62\,\%$ of parameters outside the $3\sigma$ level.

The differences between the parameter values derived in these two studies might be caused by the different quality metrics that are used. In this study, we use the likelihood defined in Sect. \ref{sec:bayesian_analy}, while DIANA uses a wavelength weighted version of $\chi$ (DIANA-$\chi$). We want to check if the fits found in this study are not just different from DIANA, but also of higher quality. To do that, we use DIANA-$\chi$ and calculate it for one representative model of the posterior distribution of every object. For this, we use the so-called median probability model, which we define as the model for which the parameters are closest to the median posterior values. This means that this model's most extreme parameter value is within the percentile level closest to the median value of the posterior compared to all other models of the posterior (similar to \citealp{Barbieri2004}). This model is a good representation of the posterior distribution, even though it is not the model with the maximum likelihood.

We calculate the DIANA-$\chi$ for SED predicted by the NNs of that model and also for the MCFOST model with the same parameters. Since data points shordwards of $0.5\,\rm \mu m$ are excluded from the Bayesian analysis, we exclude them from this calculation. As a comparison, we are calculating the DIANA-$\chi$ for the DIANA models also excluding the same observations. We find that the NN predicted SEDs for the median probable model results in a better DIANA-$\chi$ than the DIANA model for $17$ of the $30$ objects. We conclude that the better qualities are not due to the NNs' deviations from the MCFOST models, because the MCFOST SEDs of the same models also result in better DIANA-$\chi$  values in $17$ cases. Therefore, the optimisation strategy of DIANA did not converge to the optimal solution. This shows the advantages of exploring the parameter space with Bayesian analysis compared to simpler optimisation strategies.

The definition of the median probability model allows us to plot the density structure (see Fig.~\ref{fig:density_reproduce}). This shows that due to the fixing of some disk parameters and despite the parameter value differences, the disk structures of this work and the DIANA project show many similarities. Therefore, it is difficult to gain insights if one of the two disk structures is confirmed by other observations. In the next section, we evaluate the evidence for different disk structures (single zone and two-zone models). These structures are different enough that other observations can be used to distinguish between them.

\subsection{Consistent fits for all objects\label{sec:bayes_fact}}

After comparing our method to previous fits, we start to consistently fit all objects using a fully homogenised setup for both a single and a two-zone model. This effort was not possible in DIANA due to the computational cost. For doing so, we assume the stellar parameters listed in Table \ref{tab:objects} and the inclination used in DIANA, which was always taken from literature values. This allows us to incorporate external knowledge for parameters that are hard to determine based on SEDs alone and for which we have known values for every object. All other parameters are free with the grid distribution imposed as a prior (Table \ref{tab:parameters}). This results in $60$ fits with $13$ and $21$ free parameters for every object's single and two-zone model, respectively. The single zone parameter values with their uncertainties are shown in Table~\ref{tab:new_fits_single}, while the two-zone fit results are displayed in Table \ref{tab:new_fits_two}.

\begin{table}[t]
\caption{Bayes factor between single and two-zone fits for all objects.}
    \label{tab:bayes_fact}     
\centering 
\begin{tabular}{l|l|l|l|l}
Object & $B_{12}$\tablefootmark{1} & $N$\tablefootmark{2} & Evidence\tablefootmark{3}& $N_{\rm DIANA}$\tablefootmark{4} \\ \hline
 HD\,97048 & $70.60$ &2 &very strong &$2$ \\ 
 HD\,100546 & $168.65$ &2 &very strong &$2$ \\ 
 AB\,Aur & $301.06$ &2 &very strong &$2$ \\ 
 HD\,95881 & $6.76$ &2 &strong &$2$ \\ 
 HD\,163296 & $13.29$ &2 &very strong &$2$ \\ 
 49\,Cet & $117.92$ &2 &very strong &$1$ \\ 
 MWC\,480 & $24.87$ &2 &very strong &$1$ \\ 
 HD\,169142 & $364.56$ &2 &very strong &$2$ \\ 
 CQ\,Tau & $53.17$ &2 &very strong &$2$ \\ 
 HD\,142666 & $0.61$ &2 &none &$2$ \\ 
 HD\,135344B & $574.59$ &2 &very strong &$2$ \\ 
 V\,1149\,Sco & $134.87$ &2 &very strong &$2$ \\ 
 PDS\,66 & $71.67$ &2 &very strong &$2$ \\ 
 Lk\,Ca\,15 & $41.09$ &2 &very strong &$2$ \\ 
 RY\,Lup & $24.64$ &2 &very strong &$2$ \\ 
 USco\,J1604-2130 & $833.11$ &2 &very strong &$2$ \\ 
 CI\,Tau & $-0.75$ &1 &none &$2$ \\ 
 TW\,Cha & $-2.18$ &1 &weak &$2$ \\ 
 TW\,Hya & $35.16$ &2 &very strong &$2$ \\ 
 RU\,Lup & $35.23$ &2 &very strong &$1$ \\ 
 AA\,Tau & $4.59$ &2 &moderate &$2$ \\ 
 GM\,Aur & $6.62$ &2 &strong &$2$ \\ 
 DN\,Tau & $-5.38$ &1 &strong &$1$ \\ 
 BP\,Tau & $11.82$ &2 &very strong &$2$ \\ 
 DF\,Tau & $-0.02$ &1 &none &$1$ \\ 
 DO\,Tau & $11.39$ &2 &very strong &$1$ \\ 
 DM\,Tau & $19.70$ &2 &very strong &$2$ \\ 
 CY\,Tau & $48.57$ &2 &very strong &$2$ \\ 
 FT\,Tau & $41.07$ &2 &very strong &$1$ \\ 
 RECX\,15 & $13.36$ &2 &very strong &$1$ \\ 
\end{tabular}
\tablefoot{ \\
\tablefoottext{1}{Bayes factor between single and two-zone fits.}\\
\tablefoottext{2}{Prefered number of zones from our fitting}\\
\tablefoottext{3}{Interpretation of $B_{12}$ based on \cite{Trotta2008}.\\
\tablefoottext{4}{Number of zones used in DIANA}}
}
\end{table}

This systematic approach allows us for the first time to evaluate the benefit of discontinuous disk models over simpler single zone models. We evaluate the significance by calculating the Bayes factor ($B_{12}$) for every object between the two-zone model ($M_{2}$) and the single zone model ($M_{1}$). We calculate this factor as the difference of the Nested Sampling Global Log-Evidence for single ($\log(E(M_1)$) and two-zone models ($\log(E(M_2)$). According to \cite{Trotta2008} $B_{12}<1$, $1<B_{12}<2.5$, $2.5<B_{12}<5$, $5<B_{12}<11$, and $11<B_{12}$ correspond to no evidence, weak evidence, moderate evidence, strong evidence, and very strong evidence for a two-zone model over the single zone model, respectively. Negative values mean that single zone models are preferred with the corresponding evidence. 

Table \ref{tab:bayes_fact} list the Bayes factors for all objects with their interpretation. For $22$ of the $30$ objects, we find very strong evidence that a two-zone model reproduces the SED observations better. While some objects return inconclusive results, only $4$ objects return negative Bayes factors, corresponding to evidence of a single zone model over a two-zone model. This dominance of two-zone models is also reflected in DIANA, where $22$ objects used a two-zone fit. Nevertheless, we find very strong evidence for two-zone models for objects (49\,Cet, MWC\,480, RU\,Lup, DO\,Tau, FT\,Tau, and RECX\,15) that are fitted with a single zone model in the DIANA project. On the other hand, we find weak evidence for a single zone model for TW\,Cha that is fitted with two-zones in DIANA. 

For the objects with a complexity mismatch to DIANA, many disk parameters are similar between their single and two-zone fits (Comparing Table~\ref{tab:new_fits_single} to Table~\ref{tab:new_fits_two}). This means, that even though SED fitting is very degenerate, certain parameters (e.g. disk mass and innermost radius) can be determined relatively well independently of the exact model.

This is also reflected in the single and two-zone density structures of the median probability models (see Fig.~\ref{fig:density_new_data}). For many objects, the single zone and two-zone density distribution has a similar radial and vertical structure. This is very prominent for objects that do not have very strong evidence for one of the two structures. One example is CI\,Tau, which shows no evidence for either model complexities. The two density structures differ only in the two-zone model's gap at about $10\,\rm AU$. Other examples for high similarities between the single and two-zone models are HD\,142666 and AA\,Tau for which the two-zone model is preferred even though not strongly.

The resulting disk structures can be roughly compared with images of the objects. Even though we cannot expect the structures derived by SED fitting to reproduce the images well, we are interested if the preference of any disk structure can be confirmed by image data, especially if this study and DIANA prefer different disk complexities.

We find very strong evidence for MWC\,480 to be described by a two-zone model, while DIANA used a single zone model to fit the object. Images by the Atacama Large Millimeter/submillimeter Array (ALMA) at $1.33\,\rm mm$ show a prominent disk gap at about $70\,\rm AU$ with a width of about $30\,\rm AU$ \citep{Long2018}.
This is roughly consistent with the gap in the two-zone model, even though this gap is closer in (inner radius of the outer zone of about $45\,\rm AU$). Therefore, we conclude that the evidence for a two-zone model to fit the SED is confirmed by image data in this case.

The images of CI\,Tau and FT\,Tau of the same study \citep{Long2018} show four rings with corresponding gaps and one gap, respectively. Compared to MWC\,480 these gaps have a much lower intensity contrast, which makes both disk structures plausible and does not give new insight in the complexity mismatch between our study and DIANA.

Figure~\ref{fig:sed_fits} shows the resulting SEDs of the posterior distribution. The SED contours are derived by using all models from the posterior, and calculating the $1\sigma$ ($16$th and $84$ percentile), $2\sigma$ ($2.5$th and $97.5$th percentile), and $3\sigma$ ($0.05$th and $99.95$th percentile) level for every wavelength point.
The single zone or two-zone fit is selected for every object based on their Bayes factor. 

For all objects, either the single or the two-zone fit does reproduce the observations reasonably well. We point out a few notable cases. For two objects, the UV fits does not overlap the observations. RU\,Lup's UV is not easily described by a power law that intersects with the stellar spectrum. Therefore, a power law that reproduces the UV flux is chosen. For DO\,Tau the reddened power law underestimates the UV flux compared to the observations. Due to DO\,Tau's strong reddening, it is not possible to find a power law within the allowed range of $f_{\rm UV}$ and $p_{\rm UV}$ that overlaps with the observation and intersects the stellar spectrum.

 The mid-IR wavelength region for every fit is shown in Fig.~\ref{fig:midir_fits}. The silicate feature is well-fitted for all objects keeping in mind that a single parameter is used to describe the dust composition. For V\,1149\,Sco, Lk\,Ca\,15, RY\,Lup, TW\,Cha, and GM\,Aur it seems, despite the overlap, the model's silicate feature peaks at a wavelength shorter than the peak wavelength of the observed feature. This hints that another dust composition could be a better representation for these objects, since this shift can be reproduced by changing the dust composition (see Fig. 4 in \citealp{min2007}).

Additionally, we find that most PAH features are reproduced well by the models. Especially the strong features of HD\,97048 and HD\,169142 are well modelled and objects without PAH features are fitted by models without features.

\subsection{Determination of the dust mass\label{sec:dust_mass}}

One of the most fundamental disk parameters is the dust mass of the disk. While full RT models that are used for SED fitting derive dust masses (even though without uncertainties), most often the dust mass is inferred from single flux measurements (e.g. \citealp{Hildebrand1983,Andrews2005}). The used analytical relation relies on many assumptions and literature values, which makes the relation often unreliable when comparing with full radiative transfer models \citep[e.g.][]{Woitke2019,Ballering2019}.The dust mass $M_{\rm dust}^{\rm ana}$ is connected to the observed flux ($F_{\nu}$) by:
\begin{align}
    M_{\mathrm{dust}}^{\mathrm{ana}}= \frac{F_{\nu}d^2}{\kappa_{\nu}^{\mathrm{ana}} B_{\nu}(T_{\mathrm{dust}}^{\mathrm{ana}})} = \frac{F_{\nu}d^2 c^2}{\kappa_{\nu}^{ana} \cdot 2 \nu^2 k T_{\mathrm{dust}}^{\mathrm{ana}}} \label{eq:massdust_approx}.
\end{align}
The relation uses a flux measurement at a frequency $\nu$, the distance ($d$), the dust opacity $\kappa_{\nu}\,\rm[cm^2/g(dust)]$, and the Planck function ($B_{\nu}$) at the average dust temperature $T_{\mathrm{dust}}^{\mathrm{ana}}$. Assuming that the dust is optically thin and emits in the Rayleight-Jeans limit at the observed wavelength, the Rayleight-Jeans law approximates the Planck function using the speed of light $c$, the Boltzmann constant $k$, and the average dust temperature $T_{\mathrm{dust}}^{\mathrm{ana}}$.
Normally, flux measurements in the millimetre wavelength regime are used to ensure that the assumptions hold true.
For comparison, often the same values are assumed in the literature for the average dust temperature and dust opacity (e.g. $20\,\mathrm{K}$ and $3.5\,\rm cm^2/g$ at $850\,\rm \mu m$ \citealp{Andrews2005}). 

\begin{figure}
    \centering
    \includegraphics[width=\linewidth]{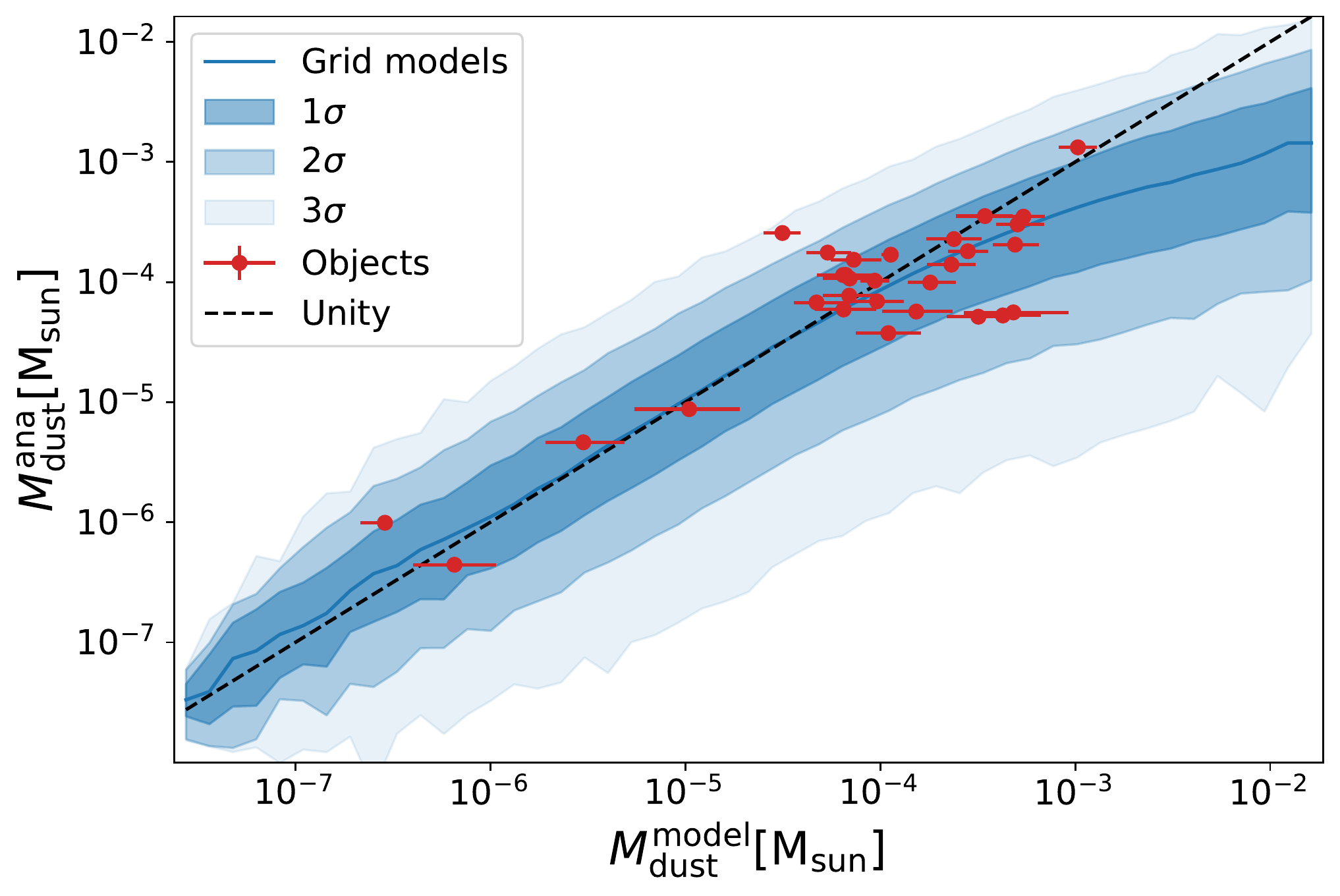}
    \caption{Calculated dust mass with Eq. \ref{eq:massdust_approx} using the flux at $850\,\mathrm{\mu m}$, $\kappa_{\rm \nu}^{\rm ana}=3.5\,\rm cm^2/g$, and $T_{\mathrm{dust}}^{\mathrm{ana}} =20\,\rm K$ in comparison with the model dust mass. The blue distribution shows all single and two-zone models that were used to train and test the NNs. The distribution is derived by dividing the value range of $M_{\rm dust}^{\rm model}$ into $50$ equal size bins in log space. The coloured lines show the median value for all models within their respective bins, the strongly, medium, and lightly shaded areas display the $1\sigma$, $2\sigma$, and $3\sigma$ percentiles of each bin. Over plotted are the object fits from Sect. \ref{sec:bayes_fact}. The model values denote the result of the posterior distribution, with the error bars representing the $1\sigma$ contour. The analytical dust masses are calculated in the same way as done for all grid models, with the fluxes at $850\,\mathrm{\mu m}$ of the respective models.}
    \label{fig:mass_estimation} 
\end{figure}

We show the analytical dust mass of Eq. \ref{eq:massdust_approx} in comparison to the model dust mass in Fig.~\ref{fig:mass_estimation}. The blue distribution shows all single and two-zone models that were produced during the grid creation. It is clearly visible that the distribution follows the unity line for about $M_{\rm dust}^{\rm model}<10^{-4}\,\rm M_\sun$. For higher masses, the analytical expression systematically underpredicts the dust masses. This suggests that the assumptions of Eq. \ref{eq:massdust_approx} break down. In particular, we suspect that for these high dust masses, disks become partly optically thick. This explains why an increase in model dust mass is not reflected any more in an increase in flux and therefore the analytical dust mass. This finding is consistent with a parameter study from \cite{Liu2022}, which finds that the analytical method can underestimate the dust mass by a factor of up to hundreds for RT models.

Over plotted in Fig.~\ref{fig:mass_estimation} are the fits for the sample (Sect. \ref{sec:bayes_fact}). For every object, the single or two-zone model is selected based on the Bayes factor (Sect. \ref{sec:bayes_fact}). The model masses and their uncertainties are directly obtained from the posteriors, with the error bars showing the $1\sigma$ level (based on its percentiles). The analytical dust masses are calculated using the interpolated model flux at $850\,\mathrm{\mu m}$, Eq. \ref{eq:massdust_approx}, $\kappa_{\rm \nu}^{\rm ana}=3.5\,\rm cm^2/g$, and $T_{\mathrm{dust}}^{\mathrm{ana}} =20\,\rm K$. We note that the errors for the analytical dust masses, propagated from the flux uncertainties, are smaller than the used markers in Fig.~\ref{fig:mass_estimation}. It can be seen that the majority of objects fall within the mass range in which the analytic expression starts to break down. Therefore, the dust mass is underestimated by the analytic formula for many high-mass objects.

In a next step, we are evaluating the potential to estimate dust masses based on a limited number of mm-flux measurements without any analytical assumptions. 
We selected $100$ random single zone models and used their mm-points as mock observations. The selected wavelength (combinations of $890\,\rm \mu m$, $1.3\,\rm mm$, and $3\,\rm mm$) are seen in the first column of Table \ref{tab:mock_obs}. The fluxes at these wavelengths are given a mock uncertainty of $1\%$, $5\%$, and $10\%$ (first row of Table~\ref{tab:mock_obs}). These mock observations are then fitted with single zone models using a full Bayesian analysis for which the stellar parameters and the inclination are fixed to their correct model values.
We used the resulting posterior distribution to estimate typical uncertainties for the disk mass. We compare the $84$th ($x_{+1\sigma}$) and $16$th percentile ($x_{-1\sigma}$) to derive a typical uncertainty factor $f_{\rm err}$:
\begin{align}
    f_{\rm err}= \sqrt{x_{+1\sigma}/x_{-1\sigma}} \label{eq:uncertainty_fact}.
\end{align}
This factor is given in Table \ref{tab:mock_obs} for the different combinations of mock observational wavelength and uncertainties. The table shows the median mass uncertainty factor for the sample of $100$ mock observations with the $84$th and $16$th percentile of the distribution given as uncertainties. Generally, the mass uncertainty decreases with more observations and with smaller flux uncertainties, which is to be expected. For a single observation at $890\,\rm \mu m$ with a flux uncertainty of $10\,\%$, the mass can be constrained by a median factor of $2.11$, while three observations at $890\,\rm \mu m$, $1.3\,\rm mm$, and $3\,\rm mm$, with uncertainties of $1\,\%$ result in a median mass uncertainty factor of $1.67$. Comparing only individual flux measurements, it comes clear that a measurement at longer wavelength can constrain the mass better than a measurement with the same uncertainty at shorter wavelength. This hints that parameters other than the disk mass are affecting the SED at longer wavelength less, which makes the fitting less degenerate and the mass determination more precise.

\begin{table}[h]
\caption{Mass uncertainty factors for mock observations.}
    \label{tab:mock_obs}     
\centering 
\begin{tabular}{l|l|l|l}
& \multicolumn{3}{c}{$\sigma_{\rm F}$\tablefootmark{(1)}} \\
$\lambda\tablefootmark{(2)} [\rm \mu m ]$ & $1\,\%$ & $5\,\%$ & $10\,\%$ \\\hline 
 & & & \\[-1.9ex]
\hline
 & & & \\[-1.9ex]
$890$ & $2.03^{0.2}_{-0.15}$ &$2.06^{0.18}_{-0.17}$ &$2.11^{0.21}_{-0.19}$\\  & & & \\[-1.9ex] \hline
 & & & \\[-1.9ex]
$1300$ & $1.9^{0.15}_{-0.12}$ &$1.9^{0.2}_{-0.09}$ &$1.96^{0.17}_{-0.13}$\\  & & & \\[-1.9ex] \hline
 & & & \\[-1.9ex]
$3000$ & $1.83^{0.09}_{-0.07}$ &$1.85^{0.08}_{-0.07}$ &$1.9^{0.1}_{-0.08}$\\  & & & \\[-1.9ex] \hline
 & & & \\[-1.9ex]
$890$, $1300$ & $1.74^{0.15}_{-0.09}$ &$1.8^{0.23}_{-0.11}$ &$1.88^{0.24}_{-0.12}$\\  & & & \\[-1.9ex] \hline
 & & & \\[-1.9ex]
$1300$, $3000$ & $1.71^{0.09}_{-0.08}$ &$1.74^{0.1}_{-0.09}$ &$1.79^{0.15}_{-0.07}$\\  & & & \\[-1.9ex] \hline
 & & & \\[-1.9ex]
$890$, $1300$, $3000$ & $1.67^{0.1}_{-0.07}$ &$1.7^{0.13}_{-0.07}$ &$1.76^{0.14}_{-0.08}$
\end{tabular}

\tablefoot{ \\
\tablefoottext{1}{$\sigma_{\rm F}$: relative uncertainty of the observation} \\
\tablefoottext{2}{$\lambda$: wavelengths of the SED measurements}
}
\end{table}

\section{Discussion\label{sec:discussion}}

\subsection{Related parameters in the sample}

\begin{figure*}[ht]
    \centering
    \includegraphics[width=0.33\linewidth]{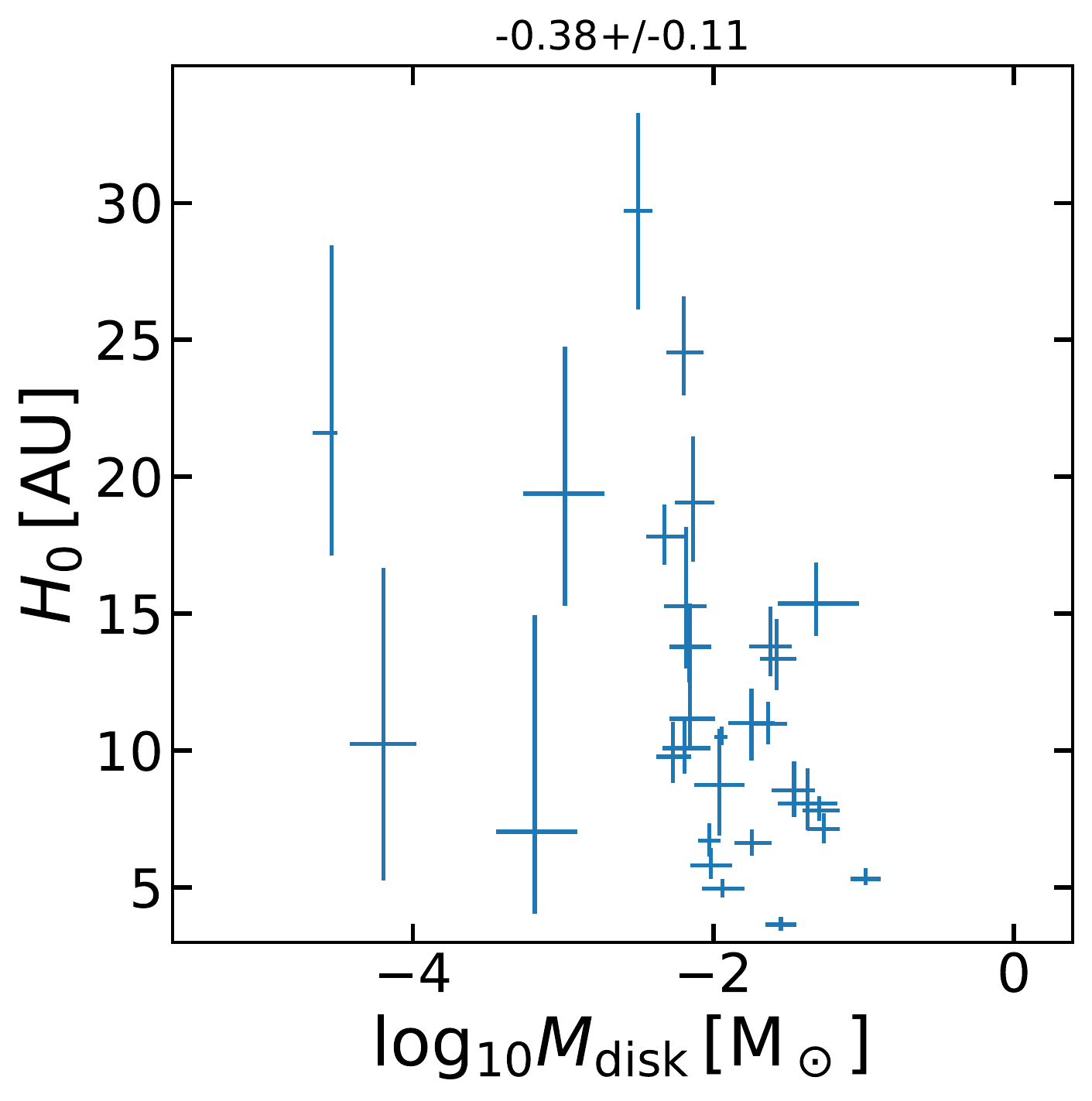}
    \includegraphics[width=0.33\linewidth]{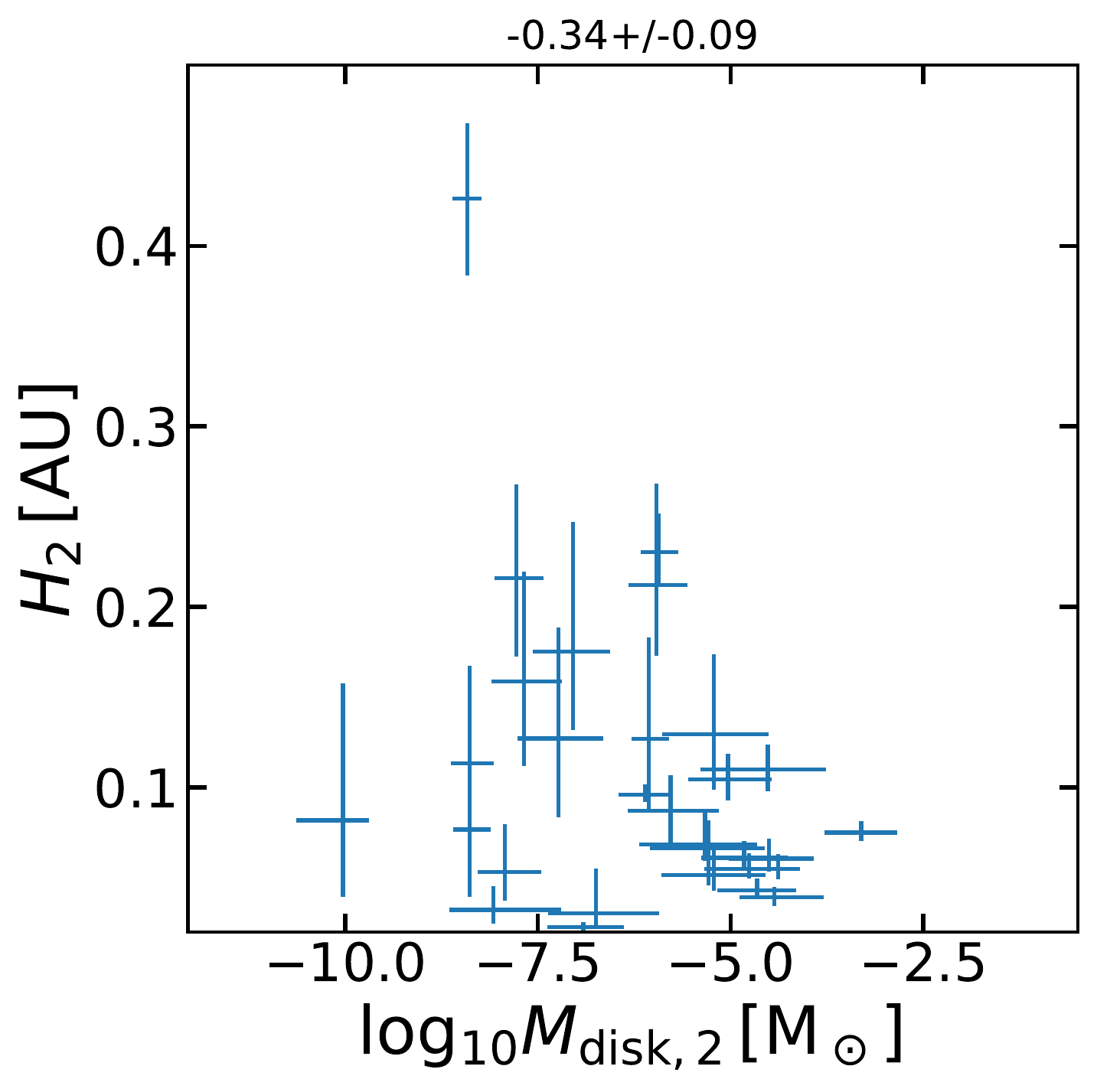}
    \includegraphics[width=0.33\linewidth]{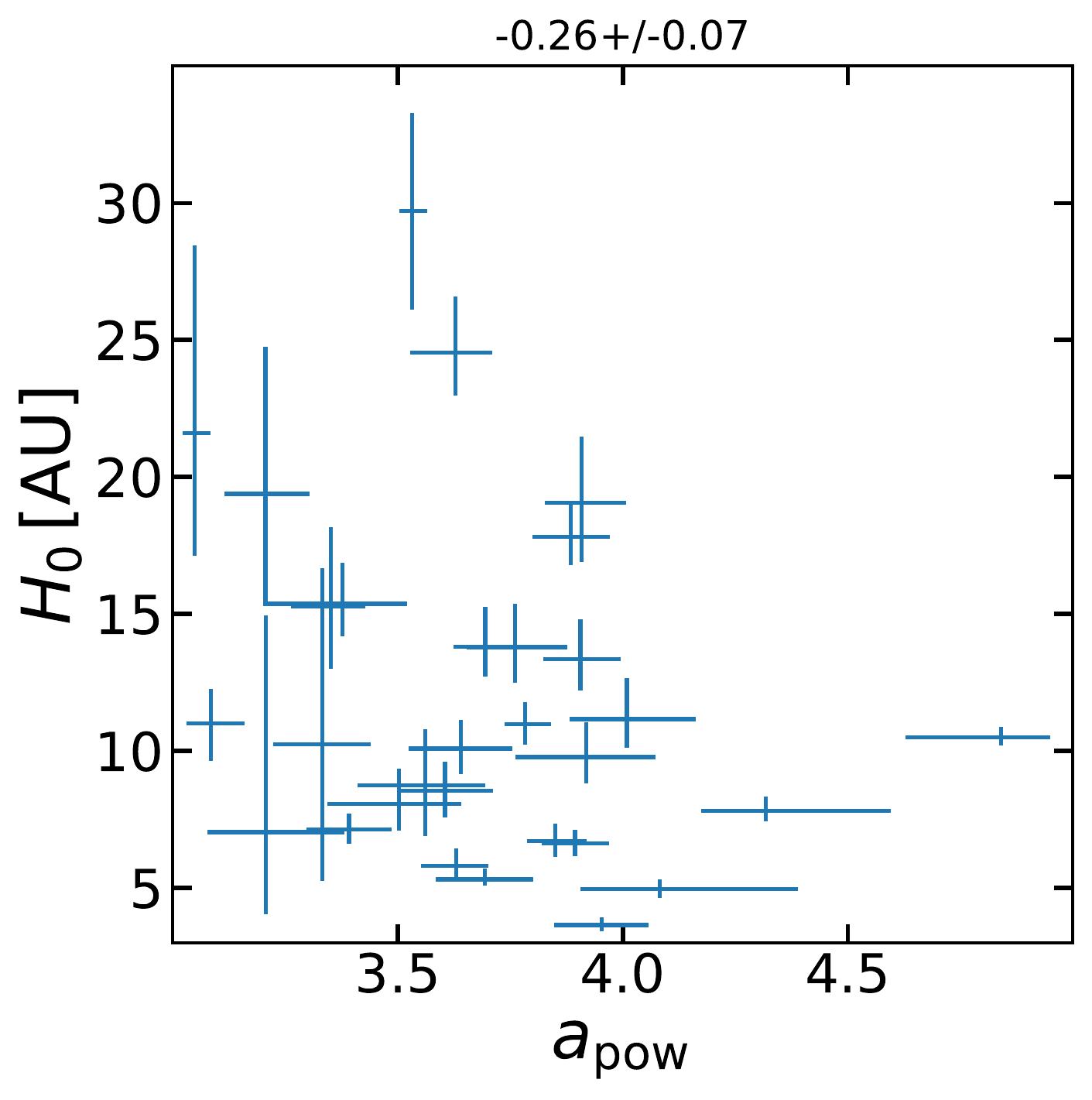}
\caption{Dependencies between selected physical two-zone disk parameters of all objects. Points with error bars denote the median values and the $84^{\rm th}$ and $16^{\rm th}$ percentiles resulting from the Bayesian analysis of two-zone fitted objects. The Pearson correlation coefficient is shown in the title of every plot.}
\label{fig:trends_two}
\end{figure*}

In this section, we examine the relations between different model parameters. Comparing all $30$ objects, we analyse if a parameter difference between objects coincides with a change of another parameter. This gives insights into what kind of disks exist. Since the two-zone models are preferred for most object, we use these fits (Table~\ref{tab:new_fits_two}) for all objects to examine parameter relations. The values of selected parameters with their $1\sigma$ uncertainties are displayed in Fig.~\ref{fig:trends_two}. The displayed parameter combinations are chosen based on the strength of their correlation and their physical importance. This correlation is quantified using the Pearson correlation coefficient, which is determined in an iterative process. The posterior of every object is projected on the to examine parameter combination. Then a random point from the posterior of every object is chosen and the correlation coefficient for this set of points is calculated. This process is repeated for a total of $10\,000$ sets of parameter values from every object's posterior. The mean coefficient values with their standard deviation are displayed on top of every panel in Fig. \ref{fig:trends_two}.

The scale height ($H_0$) and the disk mass ($M_{\rm disk}$) shows a Pearson coefficient of $-0.38\pm0.11$. The negative value corresponds to more massive disks having smaller scale heights. The explanation for this dependency is that higher mass disks are optically thicker, resulting in lower midplane temperatures, which in hydrostatic equilibrium corresponds to flatter disks. It seems impossible to have both a massive disk and a large scale height, because such disks would extend vertically very high before they become optically thin.

This relation is also true for the inner disk zone, with a Pearson coefficient of $-0.34\pm0.09$ between the disk mass $M_{\rm disk,2}$ and scale height $H_2$ of the inner zone. We therefore argue that this correlation has a real physical origin in the disk structure, as described above.

The third panel in Fig. \ref{fig:trends_two} shows the correlation between the dust size power law exponent $a_{\rm pow}$ and the scale height of the outer zone. High $a_{\rm pow}$ correspond to many small grains and only a few large once. The panel shows that no object with high values in $H_0$ and $a_{\rm pow}$ exist in our sample (correlation coefficient of $-0.26\pm0.07$). We argue similarly to the previous correlation, that small grains have higher opacities at shorter wavelength, which makes the disk more optically thick, decreasing the temperature and therefore the scale height.

The parameter combinations that are not displayed show some similar patterns. Firstly, some parameter combinations show correlations with high correlation coefficients, but they are enforced by our modelling process. For example, the inner radius ($R_{\rm in}$) of the outer zone strongly correlates with the outer radius of the inner zone ($R_{\rm out,2}$). This is a consequence of the constraint $R_{\rm in}\geq R_{\rm out,2}$. Additionally, the stellar temperature strongly correlates with the stellar luminosity, which is a consequence of stellar evolution. Secondly, some parameter combinations have error bars for both parameters that are larger than the value differences between objects. An example of such a combination is the PAH abundance ($f_{\rm PAH}$) and its charge ratio ($\rm PAH_{\rm charged}$). This prohibits any strong correlation. Lastly, some parameter combinations have error bars that are smaller than the difference between objects, but there is nevertheless no clear correlation visible. This is, for example, the case for the mass of the disk and the dust composition. It is not always clear if a correlation is significant or not, therefore we decided to only show the most convincing ones.

\subsection{Uncertainties of all disk parameters\label{sect:uncertain}}

\begin{figure}
    \centering
    \includegraphics[width=\linewidth]{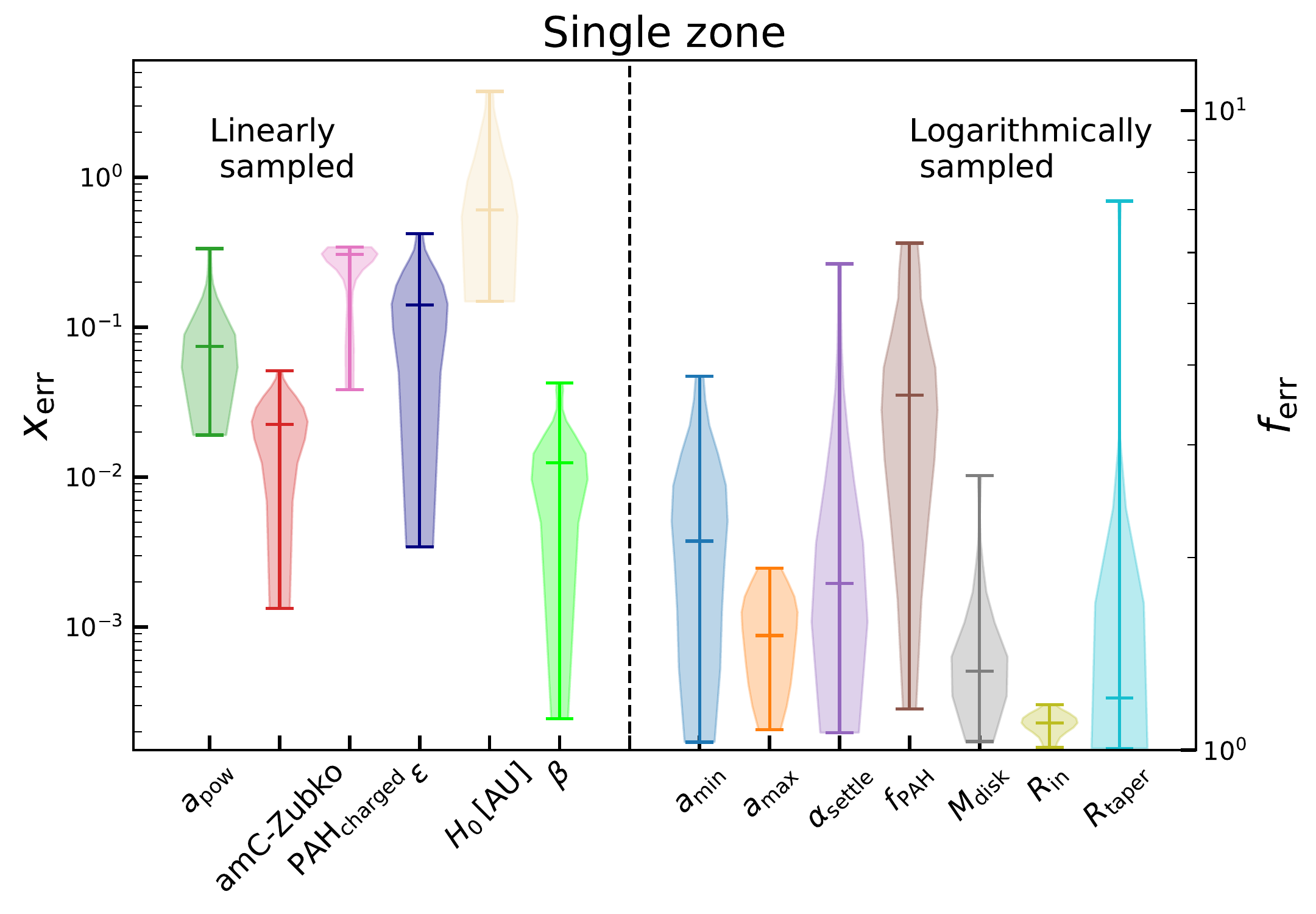}
    \caption{Violin plots showing the uncertainty distribution for all single zone parameters (Table~\ref{tab:new_fits_single}). For the left part of the plot (left of the dashed black line; linearly sampled parameters) the uncertainties are given in absolute values $x_{\rm err}$ according to Eq.~\ref{eq:uncertainty_abs} corresponding to the left axis. For the right part of the plot (logarithmically sampled parameters), the uncertainties are given as factors $f_{\rm err}$ (Eq.~\ref{eq:uncertainty_fact}) corresponding to the right axis. Every violin depicts the distribution of uncertainties for an individual parameter. The upper, middle, and lower line encode the maximal, median, and minimal uncertainty in the sample, respectively.}
    \label{fig:uncertainty_single} 
\end{figure}

\begin{figure*}[th]
    \centering
    \includegraphics[width=\linewidth]{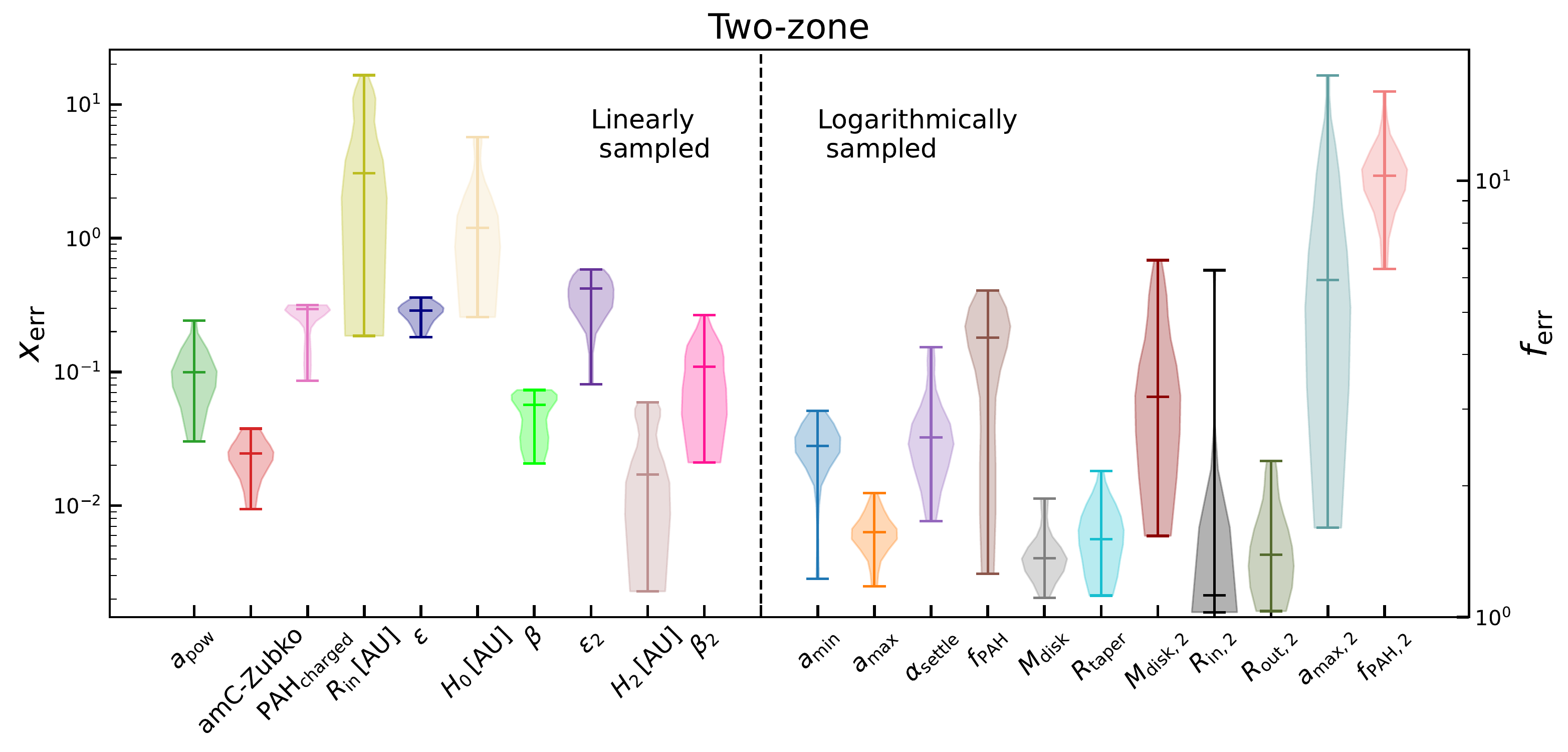}
    \caption{Violin plots showing the uncertainty distribution for all two-zone parameters (Table~\ref{tab:new_fits_two}). For the left part of the plot (left of the dashed black line; linearly sampled parameters) the uncertainties are given in absolute values $x_{\rm err}$ according to Eq.~\ref{eq:uncertainty_abs} corresponding to the left axis. For the right part of the plot (logarithmically sampled parameters), the uncertainties are given as factors $f_{\rm err}$ (Eq.~\ref{eq:uncertainty_fact}) corresponding to the right axis. Every violin depicts the distribution of uncertainties for an individual parameter. The upper, middle, and lower line encode the maximal, median, and minimal uncertainty in the sample, respectively.}
    \label{fig:uncertainty_two} 
\end{figure*}

 One main objective of this study is to calculate the uncertainties for parameters derived from SED fitting, to provide general uncertainty estimates for SED fits. We determine the uncertainties of all parameters either in absolute values or in relative factors. Absolute values are used for parameters that were sampled on a linear scale, while factors are used for logarithmically sampled parameter. In both cases, we use the $84\%$ ($x_{+1\sigma}$) and the $16\%$ ($x_{-1\sigma}$) percentile level of the posterior probability distribution to derive the uncertainty.

For linearly sampled parameters, we define the absolute uncertainty $x_{\rm err}$ as:
\begin{align}
    x_{\rm err}= \frac{x_{+1\sigma}-x_{-1\sigma}}{2}\label{eq:uncertainty_abs}.
\end{align}
For example, the flaring index of AB\,Aur derived from a two-zone fit is $1.371^{+0.016}_{-0.025}$ (see Table \ref{tab:new_fits_two}), which according to Eq. \ref{eq:uncertainty_abs} results in an absolute uncertainty $x_{\rm err}$ of $0.0205$.

For logarithmically sampled parameters, the uncertainty is given as a factor according to Eq. \ref{eq:uncertainty_fact}. The same fit of AB\,Aur derives a disk mass of $0.0113^{\times1.1}_{\div1.1}\,{\rm M}_\odot$, which results in an uncertainty factor of $\sqrt{1.1\times 1.1} = 1.1$.

Figure~\ref{fig:uncertainty_single} and \ref{fig:uncertainty_two} summarise the uncertainty distribution using violin plots for all single zone fits (Table \ref{tab:new_fits_single} and two-zone fits (Table~\ref{tab:new_fits_two}), respectively. The upper, middle, and lower lines encode the maximal, median, and minimal uncertainties of all fits (Sect. \ref{sec:bayes_fact}) for the parameters named on the x-axis. The width of the shaded area displays the distribution of all values. 

For the single zone fits (Fig.~\ref{fig:uncertainty_single}), we first examine the parameters with absolute errors. The amount of amorphous carbon (amc-Zubko) has uncertainties between $0.001$ and $0.05$ with a median value of $0.02$. The PAH charge ratio ($\rm PAH_{\rm charged}$) is among the worst constrained parameters compared to their allowed range. For many objects, no PAH features are visible, which explains an absolute uncertainty of the charge ratio of up to $0.34$ with a median value of $0.31$, which is large when considering the allowed parameter range of $1$. For the best constrained charge ratio, the uncertainty drops to $0.038$, which shows to what level it is possible to constrain this parameter if strong features are available. The scale height $H_0$ is constrained by between $0.15\,\rm AU$ and $3.7\,\rm AU$ with a median uncertainty of $0.61\,\rm AU$, which is small compared to the allowed parameter range of $32\,\rm AU$. 
The best constrained linearly sampled parameter with respect to its allowed range is the flaring index ($\beta$) which can have an uncertainty as low as $2.4\cdot 10^{-4}$ even though the range goes up to $0.5$.

From the set of logarithmically sampled parameters, the inner radius ($R_{\rm in}$) is the best constrained parameter. It has a maximum and median uncertainty factor in our sample of $1.18$ and $1.10$, respectively. 
Interestingly, the disk mass is determined by a factor of $1.33$ (median) with the distribution ranging from $1.03$ up to $2.69$. This means that even though five orders of magnitudes of disk masses were allowed for every fit, SED fitting can constrain the mass very well. This is consistent with \cite{Ribas2020}, who also find the disk mass well constrained for all objects. We note that the disk mass uncertainties from SED fitting are only slightly smaller than the once derived from fitting individual flux measurements (Sect. \ref{sec:dust_mass}). The worst constrained logarithmically sampled parameters are the viscosity parameter ($\alpha_{\rm settle}$) and the amount of PAHs ($f_{\rm PAH}$) which have maximal uncertainty factors of $5.8$ and $6.2$, respectively. The difference between these parameters is that the settling has a lower median ($1.82$ compared to $3.59$) and minimal values ($1.07$ compared to $1.16$). The minimal ($a_{\rm min}$) and maximal dust size ($a_{\rm max}$) are constrained by median factors of $2.1$ and $1.5$, respectively. The parameter constrained by the smallest factor is the best constrained fit of the taper radius ($R_{\rm taper}$), which is $1.007$, while the worst constrained fit has the largest factor ($7.22$) of all parameters.

Overall, the parameter uncertainties vary greatly between different objects. This shows that different numbers of observations, different type of objects, or specific disk properties influence how well disk parameters can be determined. However, some parameters are better constrained than others.

Giving more degrees of freedom by adding multiple parameters for a two-zone model (Fig.~\ref{fig:uncertainty_two}) increases the uncertainties for parameters that are also used for single zone fits by a small factor. For example, the disk mass of the outer zone is constrained within a factor between $1.11$ and $1.86$ (median of $1.37$), which is a slightly larger median value than the respective value for the single zone fits (Fig.~\ref{fig:uncertainty_single}). 
For the set of parameters describing the inner disk zone, the mass ($M_{\rm disk,2}$), maximum dust size ($a_{\rm max,2}$), and the PAH abundance ($f_{\rm PAH,2}$) are most uncertain with median (maximal) factors of $3.20$ ($6.58$), $5.93$ ($17.45$), and $10.28$ ($16.03$), respectively. While the mass and maximum dust size are constrained by small factors for individual objects, the PAH abundance of the inner zone is never better constrained than a factor of $6.28$.

We note that a fraction of the parameter uncertainties originate from the assumed uncertainty for the NNs' predictions. Therefore, a Bayesian analysis using the radiative model for every iteration will decrease the parameter uncertainties. The used prior will also affect the posterior. We rerun the Bayesian analysis with flat priors to estimate the influence of them. The uncertainties increase only slightly, showing that the given measurements predominantly determine the posterior. 

The uncertainties derived in this study are roughly consistent with uncertainties estimated by \cite{Woitke2019} for three objects. For example, their estimated uncertainties for the dust parameter ($a_{\rm min}$, $a_{\rm max}$, $a_{\rm pow}$, $\alpha_{\rm settle}$, and amC-Zubko) are well within the distributions displayed in Fig.~\ref{fig:uncertainty_two}.

\subsection{Degeneracies}

\begin{figure*}[th]
    \centering
    \includegraphics[width=14cm]{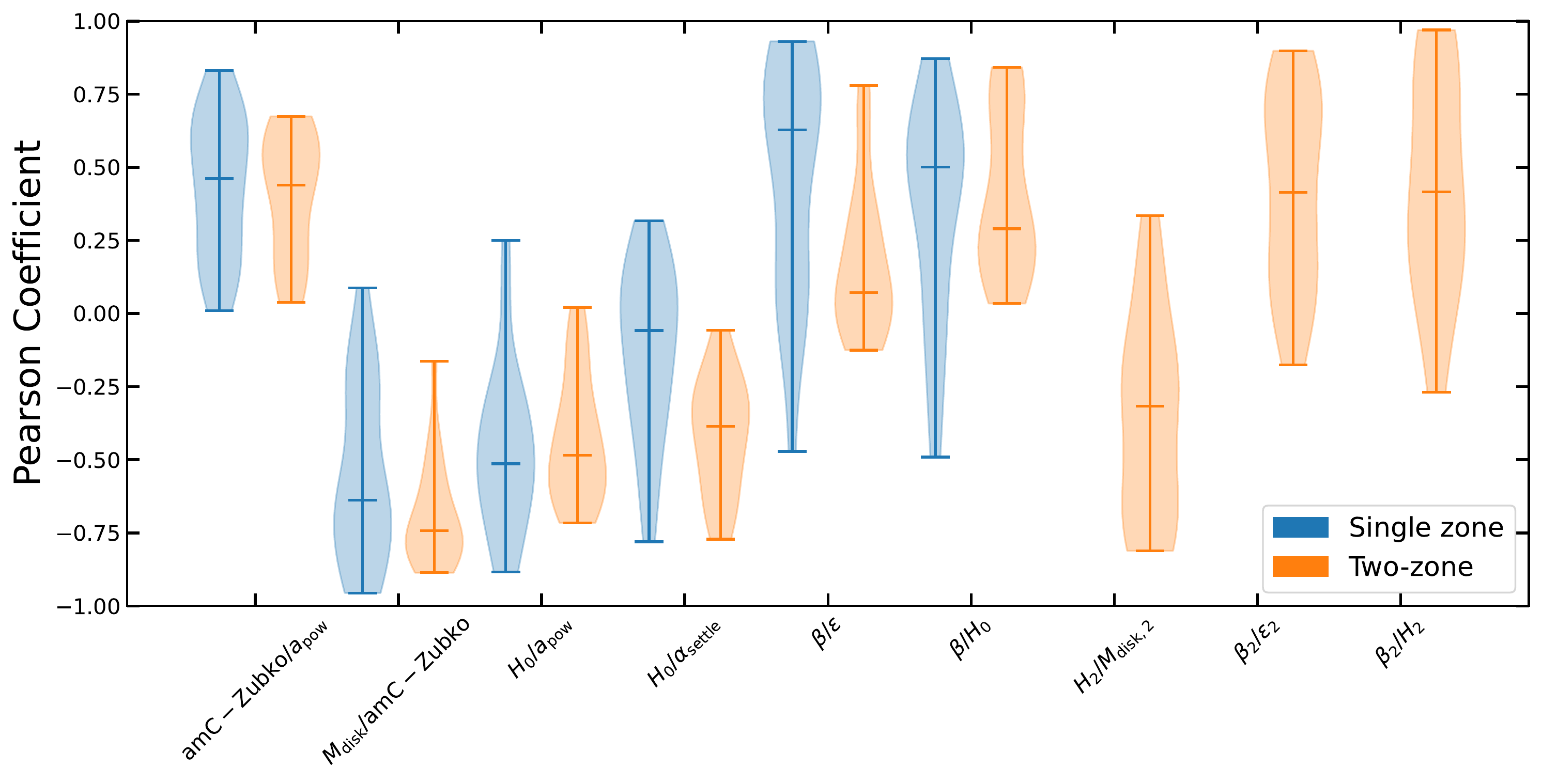}\\[-2mm]   
    \caption{Pearson correlation coefficients for selected combinations of model parameters. Values close to $+1$ and $-1$ denote strong positive and negative correlations, respectively, whereas 0 means no correlation.  Every violin depicts the distribution of correlation coefficients for all single (blue) and two-zone (orange) models (Sect.~\ref{sec:bayes_fact}). The upper, middle, and lower bars show the maximal, median, and minimal value, respectively.}
    \label{fig:degeneracies} 
\end{figure*}

In this section, we examine the degeneracies in SED-fitting that we have identified based on our full Bayesian analysis (Sect.~\ref{sec:bayes_fact}). While SED-fitting is well-known to be degenerate, quantifying this effect is difficult due to the high computational cost.

We quantify a degeneracy between two parameters by the Pearson correlation coefficient of the joint posterior distributions using all single and two-zone models introduced in Sect.~\ref{sec:bayes_fact}. In total, there are $78$ single zone and $210$ two-zone parameter relations to explore that represent all combinations of the $13$ single zone and $21$ two-zone parameters that were fitted. We display parameter combinations that show large positive or negative Pearson coefficients (mean absolute value larger than $0.35$) in Fig.~\ref{fig:degeneracies}.
The upper, middle, and lower bars encode the maximum, median, and minimum values for a certain parameter combination for all single (blue) and two-zone (orange) fits. The width of the shaded area corresponds to the value distribution. If a simultaneous increase in both parameters results in similar SEDs, the Pearson coefficient is positive, whereas, if an increase of one parameter can be counteracted by a decrease of another, the Pearson coefficient is negative. If no correlation is found, the Pearson coefficient is close to $0$, and the two parameters are considered not to be degenerate.

The largest degeneracy is found between the disk mass, $M_{\rm disk}$, and the volume fraction of amorphous carbon in the dust grain material, amC-Zubko. This degeneracy shows that a higher disk mass in combination with less amorphous carbon can result in similar SEDs. Clearly, an increase of either of these two parameters results in more millimetre fluxes. However, Fig.~\ref{fig:degeneracies} also shows that there are many objects where this Pearson coefficient is quite small. We explain this behaviour by the inclusion of mid-IR data, which measures the strength of the main silicate features at $10\,\rm\mu m$ and $20\,\rm \mu m$. For objects which exhibit strong silicate emission features, the use of too much amorphous carbon results in too flat mid-IR dust opacities under the assumption that the dust composition does not vary as a function of altitude nor radius. This is a good example that some well-known degeneracies get smaller when additional data is taken into account.

For strong silicate emission features, we need the dust absorption opacity to be dominated by small ($\mu$m-sized) particles, for example, the power law exponent of the dust size distribution ($a_{\rm pow}$) must be large.  Otherwise, if $a_{\rm pow}$ is small, the grey opacities of the large dust particles flatten out the dust opacity at mid-IR wavelengths. This explains why $a_{\rm pow}$ is found to be positively degenerate with the amorphous carbon fraction amC-Zubko, which has a similar effect on the dust opacity in the mid-IR when increased.

The most degenerate of our disk parameters is the scale height of either disk zone ($H_0$ and $H_2$), which generally has a big impact on the SED. The scale height is found to be degenerate with $a_{\rm pow}$, with the dust settling parameter $\alpha_{\rm settle}$, with the disk flaring index $\beta$, and with the mass of the inner disk zone $M_{\rm disk,2}$ in case of a two-zone models. All these degeneracies have similar physical explanations.  Larger scale heights, less dust settling, more disk mass and a steeper dust size power law all result in more grains high up in the disk, which affects the SED in similar ways.

Stronger disk flaring $\beta$ reduces the scale heights inside the reference radius ($100\,\rm AU$ for $H_0$, and $1\,\rm AU$ for $H_2$) and increases the scale heights outside of that radius. The positive Pearson coefficient between $\beta$ and the scale height itself suggests that the inner parts of the disk are more important for the SED than the outer parts.
The flaring indices $\beta$ of both zones are also degenerate with the respective surface density exponent $\epsilon$. Increasing either of these parameters changes the slope of the SED. An increase of $\epsilon$ and a decrease of $\beta$ both increase the fluxes at short wavelengths ($\lambda\!\la\!10\,mu$m) and decrease the fluxes at long wavelengths ($\lambda\!\ga\!100\,\mu$m).
However, the spread of the Pearson coefficients for these correlations is already quite large, indicating that the sign of the correlation can actually change with circumstances, depending on, for example, whether or not the outer disk zone is situated in the shadow cast by the inner disk.

Generally, parameters that exist for single and two-zone models show similar degeneracies. The degeneracies seen for the single zone models are mirrored by similar degeneracies for inner zone parameters in two-zone models. For example, is the degeneracy of $\beta$ with $\epsilon$ and $H_0$ similar to the degeneracy of $\beta_2$ with $\epsilon_2$ and $H_2$.

All found degeneracies are between parameters that have a large impact on the SED. Parameters that do not affect the SED strongly are poorly constrained, which makes them less degenerate.

\cite{Alonso-Albi2009} conclude that the slope of the SED at mm-wavelengths is for optically thin disks degenerate with the maximum dust size and the exponent of the dust size distribution. We do not see this degeneracy back in our SED fitting, which means that flux measurements at different wavelengths can break it.

\subsection{Limitations of this method}

\begin{figure}
    \centering
    \includegraphics[width=\linewidth]{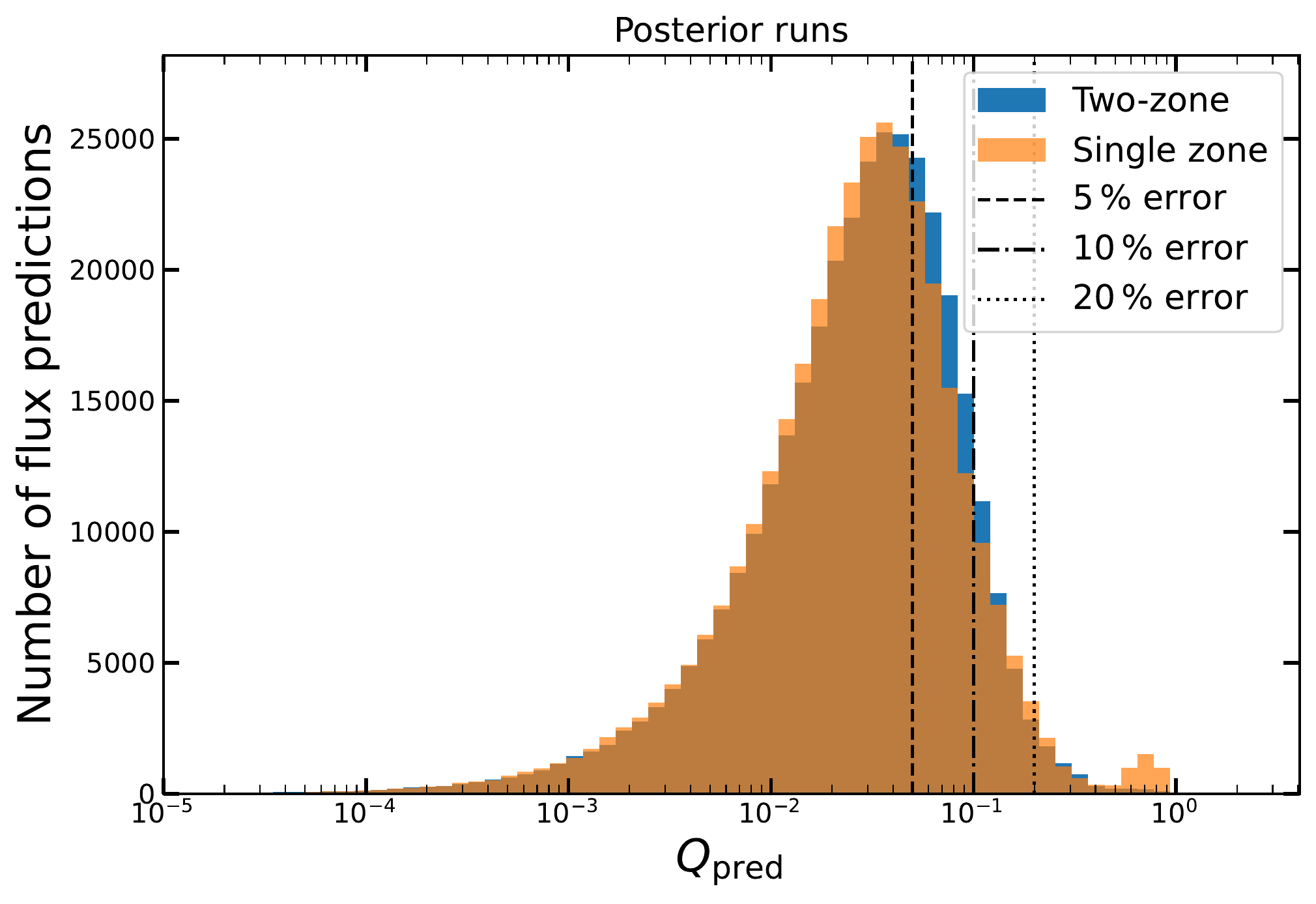}
    \caption{Quality distribution according to Eq.~(\ref{eq:quality_indi}) of the flux predictions at wavelength longer than $0.5\,\rm \mu m$ for $100$ random models from the single (orange) and two-zone (blue) posterior of every object. The histogram shows how many models fall in each quality bin. The dashed, dashed-dotted, and dotted black lines show the $5\,\%$, $10\,\%$, and $20\,\%$ difference, respectively.}
    \label{fig:uncertainty_mcfostreruns} 
\end{figure}

In this section, we explore limitations and possible improvements of our method. After deriving the single zone (Table~\ref{tab:new_fits_single}) and two-zone (Table~\ref{tab:new_fits_two}) posteriors for every object, we selected $100$ random models from every posterior to determine the quality of the NNs' predictions at these points in parameter space. To calculate the prediction quality, the true SEDs must be known. Therefore, we run MCFOST of every selected model.
The measure of quality is the relative flux difference between true and predicted SEDs as introduced in Eq.~\ref{eq:quality_indi}. Figure~\ref{fig:uncertainty_mcfostreruns} show the qualities for the selected models for the single zone and two-zone posterior of every object. Since wavelength points shortwards of $0.5\,\mu$m are not used to calculate the likelihood function, we excluded these points from the analysis.

$68\%$ ($1\sigma$) of these single zone and two-zone SED fluxes were predicted by our NNs with qualities better than $4.5\%$ and $4.8\%$, respectively. Even though this is slightly worse than the achieved qualities of the test sets, $95\%$ of single zone (two-zone) predicted SEDs have qualities better than $14\%$ ($13\,\%$). These errors, introduced by the necessity to use NNs for fast SED prediction, are generally small compared to the sum of measurement uncertainties and systematic problems inherent in SED fitting, such as source variability and systematic errors using multi-instrument data.
 
The worst predictions are often caused by models with parameter values close to the most extreme allowed values. This can be especially seen for the single zone predictions that do not describe the observations well and drift to extreme parameter values. We expect these predictions to be slightly worst, since the NNs have fewer training models in these areas of parameter space. This can be seen best in the case of 49\,Cet, which is best fitted with inner radii as large as possible. The model from the two-zone posterior have an inner radius of the outer zone that is close to the maximum value of $70\,\rm AU$ allowed by the prior. For the single zone model the inner radius is sampled differently, which results for the given stellar parameters a tighter constraint of about $40\,\rm AU$. This is why the two-zone model is strongly preferred based on the Bayes factor. This extreme configuration has two additional consequences. First, the two-zone model requires an inner zone, but since the object does not require this, it is close to the minimum possible mass. Second, the single zone posterior causes the strongest disagreements when calculating MCFOST models for the models of the posterior. This is because the area of parameter space that would be preferred to fit the model (large inner radii) is forbidden. This results in the Nested sampling algorithm searching and finding the part at which the NN's predictions deviate the most from the RT runs and can explain the observations. This is reflected in the extreme choice also for many other fitted parameters of 49\,Cet.

\section{Summary and conclusion\label{sec:summary}}

We used a Bayesian framework to fit the SEDs of $30$ protoplanetary disks based on the observational data collected by the FP7 DIANA project\footnote{https://diana.iwf.oeaw.ac.at}, based on single and two-zone full RT MCFOST models, emulated by NNs. The emulation decreases the SED computation time by a factor larger than $10^5$. This webpage\footnote{https://tillkaeufer.github.io/sedpredictor} demonstrates the SED emulation process and (i) shows graphically how individual model parameters affect the SED, and (ii) can be used to compare models easily to observations getting first estimates for well-fitting disk parameters. The main conclusions of this paper are:
\begin{itemize}
    \item Our trained NNs can emulate the SEDs of single and two-zone disk models with flux errors of typically less than $5\%$. This acceleration enables us to fit a large set of objects using Bayesian analysis with or without using additional constraints on the disk properties.
    \item We find significant differences between the parameter posterior distributions obtained in this study and the previously determined values based on a genetic $\chi^2$ optimisation \citep{Woitke2019}. For single and two-zone fits, the parameter values from \cite{Woitke2019} are outside the $3\sigma$ level of the posteriors derived in this study in about $25\%$ and $22\%$ of the cases, respectively.
    \item We find significant statistical evidence that most objects are better described by a discontinuous (two-zone) disk structure compared to the continuous (single zone) models.
    \item The disk dust masses derived from single mm-flux measurements systematically underpredict the true dust masses in the models for high-mass disks (dust mass $\geq\!10^{-4}\,\rm M_\odot$). Most objects fall into this mass range. 
    \item By an additional emulator-based Bayesian analysis, we have derived the uncertainties in disk mass determination from single mm-flux measurements. We show that this uncertainty decreases with the number and wavelength of the photometric fluxes used.
    \item Our analysis shows a few clear, but not so well-known, correlations between disk parameters. In particular, high-mass disks have lower scale heights than their low-mass counterparts. This is in line with vertical hydrostatic equilibrium based on cooler midplane temperatures in the high-mass disks.
    \item We provide typical uncertainties for disk parameters derived with our method. We show that certain parameters are relatively well constrained (the dust mass of a single zone model is typically constrained by a factor of $1.3$), while other parameters are poorly constrained (the minimal dust size has an uncertainty factor of $2.1$).
    \item SED fitting is known to be highly degenerate. Using our statistical framework, we are able, for the first time, to quantify these degeneracies. The disk mass and dust composition show the clearest degeneracy. Additionally, the scale height is degenerate with many parameters, like the dust size power law exponent and the flaring index.
\end{itemize}
This study has shown that a combination of complex RT disk models with novel machine learning techniques provide a powerful tool to analyse SED data including all degeneracies. We plan to use this method on more complex thermo-chemical disk models in the future. These models take about a factor of $100$ longer than the dust RT models discussed in this paper, making the need for a faster generation of molecular emission spectra even greater.
 
\begin{acknowledgements}
      We acknowledge funding from the European Union H2020-MSCA-ITN-2019 under grant agreement no. 860470 (CHAMELEON).
\end{acknowledgements}
\bibliographystyle{aa} 
\bibliography{lib.bib} 

\appendix

\section{Hypergrids\label{app:train}}
The training of a NN is an iterative process to minimise the loss function. This corresponds in general to predictions of the NN that are more similar to the true output, in our case the true SED of a model. The minimisation is done by adjusting the weights of the NN until the loss function converges. However, there are many NN architectures and other settings, so-called hyperparameters, that change the NN or the training process. Finding good settings is a process of trail and error that involves running multiple NNs with different settings and comparing the results.

\begin{figure}[t]
    \centering
    \includegraphics[width=\linewidth]{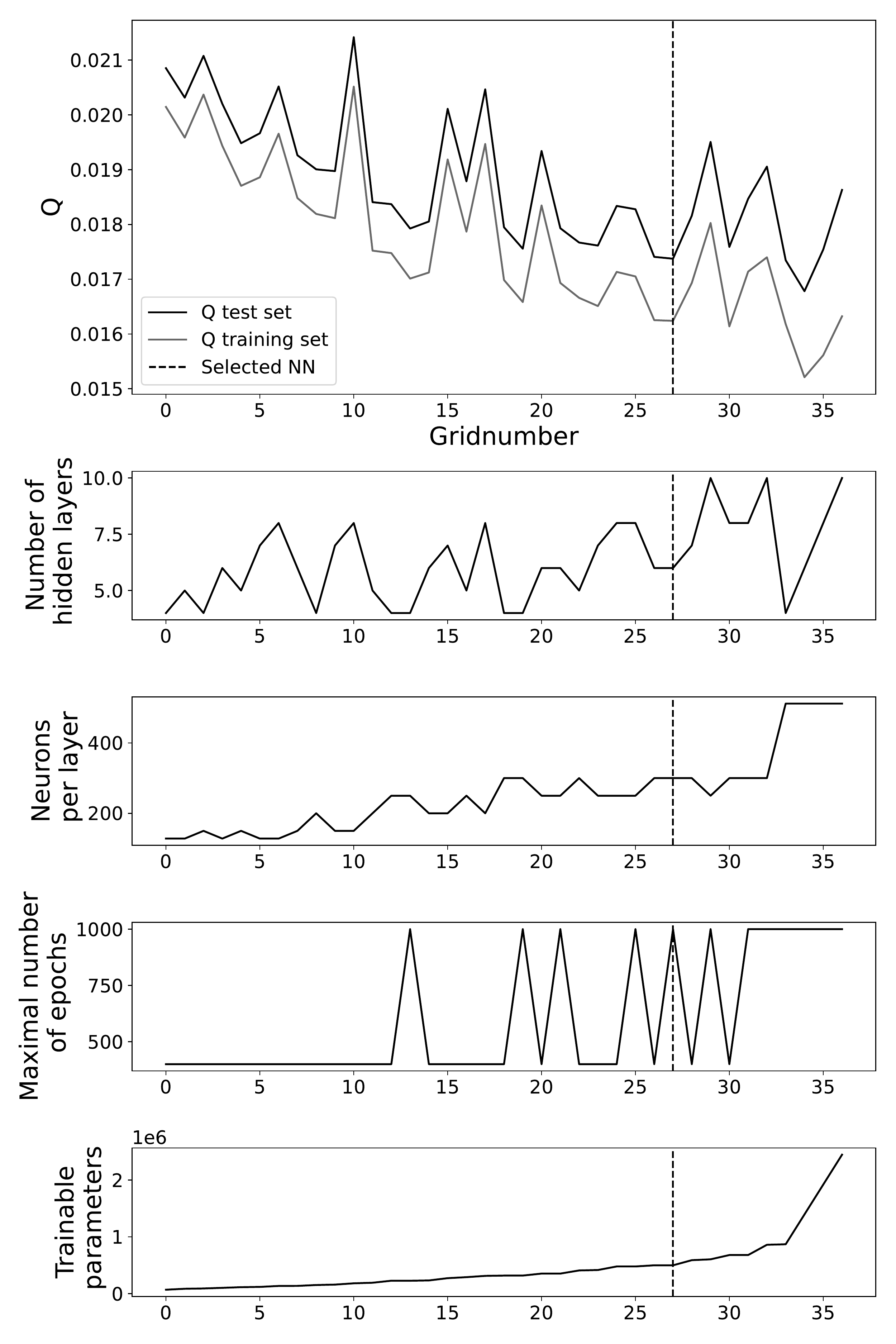}
    
    \caption{Hypergrid to determine the best architecture of single zone NNs.
    The upper panel displays the quality $Q$ of the test and training set for $37$ different NNs as explained in Eq. \ref{eq:quality}. The number of layers, neurons per layer, and the maximum number of epochs trained for every NN are shown in the lower panels. The lowest panel displays the number of trainable parameters in the NNs. The dashed line denotes the selected NN.}
    \label{fig:hyper} 
\end{figure}

\begin{figure}[ht]
    \centering
    \includegraphics[width=\linewidth]{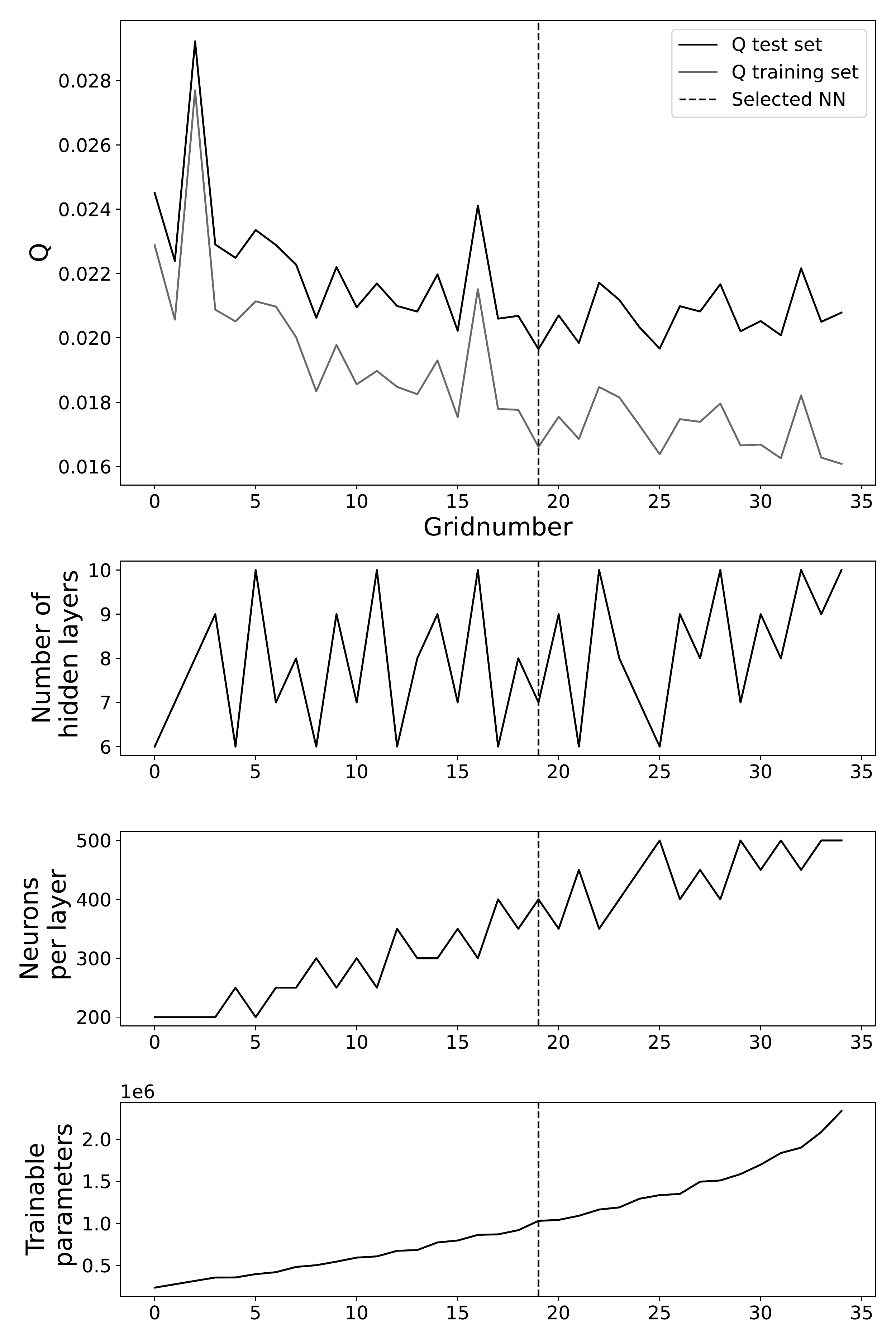}
    
    \caption{Hypergrid to determine the best architecture of two-zone NNs.
    The upper panel displays the quality $Q$ of the test and training set for $35$ different NNs as explained in Eq. \ref{eq:quality}. The number of hidden layers and the number of neurons per layer for every NN are shown in the lower panels. The lowest panel displays the number of trainable parameters in the NNs. The dashed line denotes the selected NN.}
    \label{fig:hyper_two} 
\end{figure}

We used hypergrids to determine good values for the most important hyperparameters. Therefore, we trained multiple NNs with different settings and evaluated the quality of the predictions. The measure of quality ($Q$) is the mean difference between the logarithmic predicted ($\nu F_{predict}$) and true SED fluxes ($\nu F_{true}$) averaged over all wavelength points $N_{\lambda}$ and all models in the data set $N_{model}$:

\begin{align}
    Q=\frac{1}{N_{model}} \sum_{j=1}^{N_{model}}{\sqrt{\sum_{i=1}^{N_{\lambda}}{\frac{{\left[\log_{10}{\left(\nu_{i} F_{predict,i,j}\right)}-\log_{10}{\left(\nu_{i} F_{true,i,j}\right)}\right]}^2}{N_{\lambda}}}}}\label{eq:quality}.
\end{align}

We examined $4$ hyperparameters by running over $74$ and $35$ networks for single and two-zone predictions, respectively. While two of these hyperparameters (number of hidden layers and neurons per layer) determine the architecture of the NN, two others specify scaling (standardising or normalising) of the input and output data. Evaluating the influence of it is possible because the defined quality $Q$ does not depend on the scaling of the SED, in contrast to the loss function.

The training was done on $70\%$ of the data, with the rest used to evaluate the predictions. This is the same procedure used for the final NN. To increase comparability, the split in training and test set is the same for all NNs in all hypergrids.

To evaluate the effect of the scalers we trained a set of NNs with a learning rate of $0.1$ for $400$ epochs with different number of hidden layers (5,6), neurons per layer (100,128,150,200), and scaling (normalising and standardising) the input and output data resulting in $32$ different NNs. Table \ref{tab:hyper_inout} shows the achieved average qualities for the test and training set for combinations of input and output scalers. We conclude that runs using standardised output achieve better qualities, while the different between normalising and standardising the input is small. Nevertheless, standardising the input gives slightly better predictions. Therefore, we choose to standardise input and output for all NNs, not just single zone NNs, but also for the two-zone NNs. 

\begin{table}[t]
\caption{Hypergrid for input and output scaler.}
    \label{tab:hyper_inout}     
\centering 

\begin{tabular}{r|c|c}
 & \multicolumn{2}{c}{Output scaler}   \\ 
Input scaler& Std\tablefootmark{(1)} & Norm\tablefootmark{(2)} \\ \hline
 Std\tablefootmark{(1)} & \begin{tabular}{@{}c@{}}$Q_{train}\tablefootmark{(3)}= \mathbf{0.02111}$ \\$Q_{test}\tablefootmark{(3)}=\mathbf{0.02185}$ \end{tabular}
& \begin{tabular}{@{}c@{}}$Q_{train}\tablefootmark{(3)}= 0.03691$ \\$Q_{test}\tablefootmark{(3)}=0.03762$ \end{tabular} \\ \hline
   Norm\tablefootmark{(2)} & \begin{tabular}{@{}c@{}}$Q_{train}\tablefootmark{(3)}= 0.02126$ \\$Q_{test}\tablefootmark{(3)}=0.02200$ \end{tabular} & 
   \begin{tabular}{@{}c@{}}$Q_{train}\tablefootmark{(3)}= 0.03701$ \\$Q_{test}\tablefootmark{(3)}=0.03774$ \end{tabular}\\
   \end{tabular}

\tablefoot{ \\
\tablefoottext{1}{Std: standardising of the data} \\
\tablefoottext{2}{Norm: normalising the data}\\
\tablefoottext{3}{$Q_{train,test}$: quality according to Eq. \ref{eq:quality} of the training and test set, respectively.}

}
\end{table}

We trained more single zone NNs for $400$ epochs to determine good architectures. Larger networks were additionally trained with $1000$ epochs, to ensure that they converge. To increase comparability, every training that did not improve for $10$ epochs was stopped. As seen in Fig.~\ref{fig:hyper}, we ordered the NNs by the number of trainable parameters in it. This reveals that generally larger NNs perform better on our data. However, it also becomes clear that for larger NNs, the prediction quality for the test set becomes significantly worse than the same quality for the training set, which is a sign of overfitting.

Balanced between the quality of the predictions, overfitting, and the time needed for training, we decided on using $6$ hidden layers with $300$ neurons each. This gives a quality $Q$ of $0.01618$ and $0.017347$ for the training and test set, which is close to the best quality (test score of $0.01678$) in the hypergrid. Additionally, this network takes $\sim 1\,\mathrm{min}$ to train per epoch on our machines and has with $499,340$ trainable parameters a reasonable size with respect to the size of the training set ($470,379$ training SEDs).

Fig \ref{fig:hyper_two} displays the quality of the predicted SEDs for two-zone NNs. We trained every network for $500$ epochs. Again, larger NNs perform better, but while the training score improves steadily with size, the test score does not improve further for networks larger than $\sim 1,000,000$ trainable parameters which is roughly the number of SEDs used for the training ($1,195,285$ SEDs). Therefore, we choose a NN with $7$ hidden layers with $400$ neurons per layer ($1,029,340$ trainable parameters) as the best setup for two-zone predictions. This network is with a training score of $0.01662$ close to the optimal scores of $0.016086$, while the test score of $0.019649$ is the best test score found in the hypergrid.

\section{Detailed grid creation\label{App:grid}}

After explaining the general idea of the grid creation (Sect. \ref{sec:grid_creation}), this section focusses on individual parameters and their sampling, highlighting the physical knowledge that influences parameters. 
The four stellar parameters ($M_\star$, age, $L_\star$, and $T_{\rm eff}$) cannot be varied independently, since only certain combinations are physically possible. Therefore, we took two approaches, either sampling $M_\star$ and age or $L_\star$ and $T_{\rm eff}$ and deriving the other two in a way that makes the stars consistent with pre-main-sequence tracks from \cite{Siess2000}.

\begin{figure}
    \centering
    \includegraphics[width=\linewidth]{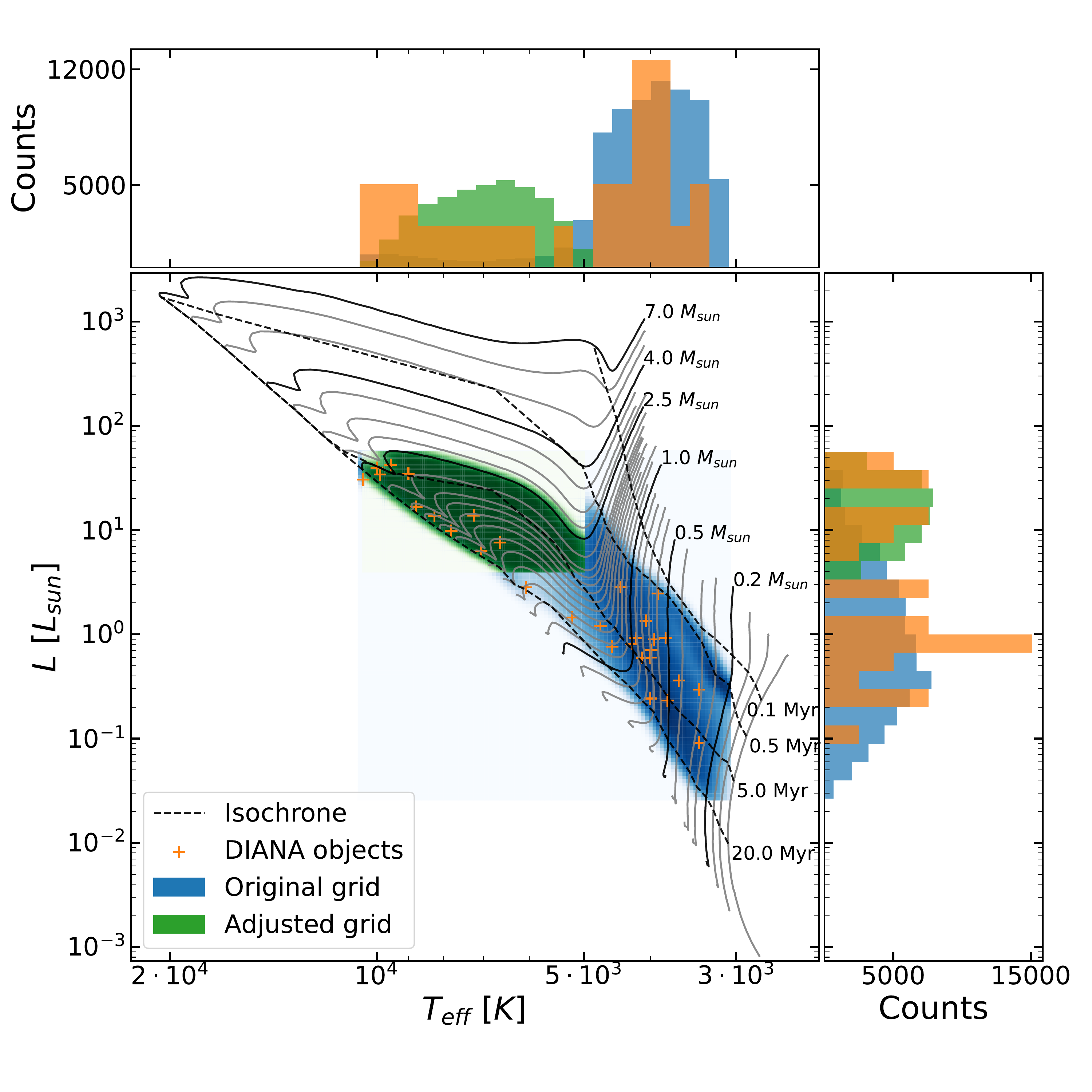}
    \caption{Hertzsprung-Russell diagram of the DIANA sample (orange markers) and all stars in the original single zone grid (blue) and the adjusted grid (green). The grey lines are the evolutionary tracks from \citep{Siess2000} for different masses, and the coloured dashed lines are isochrones. The histograms show the distribution of values for luminosity and temperature for the sample and the grid in the same colours as used in the main plot. The sample's histogram is scaled to make a comparison easier.}
    \label{fig:hrd_sampling} 
\end{figure}

For single zone models, we sampled the stellar mass and age (limits in Tab \ref{tab:parameters}). A small NN predicted the temperature and luminosity based on these parameters. This NN was trained on a subset of pre-main-sequence track points from \cite{Siess2000} and consists of $4$ hidden layers with $32$ neurons each. For a separated test set, the temperature predictions show a mean deviation of $0.18\,\%$ with a maximal uncertainty of $2.11\,\%$. The luminosity predictions have a mean error of $1.94\,\%$.
The resulting stellar distribution is shown in blue in the Hertzsprung–Russell diagram (HRD; Fig.~\ref{fig:hrd_sampling}). When comparing this distribution to the DIANA sample, it comes clear that the grid has much fewer bright and hot stars.

\begin{figure}
    \centering
    \includegraphics[width=\linewidth]{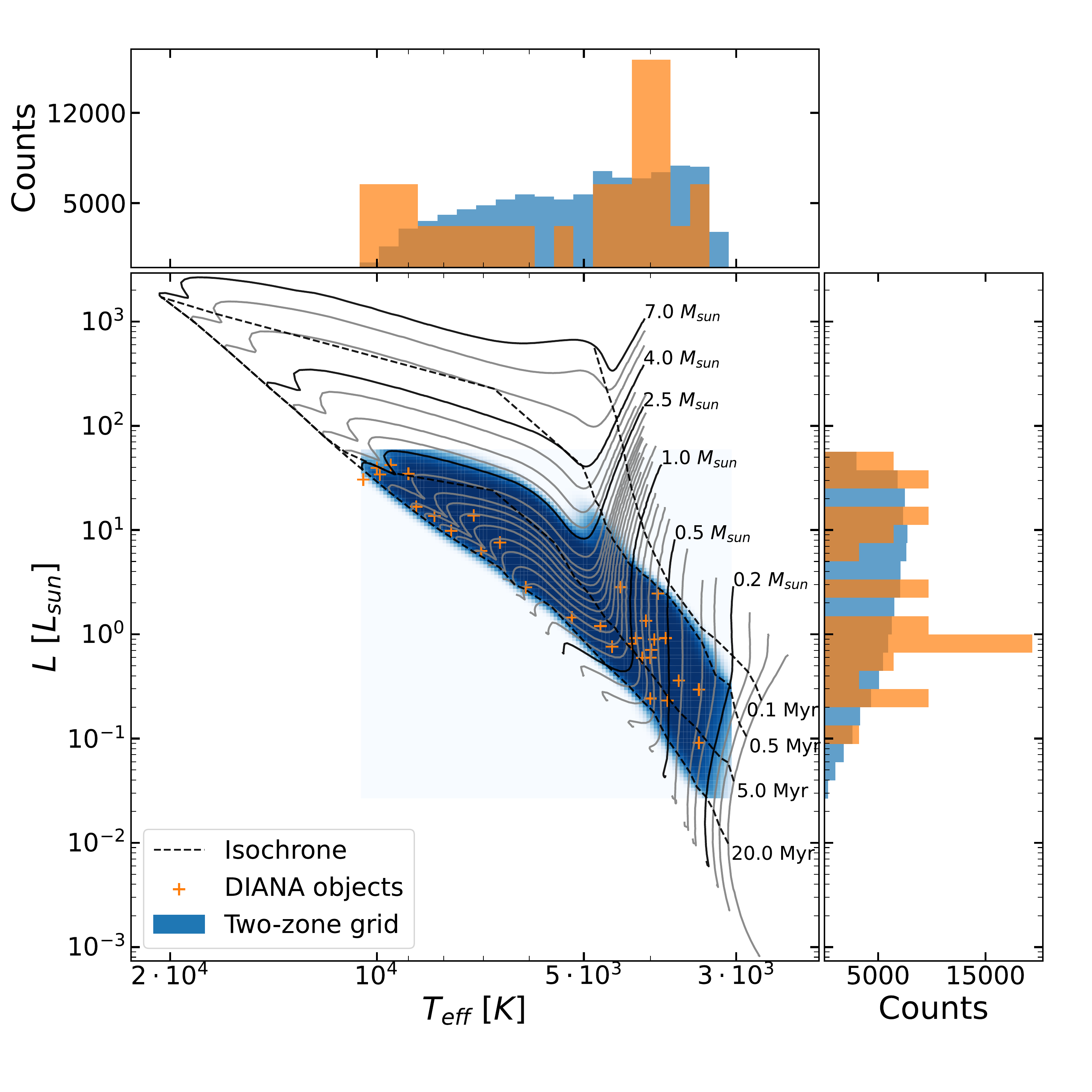}
    \caption{HRD for two-zone models. Explanation is the same as for \ref{fig:hrd_sampling}.}
    \label{fig:hrd_sampling_two} 
\end{figure}

The first tests with different NNs showed that this leads to worse prediction for stars that have such luminosities and temperatures. Therefore, another set of models was sampled uniformly in the log space of $T_{\rm eff}$ and $L_\star$ for masses less than $2.5\,\mathrm{M_\odot}$, $L_\star > 4\,\mathrm{L_\odot}$, $T_{\rm eff}>5000\,\rm K$, and ages between $0.5\,\mathrm{Myr}$ and $20\,\mathrm{Myr}$ to better populate this gap (green distribution in Fig.~\ref{fig:hrd_sampling}). 

For this adjusted grid, $M_\star$ is predicted by another NN trained on pre-main-sequence tracks using $L_\star$ and $T_{\rm eff}$. It consists of $6$ layers with $32$ neurons each. For a test set, it has a mean mass error of $0.39\,\%$ and a maximal error of $4.19\,\%$.

The combination of both grids results in a distribution similar to the stellar parameters of the DIANA target stars.
In total, $452,444$ SEDs are created using the mass and age sampling for the stars and $220,934$ to fix the lack of massive stars in the grid.

For the two-zone grid, $L_\star$ and $T_{\rm eff}$ are sampled with mass and age limits taken from the initial single zone grid, resulting in the stellar distribution shown in Fig.~\ref{fig:hrd_sampling_two}. The mass is again predicted by the same NN as mentioned above.

The mass of the disk is sampled indirectly by the ratio of it to the mass of the star. This is done to avoid massive disks around low-mass stars.

The innermost radius ($R_{\rm in}$ for single zone models and $R_{\rm in,2}$ for two-zone models) is determined by the temperature at this radius, with the idea to avoid disks that are too close to their star for dust to be stable. Therefore, we approximate the temperature ($T(R_0)$) at the inner edge of the disk for all objects in the DIANA sample using a simple analytic formula. The temperature is derived using the stellar temperature ($T_{\rm eff}$), stellar radius ($R_\star$), and inner radius of the most inner disk zone ($R_{0}$) for all objects in the DIANA sample using Eq. \ref{eq:tsub} with $a=0.2$ \citep{Woitke2015}: 

\begin{align}
    {T(R_0)}^4&=(1-a) \cdot \left(\frac{R_\star}{R_{0}}\right)^2 \cdot T_{\mathrm{eff}}^4  \label{eq:tsub}.
\end{align}

The distribution of temperatures for the sample is used to derive $x_{\rm min}$, $x_{\rm mean}$, and $x_{\rm std}$ of the sample. The maximal limit ($x_{\rm max}$) is set based on literature knowledge, to exclude unrealistically high temperatures.

\begin{figure}
    \centering
    \includegraphics[width=\linewidth]{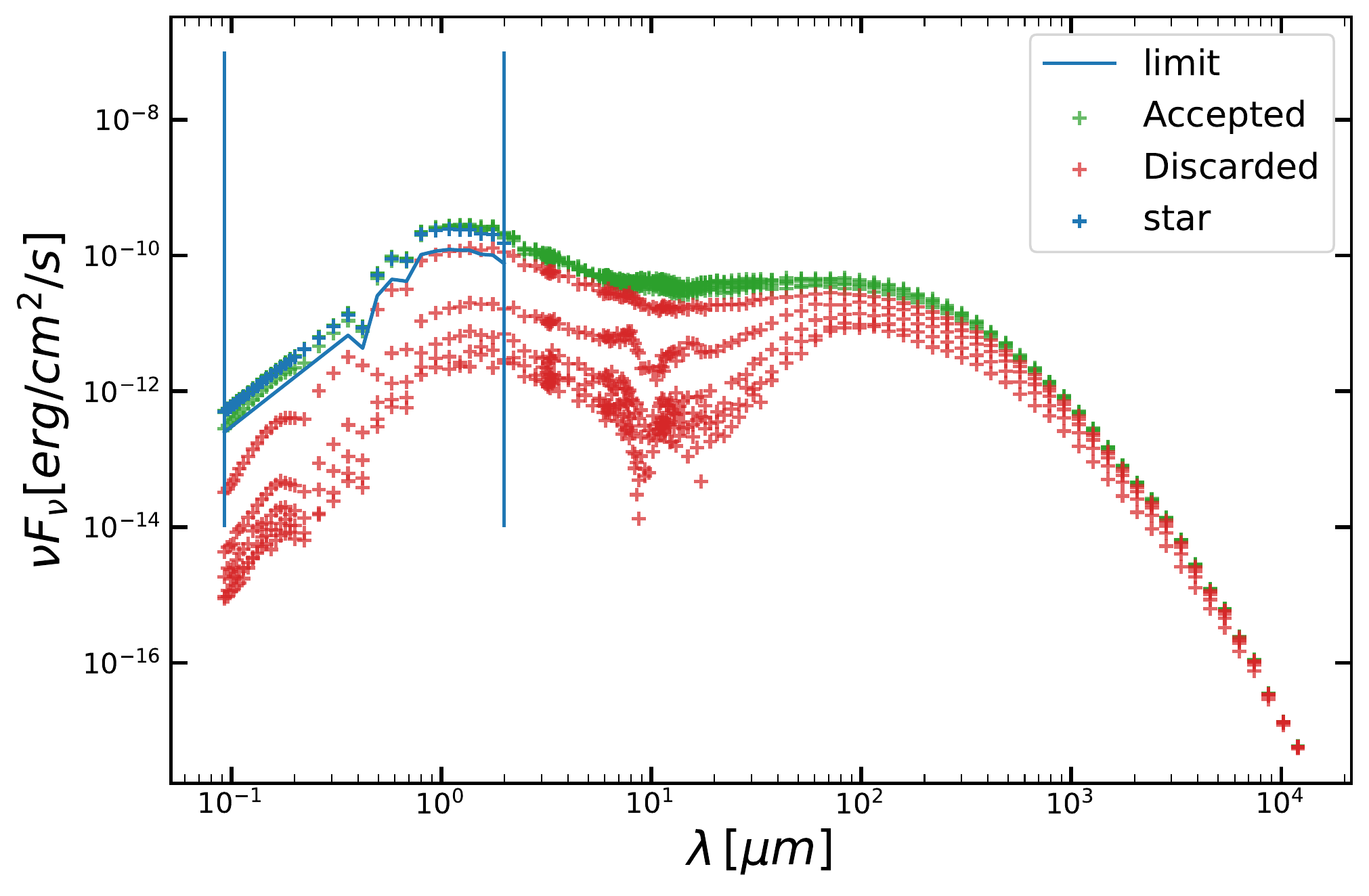}
    \caption{Example of an MCFOST model with SEDs for all 10 different inclinations. The blue points show the stellar flux at different wavelengths between $0.1\,\mathrm{\mu m}$ and $2\,\mathrm{\mu m}$ (both shown as vertical blue lines). The blue line encode the limit, which is used to distinguish non-shielded inclinations (green SEDs) that were accepted for the sample and shielded inclinations (red SEDs) that were discarded.}
    \label{fig:incl} 
\end{figure}

Using the described setup, we run all MCFOST models and created SEDs for ten different inclinations. For high inclinations (edge-on disks), often the disks block large parts of the starlight. Since these SEDs look fundamentally different (see Fig.~\ref{fig:incl}), they were excluded from our sample. We identified these shielded SEDs by having at least one flux point at a wavelength shorter than $2\mu m$ lower than half the stellar flux. It is justified to exclude them, because such objects are not present in the DIANA sample, but make the training of a NN much more challenging.

\section{Tables\label{sec:tables}}
\FloatBarrier

\begin{landscape}
\begin{table}
    \caption{Fits for all single zone objects comparable to DIANA. Notation $a(+b)$ means $a\times10^b$.}
    \label{tab:compare_diana_single}
     \centering
\resizebox{24cm}{!}{
\begin{tabular}{lllllllllll} \hline
Object & Work & $L_\star$ [$\rm L_\odot$] & $T_{\rm eff}$ [K] & $f_{\rm UV}$ & $p_{\rm UV}$ & $a_{\rm min}$ [$\rm \mu m$] & $a_{\rm max}$ [$\rm \mu m$] & $a_{\rm pow}$ & $\alpha_{\rm settle}$ & amC-Zubko[$\%$] \\ \hline 
\multirow{2}*{49\,Cet} & This work &  &   &   &   & $0.0491^{\times1.2}_{\div1.04}$ & $4.07^{\times1.1}_{\div1.01}(+3)$ & $3.04329^{+0.0012}_{-0.00030}$ & $0.079^{\times1.1}_{\div1.1}$ & $29.7^{+0.1}_{-0.1}$ \\ 
 & DIANA &$16.78$ & $8768$ & $0.001$ & $2.0$ & $0.0374$ & $1.26(+3)$ & $3.2$ & $0.000141$ & $16.0$ \\ \hline
\multirow{2}*{MWC\,480} & This work &  &   &   &   & $0.013^{\times2.4}_{\div2.6}$ & $5.5^{\times1.4}_{\div1.6}(+3)$ & $3.45^{+0.09}_{-0.09}$ & $1.24^{\times1.3}_{\div1.2}(-5)$ & $11.0^{+3}_{-2.6}$ \\ 
 & DIANA &$13.7$ & $8250$ & $0.005$ & $2.0$ & $0.0199$ & $4.28(+3)$ & $3.62$ & $7.31(-5)$ & $19.0$ \\ \hline
\multirow{2}*{RU\,Lup} & This work &  &   &   &   & $0.008^{\times2.4}_{\div2.7}$ & $6.0^{\times1.4}_{\div1.7}(+3)$ & $3.76^{+0.07}_{-0.07}$ & $0.007^{\times3.3}_{\div2.3}$ & $19.8^{+2.5}_{-2.4}$ \\ 
 & DIANA &$1.35$ & $4060$ & $0.058$ & $0.5$ & $0.051$ & $4.62(+3)$ & $3.52$ & $0.00201$ & $15.0$ \\ \hline
\multirow{2}*{DN\,Tau} & This work &  &   &   &   & $0.038^{\times2.6}_{\div3.2}$ & $3.2^{\times1.8}_{\div1.8}(+3)$ & $3.07^{+0.25}_{-0.05}$ & $0.00024^{\times2.9}_{\div2.1}$ & $20.9^{+4}_{-3.5}$ \\
\hline
\end{tabular}}
\end{table}
\end{landscape}

\begin{sidewaystable*}[p]

    \caption{Fits for all two-zone objects comparable to DIANA}
    \label{tab:compare_diana_two}
     \centering
\begin{tabular}{lllllllllll} \hline
Object & Work & $L_\star$ [$\rm L_\odot$] & $T_{\rm eff}$ [K] & $f_{\rm UV}$ & $p_{\rm UV}$ & $a_{\rm min}$ [$\rm \mu m$] & $a_{\rm max}$ [$\rm \mu m$] & $a_{\rm pow}$ & $\alpha_{\rm settle}$ & amC-Zubko[$\%$] \\ \hline 
\multirow{2}*{HD\,97048} & This work &  &   &   &   & $0.10^{\times1.6}_{\div2.5}$ & $349^{\times1.2}_{\div1.1}$ & $3.79^{+0.09}_{-0.08}$ & $0.035^{\times1.8}_{\div2.0}$ & $10.2^{+2.6}_{-2.4}$ \\ 
 & DIANA &$45.82$ & $10000$ & $0.001$ & $1.4$ & $0.054$ & $96.03$ & $3.49$ & $0.0135$ & $8.0$ \\ \hline
\multirow{2}*{HD\,100546} & This work &  &   &   &   & $0.0065^{\times2.1}_{\div2.2}$ & $6.1^{\times1.4}_{\div1.5}(+3)$ & $3.582^{+0.029}_{-0.032}$ & $0.00019^{\times1.7}_{\div1.5}$ & $27.2^{+1.8}_{-2.3}$ \\ 
 & DIANA &$33.55$ & $10470$ & $0.005$ & $2.0$ & $0.0416$ & $2.98(+3)$ & $3.34$ & $0.000117$ & $17.0$ \\ \hline
\multirow{2}*{AB\,Aur} & This work &  &   &   &   & $0.010^{\times2.4}_{\div2.7}$ & $2.8^{\times1.9}_{\div1.9}(+3)$ & $4.90^{+0.06}_{-0.13}$ & $0.047^{\times1.7}_{\div2.7}$ & $12.5^{+1.1}_{-1}$ \\ 
 & DIANA &$49.35$ & $9550$ & $0.03$ & $2.0$ & $0.0256$ & $4.55(+3)$ & $3.92$ & $0.00252$ & $8.0$ \\ \hline
\multirow{2}*{HD\,95881} & This work &  &   &   &   & $0.017^{\times3.2}_{\div3.4}$ & $7.2^{\times1.2}_{\div1.4}(+3)$ & $3.32^{+0.07}_{-0.08}$ & $0.008^{\times2.3}_{\div2.4}$ & $16.0^{+4}_{-4}$ \\
\hline
\end{tabular}
\end{sidewaystable*}

\addtocounter{table}{-1}
\begin{sidewaystable*}[p]
   
    \caption{continued.}
     \centering
\begin{tabular}{lllllllllll} \hline
Object & Work & $f_{\rm PAH}$ & $\rm PAH_{\rm charged}$ & $M_{\rm disk}$ [$\rm M_\odot$] & $R_{\rm in}$ [AU] & $R_{\rm taper}$ [AU] & $\epsilon$ & $H_0$ [AU] & $\beta$ & $M_{\rm disk,2}$ [$\rm M_\odot$] \\ \hline 
\multirow{2}*{HD\,97048} & This work &$0.77^{\times1.2}_{\div1.3}$ & $0.80^{+0.09}_{-0.09}$ & $0.079^{\times1.3}_{\div1.3}$ & $45^{+12}_{-8}$ &   &   & $5.23^{+0.13}_{-0.15}$ & $1.377^{+0.012}_{-0.015}$ & $7^{\times2.3}_{\div2.1}(-7)$ \\ 
 & DIANA &$0.417$ & $0.63$ & $0.132$ & $45.96$ & $100.0$ & $1.0$ & $5.75$ & $1.29$ & $12.33(-7)$ \\ \hline
\multirow{2}*{HD\,100546} & This work &$0.018^{\times1.4}_{\div1.5}$ & $0.66^{+0.19}_{-0.18}$ & $0.0035^{\times1.2}_{\div1.2}$ & $29.6^{+2.7}_{-3.1}$ &   &   & $30.8^{+2.9}_{-3.2}$ & $1.11^{+0.07}_{-0.07}$ & $2.6^{\times2.5}_{\div2.2}(-8)$ \\ 
 & DIANA &$0.00278$ & $0.89$ & $0.0071$ & $19.34$ & $100.0$ & $1.12$ & $32.37$ & $1.03$ & $8.72(-8)$ \\ \hline
\multirow{2}*{AB\,Aur} & This work &  &   & $0.0101^{\times1.1}_{\div1.1}$ & $8.8^{+3.1}_{-2.5}$ &   &   & $10.76^{+0.29}_{-0.30}$ & $1.372^{+0.016}_{-0.021}$ & $9^{\times1.8}_{\div1.9}(-7)$ \\ 
 & DIANA &$0.4$ & $0.9$ & $0.022$ & $62.78$ & $300.0$ & $1.0$ & $14.57$ & $1.0$ & $623.69(-7)$ \\ \hline
\multirow{2}*{HD\,95881} & This work &$0.036^{\times3.0}_{\div4.0}$ & $0.28^{+0.35}_{-0.19}$ & $3.1^{\times1.3}_{\div1.2}(-5)$ & $1.57^{+0.35}_{-0.22}$ &   &   & $16^{+4}_{-4}$ & $1.22^{+0.05}_{-0.07}$ &   \\
\hline
\end{tabular}
\end{sidewaystable*}

\addtocounter{table}{-1}

\begin{sidewaystable*}[p]
   
    \caption{continued.}
     \centering
\begin{tabular}{llllllllllllllllllll} \hline
Object & Work & $\epsilon_2$ & $H_2$ [AU] & $\beta_2$ & $R_{\rm in,2}$ [AU] & $R_{\rm out,2}$ [AU] & $a_{\rm max,2}$ [$\rm \mu m$] & $f_{\rm PAH,2}$ & $incl$ [deg] \\ \hline 
\multirow{2}*{HD\,97048} & This work &$1.62^{+0.27}_{-0.4}$ & $0.0208^{+0.0008}_{-0.0005}$ & $1.06^{+0.08}_{-0.07}$ & $0.337^{\times1.1}_{\div1.1}$ &   &   &   &   \\ 
 & DIANA &$1.42$ & $0.023$ & $1.18$ & $0.301$ & $10.0$ & $a_{\rm max}$ & $0.0001$ & $40.0$ \\ \hline
\multirow{2}*{HD\,100546} & This work &$1.1^{+0.5}_{-0.5}$ & $0.14^{+0.05}_{-0.04}$ & $0.85^{+0.20}_{-0.19}$ &   &   & $5^{\times16}_{\div30}(+1)$ &   &   \\ 
 & DIANA &$0.35$ & $0.181$ & $0.86$ & $0.55$ & $4.0$ & $0.29(+1)$ & $f_{\rm PAH}$ & $42.0$ \\ \hline
\multirow{2}*{AB\,Aur} & This work &  & $0.0959^{+0.004}_{-0.0032}$ & $0.85^{+0.06}_{-0.09}$ & $0.426^{\times1.1}_{\div1.1}$ &   &   &   &   \\ 
 & DIANA &$0.435$ & $0.0153$ & $0.88$ & $0.4$ & $R_{\rm in}$ & $a_{\rm max}$ & $f_{\rm PAH}$ & $30.0$ \\ \hline
\multirow{2}*{HD\,95881} & This work &$0.15^{+0.4}_{-0.35}$ & $0.062^{+0.004}_{-0.004}$ &   & $0.35^{\times1.2}_{\div1.1}$ &   &   &   &   \\
\hline
\end{tabular}

\end{sidewaystable*}

\begin{sidewaystable*}[p]

    \caption{Fits for all objects with single zone models with all parameters fitted}
    \label{tab:new_fits_single}
\begin{tabular}{llllllllll} \hline & & & & & & & & & \\[-1.9ex] 
Object & $L_\star$ [$\rm L_\odot$] & $T_{\rm eff}$ [K] & $f_{\rm UV}$ & $p_{\rm UV}$ & $a_{\rm min}$ [$\rm \mu m$] & $a_{\rm max}$ [$\rm \mu m$] & $a_{\rm pow}$ & $\alpha_{\rm settle}$ & amC-Zubko[$\%$] \\  & & & & & & & & & \\[-1.9ex]  \hline  & & & & & & & & & \\[-1.9ex] 
HD\,97048 &$45.817$ & $10000$ & $0.001$ & $1.4$ & $0.166^{\times1.1}_{\div1.3}$ & $3.0^{\times1.9}_{\div2.0}(+3)$ & $3.62^{+0.08}_{-0.08}$ & $0.078^{\times1.2}_{\div1.3}$ & $5.6^{+0.6}_{-0.4}$ \\  & & & & & & & & & \\[-1.9ex]  \hline  & & & & & & & & & \\[-1.9ex] 
HD\,100546 &$33.55$ & $10470$ & $0.005$ & $2.0$ & $0.0017^{\times1.5}_{\div1.4}$ & $7.0^{\times1.3}_{\div1.6}(+3)$ & $3.730^{+0.026}_{-0.021}$ & $0.0036^{\times1.4}_{\div1.5}$ & $26.1^{+1.3}_{-1}$ \\
\end{tabular}
\end{sidewaystable*}

\addtocounter{table}{-1}
\begin{sidewaystable*}[p]
   
    \caption{continued.}
\begin{tabular}{llllllllll} \hline & & & & & & & & & \\[-1.9ex] 
Object & $f_{\rm PAH}$ & $\rm PAH_{\rm charged}$ & $M_{\rm disk}$ [$\rm M_\odot$] & $R_{\rm in}$ [AU] & $R_{\rm taper}$ [AU] & $\epsilon$ & $H_0$ [AU] & $\beta$ & $incl$ [deg] \\  & & & & & & & & & \\[-1.9ex]  \hline  & & & & & & & & & \\[-1.9ex] 
HD\,97048 &$0.11^{\times1.5}_{\div1.5}$ & $0.91^{+0.06}_{-0.09}$ & $0.173^{\times1.2}_{\div1.2}$ & $0.40^{\times1.1}_{\div1.1}$ & $331^{\times1.04}_{\div1.1}$ & $0.26^{+0.14}_{-0.11}$ & $5.60^{+0.22}_{-0.24}$ & $1.228^{+0.010}_{-0.009}$ & $40.0$ \\  & & & & & & & & & \\[-1.9ex]  \hline  & & & & & & & & & \\[-1.9ex] 
HD\,100546 &$0.0010^{\times3.7}_{\div2.2}$ & $0.90^{+0.07}_{-0.15}$ & $0.0043^{\times1.2}_{\div1.2}$ & $0.2884^{\times1.02}_{\div1.01}$ & $211^{\times1.1}_{\div1.1}$ & $0.012^{+0.022}_{-0.009}$ & $20.3^{+0.6}_{-0.5}$ & $1.3968287^{+0}_{-0.0005}$ & $42.0$ \\
\end{tabular}
\end{sidewaystable*}

\begin{sidewaystable*}[p]

    \caption{Fits for all object with two-zone models with all parameters fitted}
    \label{tab:new_fits_two}
\begin{tabular}{llllllllll} \hline & & & & & & & & \\[-1.9ex] 
Object &$L_\star$ [$\rm L_\odot$] & $T_{\rm eff}$ [K] & $f_{\rm UV}$ & $p_{\rm UV}$ & $a_{\rm min}$ [$\rm \mu m$] & $a_{\rm max}$ [$\rm \mu m$] & $a_{\rm pow}$ & $\alpha_{\rm settle}$ & amC-Zubko[$\%$] \\  & & & & & & & & \\[-1.9ex]  \hline  & & & & & & & & \\[-1.9ex] 
HD\,97048 &$45.817$ & $10000$ & $0.001$ & $1.4$ & $0.033^{\times2.5}_{\div2.9}$ & $4.0^{\times1.3}_{\div1.2}(+2)$ & $3.69^{+0.11}_{-0.11}$ & $0.007^{\times2.6}_{\div2.2}$ & $9.4^{+2.3}_{-1.9}$ \\  & & & & & & & & \\[-1.9ex]  \hline  & & & & & & & & \\[-1.9ex] 
HD\,100546 &$33.55$ & $10470$ & $0.005$ & $2.0$ & $0.011^{\times2.1}_{\div2.5}$ & $5.3^{\times1.5}_{\div1.4}(+3)$ & $3.532^{+0.032}_{-0.029}$ & $0.00013^{\times1.7}_{\div1.6}$ & $25.6^{+2}_{-1.9}$ \\
\end{tabular}
\end{sidewaystable*}

\addtocounter{table}{-1}
\begin{sidewaystable*}[p]
   
    \caption{continued.}
\begin{tabular}{llllllllll} \hline & & & & & & & & \\[-1.9ex] 
Object &  $f_{\rm PAH}$ & $\rm PAH_{\rm charged}$ & $M_{\rm disk}$ [$\rm M_\odot$] & $R_{\rm in}$ [AU] & $R_{\rm taper}$ [AU] & $\epsilon$ & $H_0$ [AU] & $\beta$ & $M_{\rm disk,2}$ [$\rm M_\odot$] \\  & & & & & & & & \\[-1.9ex]  \hline  & & & & & & & & \\[-1.9ex] 
HD\,97048 &$0.48^{\times1.5}_{\div1.7}$ & $0.80^{+0.10}_{-0.09}$ & $0.103^{\times1.3}_{\div1.3}$ & $50^{+9}_{-8}$ & $304^{\times1.1}_{\div1.1}$ & $0.82^{+0.27}_{-0.28}$ & $5.31^{+0.4}_{-0.24}$ & $1.264^{+0.034}_{-0.030}$ & $1.2^{\times3.4}_{\div2.9}(-7)$ \\  & & & & & & & & \\[-1.9ex]  \hline  & & & & & & & & \\[-1.9ex] 
HD\,100546 &$0.012^{\times1.5}_{\div1.4}$ & $0.66^{+0.17}_{-0.16}$ & $0.0031^{\times1.2}_{\div1.2}$ & $29.9^{+3.0}_{-2.7}$ & $43^{\times1.3}_{\div1.2}$ & $0.79^{+0.24}_{-0.28}$ & $30^{+4}_{-4}$ & $1.11^{+0.07}_{-0.07}$ & $6^{\times3.9}_{\div3.4}(-8)$ \\
\end{tabular}
\end{sidewaystable*}

\addtocounter{table}{-1}
\begin{sidewaystable*}[p]
   
    \caption{continued.}
\begin{tabular}{lllllllll} \hline & & & & & & & & \\[-1.9ex] 
Object &  $\epsilon_2$ & $H_2$ [AU] & $\beta_2$ & $R_{\rm in,2}$ [AU] & $R_{\rm out,2}$ [AU] & $a_{\rm max,2}$ [$\rm \mu m$] & $f_{\rm PAH,2}$ & $incl$ [deg] \\  & & & & & & & & \\[-1.9ex]  \hline  & & & & & & & & \\[-1.9ex] 
HD\,97048&$0.7^{+0.6}_{-0.6}$ & $0.0226^{+0.0028}_{-0.0017}$ & $1.22^{+0.16}_{-0.16}$ & $0.44^{\times1.2}_{\div1.2}$ & $0.68^{\times1.2}_{\div1.1}$ & $7^{\times3.4}_{\div5.5}(+1)$ & $0.0007^{\times15}_{\div18}$ & $40.0$ \\  & & & & & & & & \\[-1.9ex]  \hline  & & & & & & & & \\[-1.9ex] 
HD\,100546&$0.7^{+0.5}_{-0.4}$ & $0.13^{+0.06}_{-0.04}$ & $0.98^{+0.22}_{-0.19}$ & $0.45^{\times1.1}_{\div1.1}$ & $3.2^{\times2.3}_{\div2.1}$ & $1.5^{\times8.6}_{\div35}(+2)$ & $0.00018^{\times9.4}_{\div12}$ & $42.0$ \\
\end{tabular}
\end{sidewaystable*}

\FloatBarrier
\onecolumn
\section{Figures}

\begin{figure*}[h]
    \centering
    \includegraphics[width=0.92\linewidth]{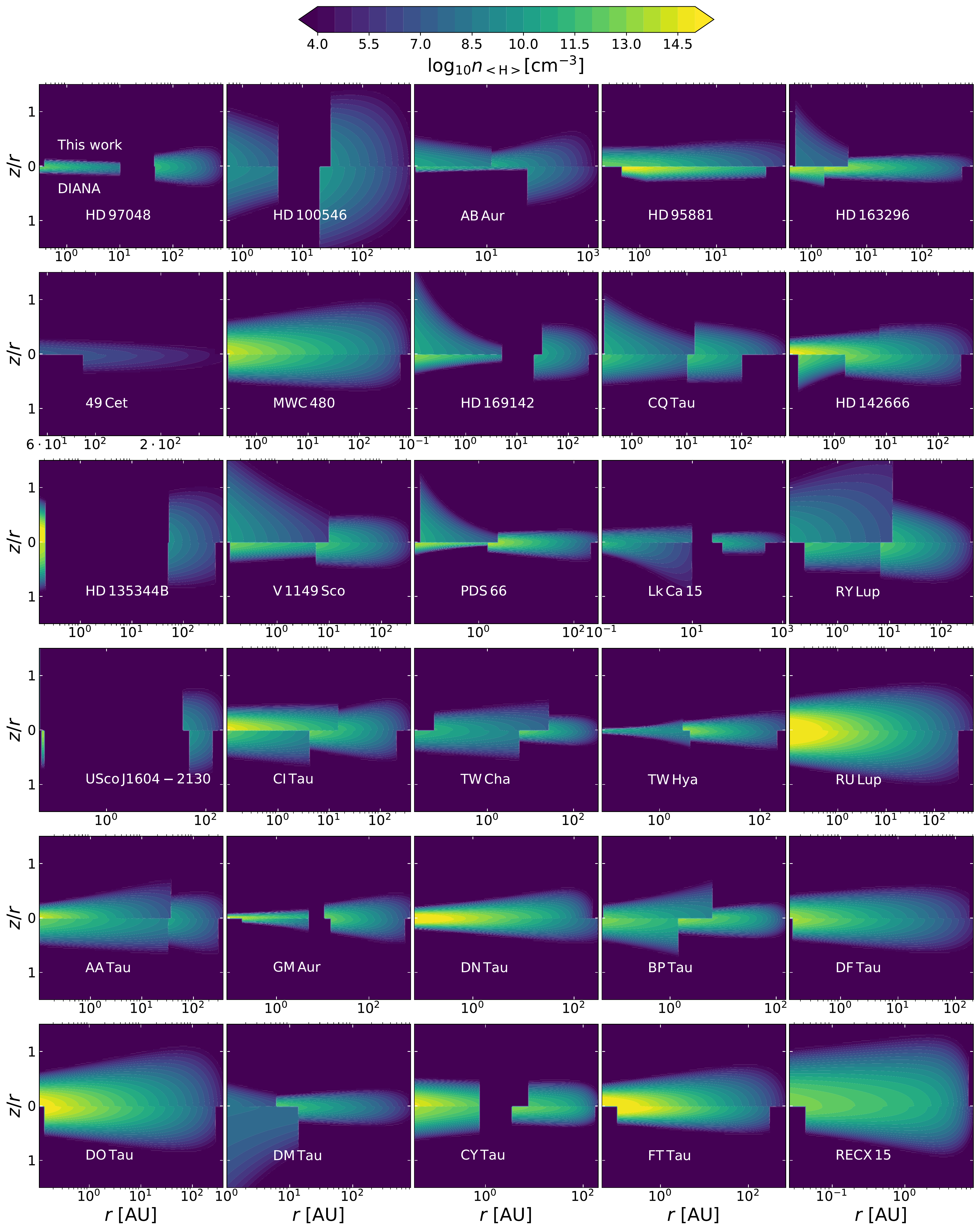}
\caption{Density plots for every object's median probable model to reproduce the DIANA results (Sect. \ref{sec:reproduce_DIANA}). Every panel shows the density structure in number densities of Hydrogen per cubic centimetre (colour bar at the top) for the objects noted in the panel. The upper and lower half of every plot depict the model from this study and DIANA, respectively. The radial axis is different for every object. The inner zones of Usco\,J1604-2130 in this work has a radial extent smaller than the radial resolution of the plots.}
\label{fig:density_reproduce}

\end{figure*}

\begin{figure*}
    \centering
    \includegraphics[width=0.92\linewidth]{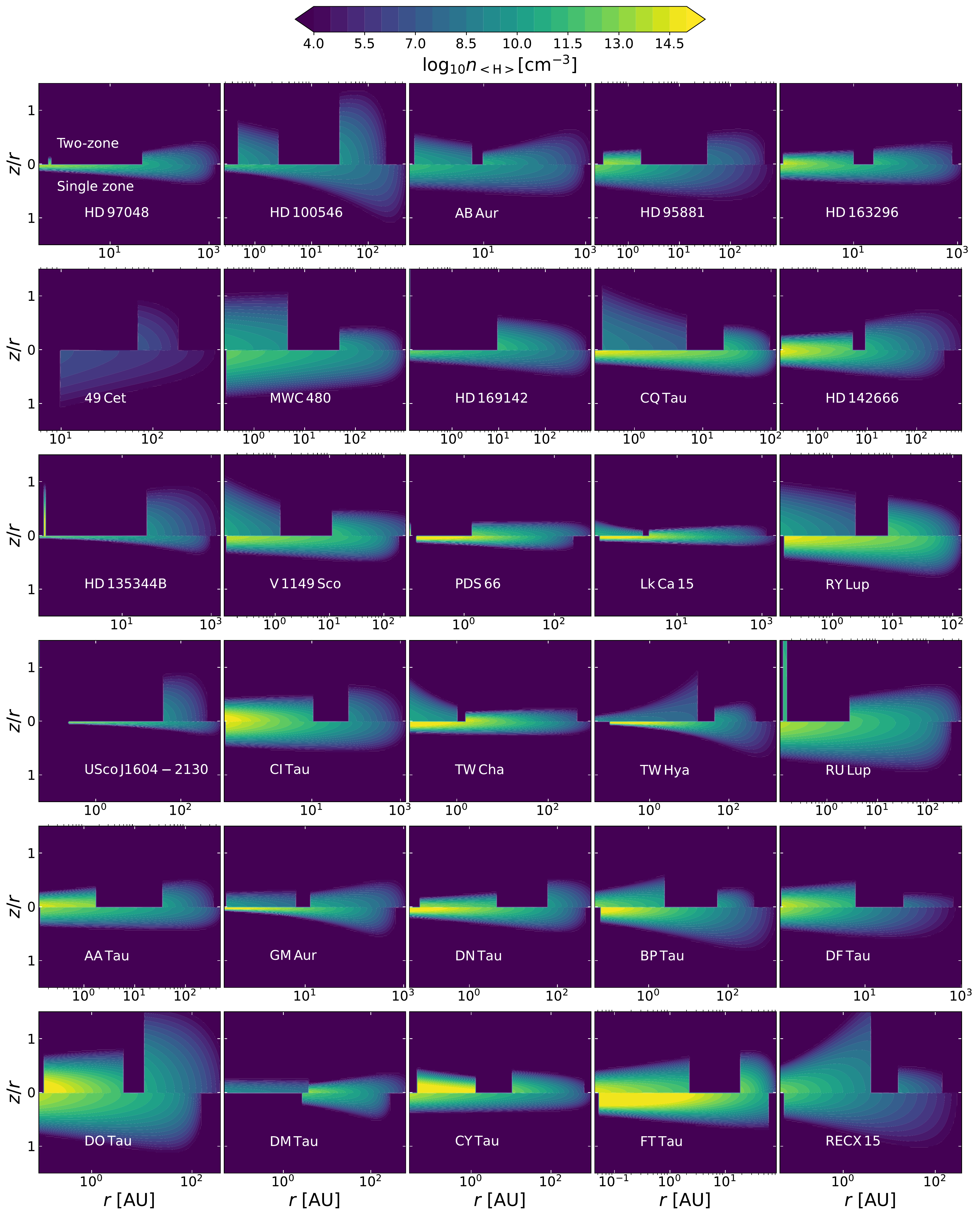}
\caption{Density plots for every object's median probable model derived in this study (Sect. \ref{sec:bayes_fact}). Every panel shows the density structure in number densities of Hydrogen per cubic centimetre (colour bar at the top) for the objects noted in the panel. The upper and lower half of every plot depict the two-zone and single zone model, respectively. The radial axis is different for every object. The inner zone of the two-zone model of 49\,Cet has densities lower than the minimum threshold of the picture. The inner zones of HD\,169142 and Usco\,J1604-2130have a radial extent smaller than the radial resolution of the plots.
}\label{fig:density_new_data}
\end{figure*}

\begin{figure*}[!hp]
\centering
    \includegraphics[width=0.49\linewidth]{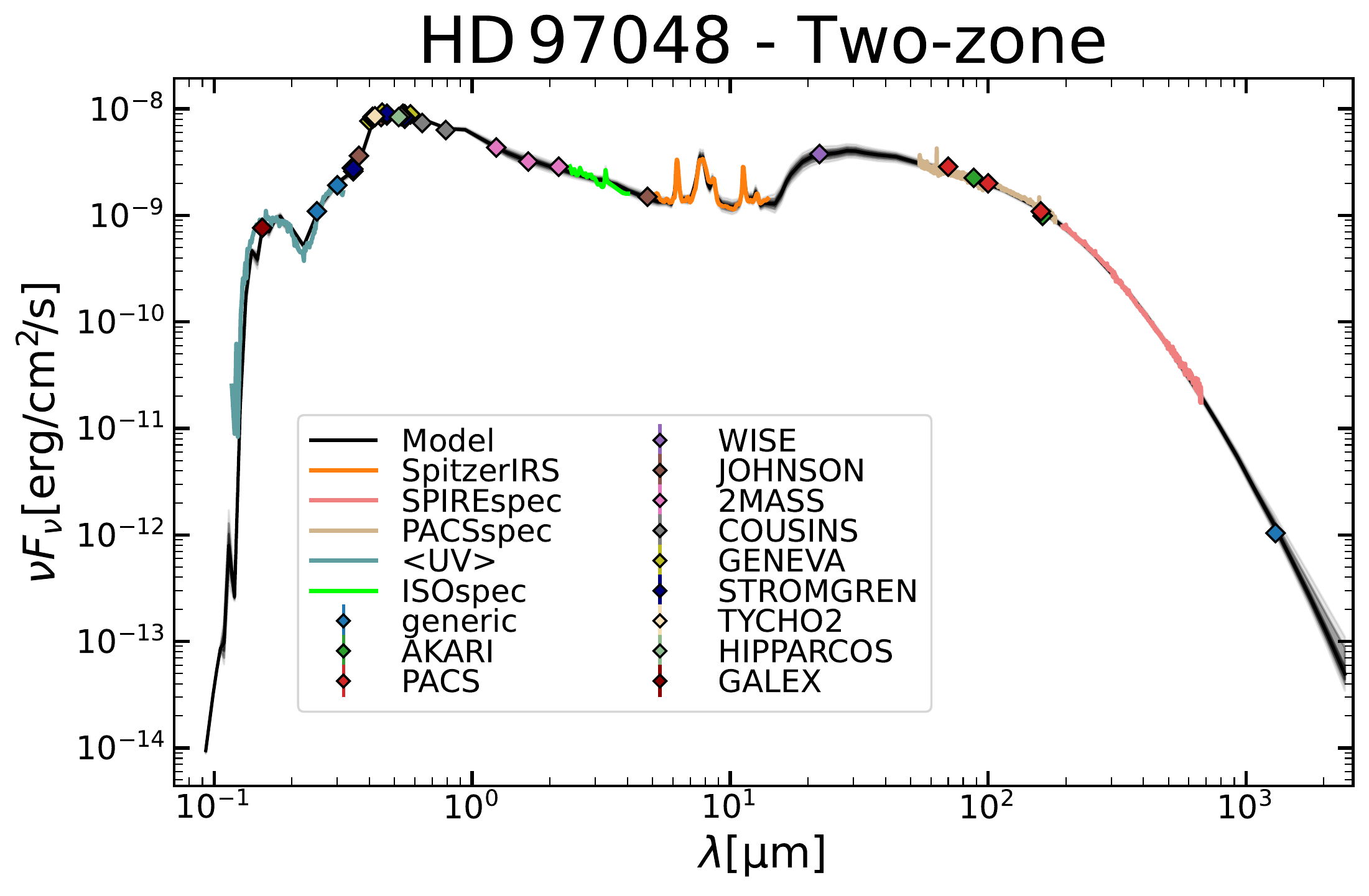}
    \includegraphics[width=0.49\linewidth]{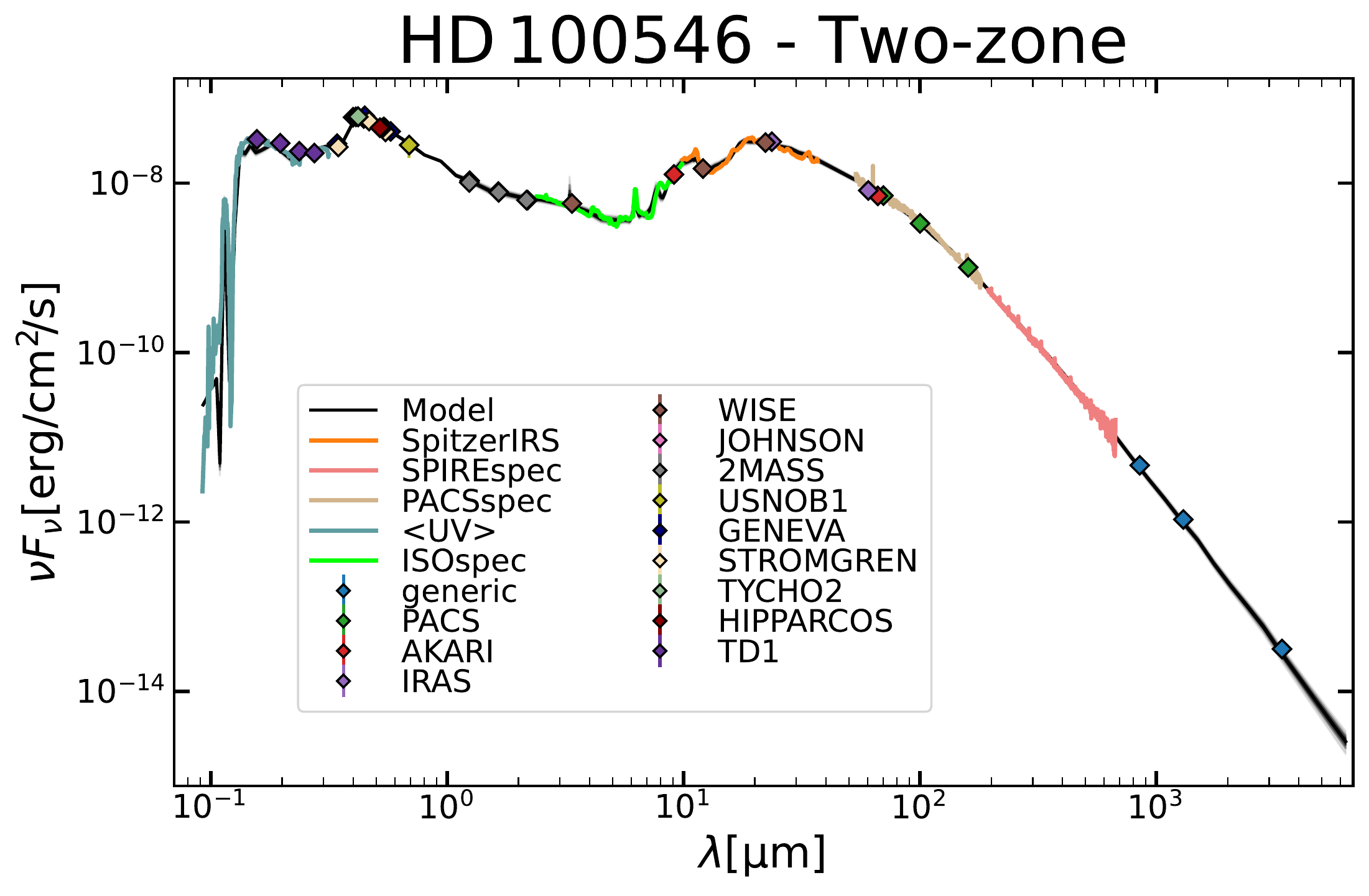}\\
    \includegraphics[width=0.49\linewidth]{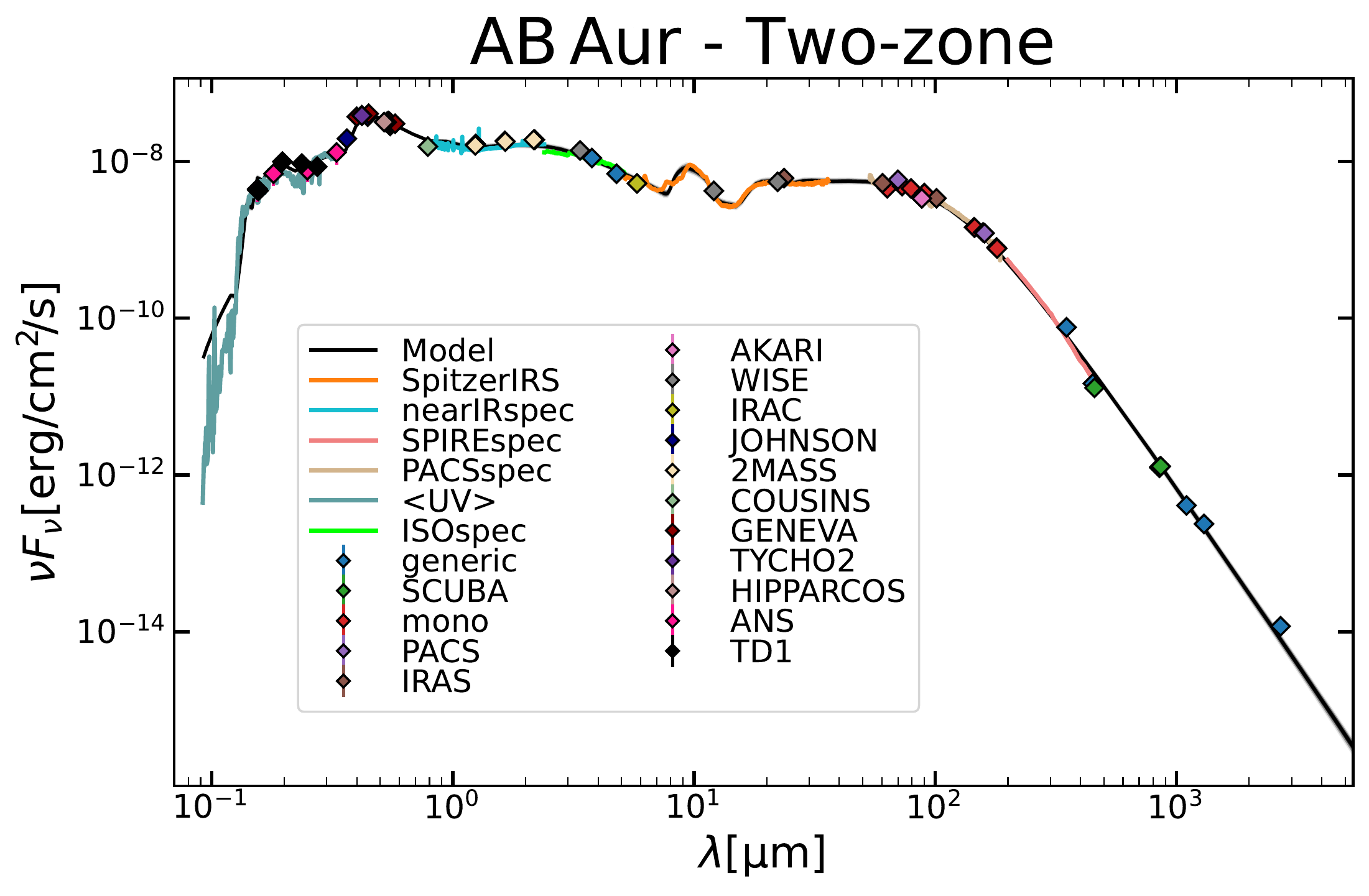}
    \includegraphics[width=0.49\linewidth]{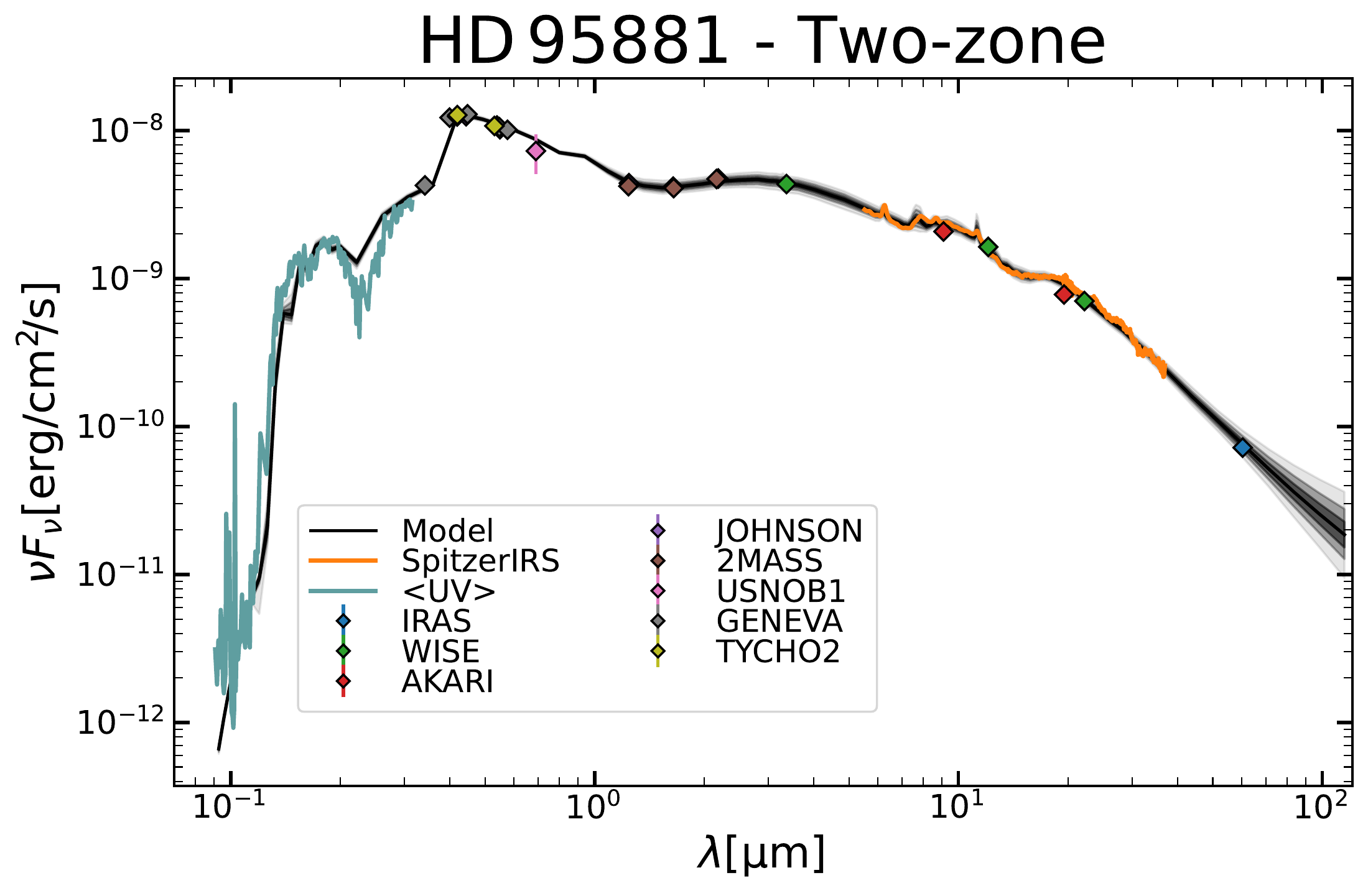}\\
    \includegraphics[width=0.49\linewidth]{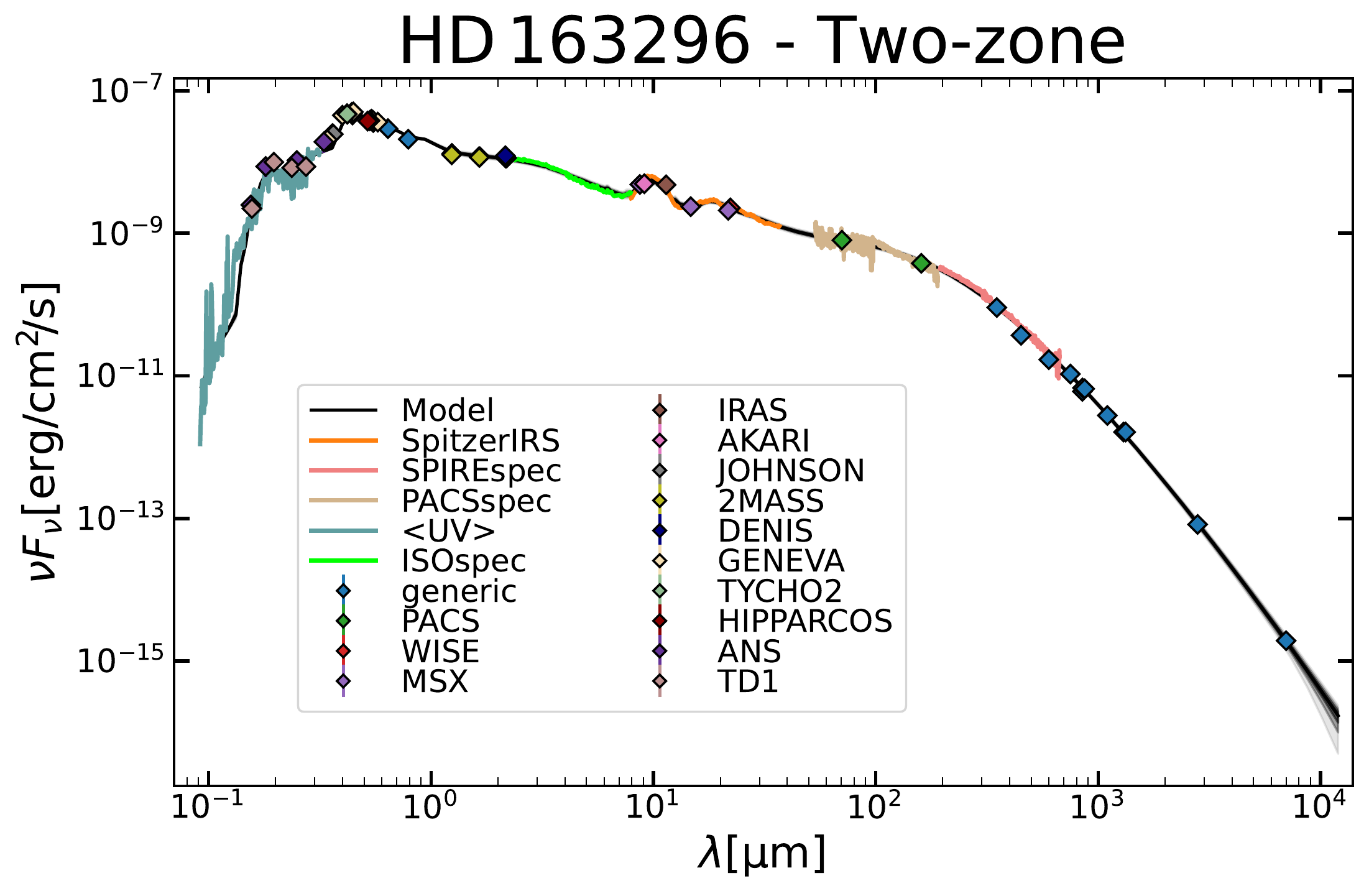}
    \includegraphics[width=0.49\linewidth]{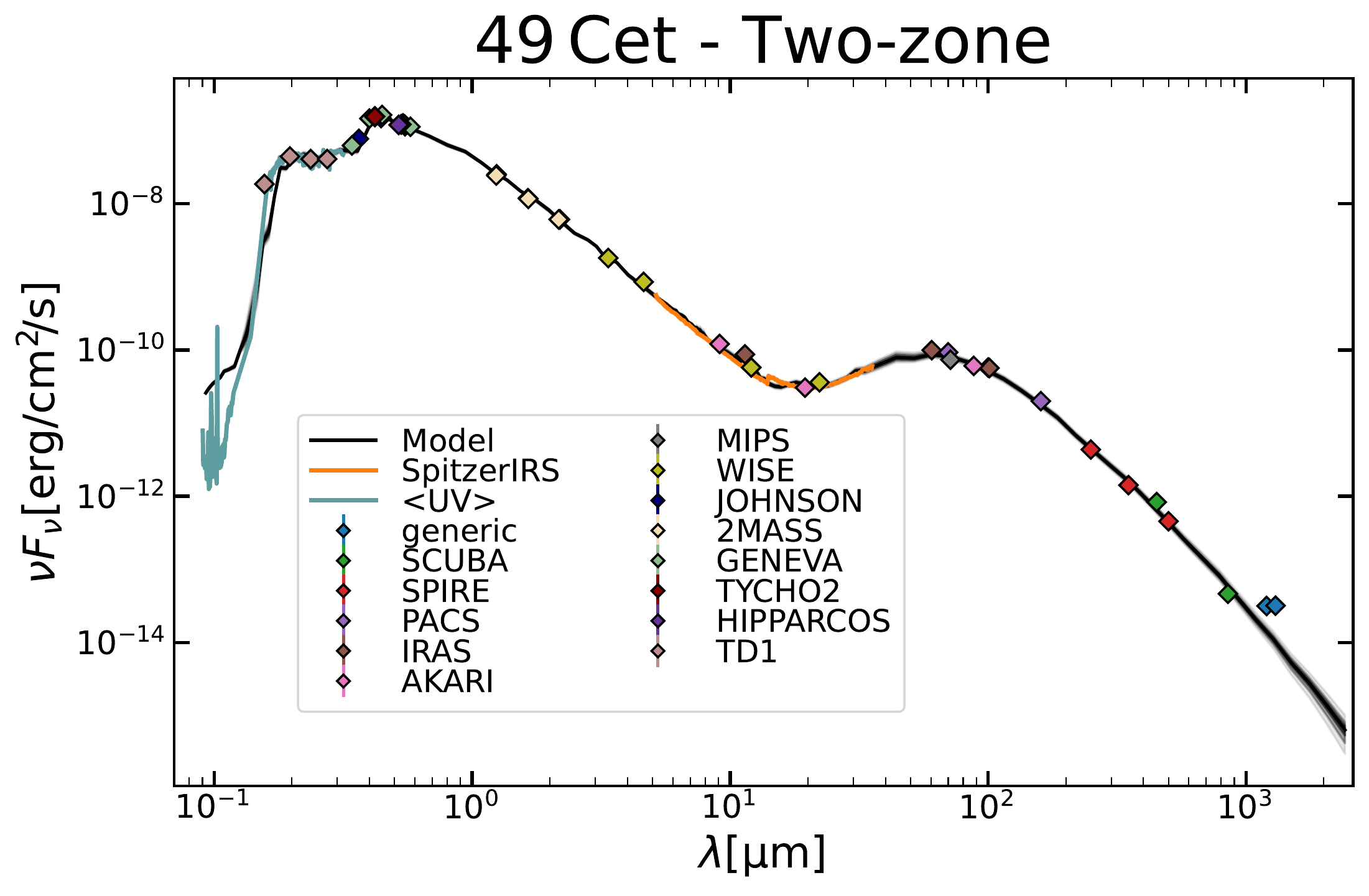}
\caption{SEDs for all objects. The coloured points and lines indicate the observation listed in the legend. The SEDs from the posterior distribution are shown in black. The line denotes the median of all model SED, with the dark, medium, and light black areas denoting the $68\,\%$, $95\,\%$, and $99.9\,\%$ percentiles, respectively. The name of the model and if it is fitted with a single or two-zone model is given in the title.}
\label{fig:sed_fits}
\end{figure*}

\addtocounter{figure}{-1}
\begin{figure*}[!hp]
\centering
    \includegraphics[width=0.49\linewidth]{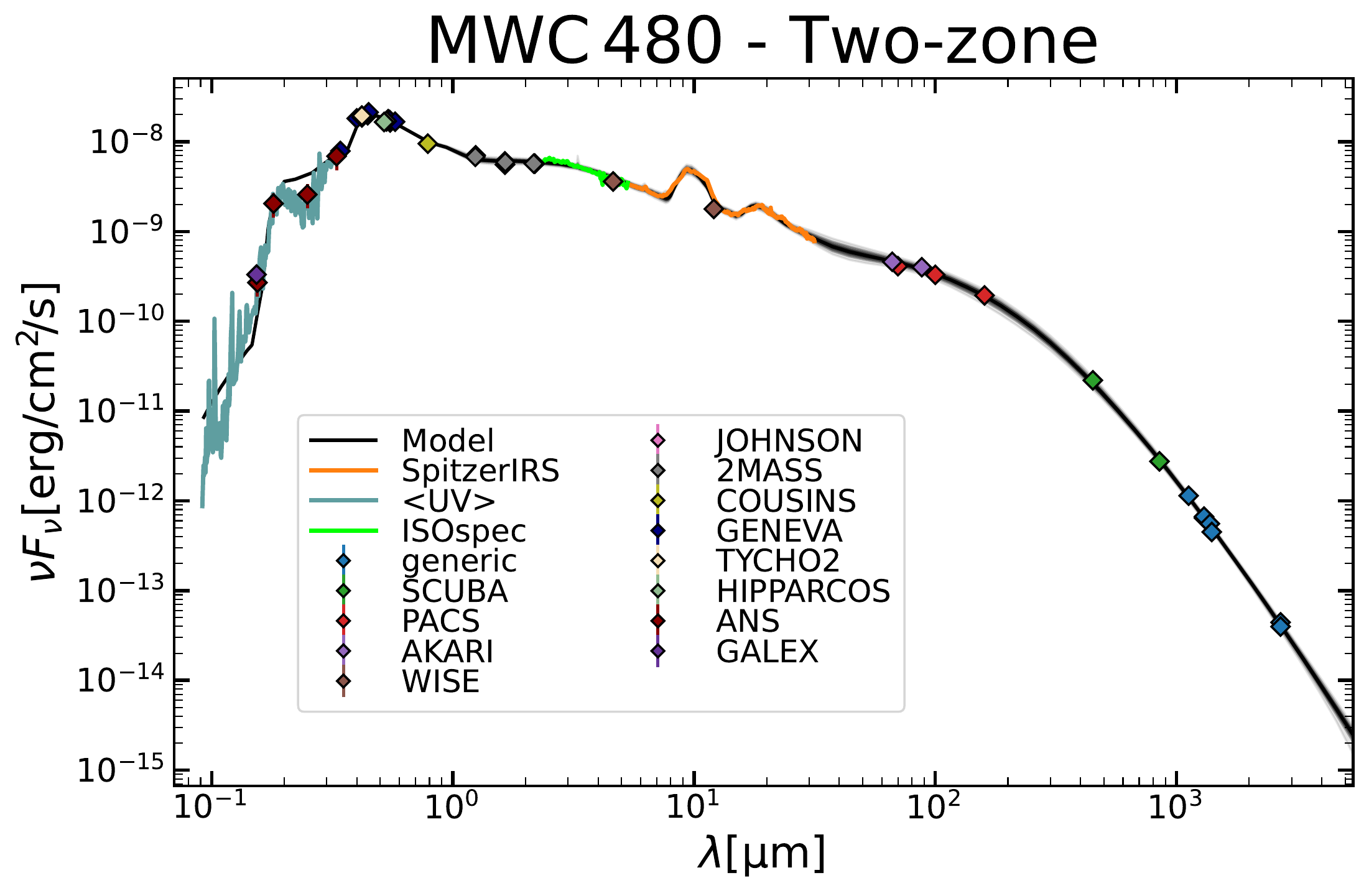}
    \includegraphics[width=0.49\linewidth]{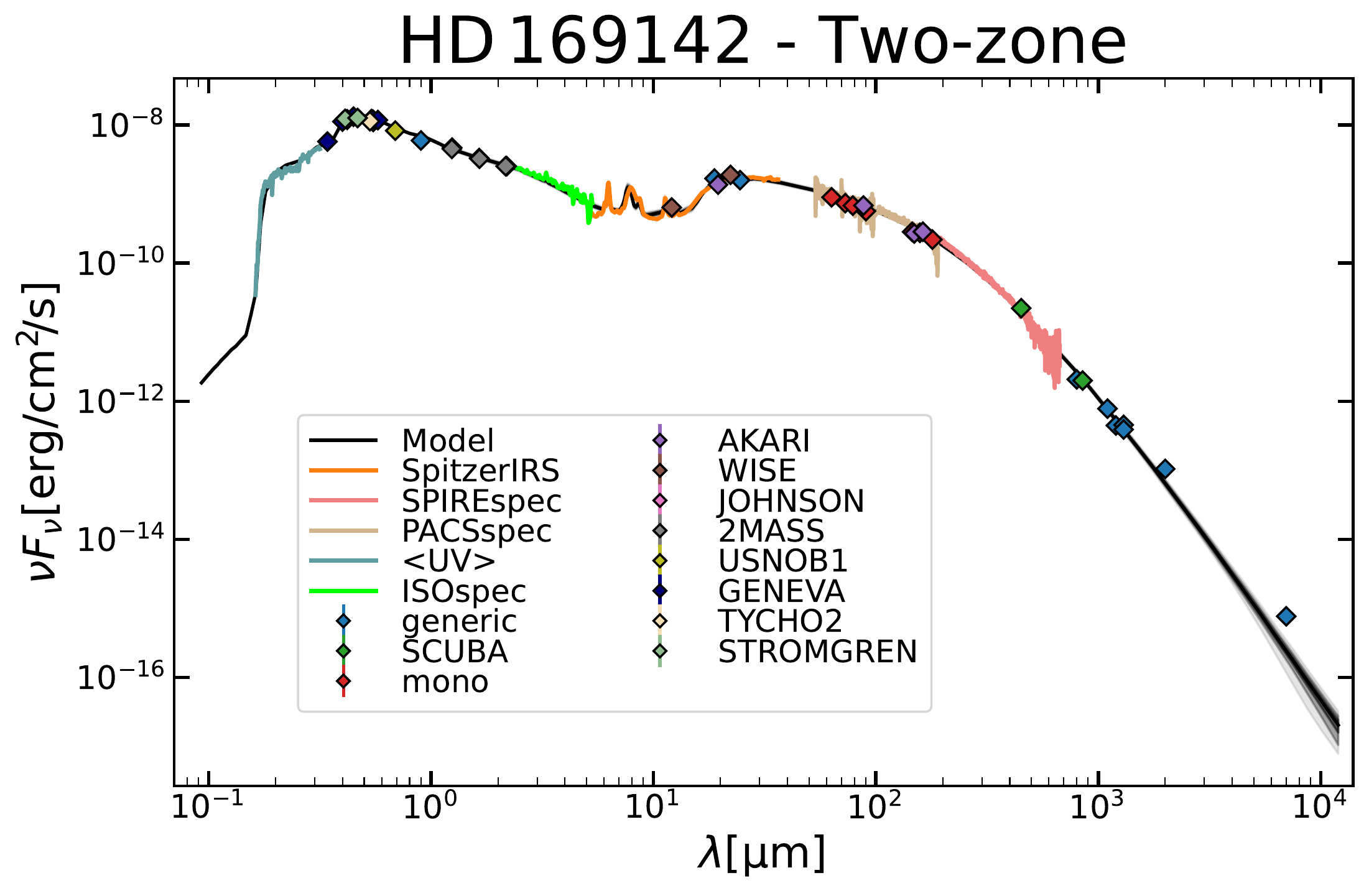}\\
    \includegraphics[width=0.49\linewidth]{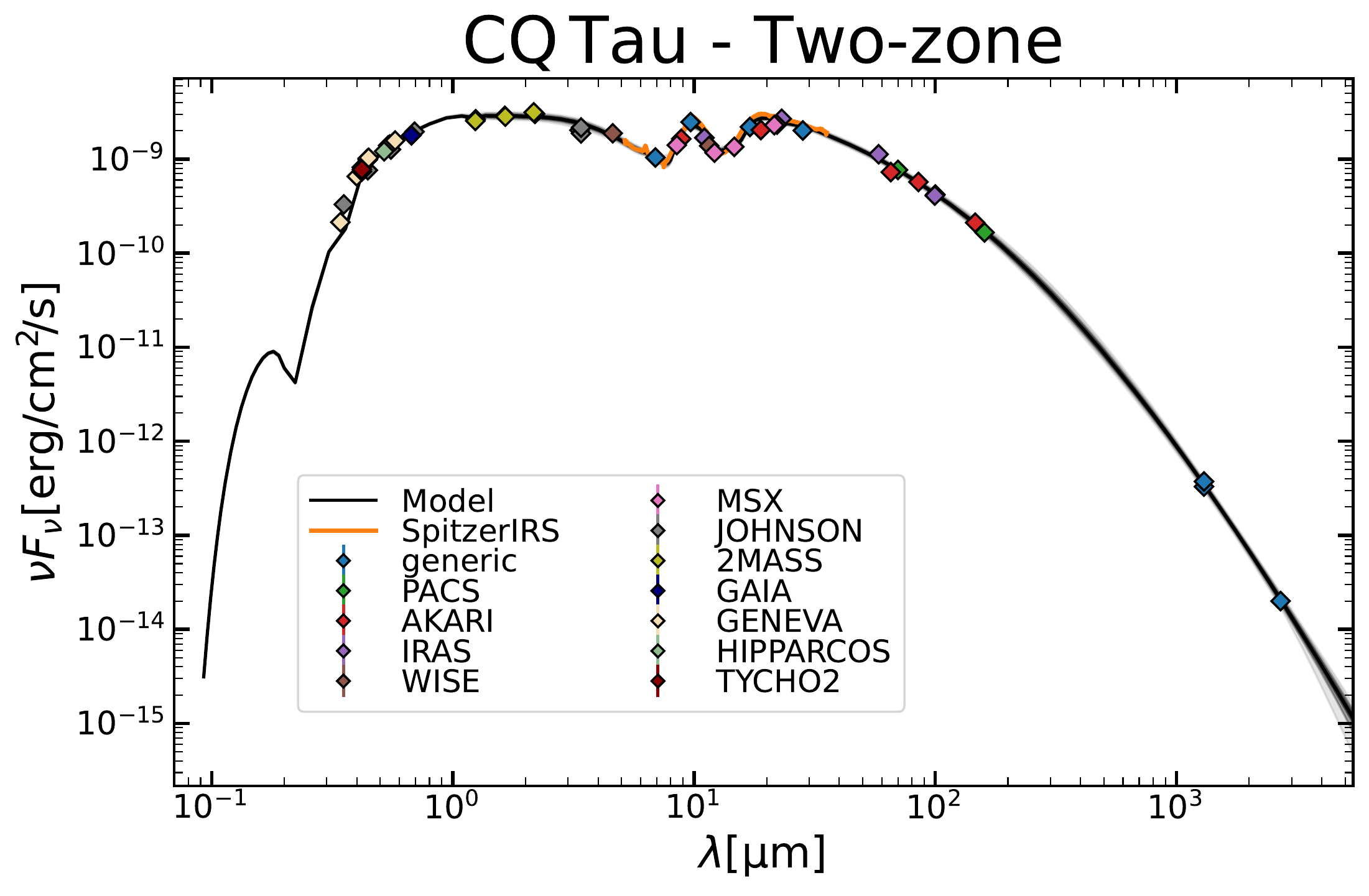}
    \includegraphics[width=0.49\linewidth]{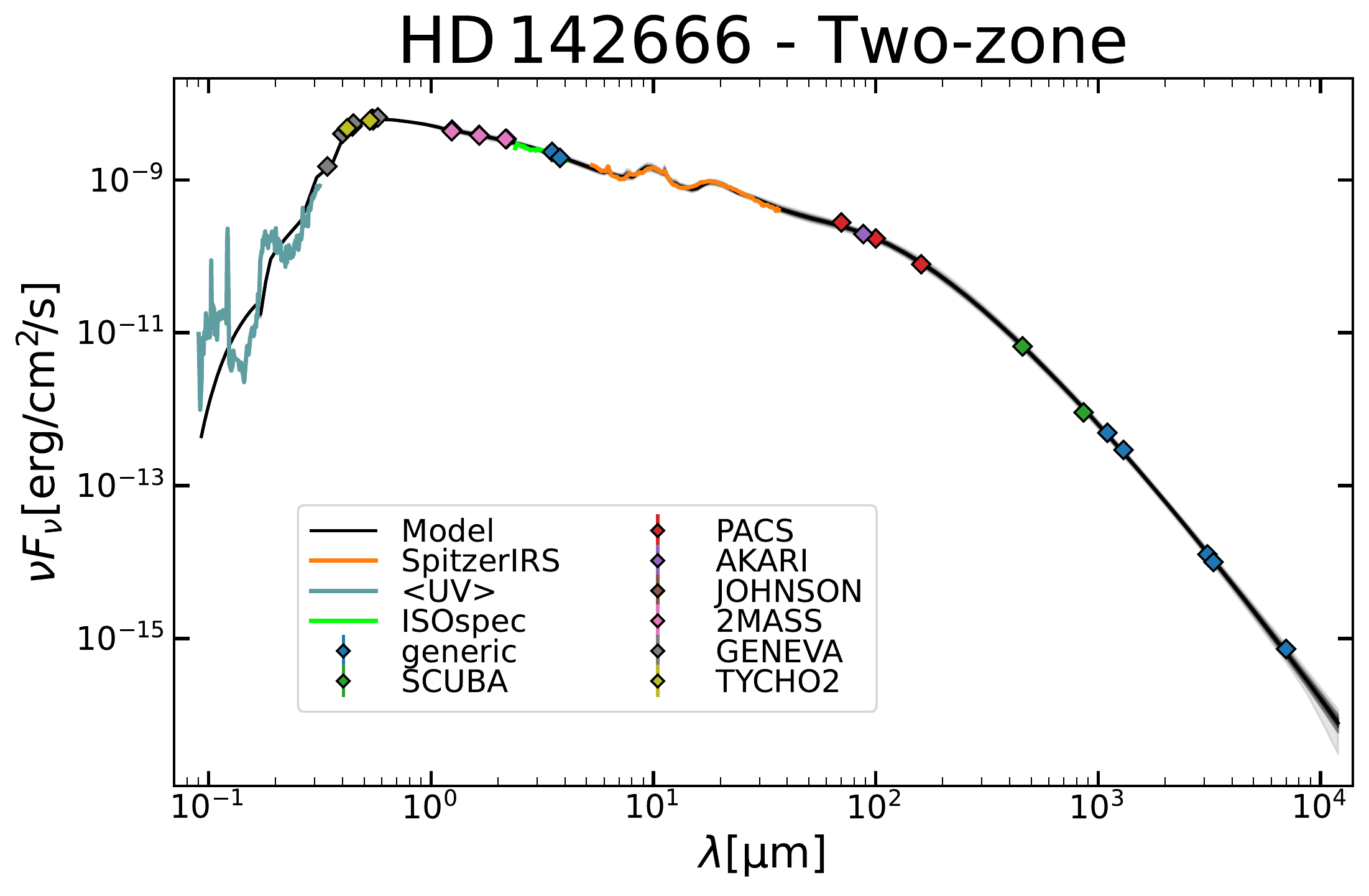}\\
    \includegraphics[width=0.49\linewidth]{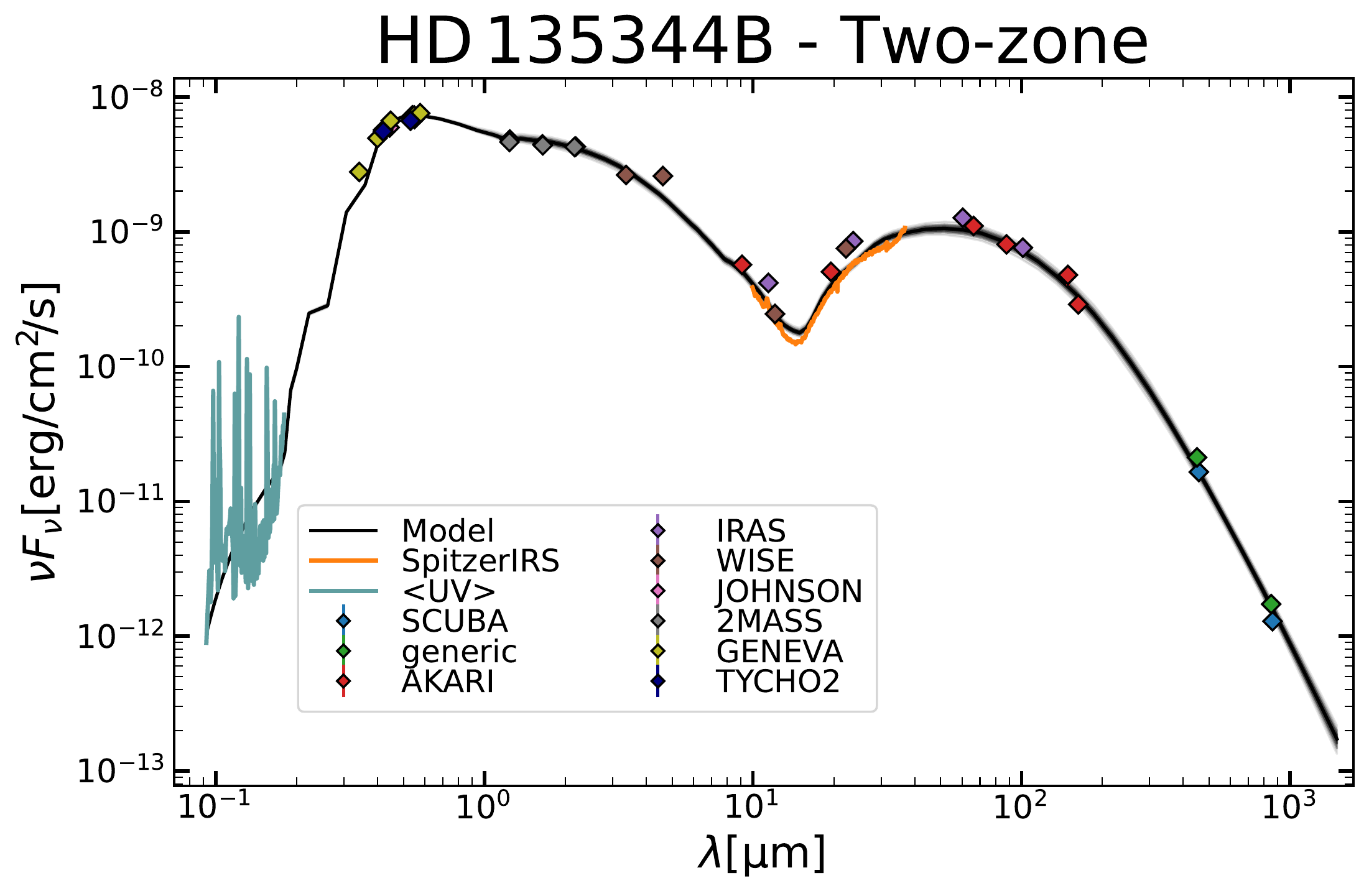}
    \includegraphics[width=0.49\linewidth]{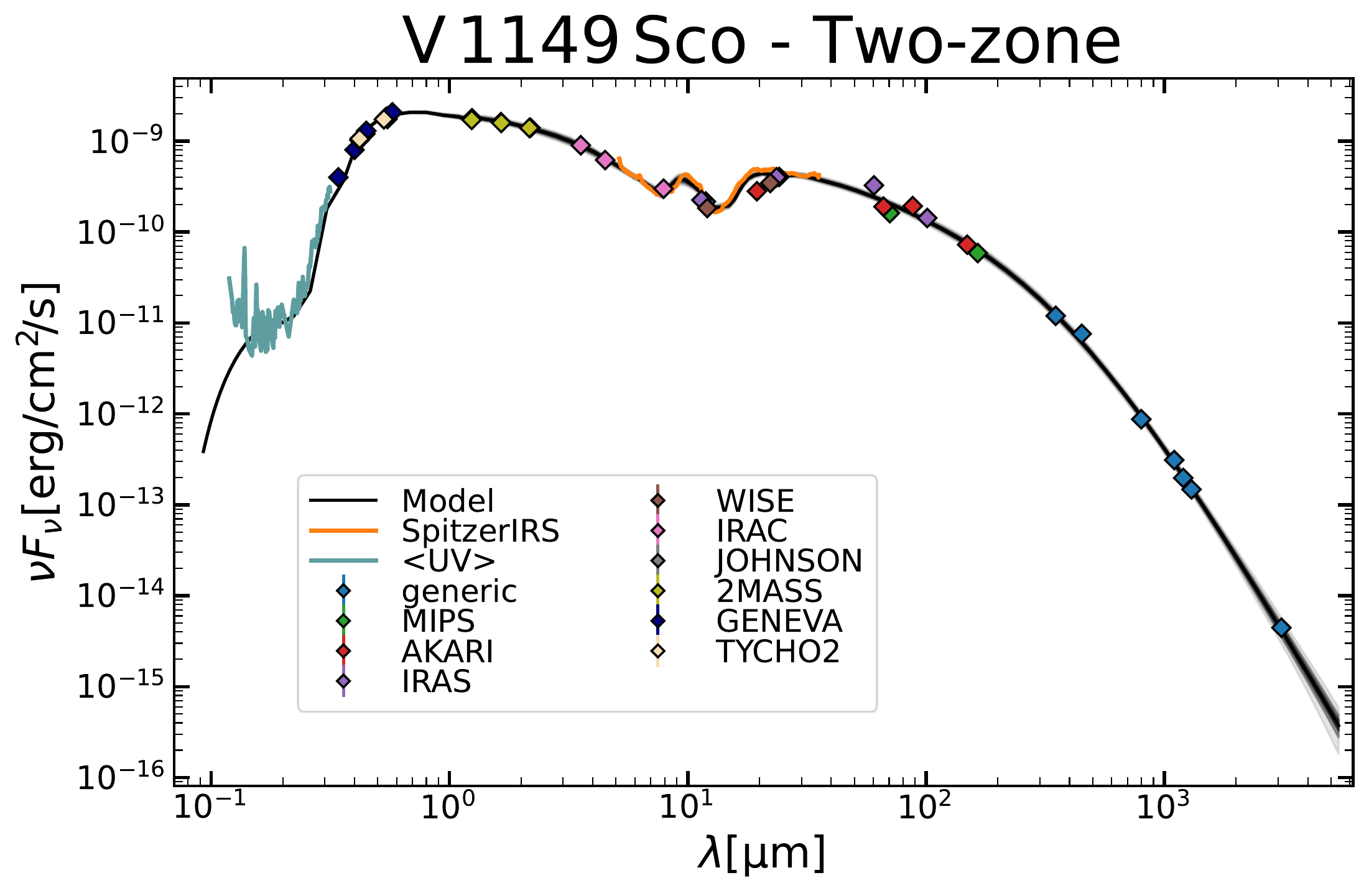}\\
    \includegraphics[width=0.49\linewidth]{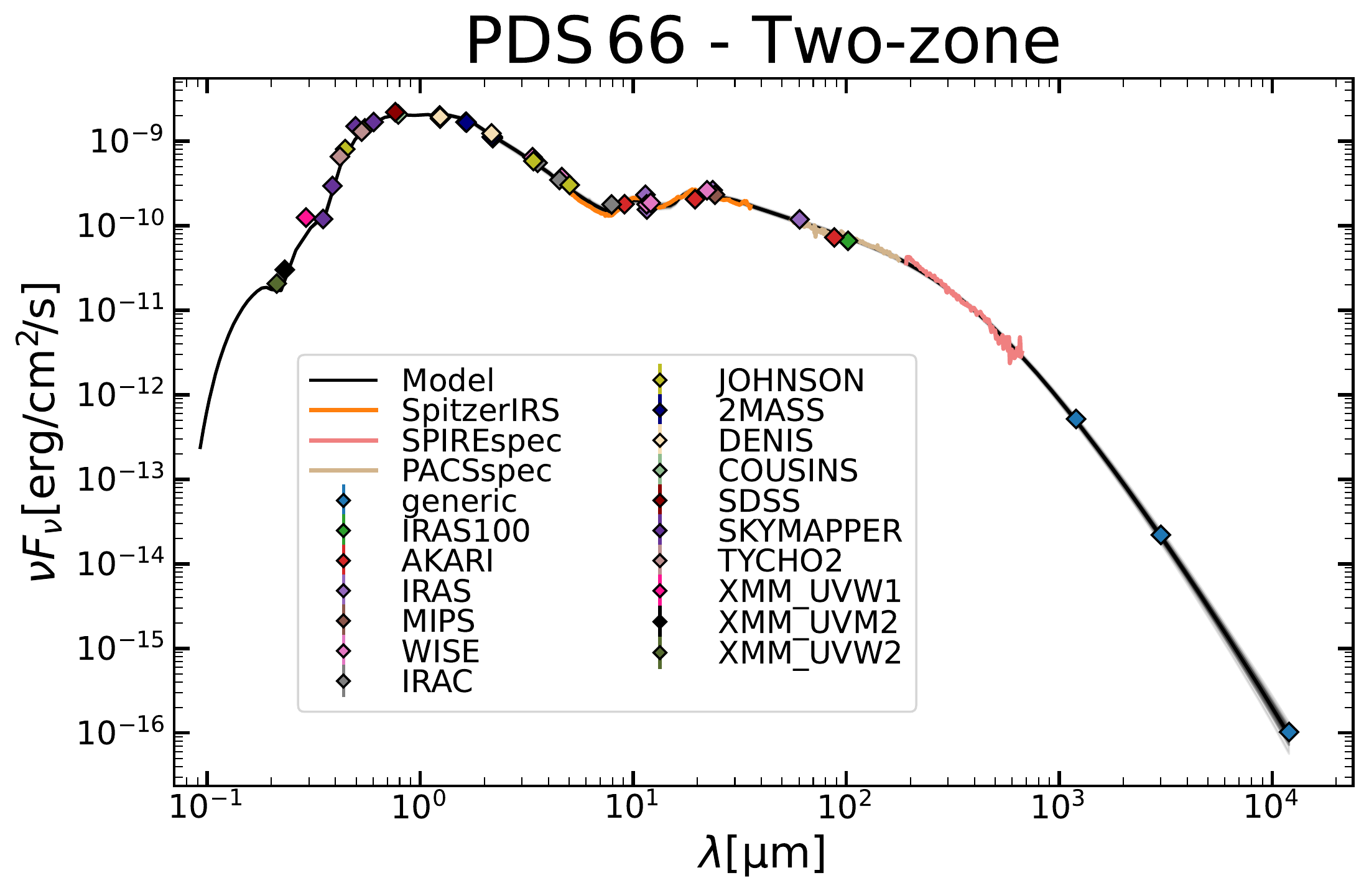}
    \includegraphics[width=0.49\linewidth]{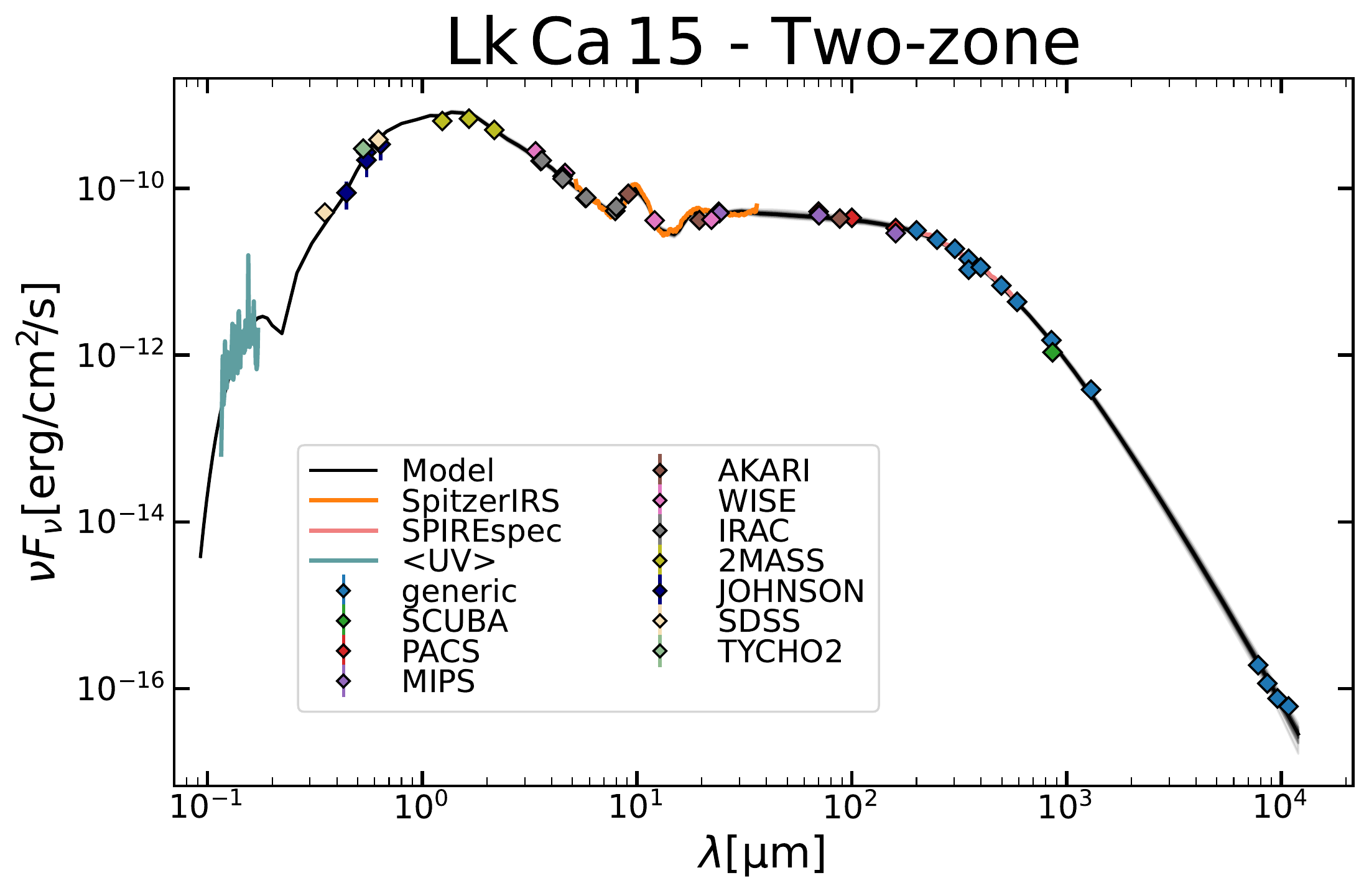}

\caption{continued.}
\end{figure*}

\addtocounter{figure}{-1}
\begin{figure*}[!hp]

    \centering
    \includegraphics[width=0.49\linewidth]{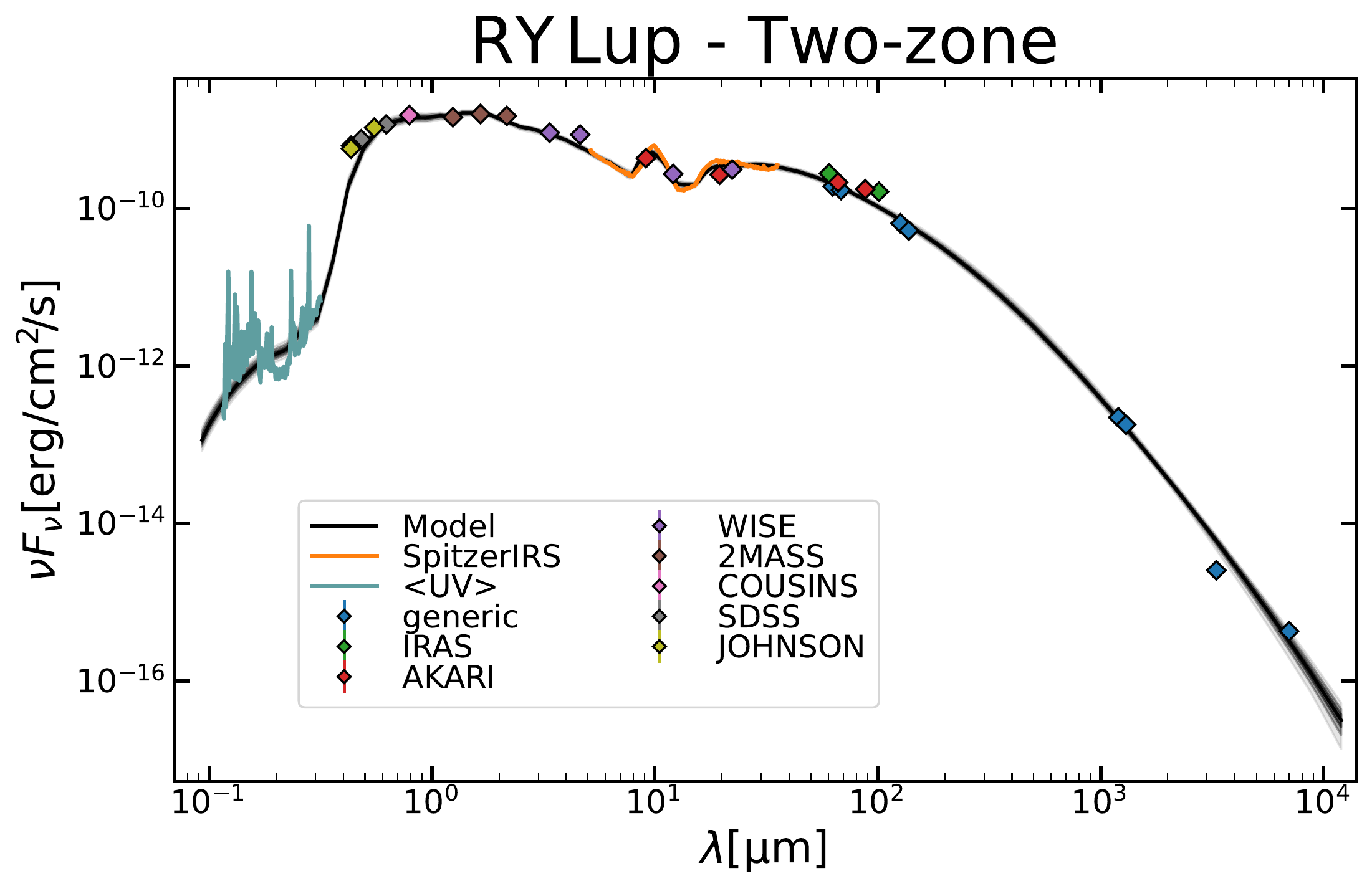}
    \includegraphics[width=0.49\linewidth]{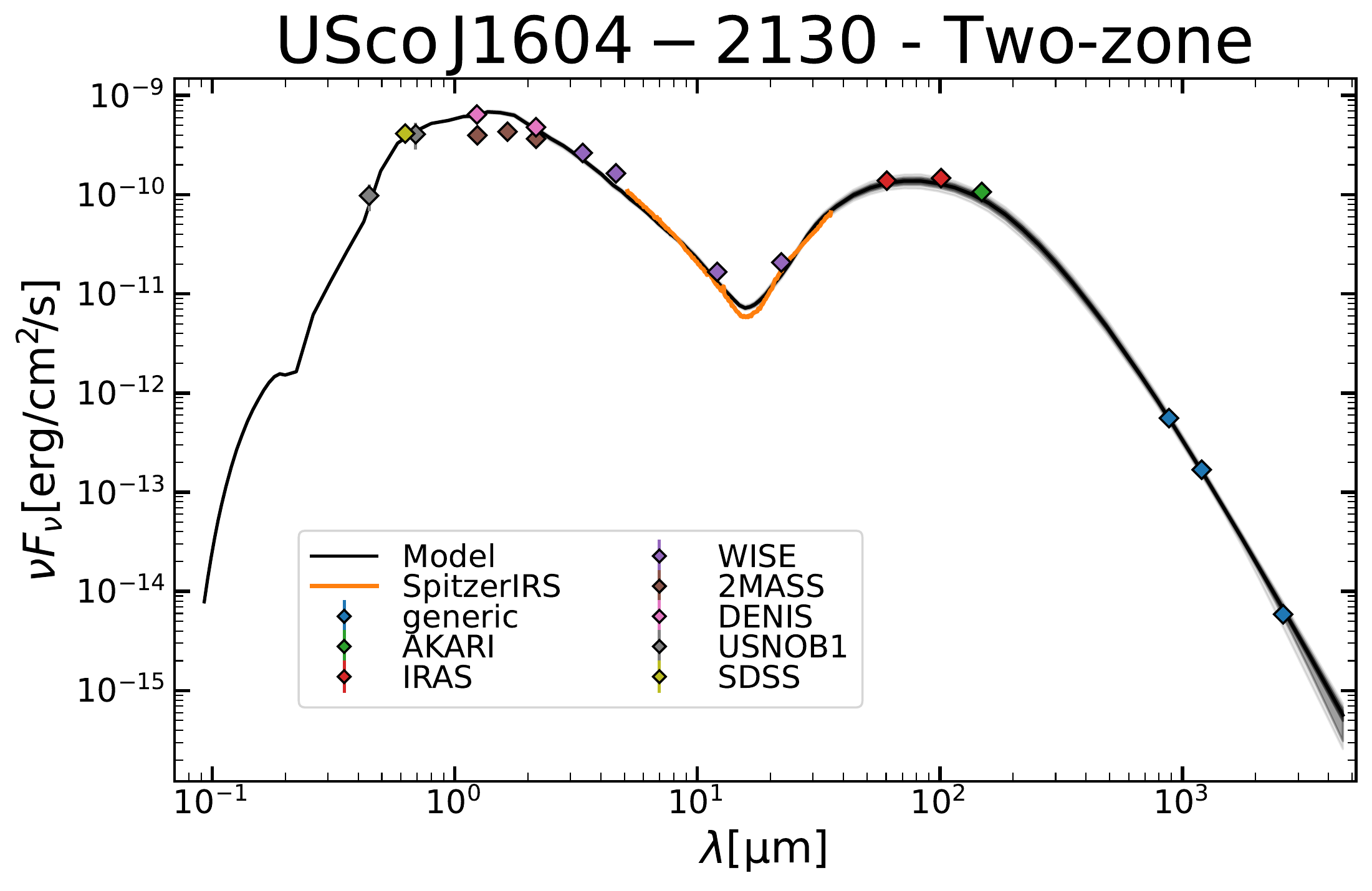}\\
    \includegraphics[width=0.49\linewidth]{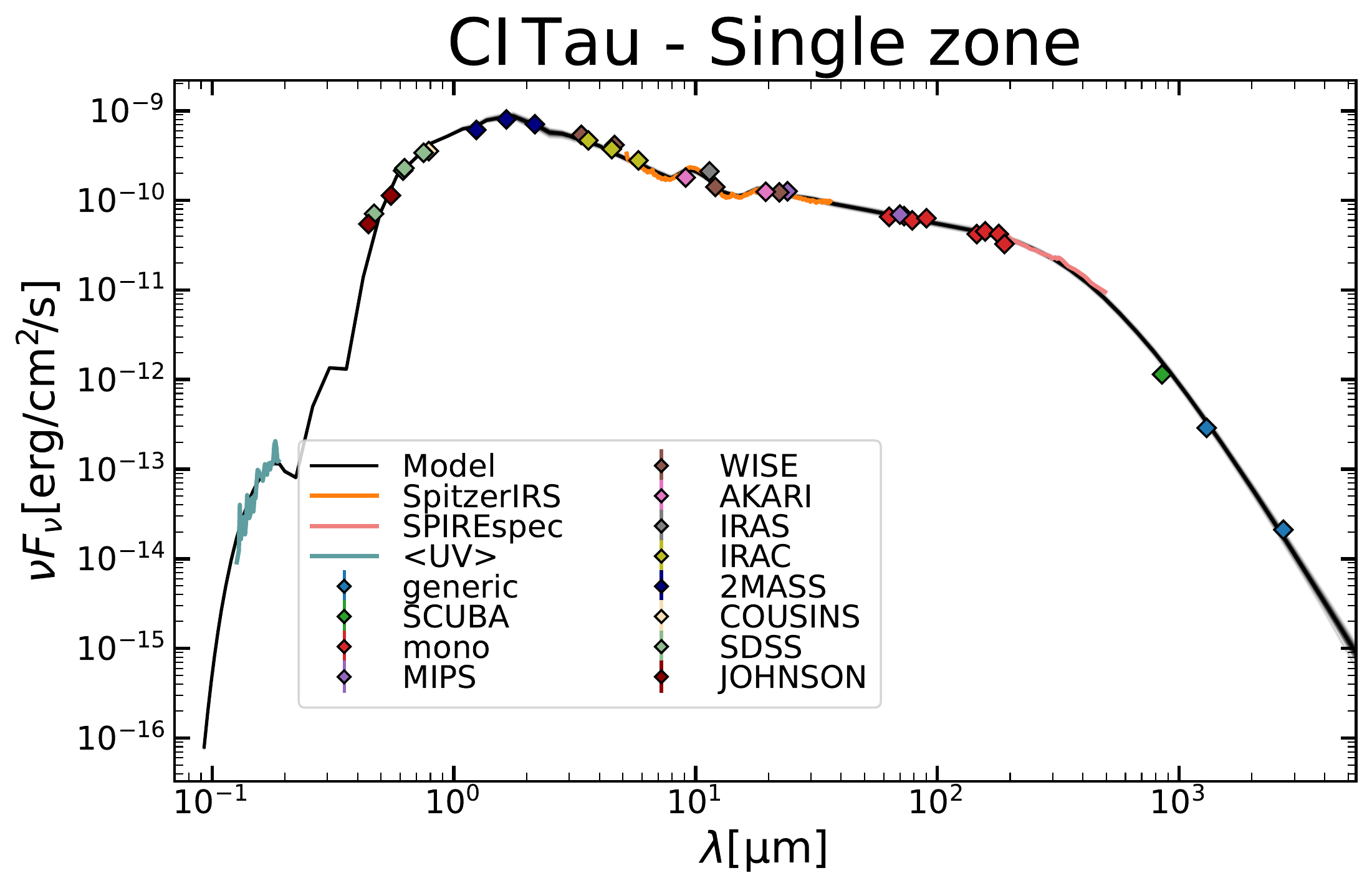}
    \includegraphics[width=0.49\linewidth]{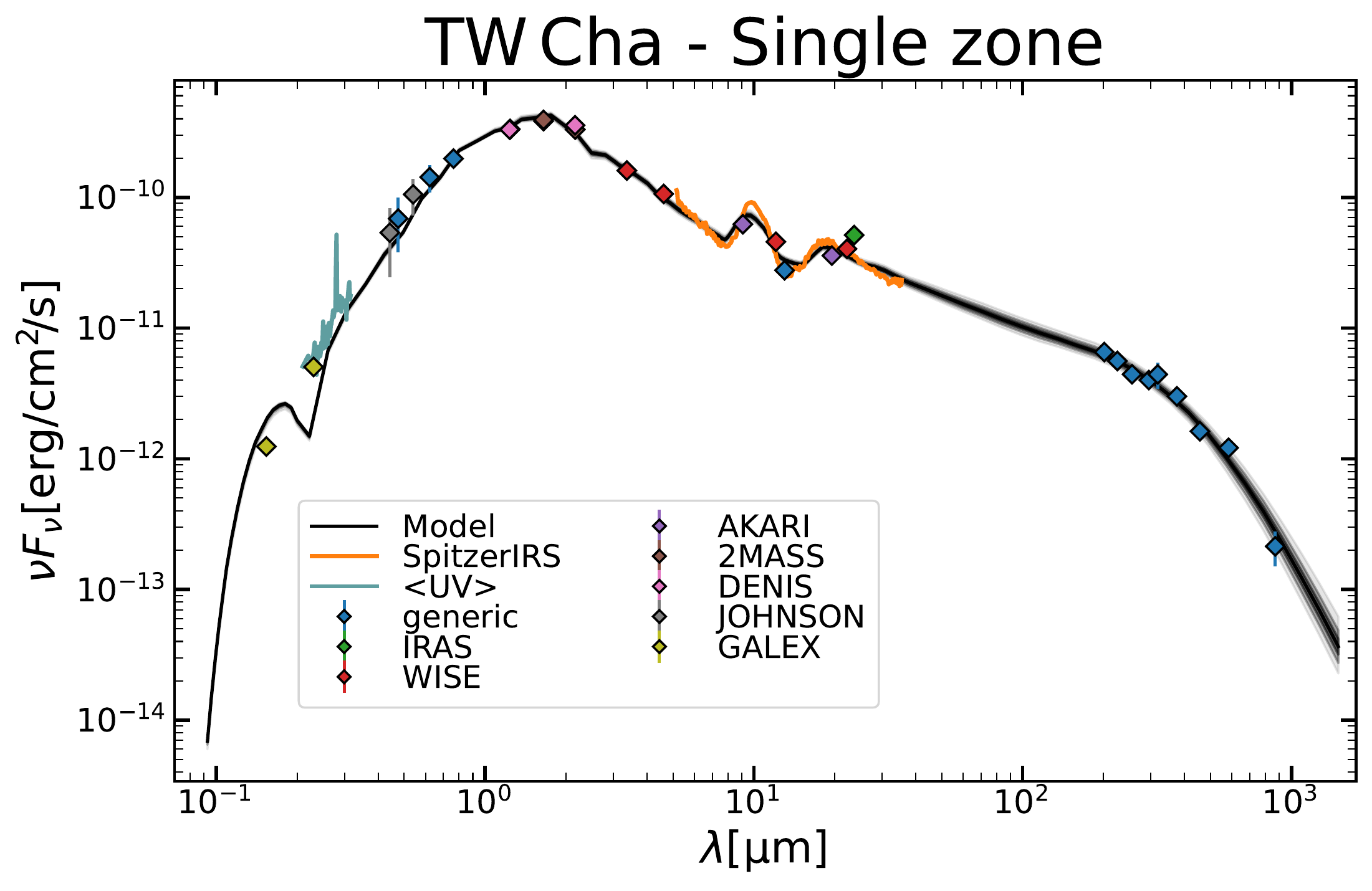}\\
     \includegraphics[width=0.49\linewidth]{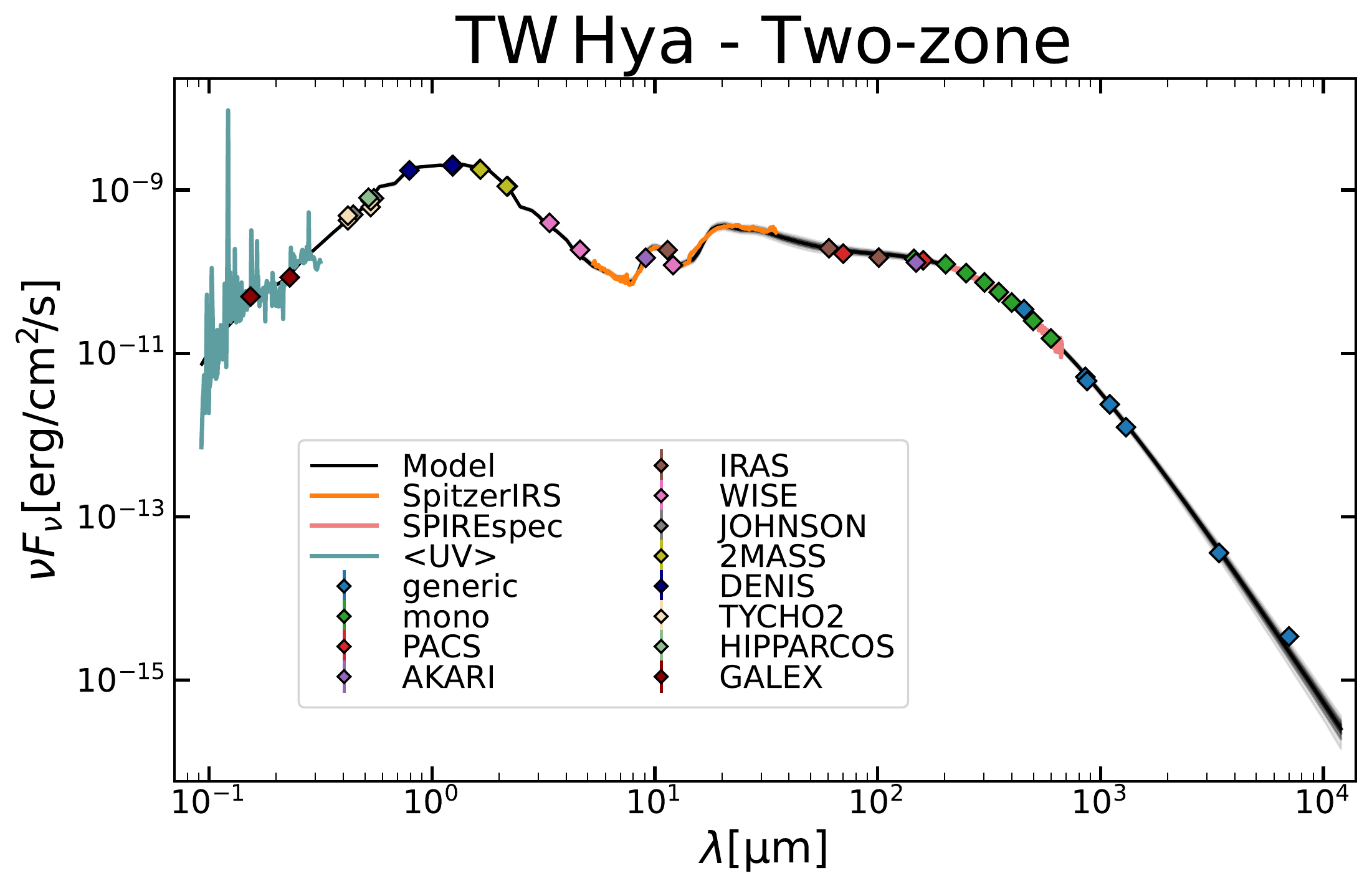}
    \includegraphics[width=0.49\linewidth]{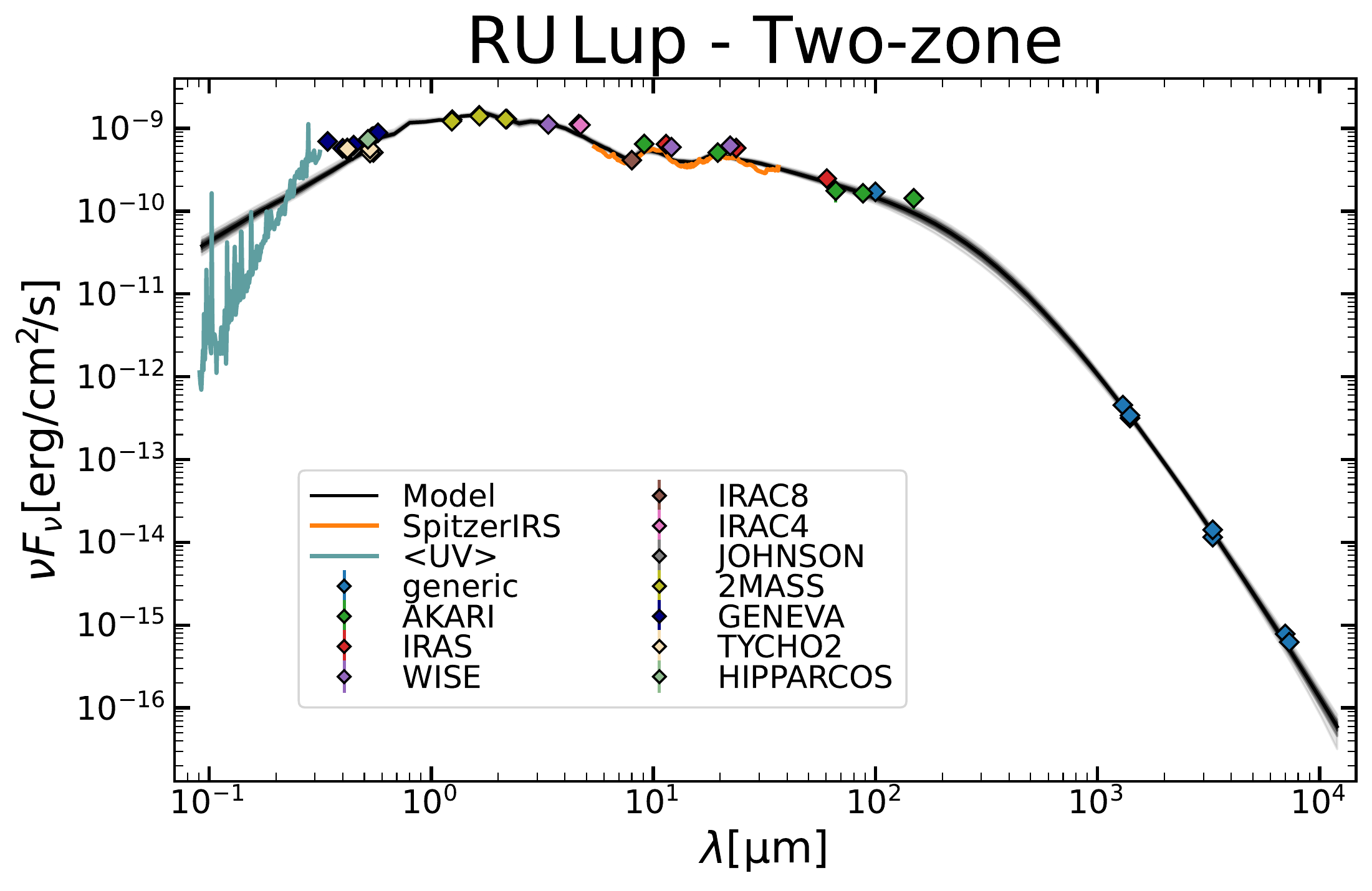}\\
    \includegraphics[width=0.49\linewidth]{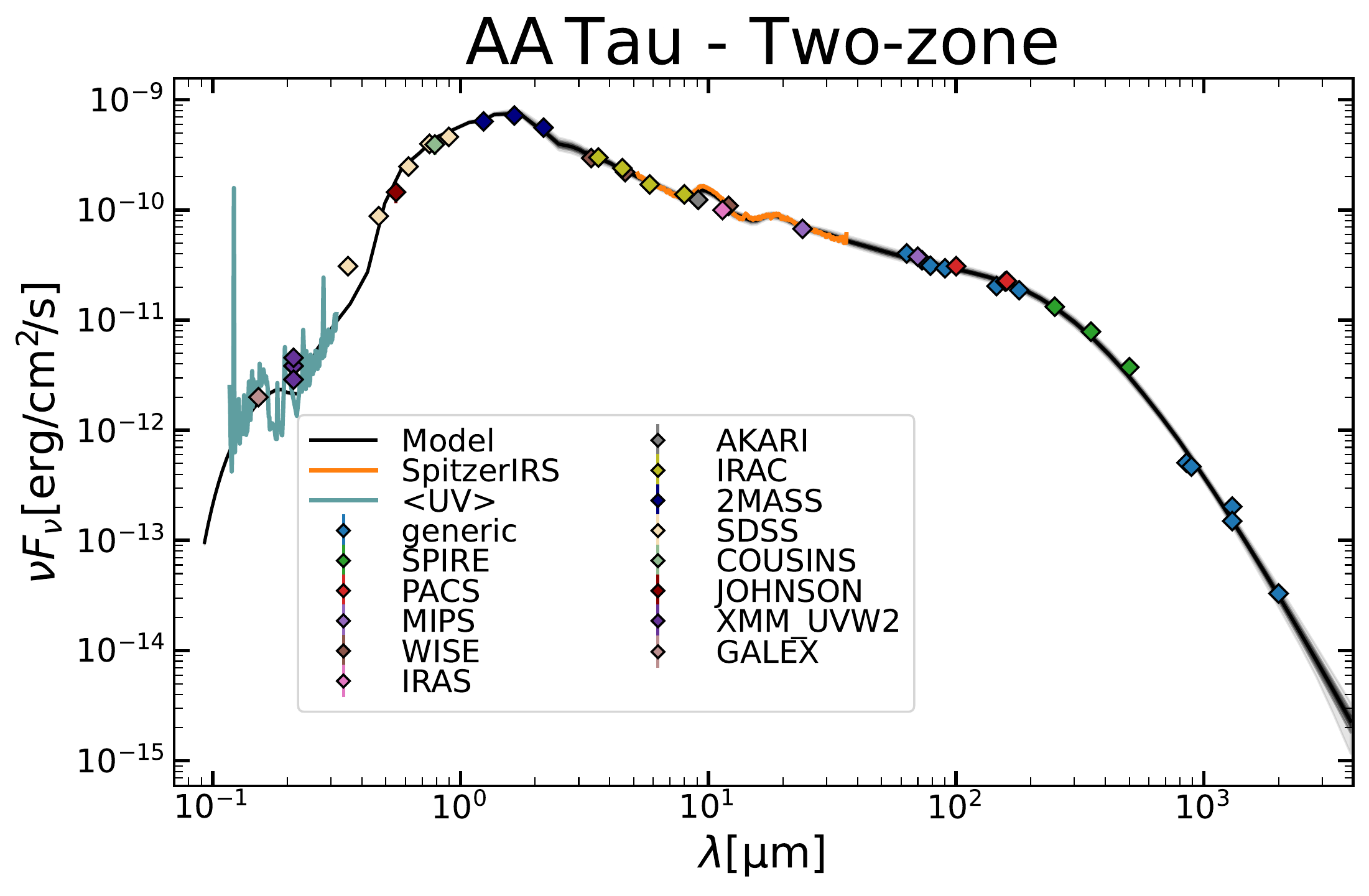}
    \includegraphics[width=0.49\linewidth]{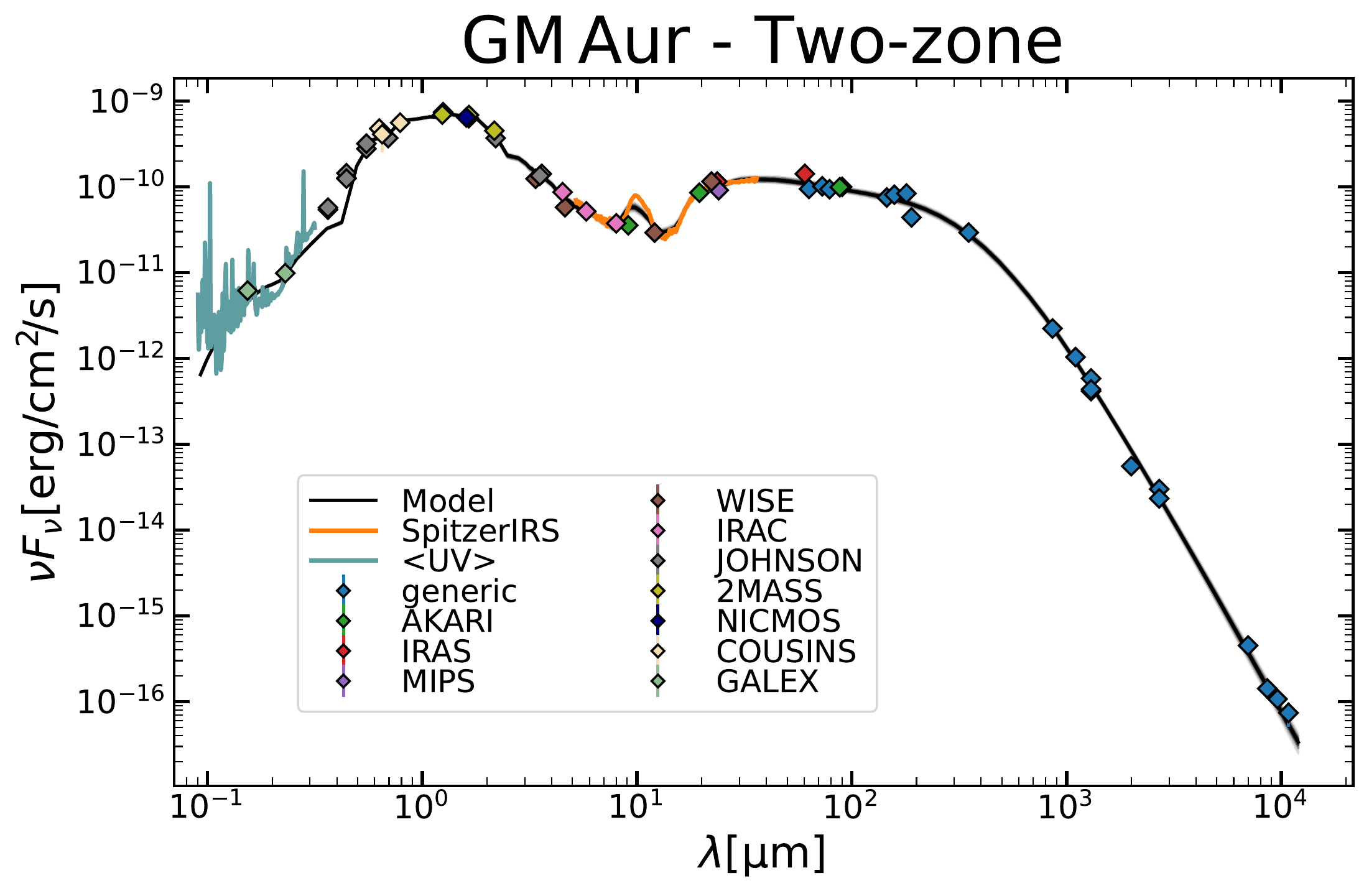}

\caption{continued.}
\end{figure*}

\addtocounter{figure}{-1}
\begin{figure*}[!hp]
\centering
    \includegraphics[width=0.49\linewidth]{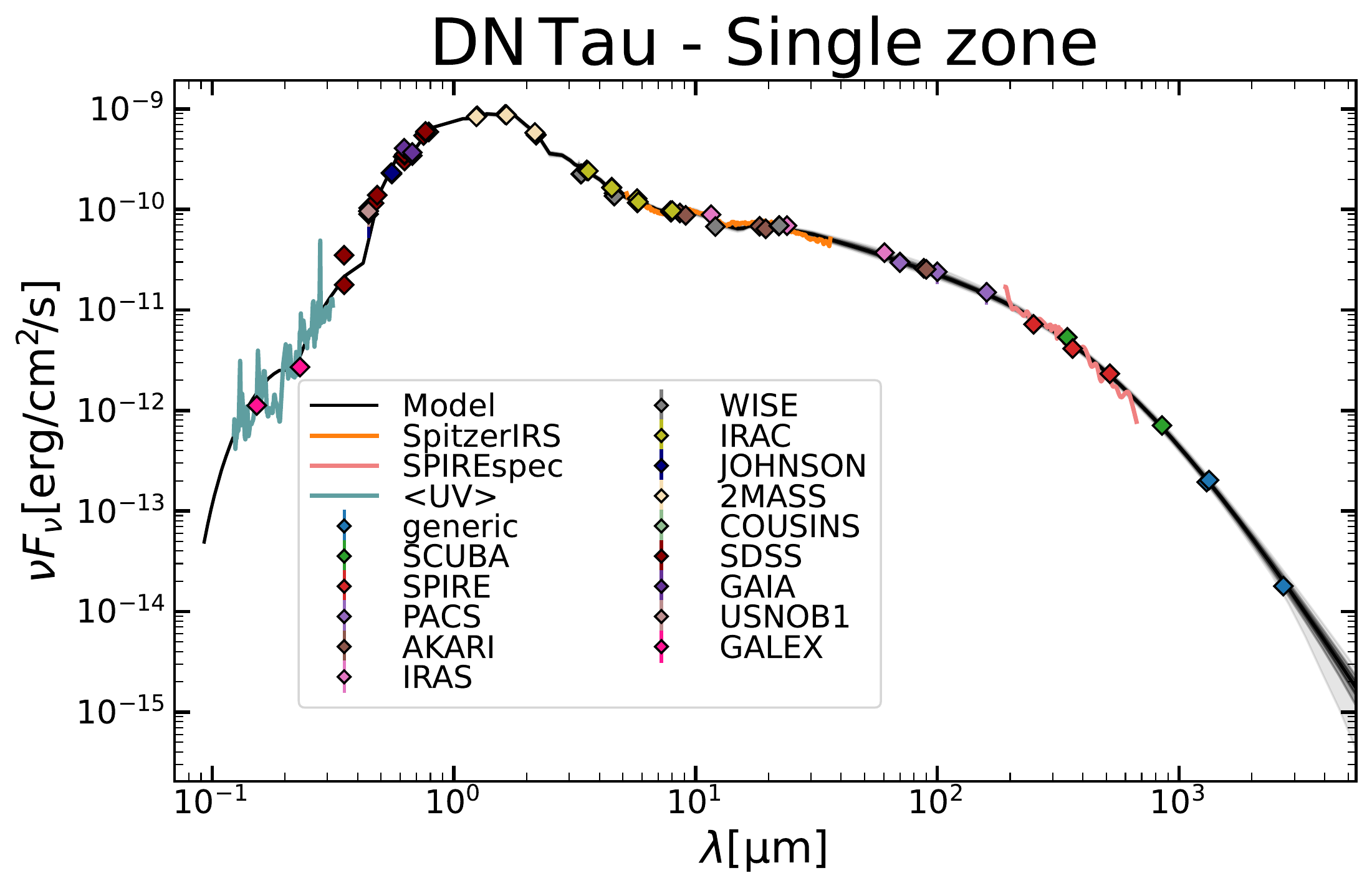}
    \includegraphics[width=0.49\linewidth]{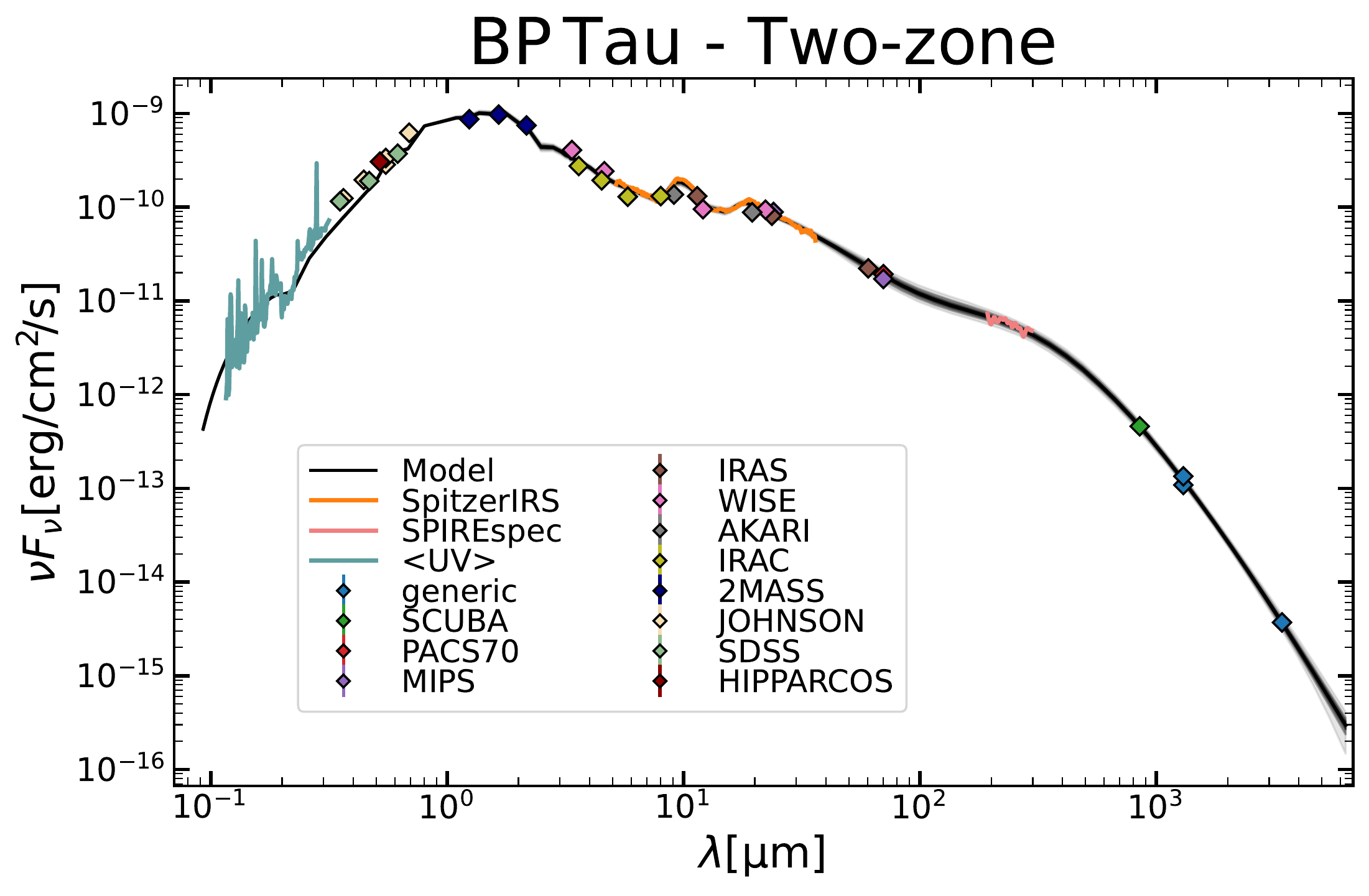}\\
    \includegraphics[width=0.49\linewidth]{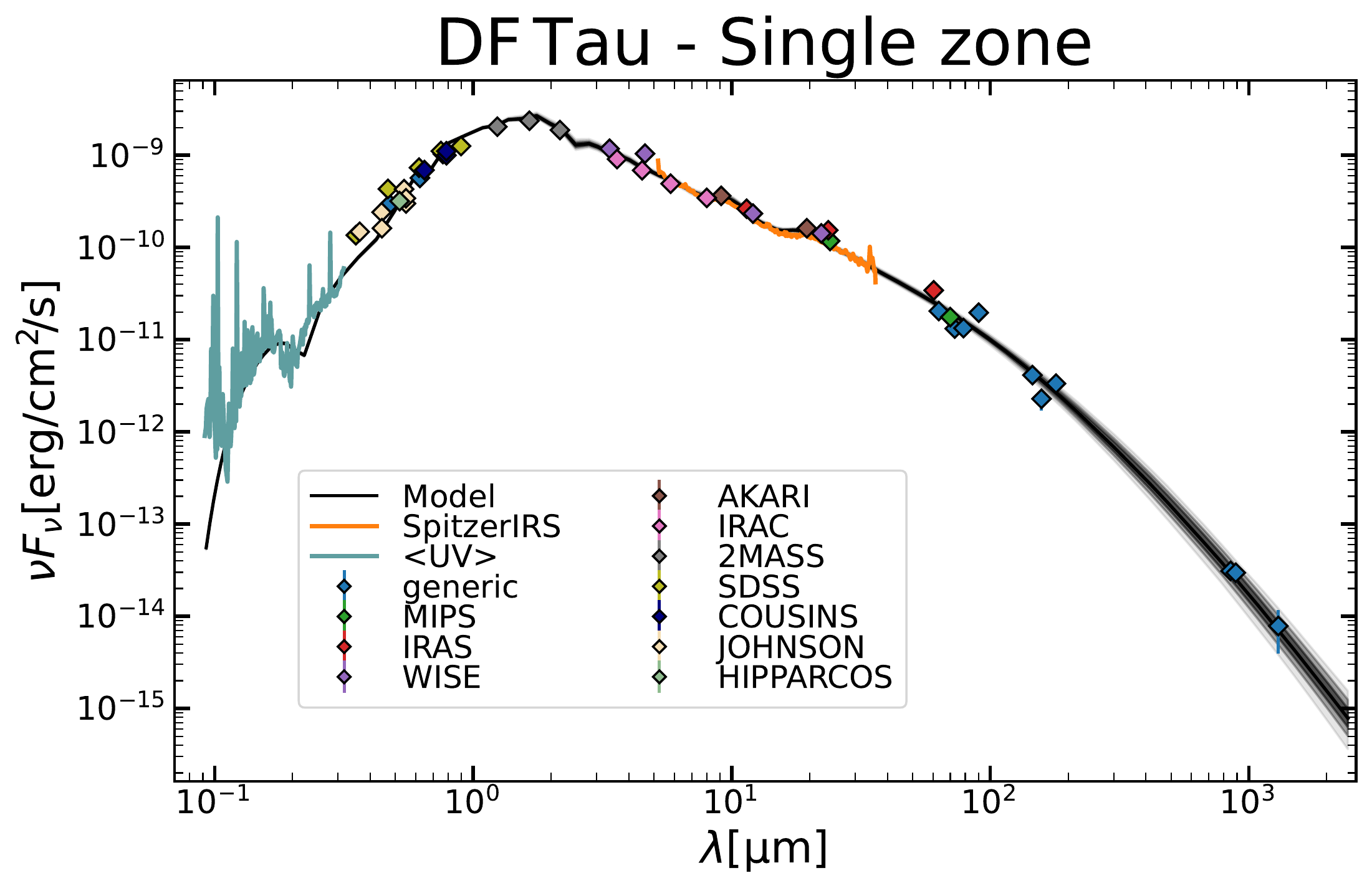}
    \includegraphics[width=0.49\linewidth]{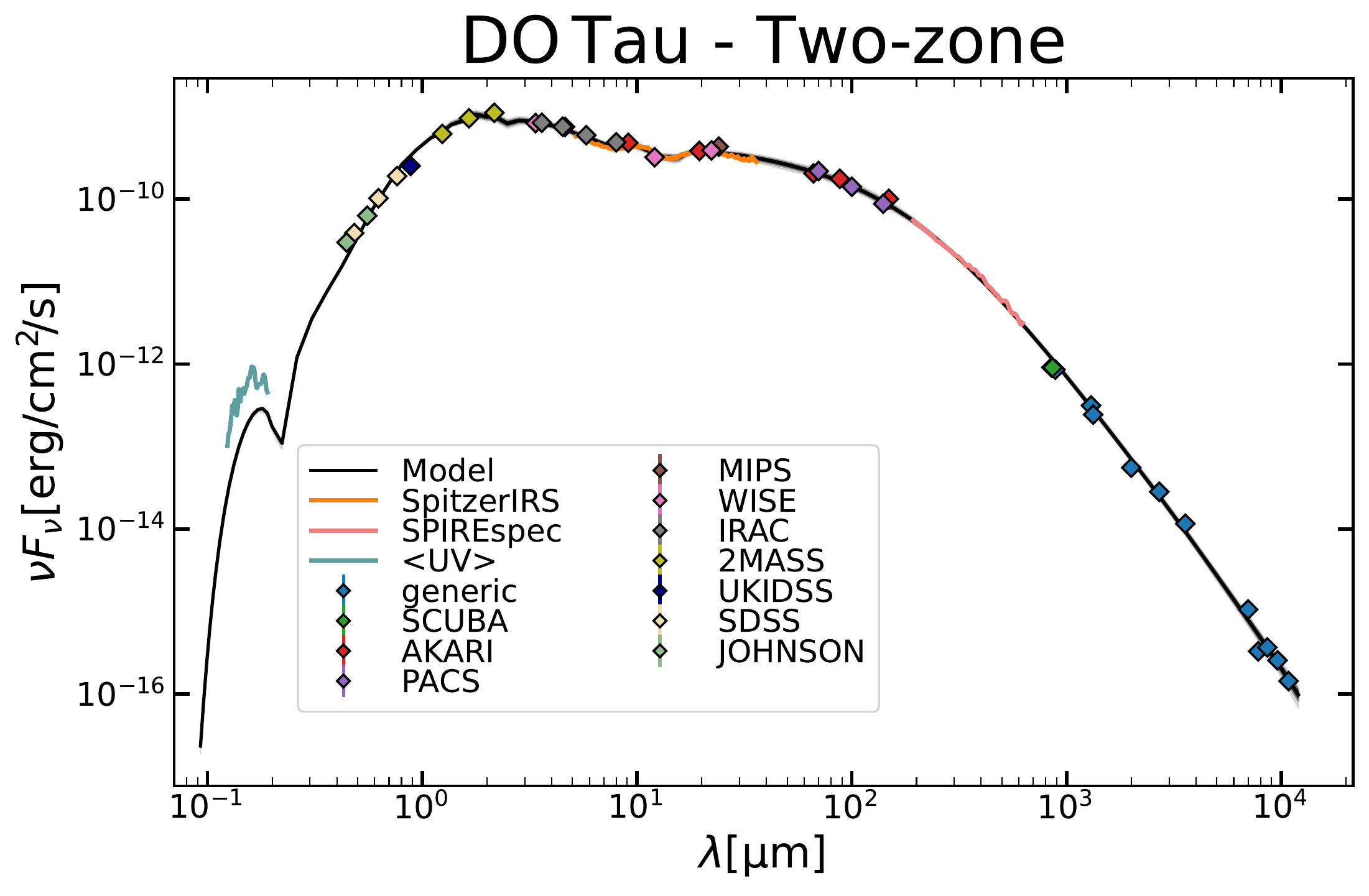}\\
    \includegraphics[width=0.49\linewidth]{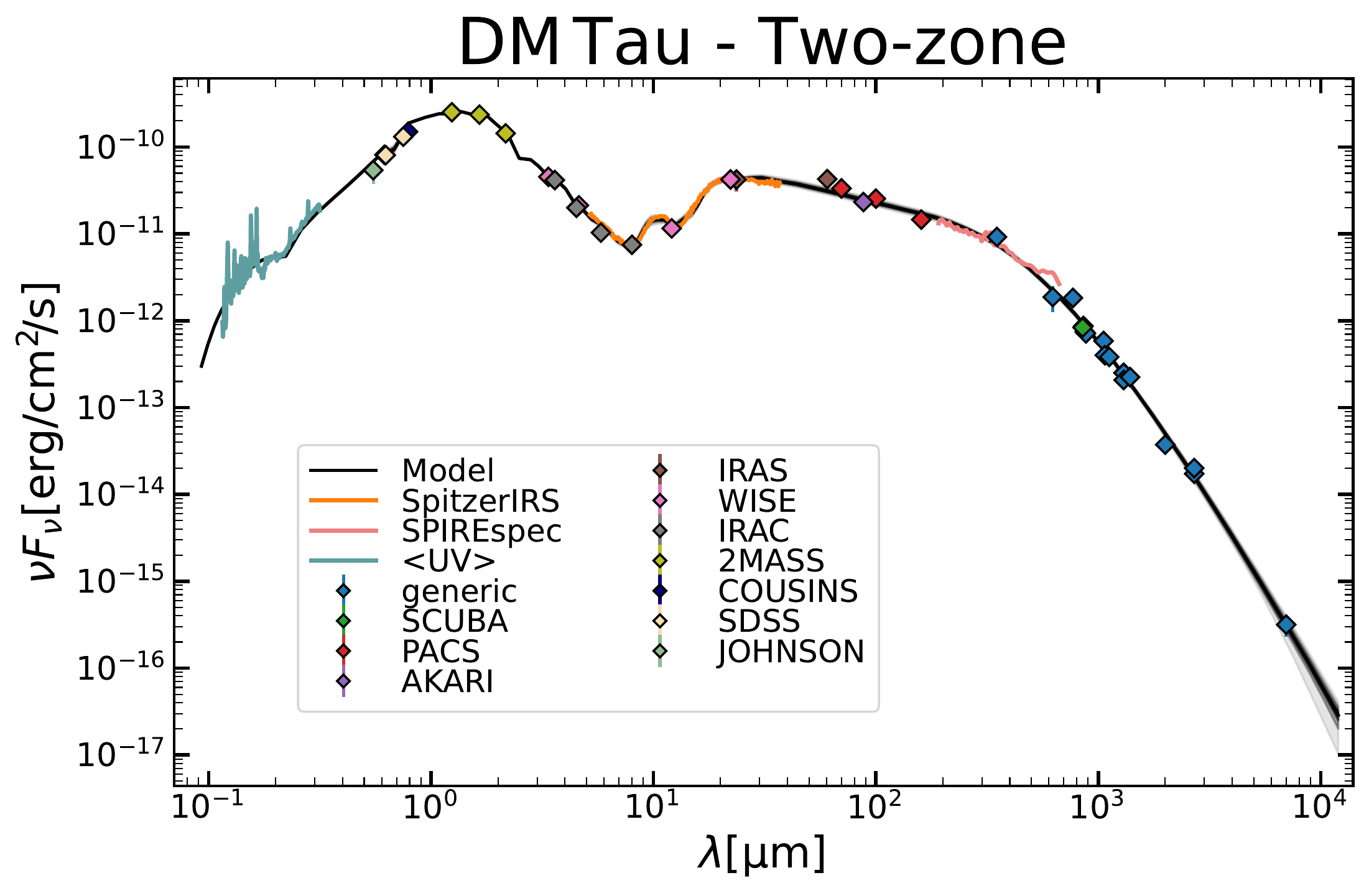}
    \includegraphics[width=0.49\linewidth]{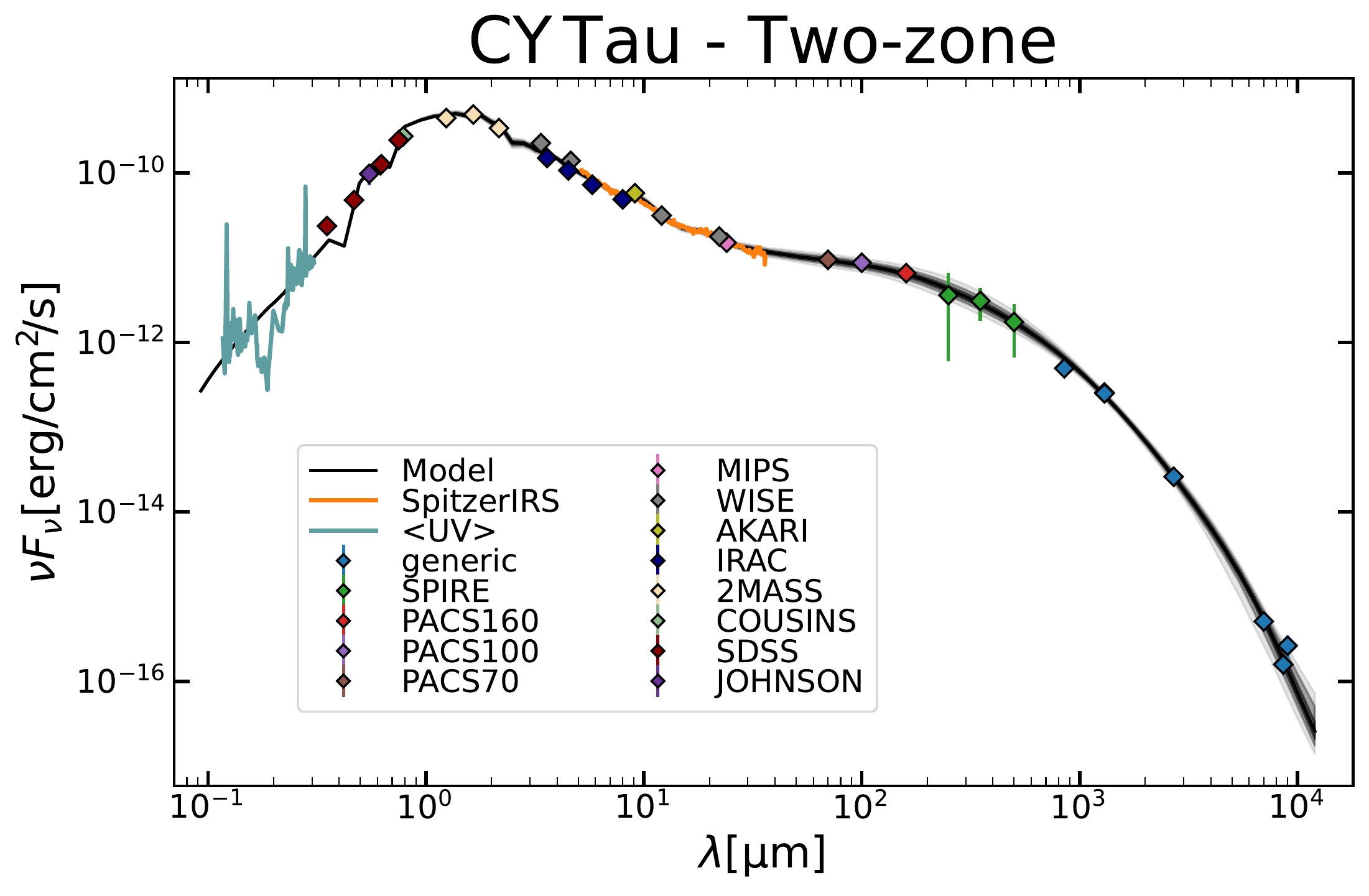}\\
    \includegraphics[width=0.49\linewidth]{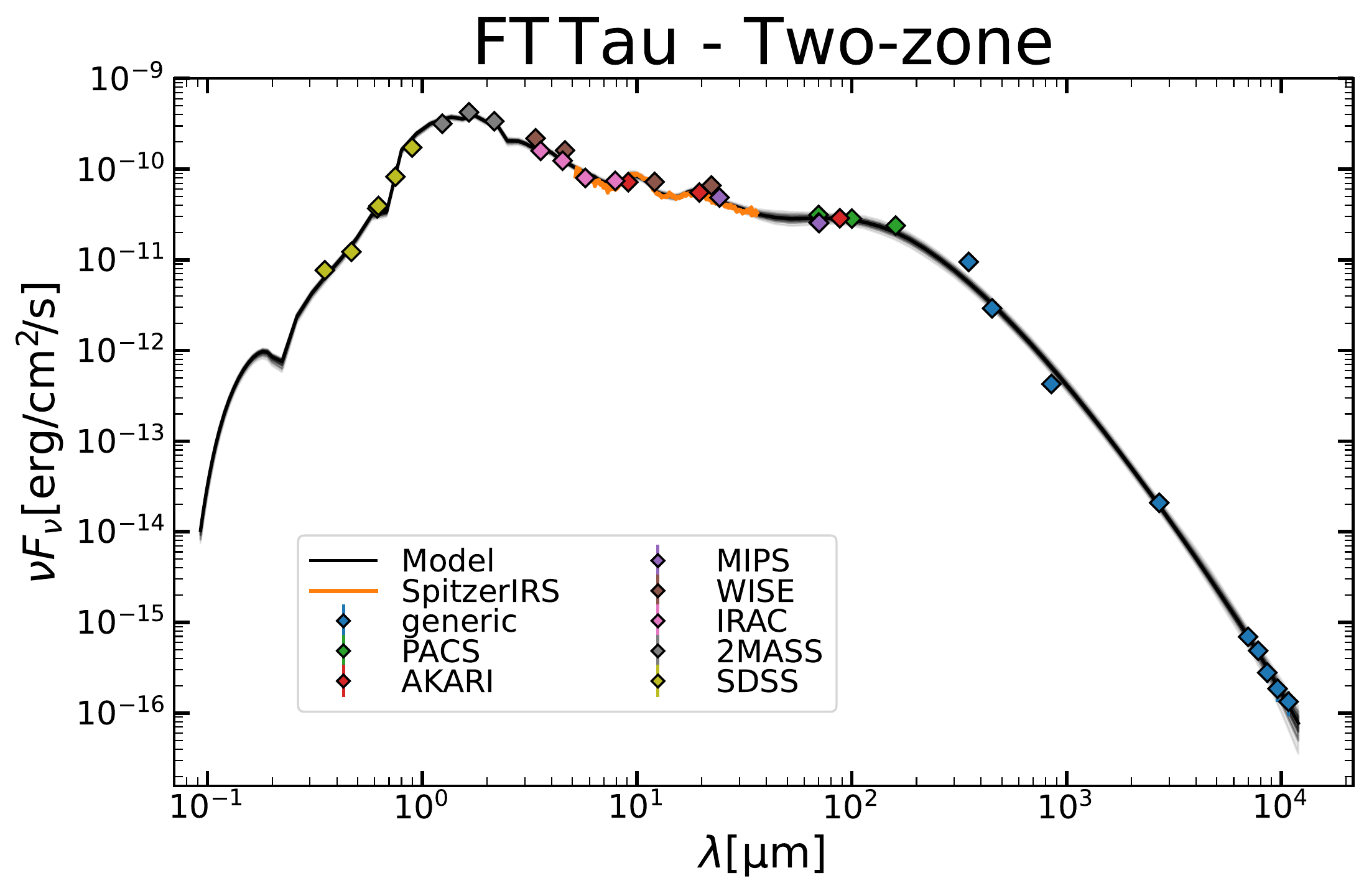}
    \includegraphics[width=0.49\linewidth]{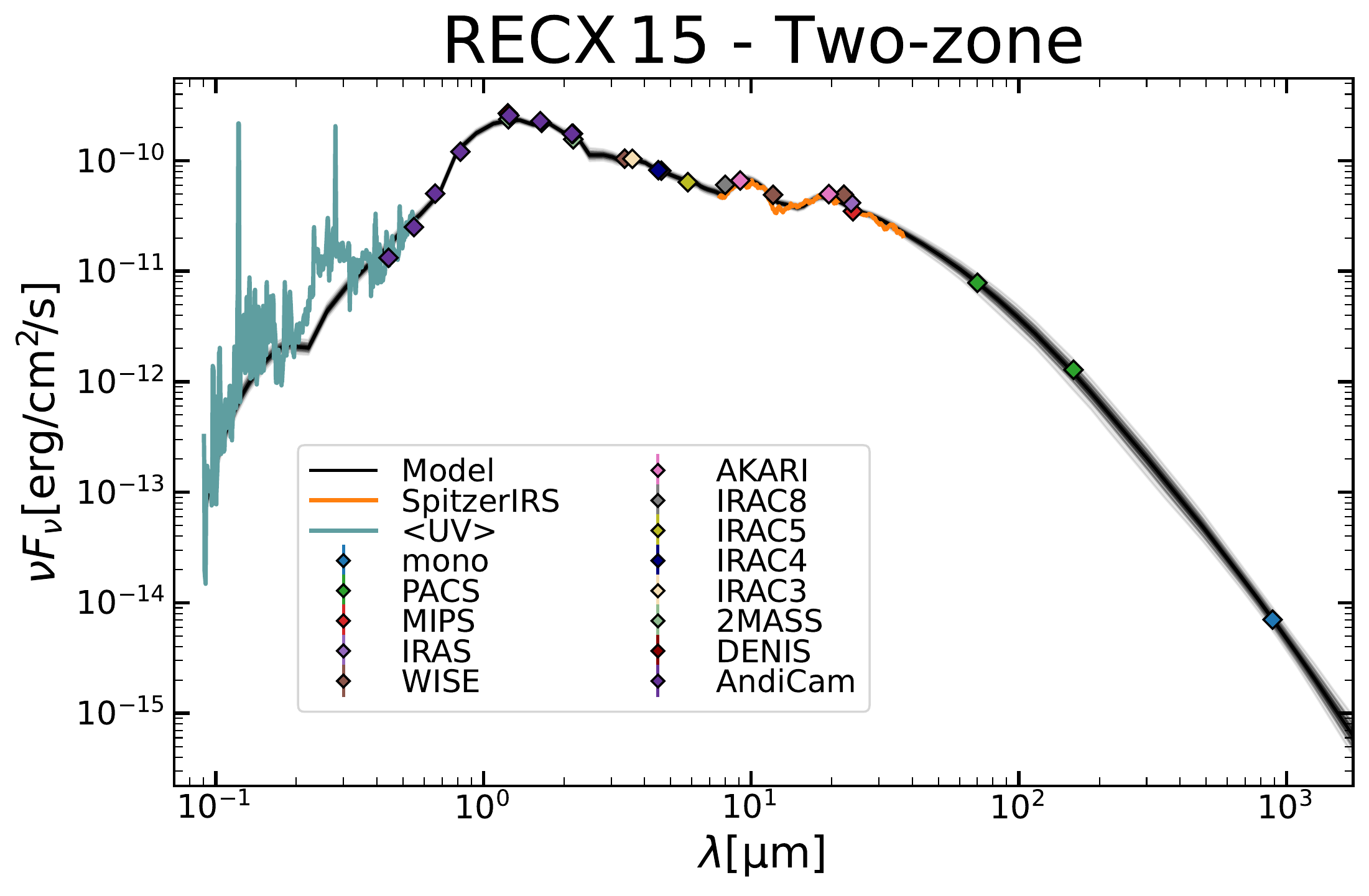}

\caption{continued.}
\end{figure*}

\begin{figure*}[!hp]
\centering
    \includegraphics[width=0.49\linewidth]{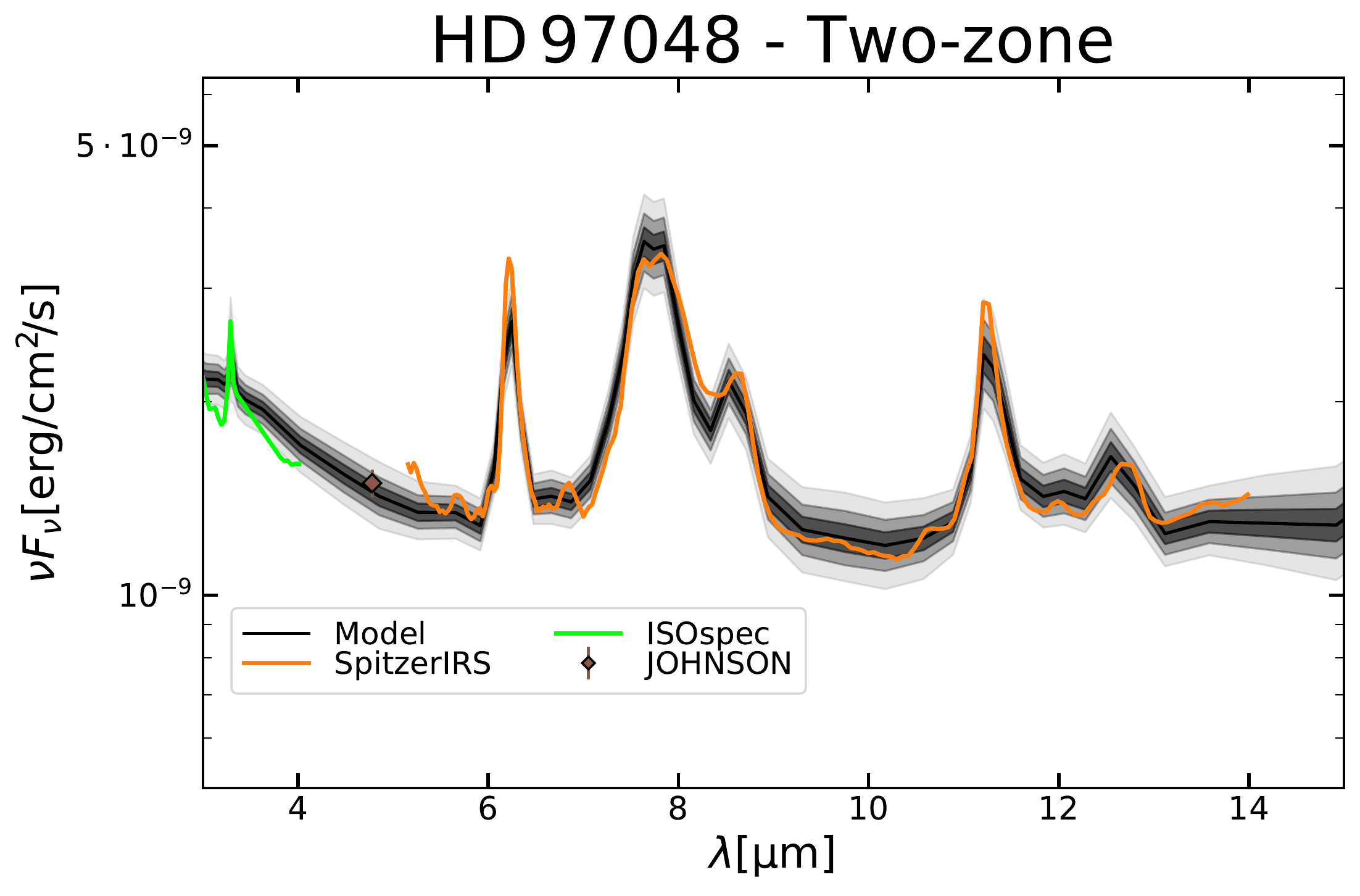}
    \includegraphics[width=0.49\linewidth]{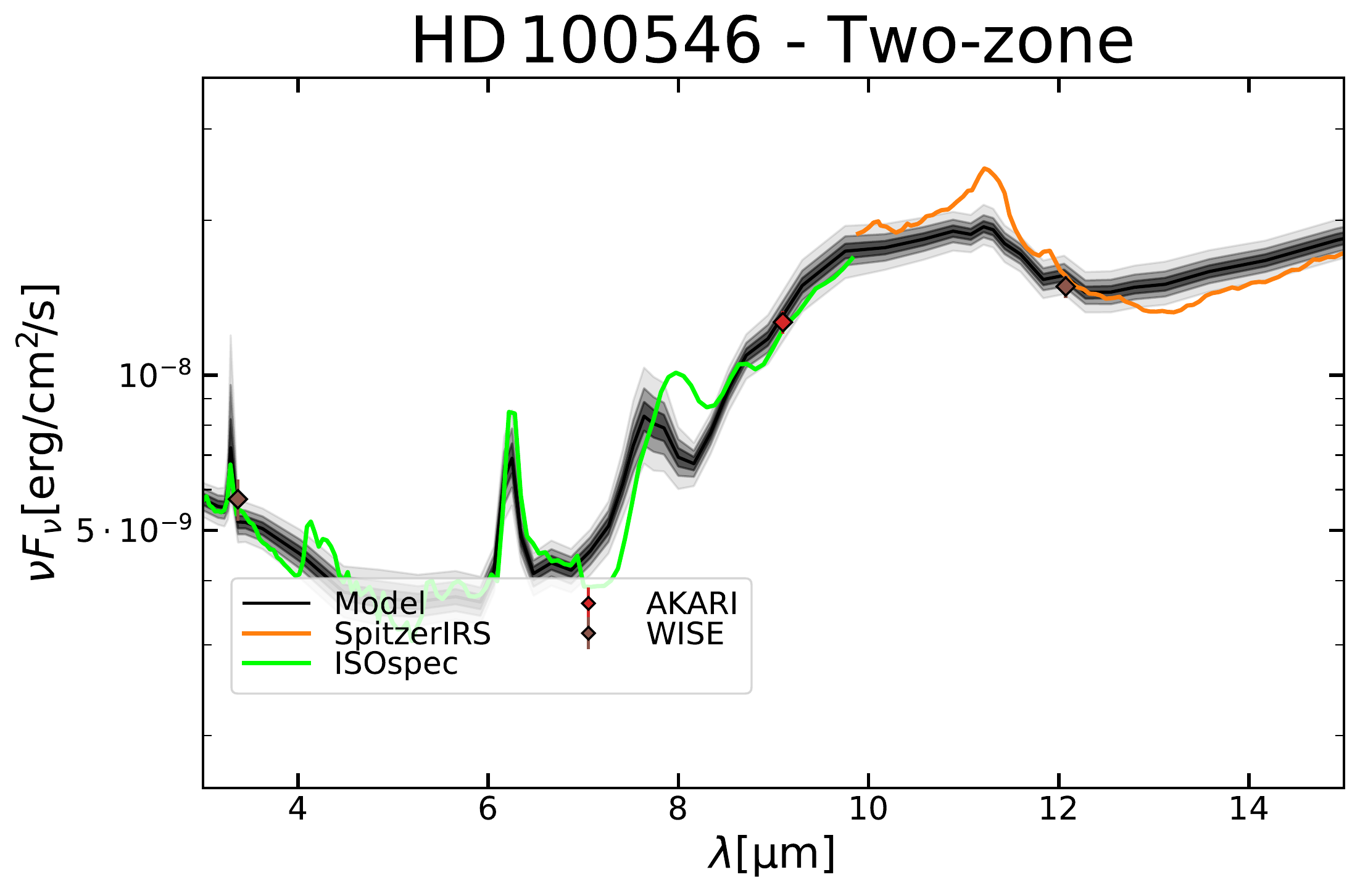}\\
    \includegraphics[width=0.49\linewidth]{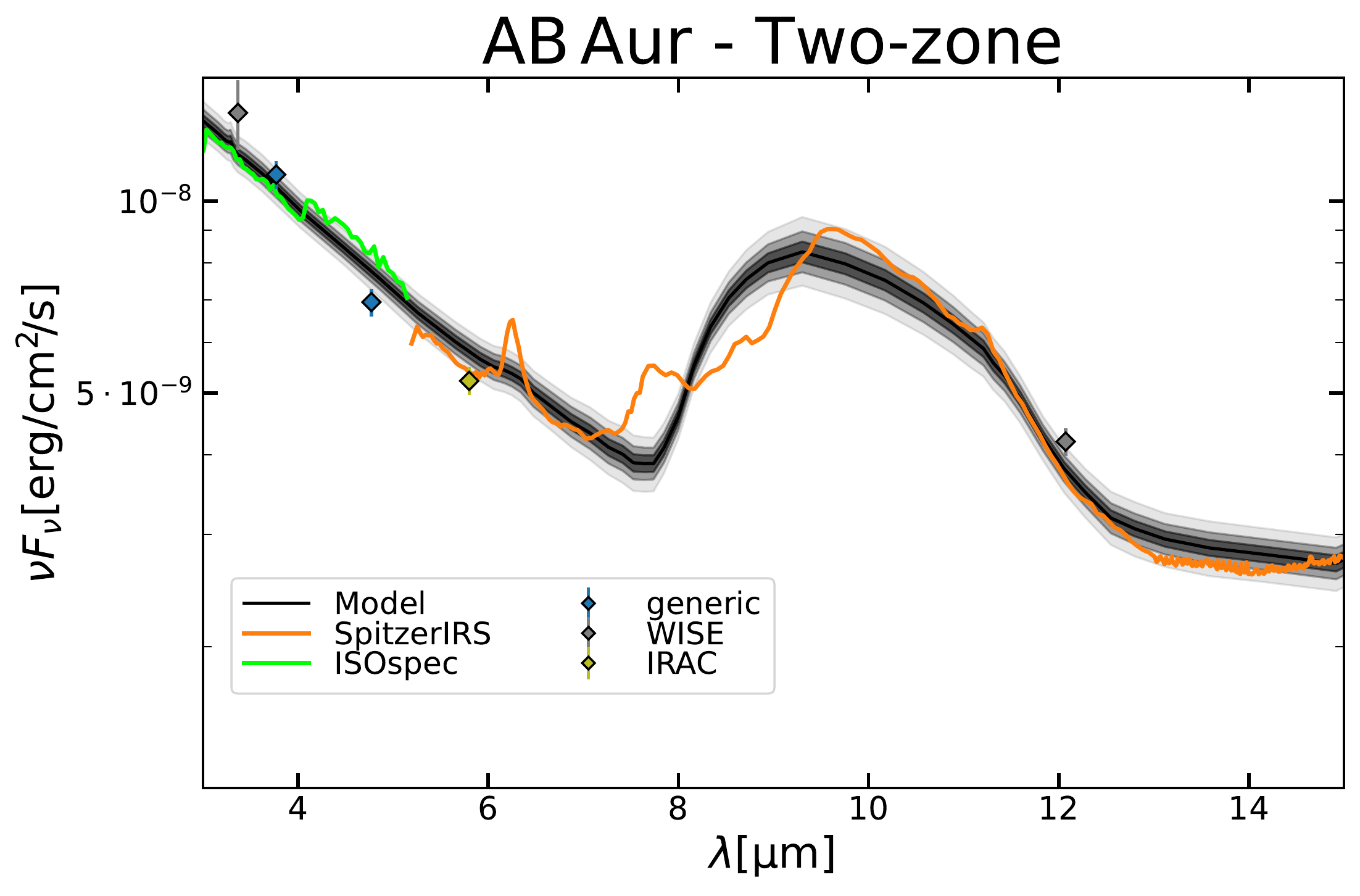}
    \includegraphics[width=0.49\linewidth]{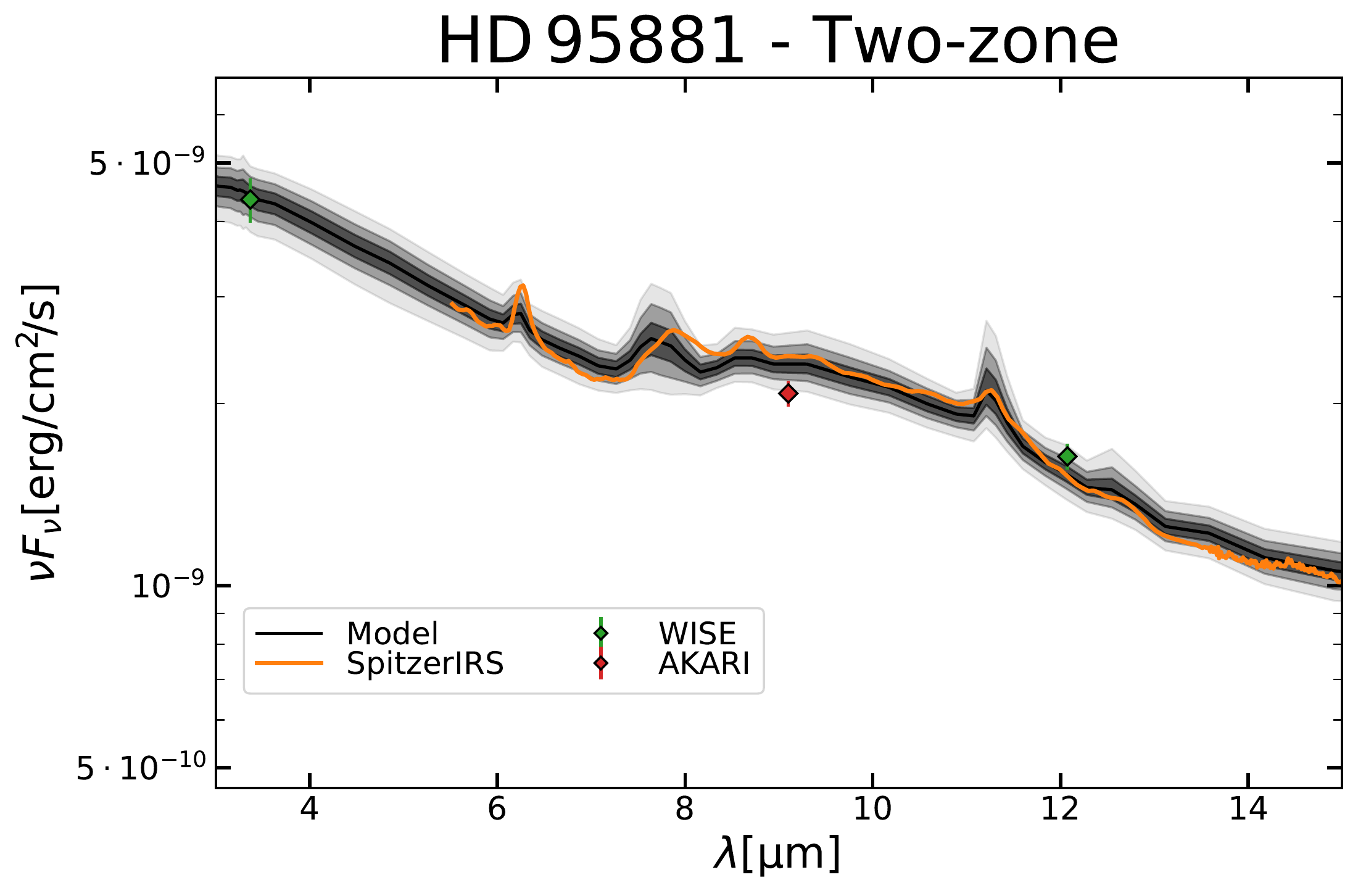}\\
    \includegraphics[width=0.49\linewidth]{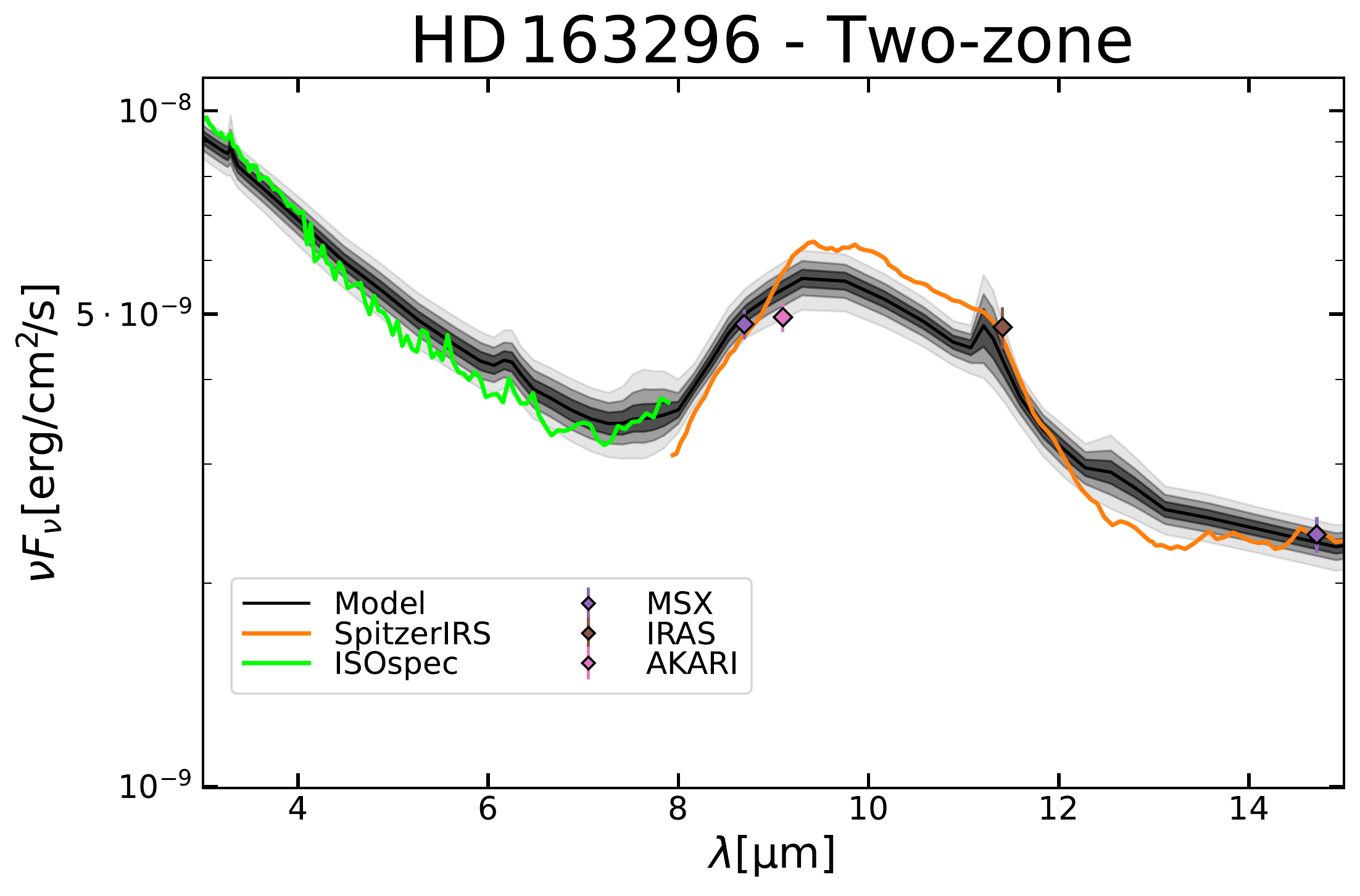}
    \includegraphics[width=0.49\linewidth]{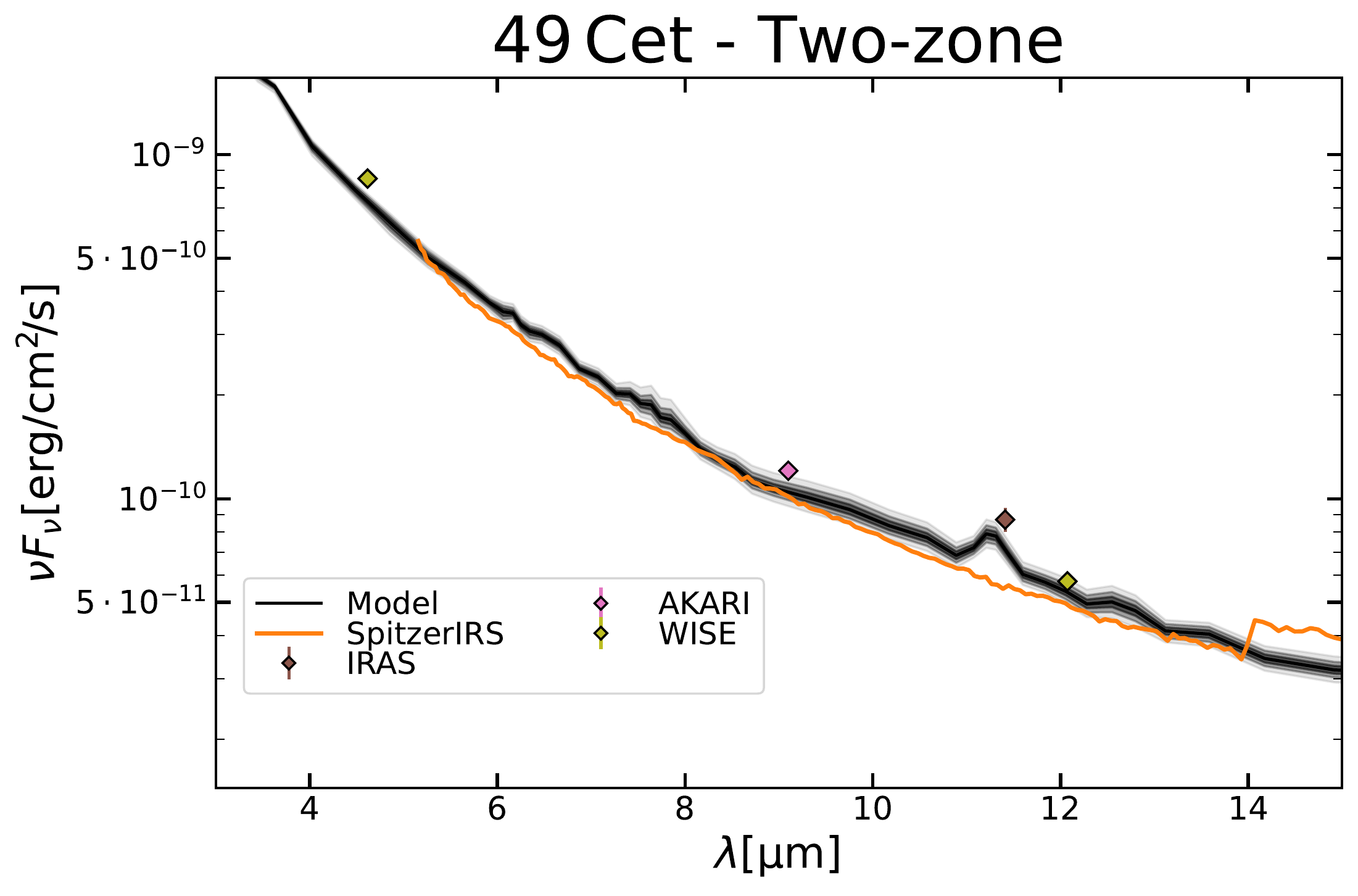}
\caption{Mid-IR region of the SED for all objects. The coloured points and lines indicate the observation listed in the legend. The SEDs from the posterior distribution are shown in black. The line denotes the median of all model SED, with the dark, medium, and light black areas denoting the $68\,\%$, $95\,\%$, and $99.9\,\%$ percentiles, respectively. The name of the model and if it is fitted with a single or two-zone model is given in the title. }
\label{fig:midir_fits}
\end{figure*}

\addtocounter{figure}{-1}
\begin{figure*}[!hp]
\centering
    \includegraphics[width=0.49\linewidth]{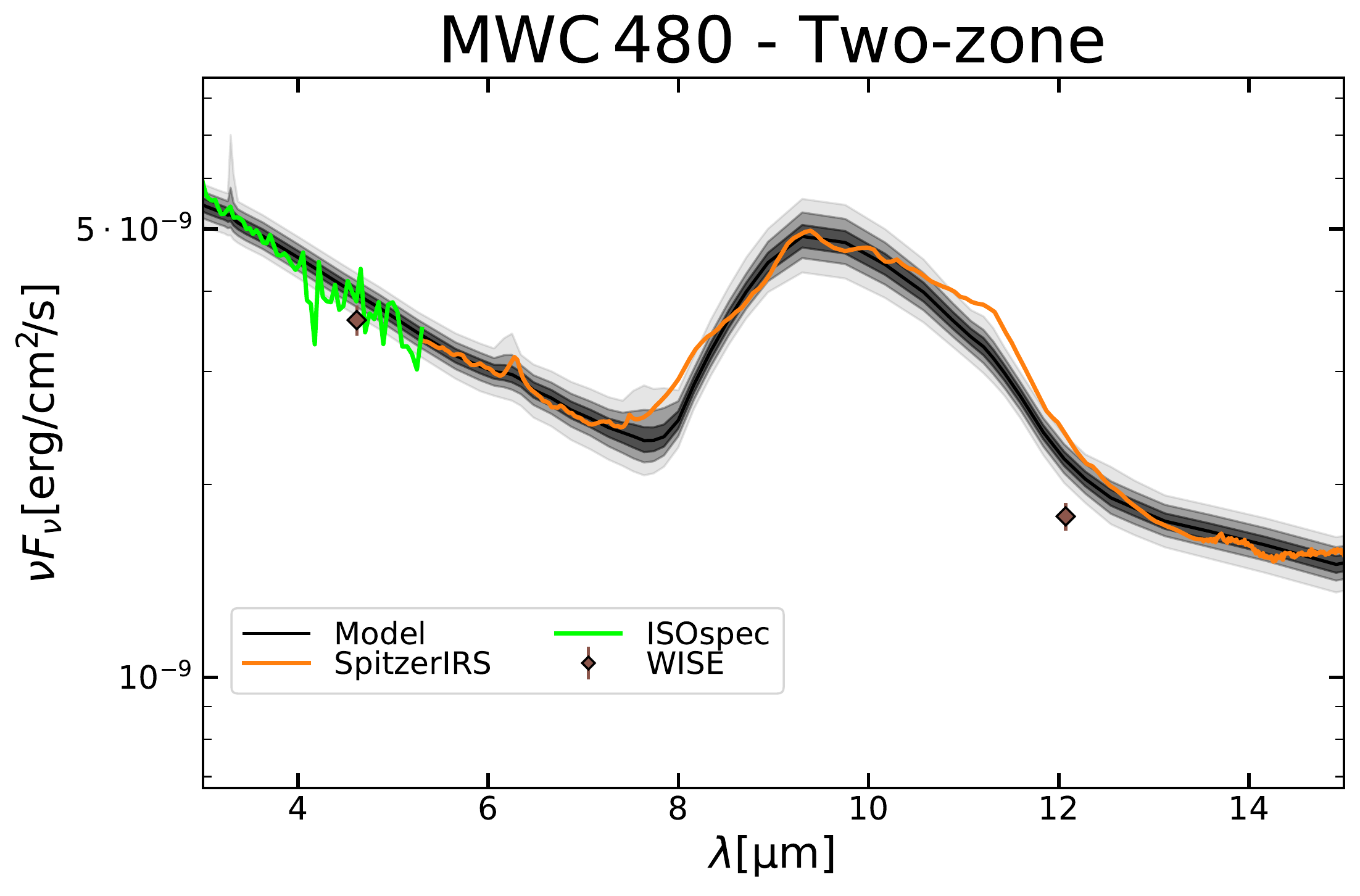}
    \includegraphics[width=0.49\linewidth]{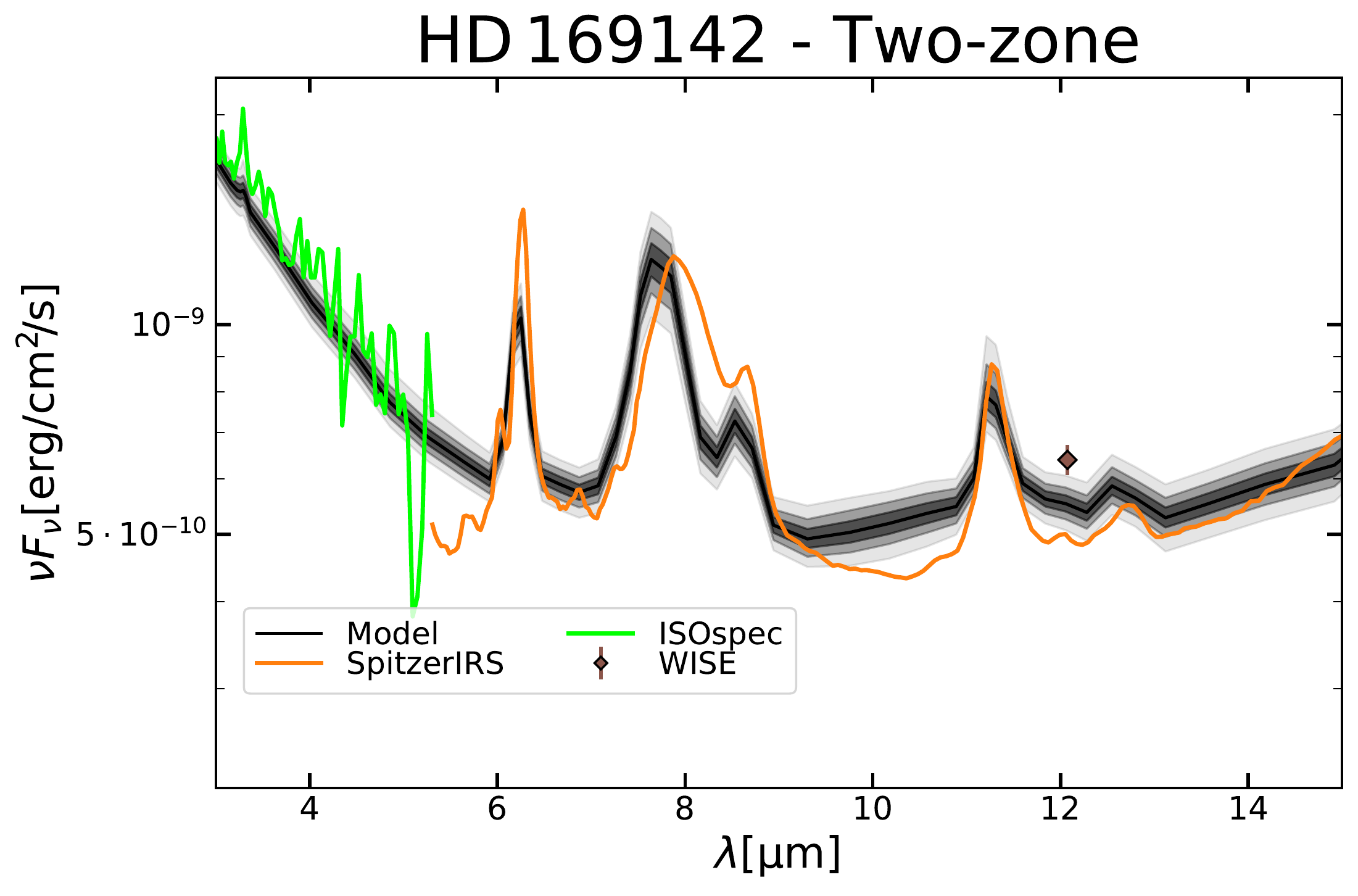}\\
    \includegraphics[width=0.49\linewidth]{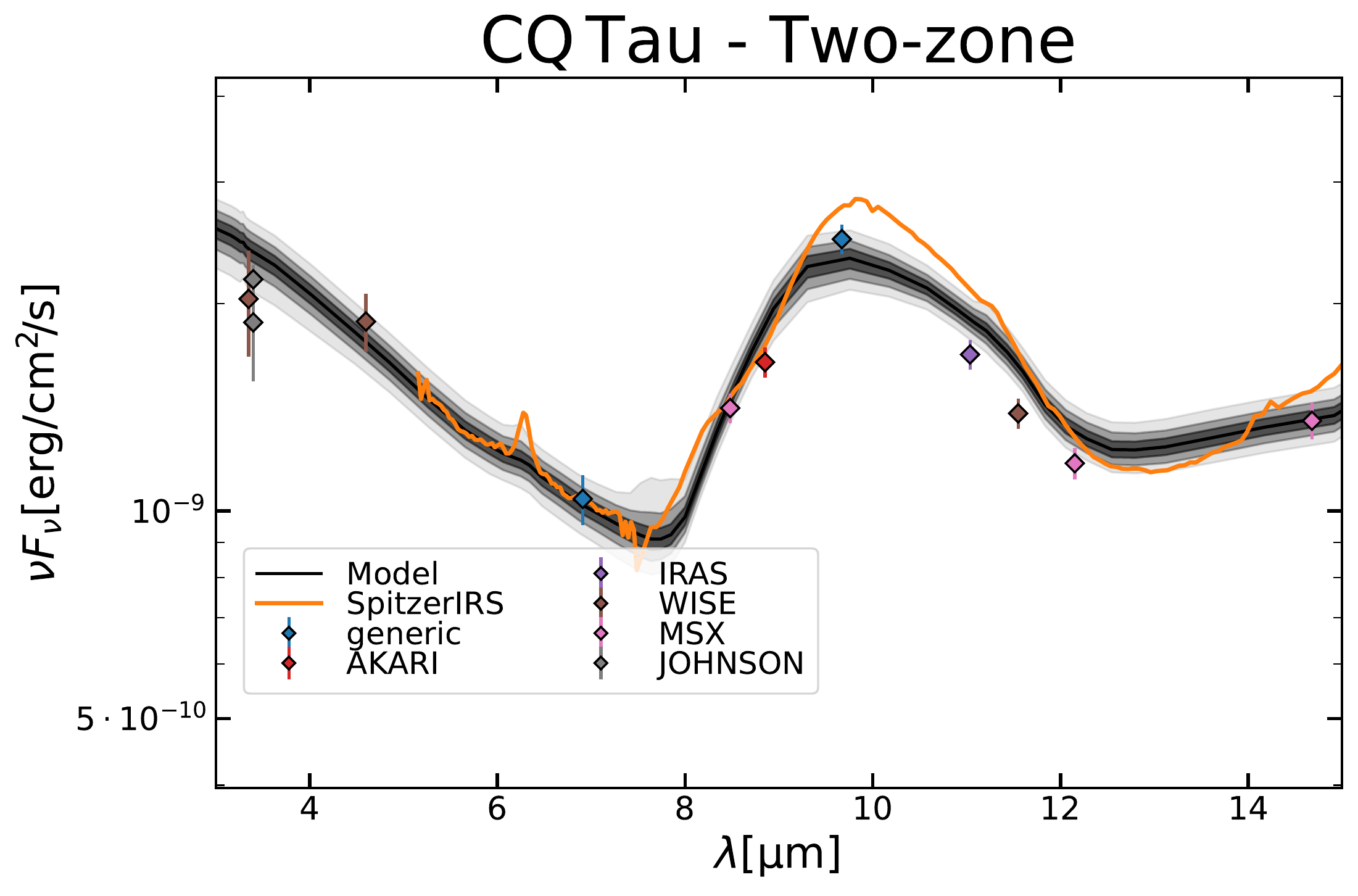}
    \includegraphics[width=0.49\linewidth]{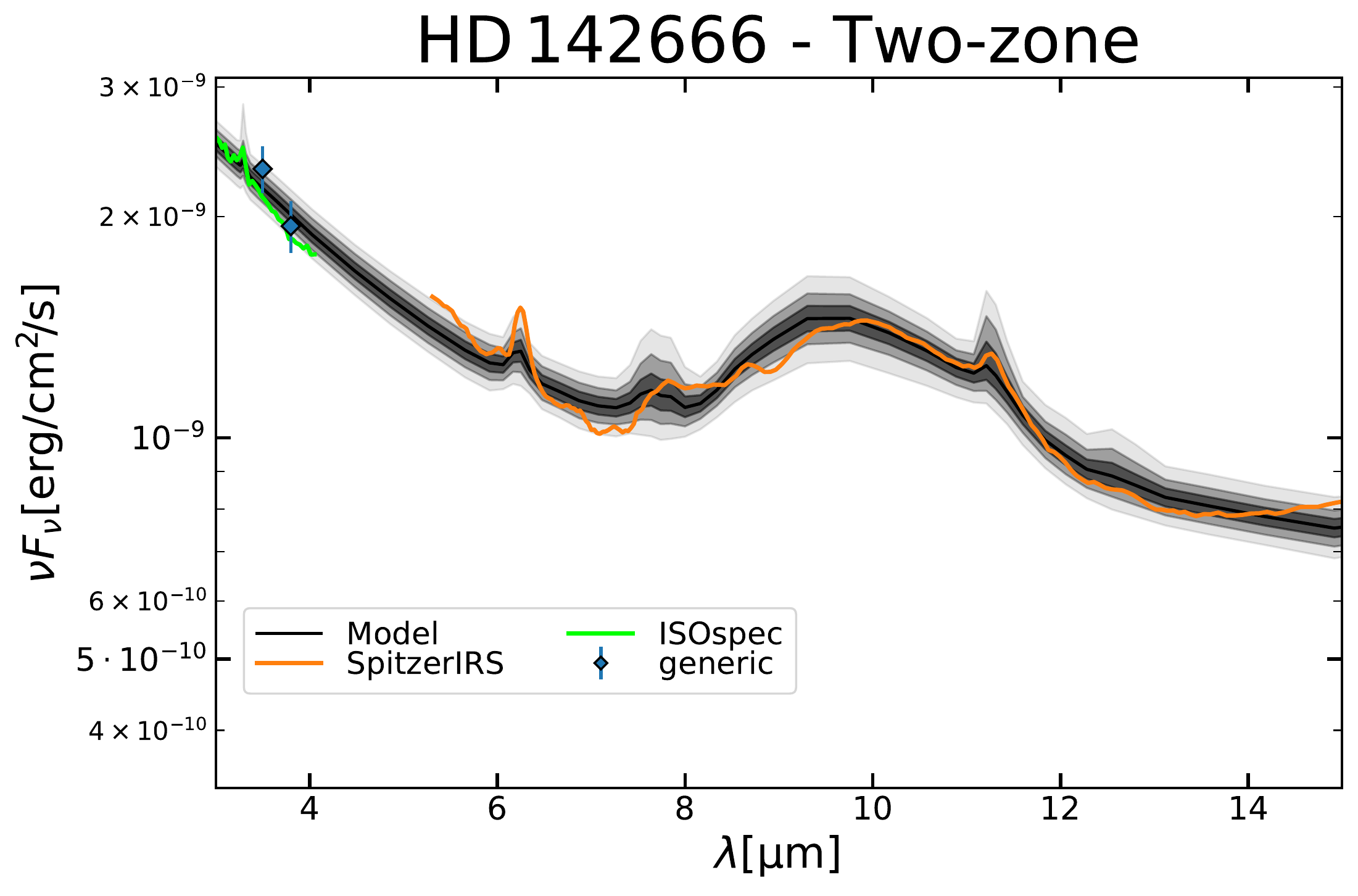}\\
    \includegraphics[width=0.49\linewidth]{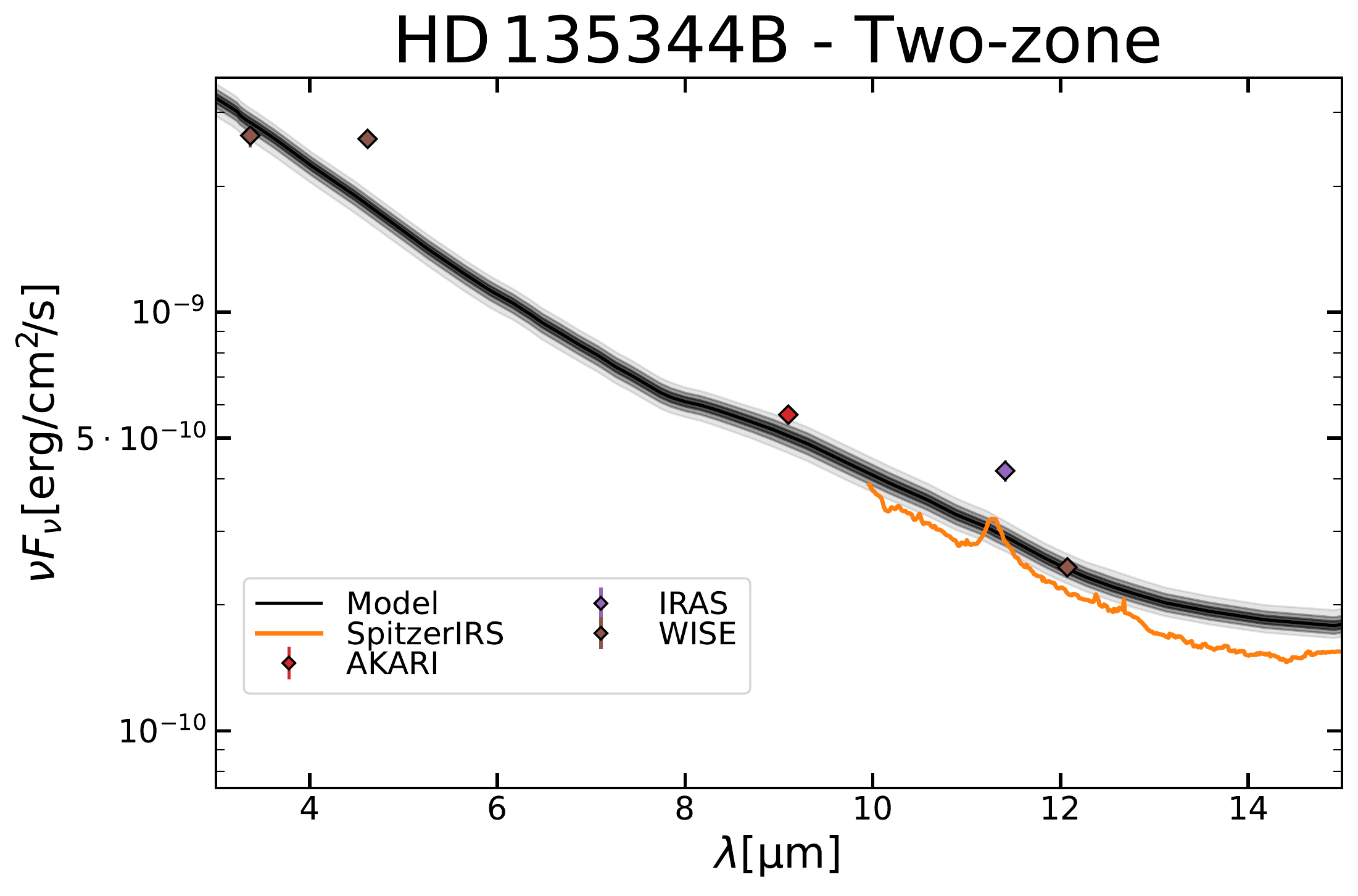}
    \includegraphics[width=0.49\linewidth]{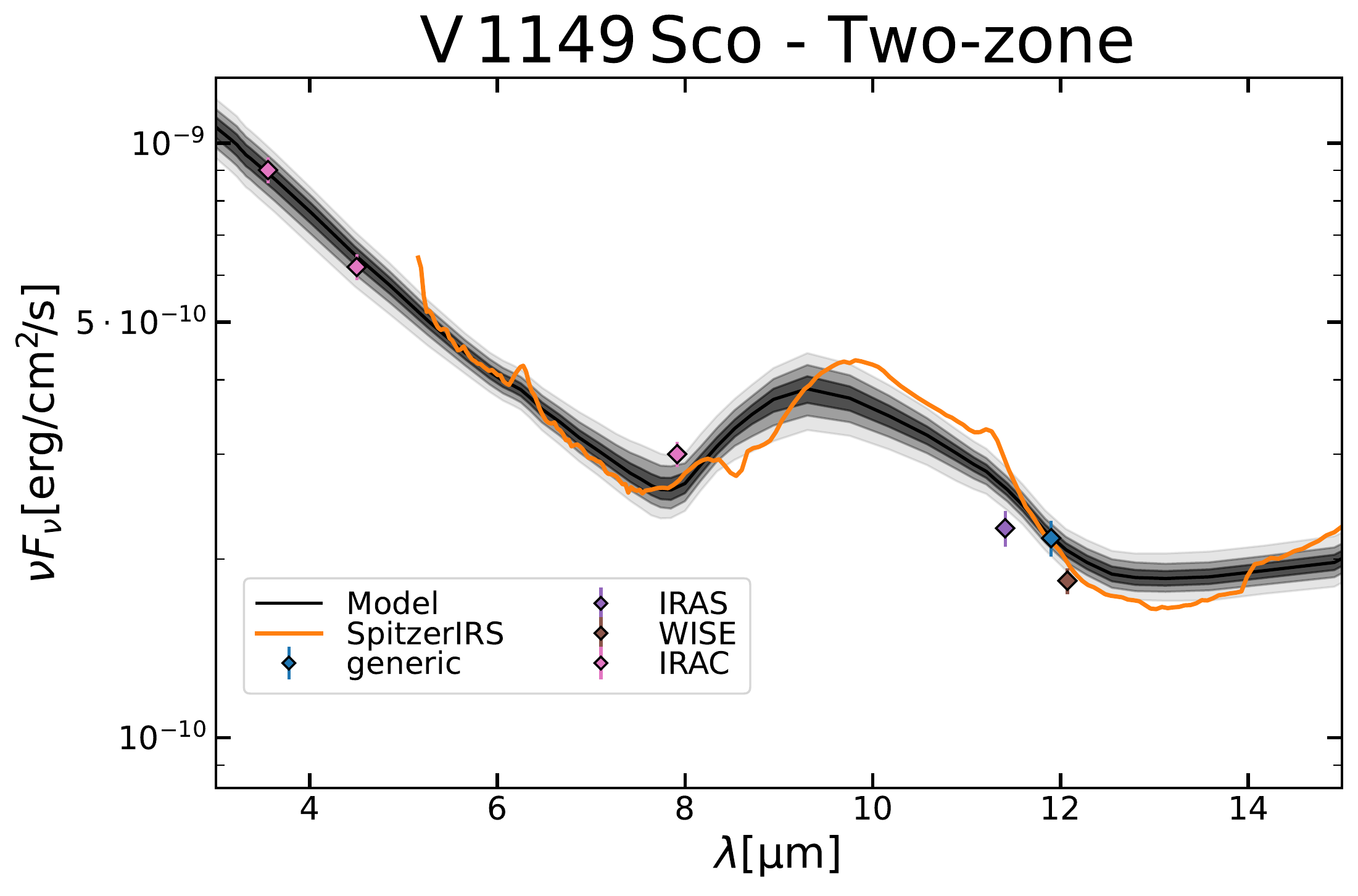}\\
    \includegraphics[width=0.49\linewidth]{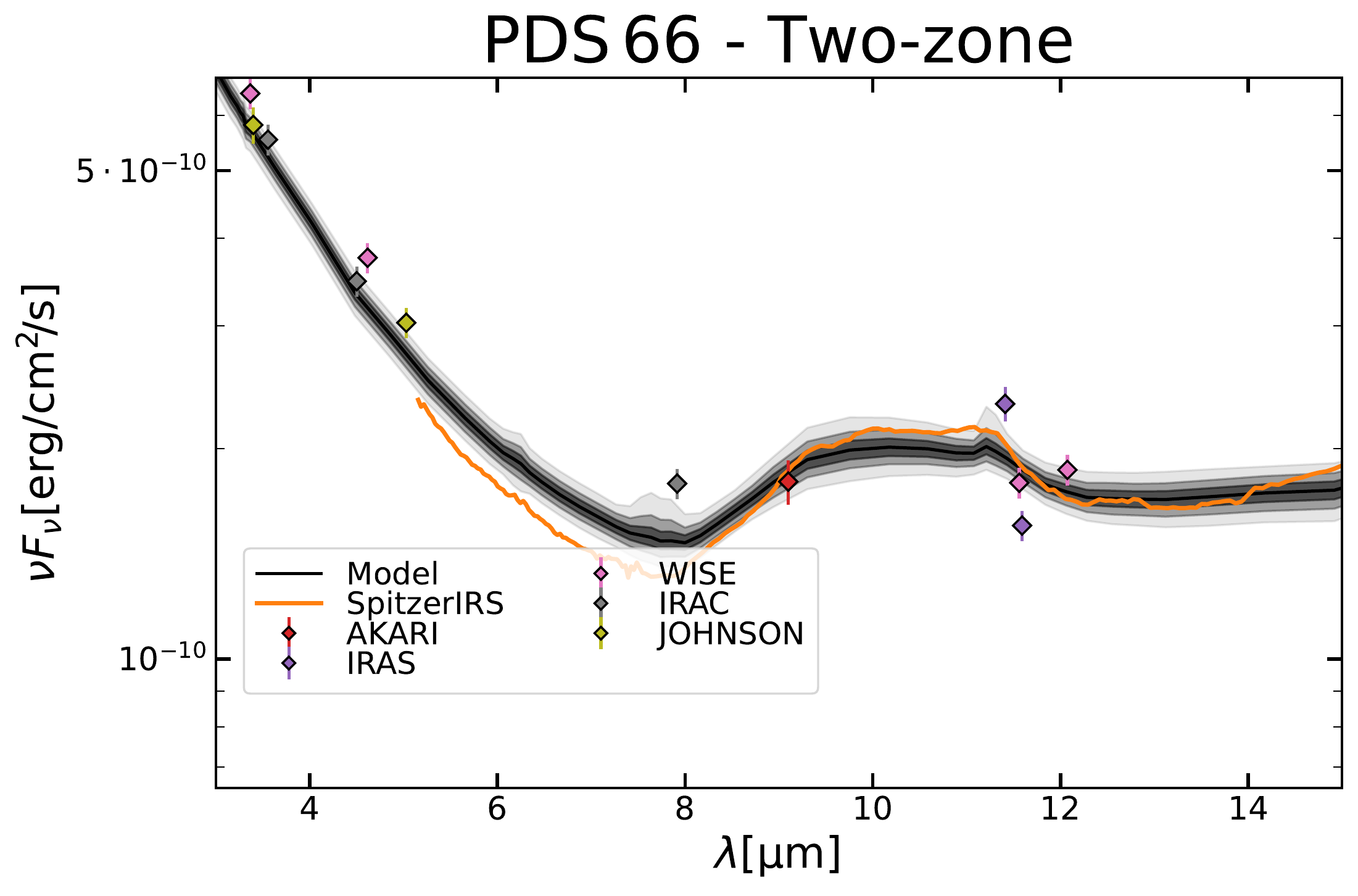}
    \includegraphics[width=0.49\linewidth]{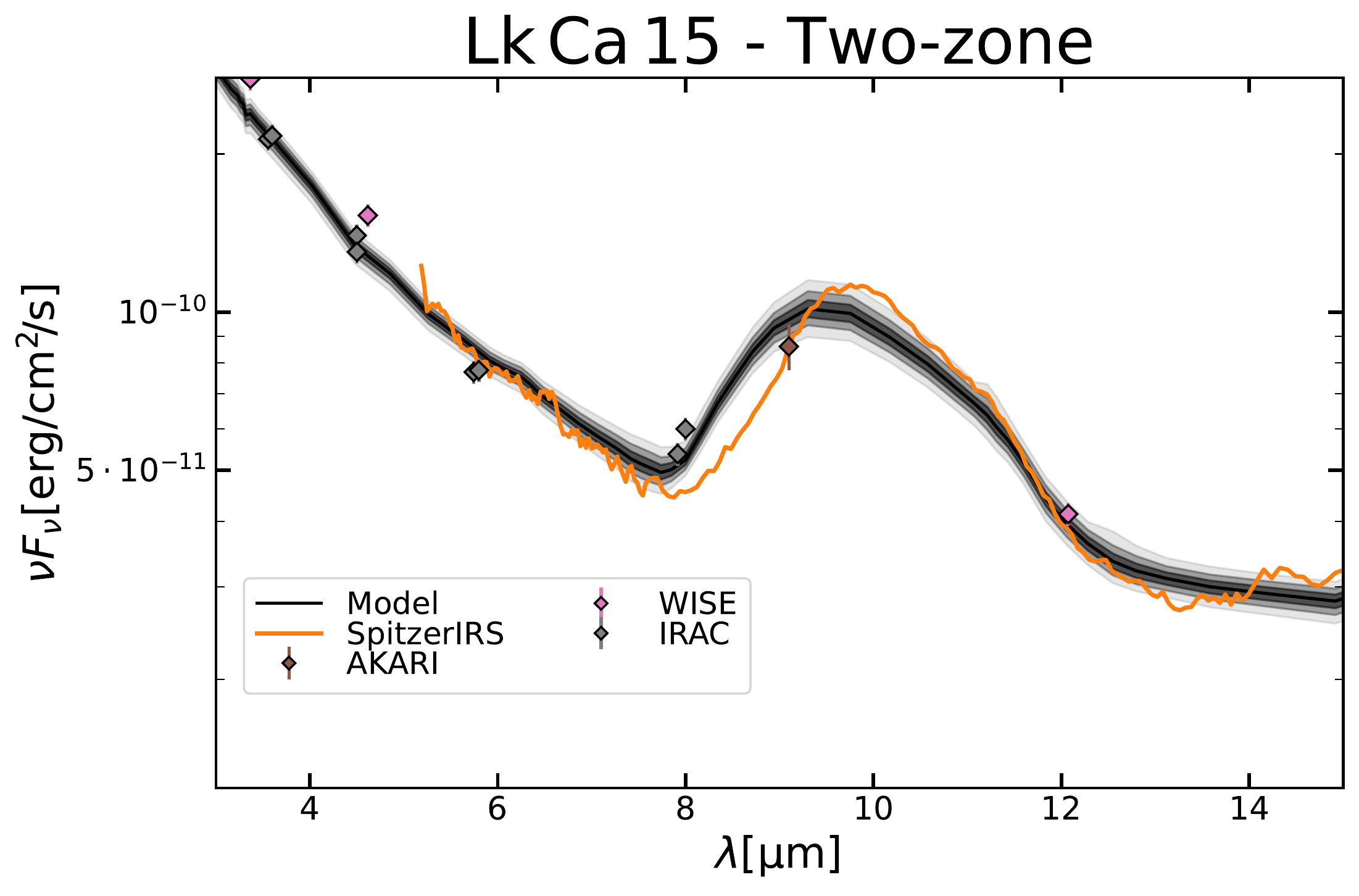}

\caption{continued.}
\end{figure*}

\addtocounter{figure}{-1}
\begin{figure*}[!hp]

    \centering
    \includegraphics[width=0.49\linewidth]{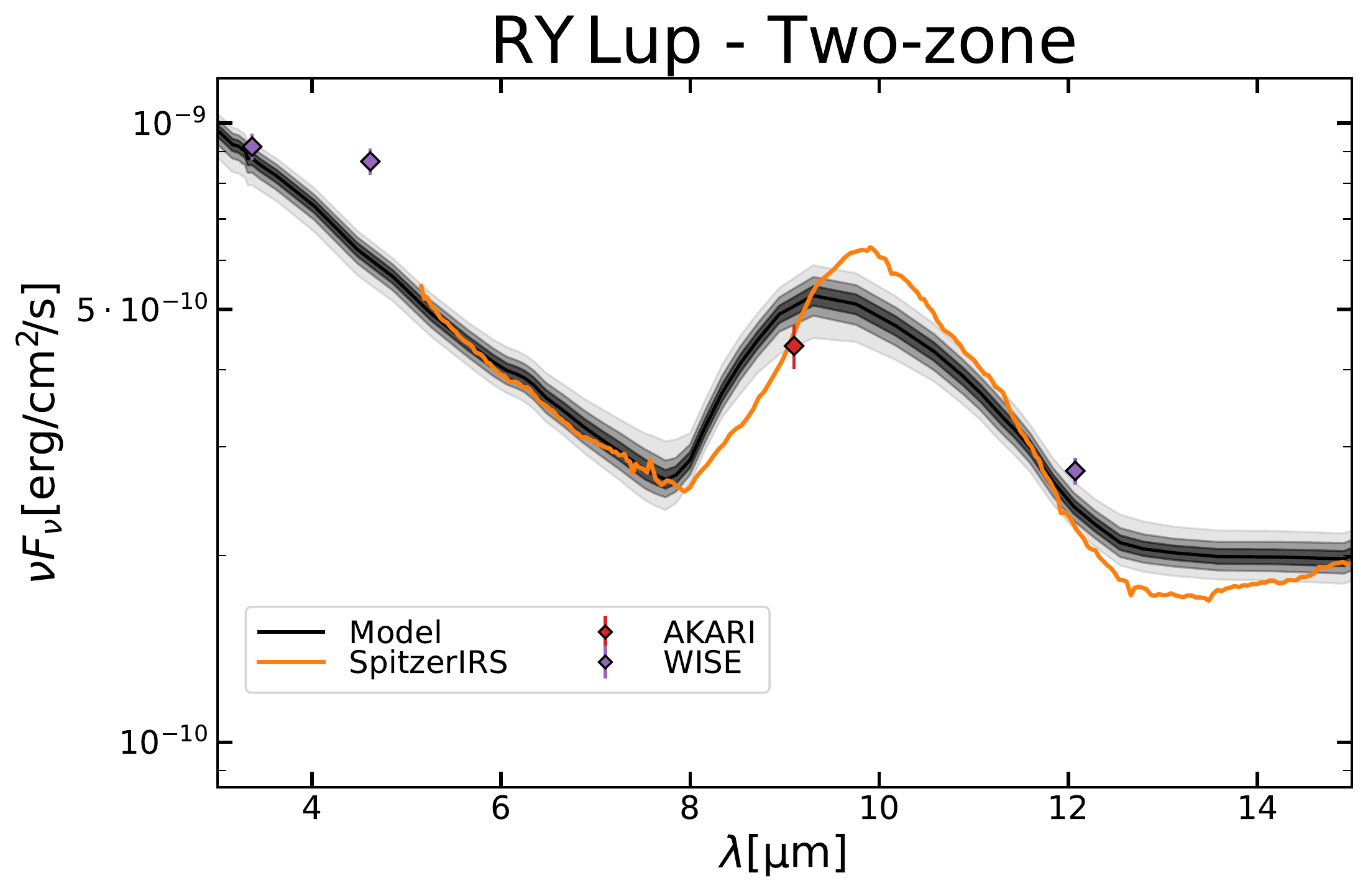}
     \includegraphics[width=0.49\linewidth]{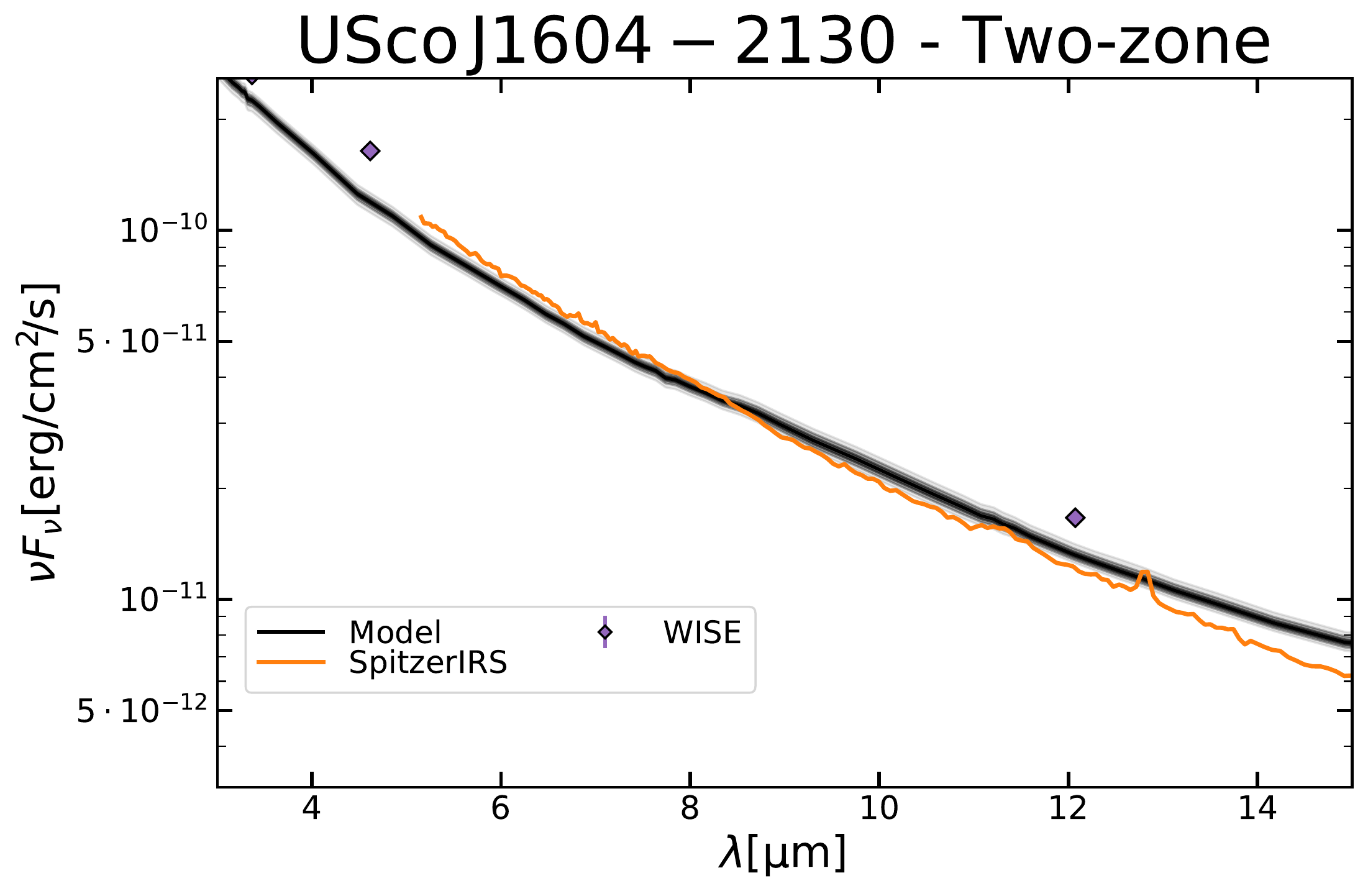}\\
    \includegraphics[width=0.49\linewidth]{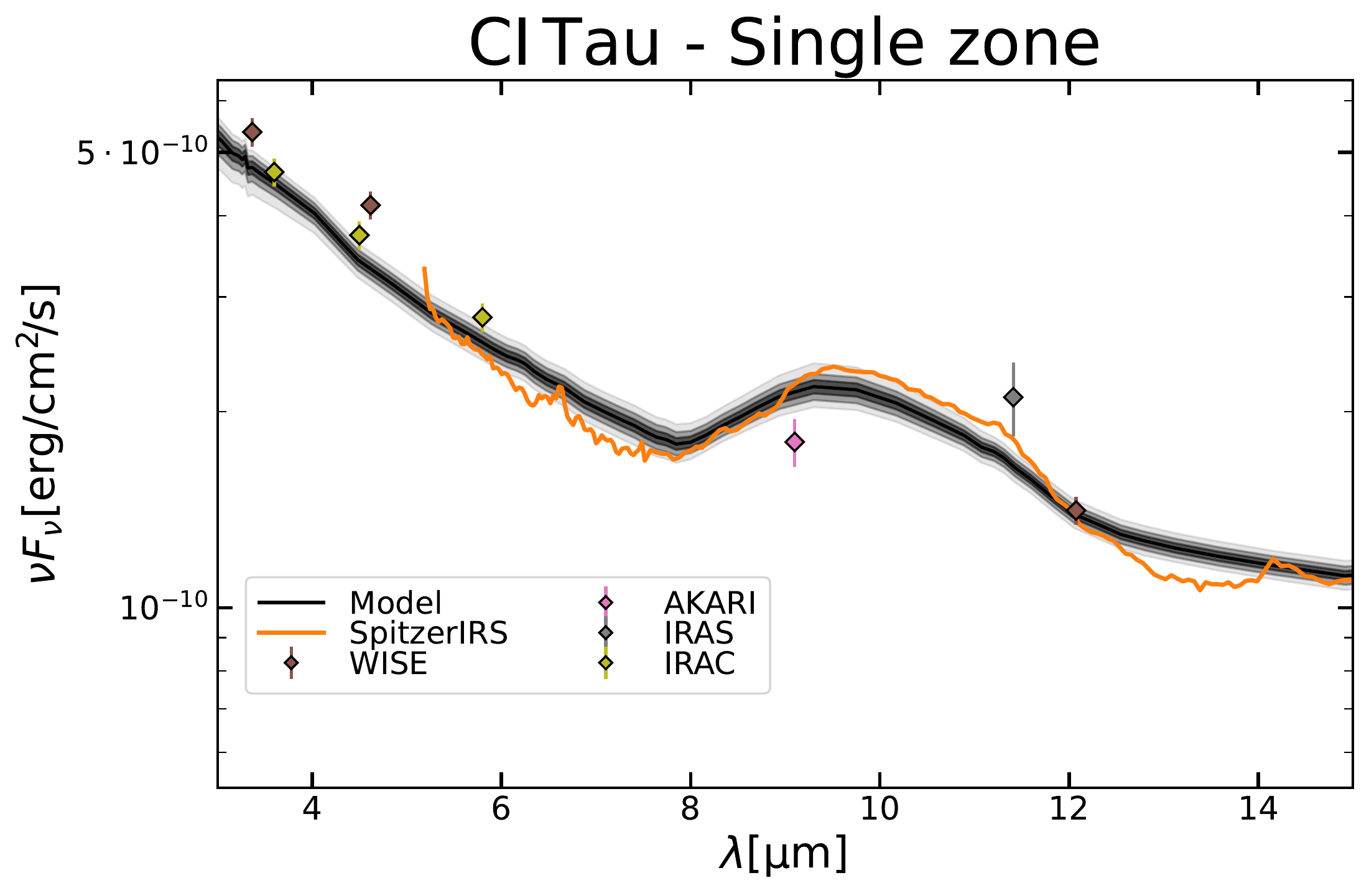}
    \includegraphics[width=0.49\linewidth]{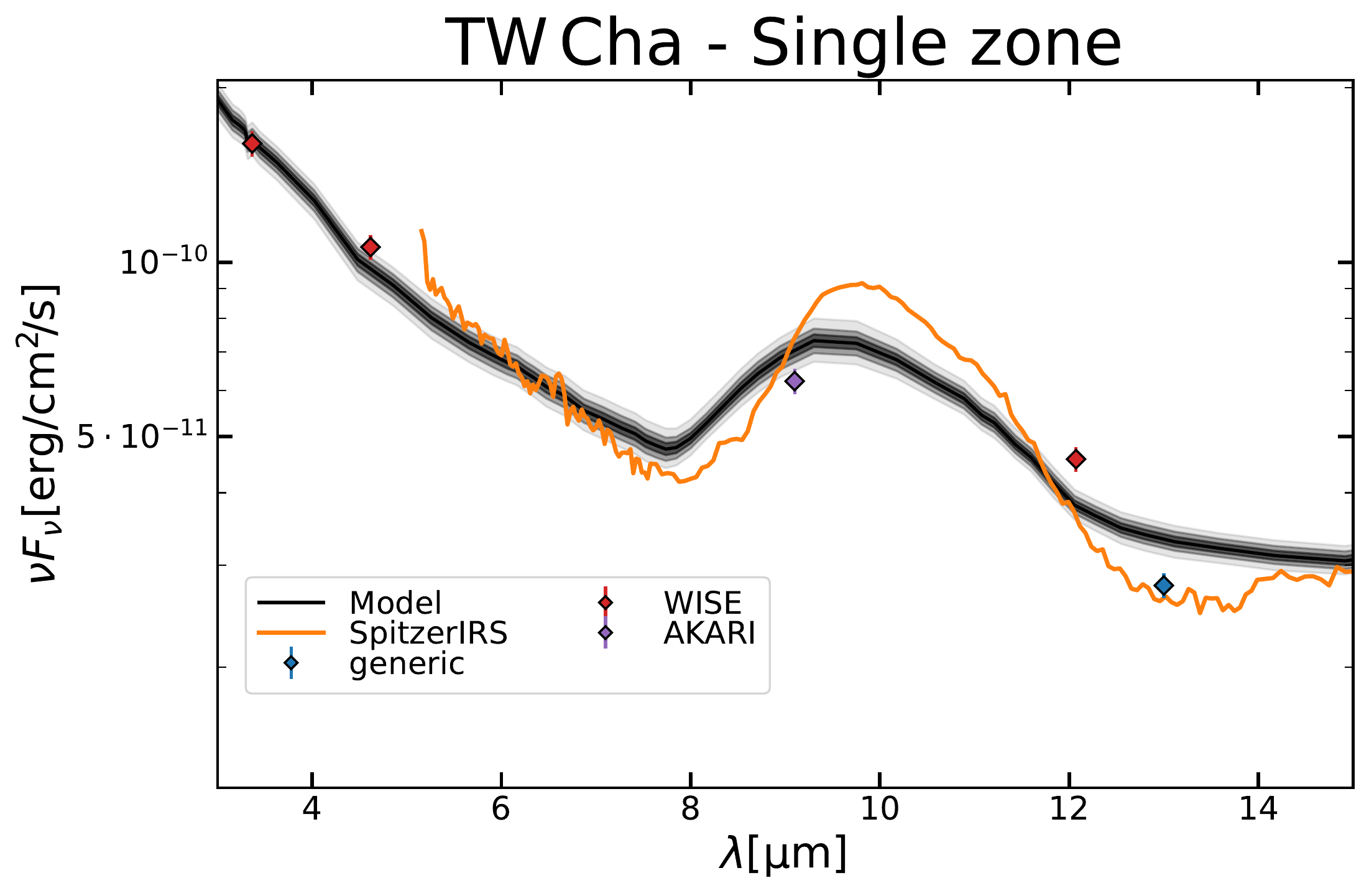}\\
     \includegraphics[width=0.49\linewidth]{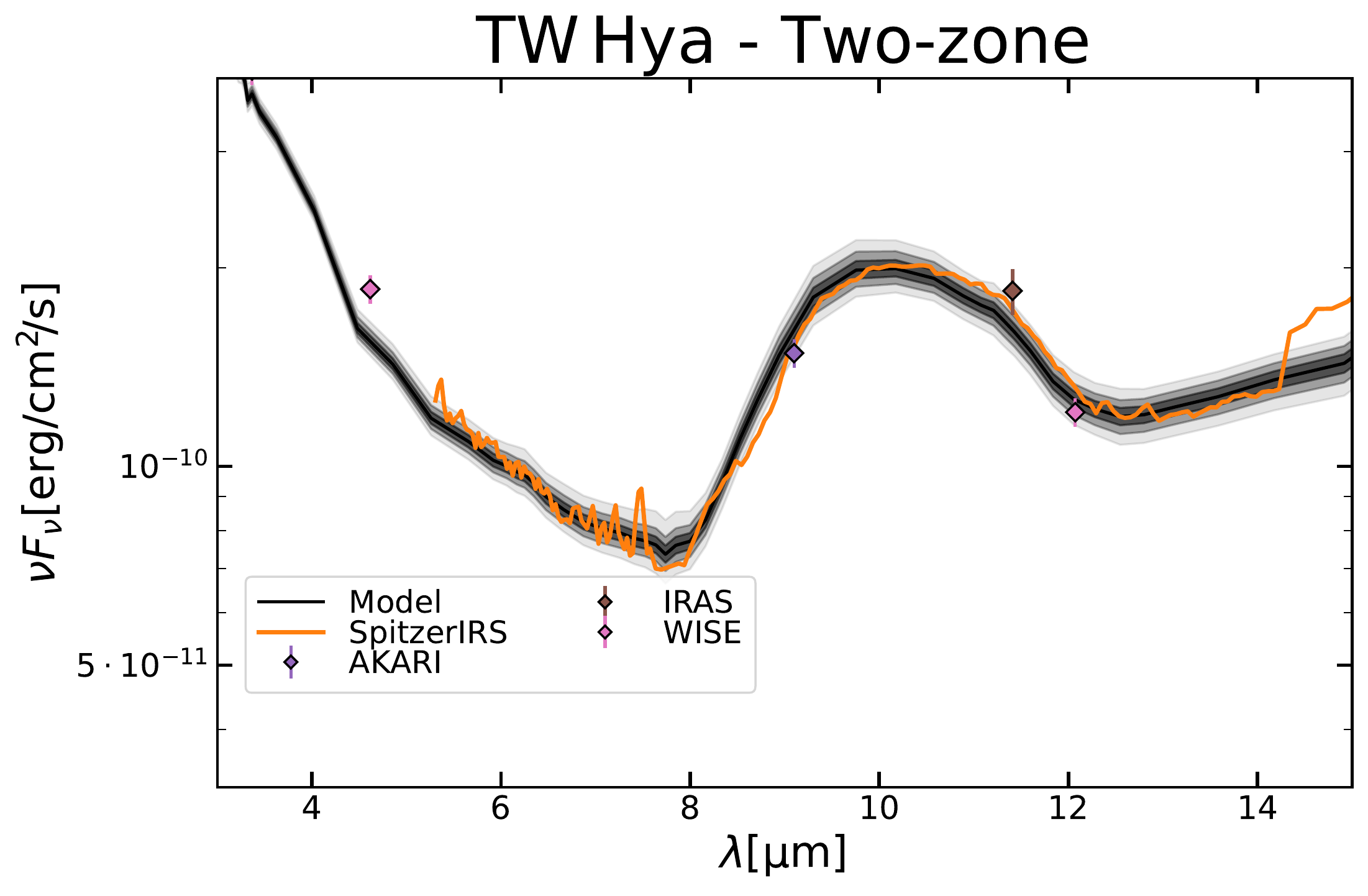}
    \includegraphics[width=0.49\linewidth]{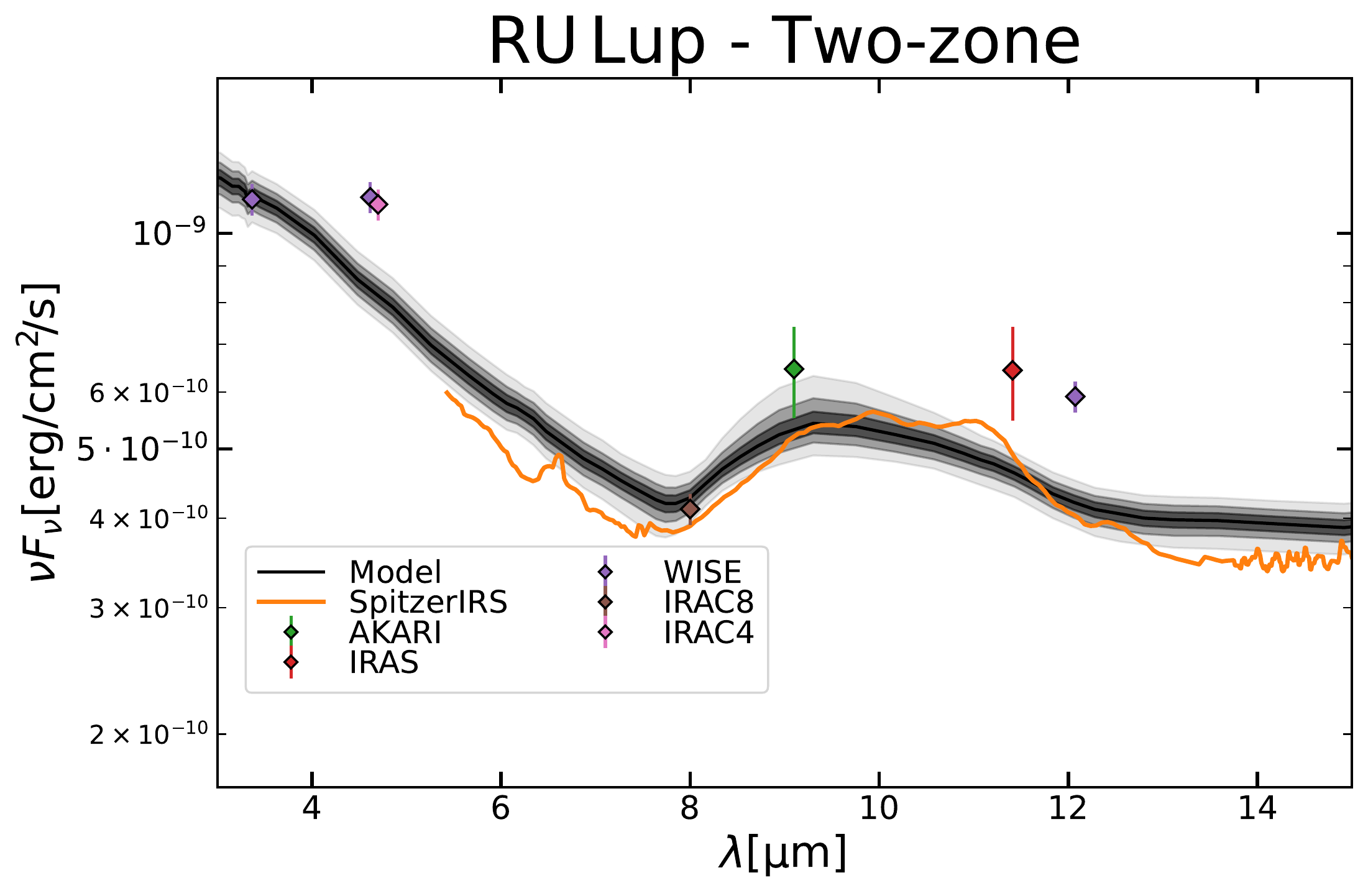}\\
    \includegraphics[width=0.49\linewidth]{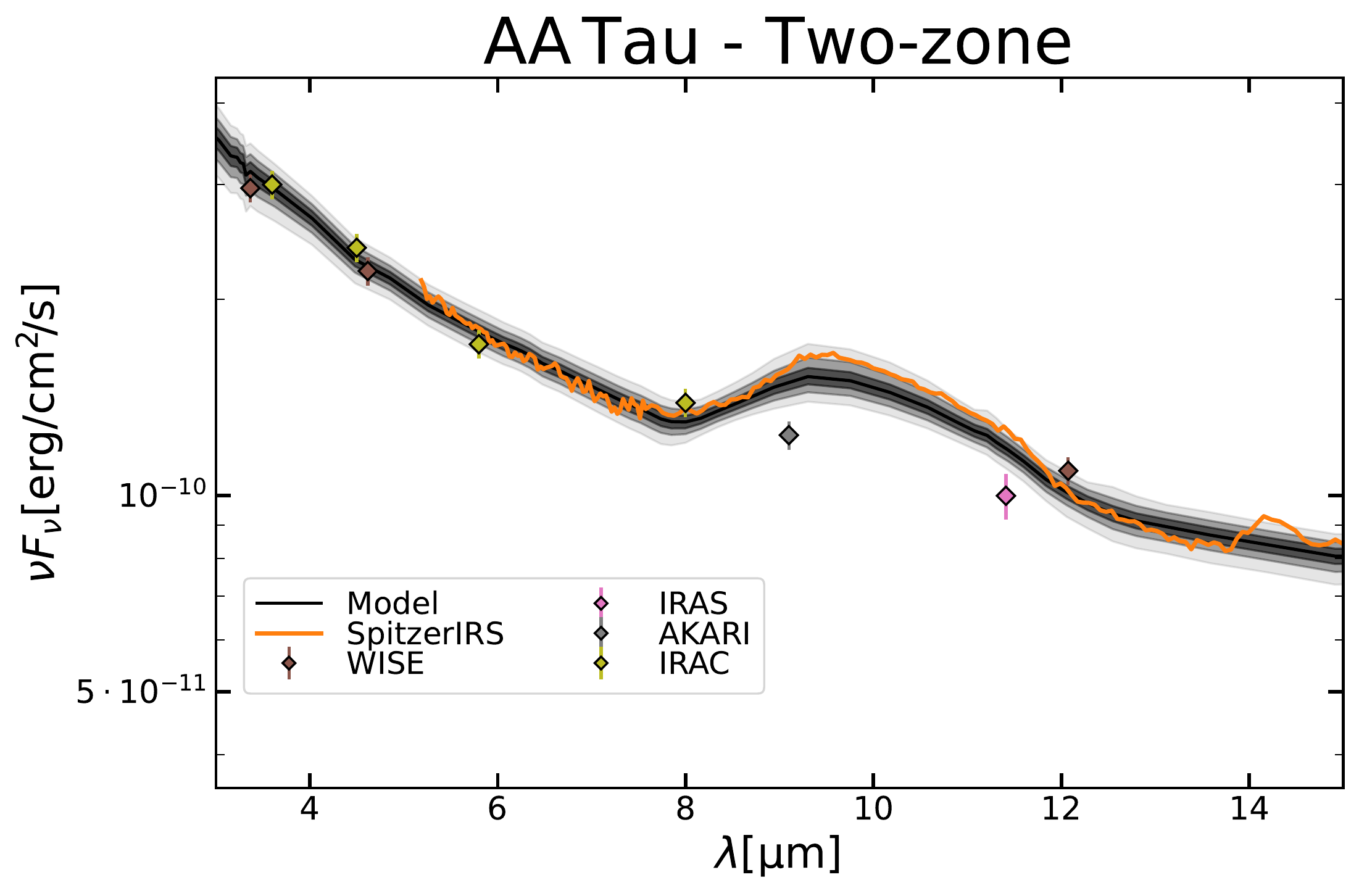}
    \includegraphics[width=0.49\linewidth]{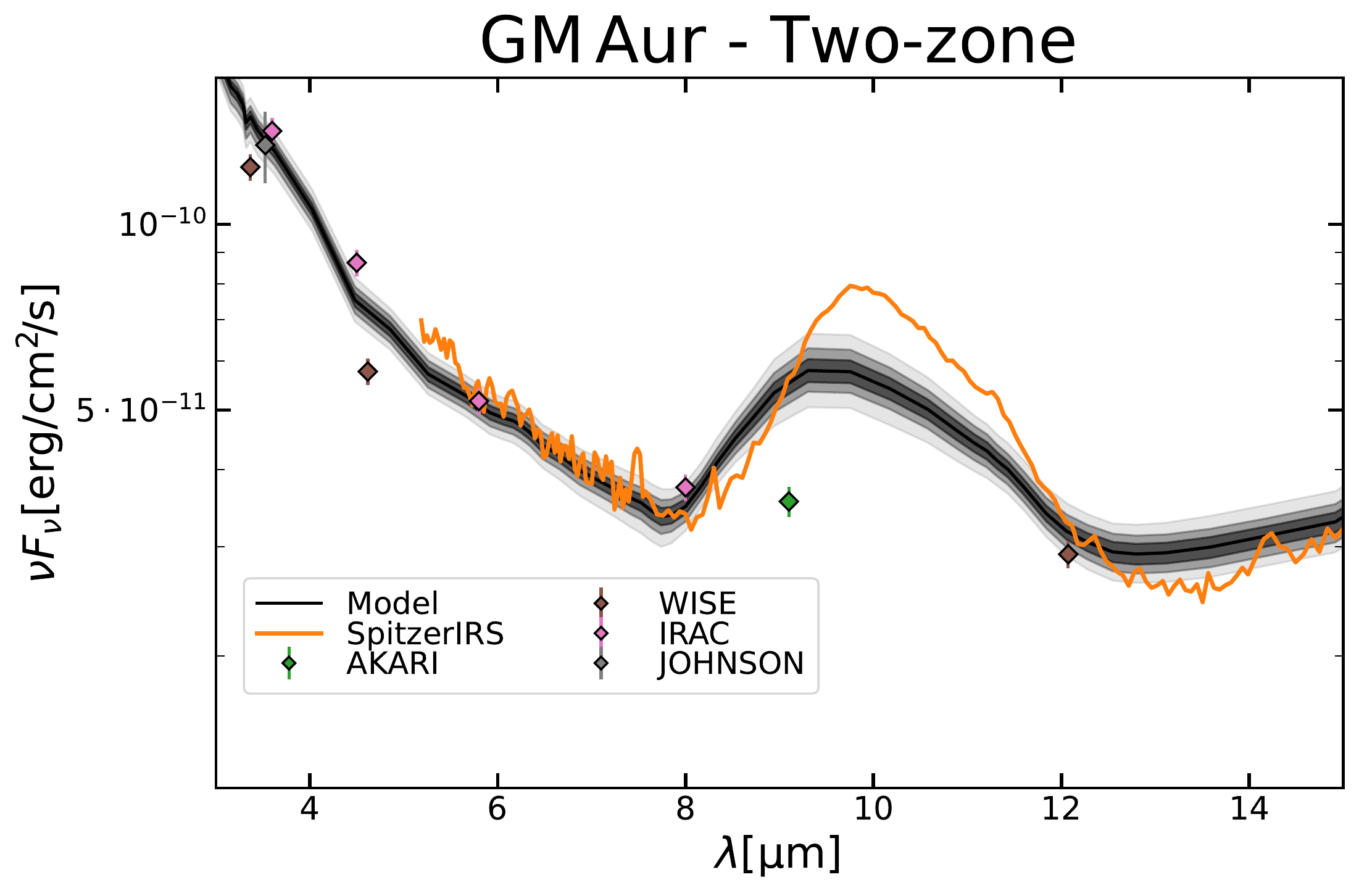}

\caption{continued.}
\end{figure*}

\addtocounter{figure}{-1}
\begin{figure*}[!hp]
\centering
    \includegraphics[width=0.49\linewidth]{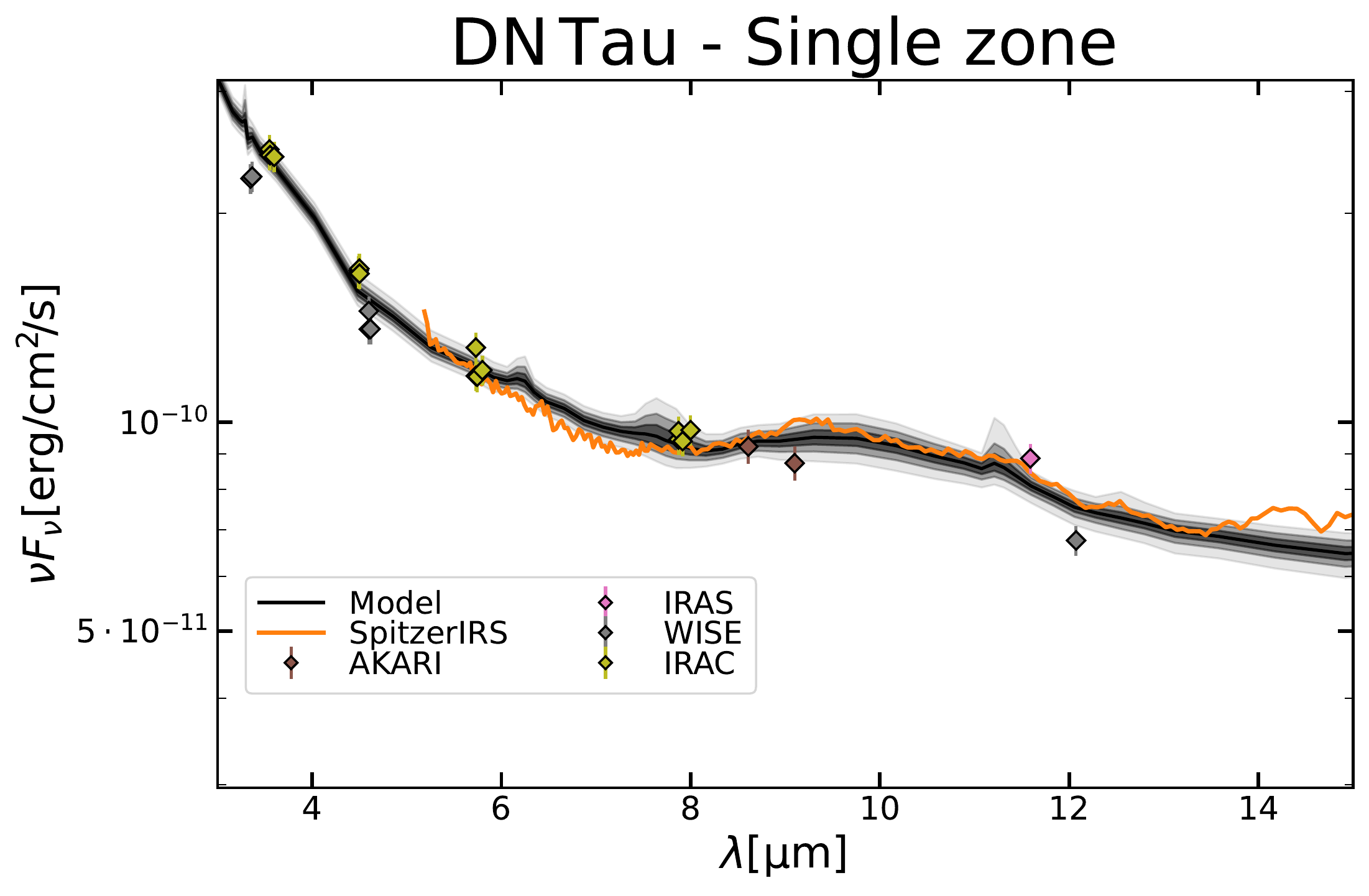}
    \includegraphics[width=0.49\linewidth]{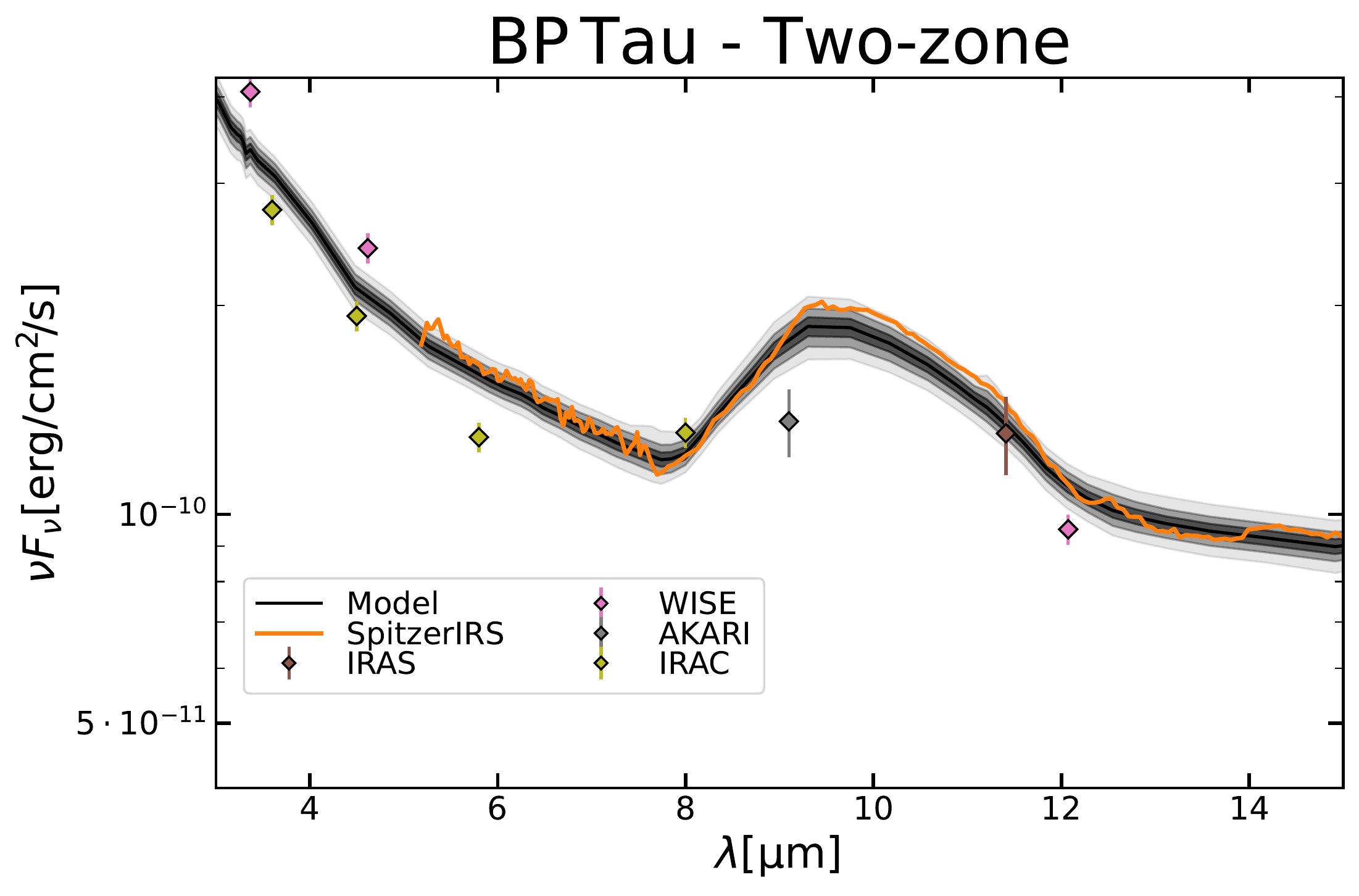}\\
    \includegraphics[width=0.49\linewidth]{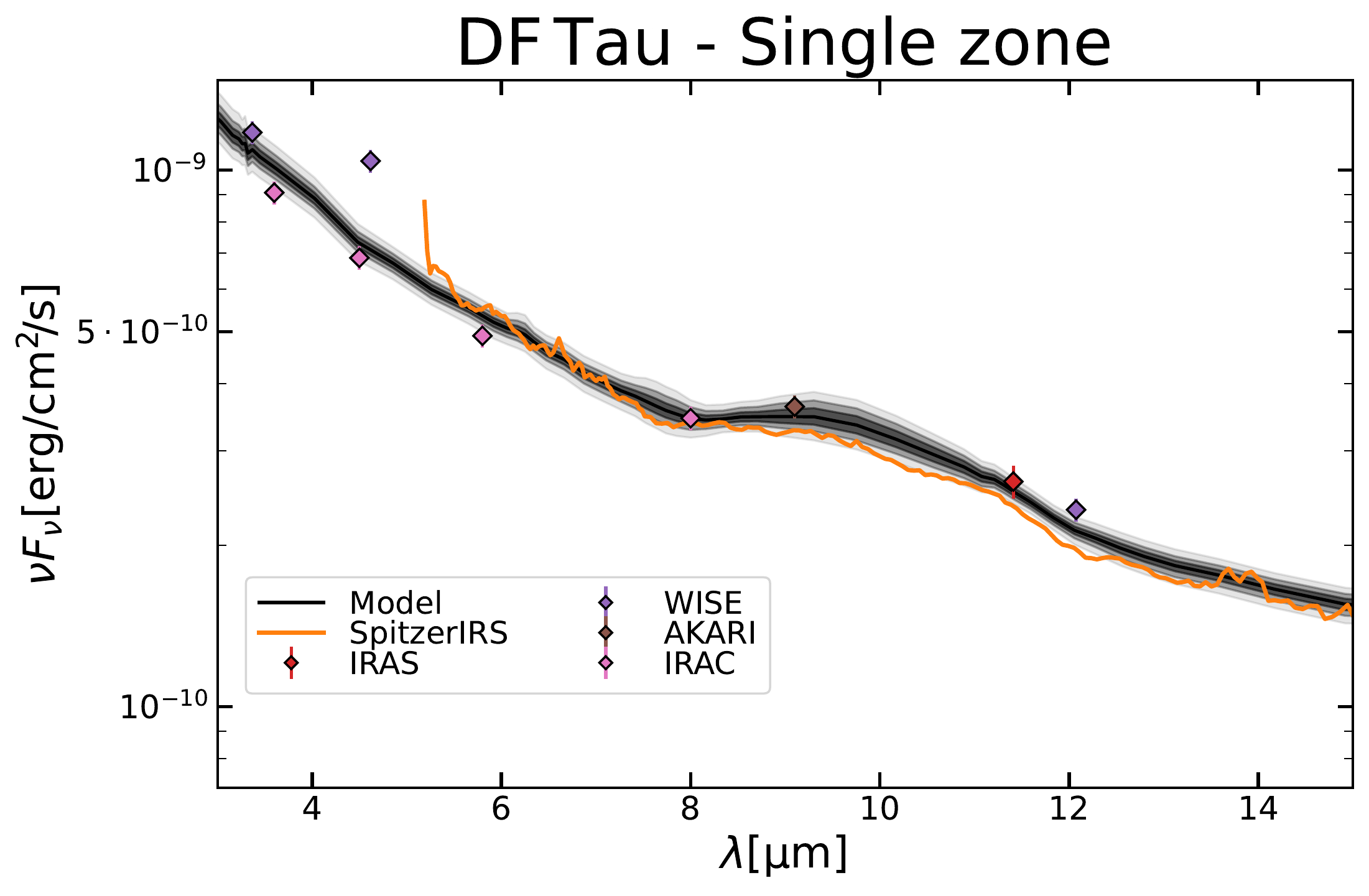}
    \includegraphics[width=0.49\linewidth]{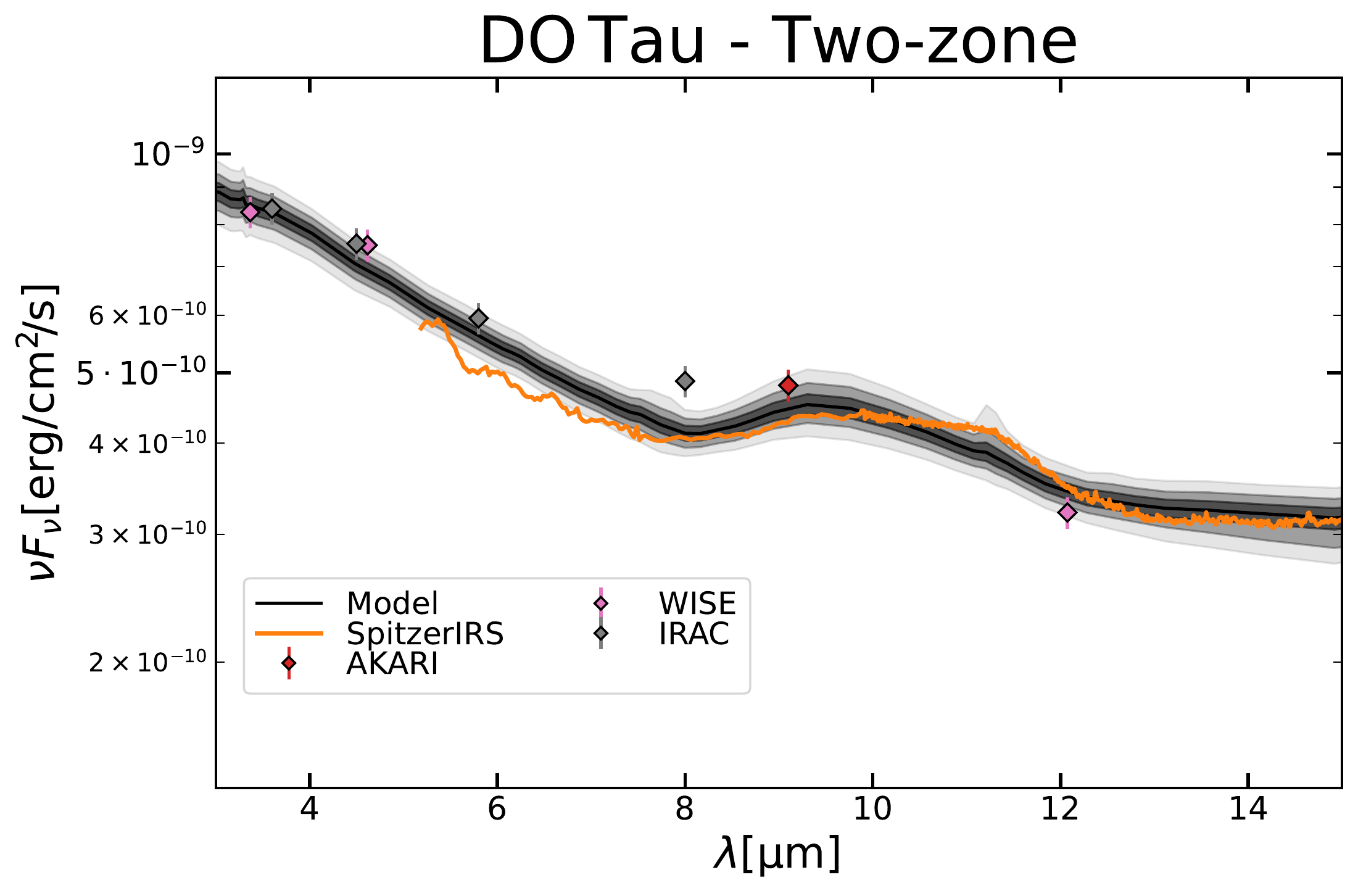}\\
    \includegraphics[width=0.49\linewidth]{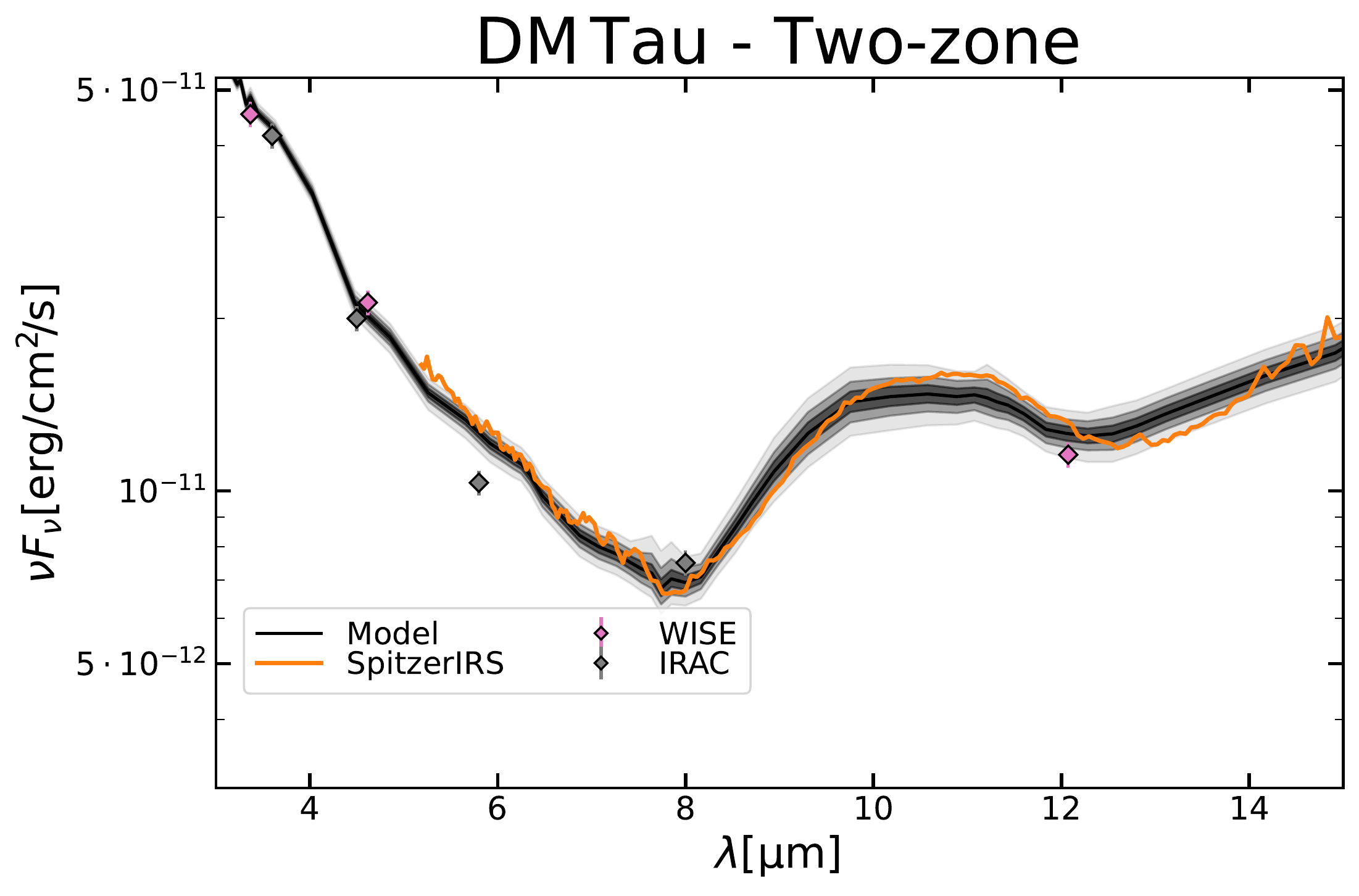}
    \includegraphics[width=0.49\linewidth]{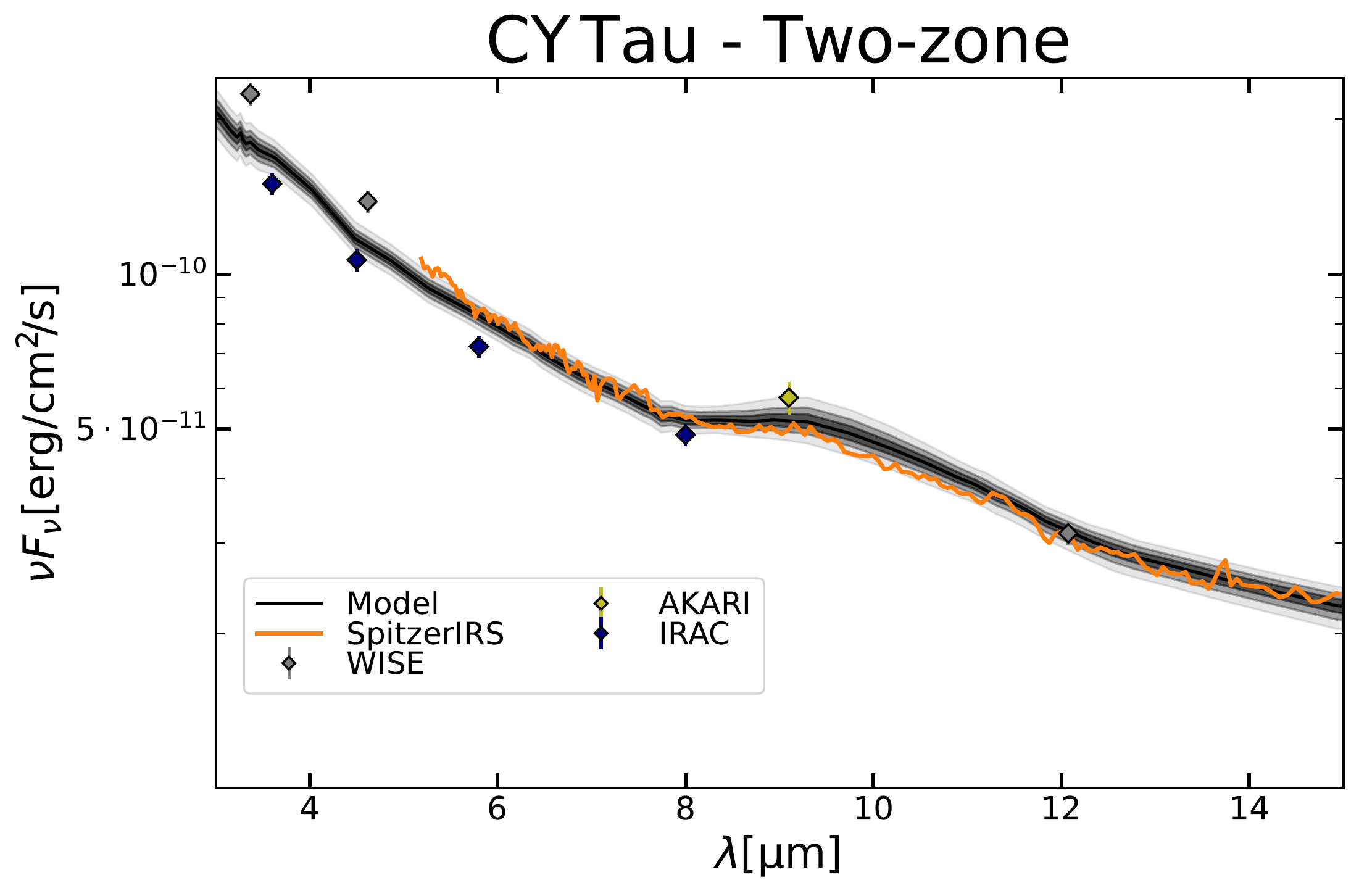}\\
    \includegraphics[width=0.49\linewidth]{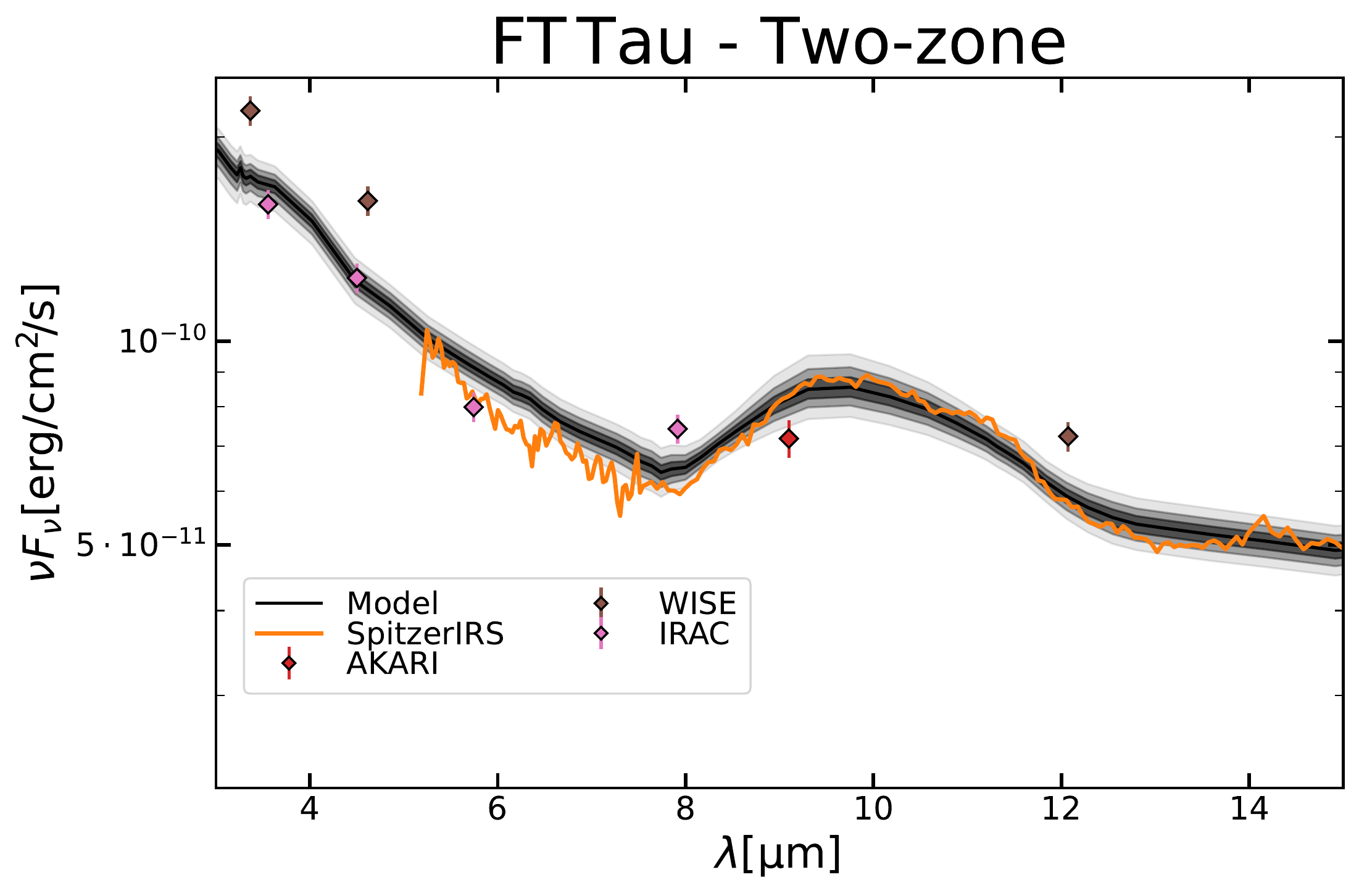}
    \includegraphics[width=0.49\linewidth]{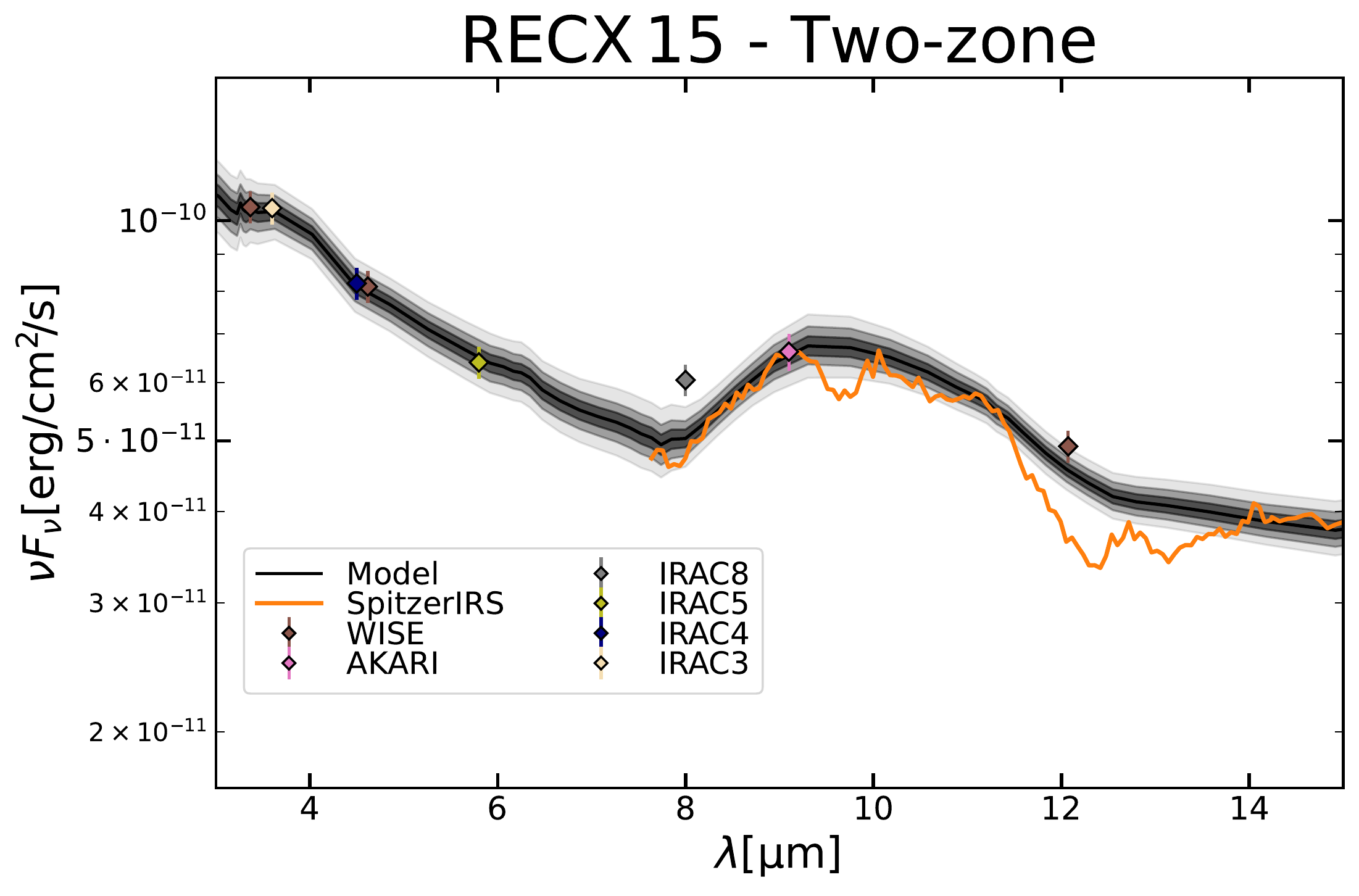}

\caption{continued.}
\end{figure*}

\end{document}